\documentclass[prd, onecolumn, preprintnumbers, superscriptaddress,
nofootinbib, notitlepage, floatfix]{revtex4-1}
\usepackage{amssymb}
\usepackage{amsmath}
\usepackage{graphicx}
\usepackage{color}
\usepackage{slashed}
\usepackage{hyperref}

\newcommand{\xm}[1]{\left\langle x^{#1}\right\rangle}

\begin{document}

\preprint{MSUHEP-20-007}

\title{Isovector parton distribution functions of the proton on a superfine lattice}

\author{Zhouyou Fan}
\affiliation{Department of Physics and Astronomy, Michigan State University, East Lansing, MI 48824, USA}
\author{Xiang Gao}
\email[]{xgao@bnl.gov}
\affiliation{Physics Department, Brookhaven National Laboratory, Upton, NY 11973, USA}
\affiliation{Physics Department, Tsinghua University, Beijing 100084, China}
\author{Ruizi Li}
\affiliation{Department of Physics and Astronomy, Michigan State University, East Lansing, MI 48824, USA}
\author{Huey-Wen Lin}
\affiliation{Department of Physics and Astronomy, Michigan State University, East Lansing, MI 48824, USA}
\affiliation{Department of Computational Mathematics, Michigan State University, East Lansing, MI 48824, USA}
\author{Nikhil Karthik}
\affiliation{Physics Department, Brookhaven National Laboratory, Upton, NY 11973, USA}
\author{Swagato Mukherjee}
\affiliation{Physics Department, Brookhaven National Laboratory, Upton, NY 11973, USA}
\author{Peter Petreczky}
\affiliation{Physics Department, Brookhaven National Laboratory, Upton, NY 11973, USA}
\author{Sergey Syritsyn}
\affiliation{RIKEN-BNL Research Center, Brookhaven National Laboratory, Upton, NY, 11973, USA}
\affiliation{Department of Physics and Astronomy, Stony Brook University, Stony Brook, NY 11794, USA}
\author{Yi-Bo Yang}
\affiliation{CAS Key Laboratory of Theoretical Physics, Institute of Theoretical Physics, Chinese Academy of Sciences, Beijing 100190, China}
\affiliation{Department of Physics and Astronomy, Michigan State University, East Lansing, MI 48824, USA}
\author{Rui Zhang}
\affiliation{Department of Physics and Astronomy, Michigan State University, East Lansing, MI 48824, USA}
\affiliation{Department of Computational Mathematics, Michigan State University, East Lansing, MI 48824, USA}

\date{\today}

\begin{abstract}
We study isovector unpolarized and helicity  parton distribution functions (PDF) of the
proton within the framework of Large Momentum Effective Theory. We use a gauge
ensemble, generated by the MILC Collaboration, with a superfine lattice spacing of
$0.042$~fm and a pion mass of $310$~MeV, enabling us to simultaneously reach
sub-fermi spatial separations and larger nucleon momenta. We compare the spatial
dependence of quasi-PDF matrix elements in different renormalization schemes with the
corresponding results of the global fits, obtained using 1-loop perturbative
matching. We present determinations of the first four moments of the
unpolarized and helicity PDFs of proton from the Ioffe-time dependence of the
isovector matrix elements, obtained by employing  a ratio-based renormalization scheme.
\end{abstract}

\maketitle

\section{Introduction}
Decades of deep inelastic scattering (DIS) and semi-inclusive DIS (SIDIS) data over
wide kinematic ranges have provided us insight into the structure of nucleon.
Significant progress also has been made in recent years. For example, the
determination of the polarized gluon distribution at
small-$x$~\cite{deFlorian:2014yva} based on the inclusive jet and pion production
data from  polarized $p$-$p$ collisions at the Relativistic Heavy-Ion Collider
(RHIC)~\cite{Adamczyk:2014ozi,Adare:2014hsq,Adare:2015ozj} and double spin asymmetries
from open-charm muon production at COMPASS~\cite{Adolph:2012ca}, and the constraints
on the polarization of sea quarks and antiquarks with longitudinal single-spin
asymmetries in $W^\pm$-boson production~\cite{Adamczyk:2014xyw,Adare:2015gsd}. In the
future, the kinematic coverage of nucleon PDFs will be be greatly extended by the
data from  from the Jefferson Lab 12-GeV program~\cite{Dudek:2012vr} and the
Electron-Ion Collider (EIC)~\cite{Accardi:2012qut}. On the energy frontier, nucleon
PDF not only was a critical input for the discovery of the Higgs boson at the Large
Hadron Collider (LHC)~\cite{CMS:2012nga,ATLAS:2012oga}, but also is expected to play
critical roles in determining the Standard-Model backgrounds during LHC's search for
physics beyond the Standard Model in future Runs~3--5.

Despite great progress in the experimental and phenomenological sides,
non-perturbative determinations of the PDFs starting from the microscopic theory of
quantum chromodynamics (QCD) remains a challenge. To obtain the quark PDF one has to
calculate the matrix element with the quark fields are separated along the lightcone
between the hadronic states. Due to the lightcone separation, straightforward
calculation of PDF is not possible using lattice QCD, a technique based on Euclidean
time formulation. One can bypass this obstacle by calculating a similar matrix
element with spatially separated quark fields at equal time within highly boosted
hadron states, which defines the so-called quasi-PDF
(qPDF)~\cite{Ji:2013dva,Ji:2014gla}. For large hadron momenta this matrix element can
be related to PDF~\cite{Ji:2013dva,Ji:2014gla}.  The Large Momentum Effective Theory
(LaMET) provides a systematic way to relate the qPDF at large, but finite, hadron
momentum to the PDF order by order in perturbation theory~\cite{Ji:2014gla}. Related
approaches to connect PDF to matrix elements of boosted hadrons calculable in the
Euclidean time lattice computations, such as ``the good lattice
cross-section"~\cite{Ma:2014jla,Ma:2017pxb} and the
pseudo-PDF~\cite{Radyushkin:2017cyf,Orginos:2017kos}, have also been proposed.
Renormalization of the underlying boosted hadron matrix elements, usually referred as
the Ioffe-time distributions (ITD), involves Wilson-line. The multiplicative
renormalizability of the ITD to all orders of perturbation theory has been
proven~\cite{Ji:2017oey,Ishikawa:2017faj}. A practical ways to implement
renormalization on the lattice, such as the use of RI-MOM
scheme~\cite{Chen:2017mzz,Constantinou:2017sej,Alexandrou:2017huk,Stewart:2017tvs,Liu:2018uuj}
and reduced Ioffe-time distributions~\cite{Orginos:2017kos}, have been established.
Relation between different theoretical approaches also is now
understood~\cite{Izubuchi:2018srq}.  Based on these theoretical developments,
unpolarized and polarized nucleon PDFs have been calculated on the
lattice~\cite{Liu:2018uuj,Liu:2018hxv,Lin:2018qky,Chen:2018xof,Alexandrou:2019lfo,Alexandrou:2018eet,Alexandrou:2018pbm,Joo:2019jct,Joo:2020spy}.
Furthermore, lattice calculations of the valence pion PDF have also
appeared~\cite{Chen:2018fwa,Izubuchi:2019lyk,Sufian:2019bol,Joo:2019bzr,Sufian:2020vzb}.
The status of this field is well summarized in recent review papers
\cite{Zhao:2018fyu,Cichy:2018mum,Monahan:2018euv,Ji:2020ect}. All these calculations
for the nucleon, so far,  have been carried out with lattice spacing $a>0.08$~fm.

Having small lattice spacing plays a crucial role in calculation of PDF within the LaMET
framework. To suppress the target mass and higher twist corrections the hadron
momentum $P_z$ should be large. But to avoid large discretization effects one must
ensure $a P_z\ll 1$. Furthermore, to obtain lightcone-PDF from qPDF one needs
perturbative matching, which, presently, is known only up to 1-loop order.
Applicability of 1-loop perturbative matching can be guaranteed only for spatial
separations $z \Lambda_\text{QCD}\ll 1$, and therefore demands use of fine lattices.
The main goal of the present work is to study systematic of the PDF calculations
within the LaMET framework by going to the extreme limit with the use of a superfine
lattice having $a=0.042$~fm. The lattice spacing used in this study is at least twice
smaller than that used in any previous lattice calculations of the nucleon PDF. The
unpolarized and helicity PDFs of the nucleon are well constrained
through global fits to experimental results. Thus, we study the systematic of our
calculations by comparing $P_z$- and $z$-dependence of renormalized qPDF matrix
elements with the same reconstructed from the well-known phenomenological PDFs using
the LaMET framework.

The rest of the paper is organized as follows. In section II we discuss the general
features of LaMET and our lattice setup. In section III we discuss the nucleon
2-point functions for large values of $P_z$ and the determination of the energy
levels of the fast moving nucleon. Section IV is dedicated to the analysis of the
nucleon 3-point functions and the calculations of bare qPDF. Section V describes the
non-perturbative RI-MOM renormalization. The comparison of the lattice results on
qPDF with the results of global analysis of unpolarized and helicity
PDF is discussed in sections VI and VII, respectively. Different from RI-MOM
renormalization, we discuss the analysis of ratios of nucleon matrix elements in
Section VIII.  Finally, section IX contains our conclusions.

\section{Lattice setup and LaMET}
In this paper, we report the results of a lattice QCD calculation
using clover valence fermions on an ensemble of $N_f=2+1+1$ gauge
configurations with lattice spacing $a=0.042$~fm, with space-time
dimensions of $64^3\times 192$ and pion mass $M_\pi \approx 310$~MeV
in the continuum limit. The gauge configurations have been generated
using Highly Improved Staggered Quarks (HISQ)~\cite{Follana:2006rc}
by the MILC Collaboration~\cite{Bazavov:2012xda}.  The gauge links
entering the clover Wilson-Dirac operator have been smeared using
hypercubic (HYP) smearing~\cite{Hasenfratz:2001hp}.  We used
tree-level tadpole improved result for the  coefficient of the
clover term and the bare quark mass has been tuned to recover the
lowest pion mass of the staggered quarks in the
sea~\cite{Rajan:2017lxk,Bhattacharya:2015wna,Bhattacharya:2015esa,Bhattacharya:2013ehc}.
We use only one step of HYP smearing to improve the discretization
effects, since it is possible that multiple applications of smearing could alter the ultraviolet results for
the PDF.  We use multigrid algorithm~\cite{Babich:2010qb,Osborn:2010mb}
in Chroma software package~\cite{Edwards:2004sx} to perform the
inversion of the clover fermion matrix allowing us collect relatively
high statistics sample.  We collected a total of 3258 measurements
using 6 sources per configuration and 543 gauge configurations. In the following, we
elaborate on the steps of our computation.

\subsection{Nucleon two-point correlators}

The two crucial components of the lattice computation are the two-point function
and the three-point function involving boosted nucleon and
the qPDF operator. The two point function for the
nucleon boosted to spatial momentum $\mathbf{P}$ is the standard operator
\begin{equation}
    C_{\rm 2pt}(t_s)= \left \langle \hat{N}_{s'}(\mathbf{P},t_s)  \hat{N}_s^\dagger(\mathbf{P},0) \right \rangle,
    \qquad
    \hat{N}_s(\mathbf{P},t)=\sum_{\mathbf{x}}\epsilon_{abc}u^{(s)}_a(\tilde{x})\left(u^{(s)}_b(\tilde{x})^T C \gamma_5 d^{(s)}_c(\tilde{x})\right)e^{-i\mathbf{P}\cdot \mathbf{x}},
\end{equation}
where $\tilde{x}=(\mathbf{x},t)$ and $t_s$ is the source-sink separation along the Euclidean time direction.
The index `$s$' refers to the kind of quark smearing that is applied to improve the
signal-to-noise of the boosted nucleon states. We either used point quark operators $\psi(x)$ or
we used the Gaussian momentum smeared~\cite{Bali:2016lva} for the quark fields, $\psi^{(s)}(x)$ that enters
$\hat{N}_s$,
\begin{equation}
    \psi^{(s)}(\tilde{x}) = S_\text{mom}\psi(\tilde{x}) = \frac{1}{1+6\alpha}\left ( \psi(\tilde{x})
+ \alpha \sum_j U_j(\tilde{x})e^{i\mathbf{k}\cdot\hat{j}}\psi(\tilde{x}+\hat{j}) \right),
\label{eq:moms}
\end{equation}
where $\mathbf{k}$ is the momentum of the quark field, $U_j(\tilde{x})$ are the gauge links in the $\hat{j}$ direction,
and $\alpha$ is a tunable parameter as in traditional Gaussian smearing.
The quark momentum should be chosen such that the signal-to-noise ratio is optimal for the given
nucleon momentum. Naively one would expect that $|\mathbf{k}|$ should be one third of the nucleon momenta \cite{Bali:2016lva}.
For this particular study, we use $j=z$ and $k_z=4 \pi/L$, and a large Gaussian-smearing parameter $\alpha=10$.
Such a momentum source is designed to align the overlap with nucleons of the desired boost momentum,
and we are able to reach higher boost momentum for the nucleon states with reasonable signals.
In the nucleon two-point correlators, we can study multiple values of the
nucleon momentum, $\mathbf{P}=\{0,0, P_z \}$ with
\begin{equation}
P_z= n_z \frac{2\pi}{L},\qquad n_z \in [0,6],
\label{eq:moms0-6}
\end{equation}
without a significant increase in computational needs. These values
of $n_z$ from 1 to 6 correspond to $P_z=0.46, 0.92, 1.38, 1.84, 2.31$ and
$2.77$ GeV in physical units respectively.  We either used smeared
fields for both the source and sink, which we refer to as SS, or
smeared fields only for the source and point fields for the sink
which we refer to as SP in the rest of the paper.

\subsection{Nucleon three-point function}

The three point function we compute is of the form
\begin{equation}
    C_{\rm 3pt}(t_s,\tau)= \mathrm{P\!\!\!P} \left \langle  \hat{N}_{s}(\mathbf{P},t_s) O_{\Gamma}(z;\tau) \hat{N}^\dagger_{s}(\mathbf{P},0) \right \rangle
\end{equation}
where $ O_{\Gamma}(z;\tau)$ is the $u-d$ isovector qPDF operator
\begin{equation}
    O_{\Gamma}(z;\tau) =\sum_{\mathbf{x}} \overline{u}(\tilde{x}+z)\Gamma W_z(\tilde{x}+z,
    \tilde{x}) u(\tilde{x}) - \sum_{\mathbf{x}} \overline{d}(\tilde{x}+z)\Gamma W_z(\tilde{x}+z,
    \tilde{x}) d(\tilde{x}).
    \label{eq:isovector}
\end{equation}
where $\tilde{x}=(\mathbf{x},\tau)$, and $W_z$ is the straight Wilson line along the spatial $z$-direction, connecting lattice sites $\tilde{x}$ and $\tilde{x}+z$ . The Dirac
$\Gamma$ used will determine the quantum numbers of the PDF --- $\Gamma=\gamma_t$ for
the unpolarized case and $\Gamma=\gamma_z\gamma_5$ for the longitudinally polarized
case. The projector operator, $\mathrm{P\!\!\!P}$, is given by
$\mathrm{P\!\!\!P}=\frac{1+\gamma_t}{2}$ for the unpolarized case and $\mathrm{P\!\!\!P}=i\gamma_z\gamma_5\frac{1+\gamma_t}{2}$
 for the longitudinally polarized case, respectively.  We
only use smeared quark sources for the computation of $C_{\rm 3pt}$. In order to
reduce the  computational cost, we only computed the $C_{\rm 3pt}$ for two large
values  $P_z=1.84$ and $2.31$~GeV, and for source-sink separations $t_s=16a, 18a, 20a$.

\subsection{Extraction of nucleon matrix element and perturbative matching to PDF}

Using the three-point and two-point functions whose calculations are described above, we
can extract the bare matrix element
\begin{equation}
h(z,P_z,\Gamma)=\langle P_z |O_{\Gamma}(z)| P_z \rangle,
\label{ME}
\end{equation}
formally in the infinite source-sink separation $t_s$ limit of their ratio
\begin{equation}
R(z,P_z,\Gamma;\tau,t_s)=\frac{C_{\rm 3pt}(\tau,t_s)}{C_{\rm 2pt}(t_s)}.
\end{equation}
To obtain the matrix element $h(z,P_z)$ from the above ratio, we calculate the nucleon
three-point function with insertion of $O_{\Gamma}(z)$ operator at
three nucleon three-point source-sink separations, approximately $t_s=0.67, 0.76, 0.84$~fm, and
describe its $t_s$- and $\tau$-dependence through $2$- and $3$-state ansatz.
In Sec.~\ref{sec3pt}, we describe our extraction of bare matrix element from
various extrapolations in detail.

The next step of the computation is the renormalization of the bare matrix element $h$.
One possible choice for $O_\Gamma$ is $O_{\gamma_z}$. However, for this
case of $\Gamma=\gamma_z$ there is a mixing with the quark bilinear operator containing the
unit matrix, $\Gamma=1$
if Wilson fermions are used \cite{Constantinou:2017sej,Chen:2017mzz,Chen:2017mie}. This mixing is absent if
we use $\Gamma=\gamma_t$, and we will use this choice for the unpolarized PDF in this study.
One way to perform the renormalization procedure on the lattice to use RI-MOM scheme~\cite{Alexandrou:2017huk,Chen:2017mzz},
where in the
renormalized matrix element is defined as
\begin{equation}
h^R(z,P_z,\mu_R,p_z^R)= Z(z,\mu^R,p_z^R) h(z,P_z,\Gamma),
\end{equation}
The non-perturbatively determined RI-MOM renormalization constant
$Z$($z$, $\mu^R$, $p_z^R$) depends on the separation $z$, the norm of the renormalization point
$\mu^R=(p^R)^2$ and the $z$ component of renormalization point $p_z^R$.
The dependence on $p_z^R$ arises because the $z$-component of the momentum now plays a special role.
We will discuss the details of the RI-MOM renormalization in section V. We will also consider an alternate
ratio scheme that has a well defined continuum limit in Sec.~\ref{secratio}. Here, the multiplicative
renormalization factor $Z_{\rm ratio}(z)$
can be taken as the hadron matrix element at a different fixed momentum $P_z'$ i.e.,
$Z_{\rm ratio}(z)=(h(z,P_z',\Gamma))^{-1}$.

After the RI-MOM renormalization one obtains the renormalized matrix element $h_R(z,P_z,\mu_R,p_z^R)$, from
which we can define the qPDF as a function of Bjorken-$x$
\begin{align}\label{eq:qpdf}
\tilde{q}(x,P_z,\mu_R,p_z^R) \equiv \int_{-\infty}^\infty {dz\over 4\pi} e^{ixP_zz} h_R(z,P_z,\mu_R,p_z^R).
\end{align}
From this formula it is clear that $h_R(z,P_z,\mu_R,p_z^R)$ can be considered as the coordinate space qPDF.
For finite momentum $P_z$, $\tilde{q}(x,P_z,\mu_R,p_z^R)$ has support in $-\infty < x < \infty$.
Unlike the physical PDF, which is frame independent, the qPDF has a nontrivial dependence on the nucleon momentum $P_z$.
When the nucleon momentum $P_z \gg \{M, \Lambda_\text{QCD}\}$ with $M$ being the nucleon mass,
the qPDF in RI-MOM scheme can be matched to the PDF defined in $\overline{\rm MS}$-scheme, $q(x,\mu)$
through the factorization theorem~\cite{Ji:2013dva,Ji:2014gla,Izubuchi:2018srq},
\begin{equation} \label{eq:fact}
\tilde{q}(x,P_z, p^R_z,\mu_R)=\int_{-1}^1 {dy\over |y|}\: C\left({x\over y},r,\frac{yP_z}{\mu},\frac{yP_z}{p_z^R}\right) \, q(y,\mu)
+\mathcal{O}\left({M^2\over P_z^2} \right) +\mathcal{O}\left({\Lambda_{\text{QCD}}^2\over P_z^2}\right) ,
\end{equation}
where $r=(\mu_R/p_z^R)^2$ and $C$ is the perturbative matching coefficient,
$\mathcal{O}(M^2/P_z^2)$ is the target-mass correction due to the non-zero nucleon
mass, and $\mathcal{O}(\Lambda_\text{QCD}^2/P_z^2)$ stands for higher-twist
contributions. The flavor indices of $q$, $\tilde{q}$, and $C$ are implied. In what
follows we will discuss the non-singlet case, and therefore, mixing with gluon and sea-quark PDFs is absent in 
the above formula. We use 1-loop expression of the kernel $C$.
(The 1-loop matching including for the singlet case also has
been worked out in Ref. \cite{Wang:2019tgg,Zhang:2018diq}.)

The matching kernel $C(x,r,P_z/\mu,P_z/p_z^R)$ for $\Gamma=\gamma_t$
was derived in Ref. \cite{Liu:2018uuj} and depends on details of
the RI-MOM scheme. It can be written in the following form
\begin{equation}
C\left(x,r,\frac{P_z}{\mu},\frac{P_z}{p_z^R}\right)=\delta(1-x)+\left[f_{1,\Gamma} \left(x,{P_z\over \mu} \right)-
\left|\frac{P_z}{p_z^R}\right| f_{2,\Gamma,\mathcal{P}} \left( 1 + \frac{P_z}{p_z^R} (x-1), r\right) \right]_{+}.
\label{eq:C_f12}
\end{equation}
The subscript `+' stands for the plus-prescription. Both the functions,
$f_{1,\Gamma}$ and $f_{2,\Gamma,\mathcal{P}}$,  depend on the choice of the $\Gamma$
in the operator insertion~ \cite{Liu:2018uuj}. On the other hand, $f_{1,\Gamma}$ is independent
of the projection operator ($\mathcal{P}$) used in defining the RI-MOM
renormalization condition,  but $f_{2,\Gamma,\mathcal{P}}$ is different for different
choices of the RI-MOM renormalization condition~ \cite{Liu:2018uuj}. We also note
that it is also possible to convert $h_R(z,P_z,\mu_R,p_z^R)$ to $\overline{\rm
MS}$-scheme and define the corresponding qPDF $\tilde q(x,P_z,\mu)$ that then can be
directly matched to $\overline{\rm MS}$ PDF~\cite{Alexandrou:2017huk}.

To study the longitudinally polarized quark PDF one can use
$\Gamma=\gamma_z\gamma_5$ or $\Gamma=\gamma_t\gamma_5$.  In the
case $\Gamma=\gamma_z\gamma_5$ there is no mixing with quark bilinear
operators with $\Gamma=1$ \cite{Alexandrou:2017huk}.  Therefore,
we will use this choice to study the longitudinally polarized quark
PDF and qPDF.  The bare matrix element of $O_{\gamma_z \gamma_5}$
can be renormalized using RI-MOM scheme and then match to PDF in
the same manner as this was done for unpolarized. The RI-MOM
renormalization for the longitudinally polarized case will be
discussed in section V, while details of the matching procedure,
including the formulas for $f_1$ and $f_2$ functions will be give
in section VII.

\section{Analysis of the nucleon two-point function}
\begin{figure}[t]
  \centering
\includegraphics[scale=0.7]{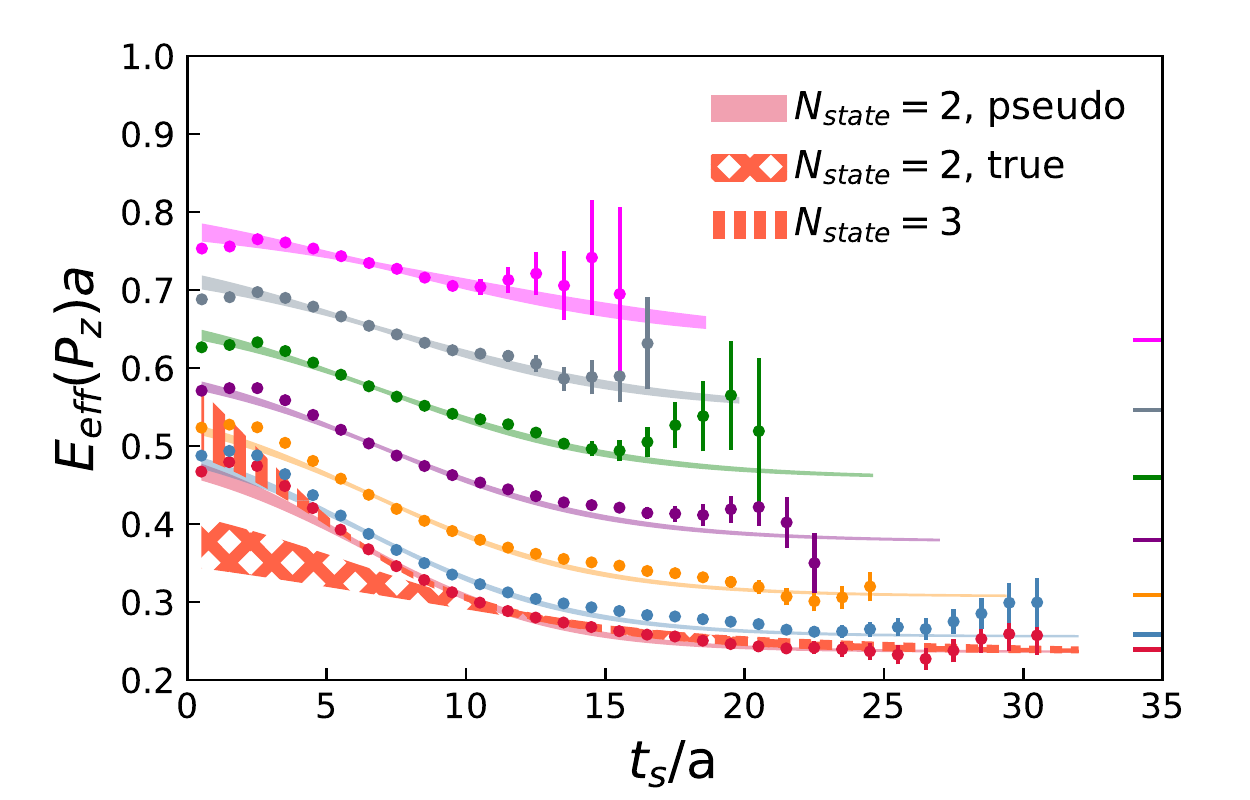}
\includegraphics[scale=0.7]{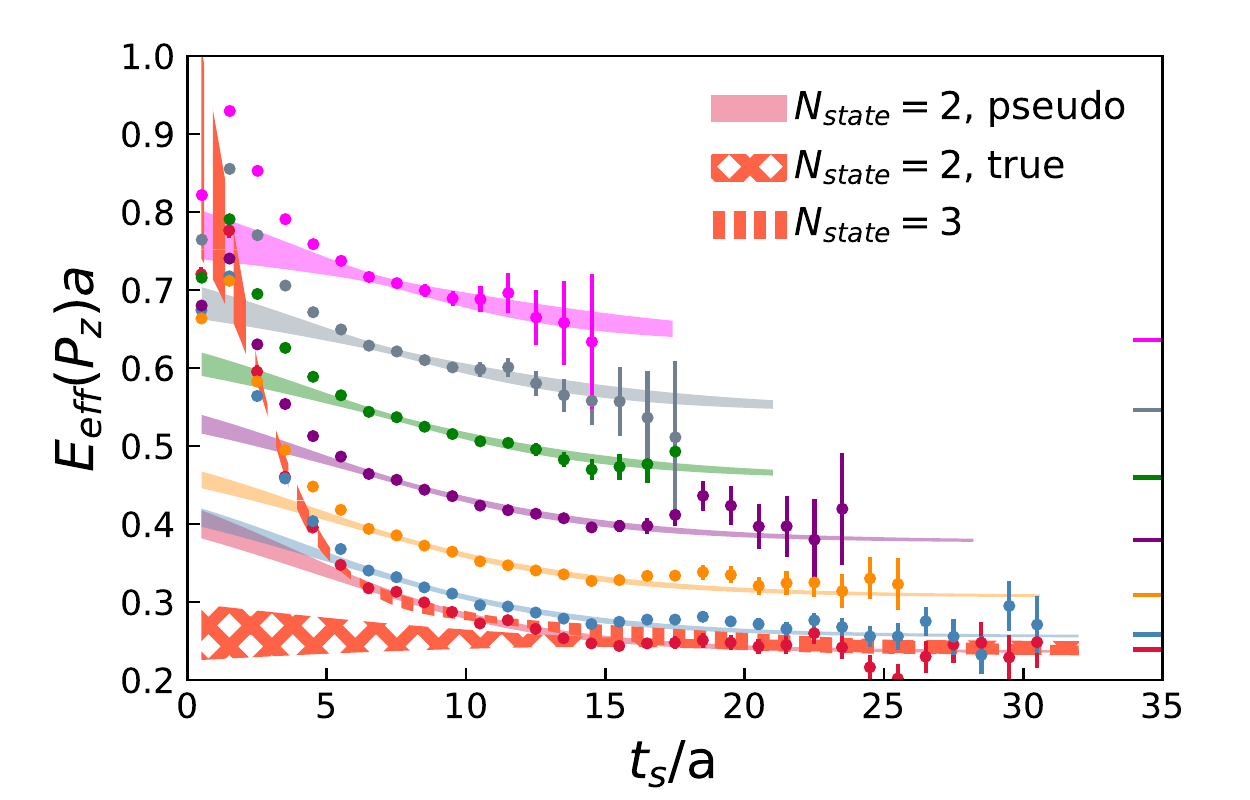}
\caption{The effective masses obtained from SP (left) and SS (right) correlators for
different momenta showed in Eq. (\ref{eq:moms0-6}). The bands come from the results of two-state
($N_\mathrm{state}=2$) and three-state ($N_\mathrm{state}=3$) fits. For
$N_\mathrm{state}=2$,  `pseudo' indicates that the effective pseudo-plateau in range
$5a < t_{\rm min} < 10a$ for the first excited state $E_1$ have been used, and `true'
indicates that the true plateau value of $E_1$ in range $ t_{\rm min} > 11a$ have
been used (see text for details).
}
\label{fig:meff}
\end{figure}

For the extraction of the qPDF matrix element of the nucleon at large momenta it is
important to understand the contribution of different energy states to the nucleon
two-point correlation function. We calculated nucleon two-point function using
smeared source and smeared sink (SS correlator), as well as smeared source and point
sink (SP correlator), for seven values of the momenta $a P_z =2 \pi/L \cdot
n_z,~n_z=0,1,2,3,4,5$ and $6$.  From the two-point correlators, $C^i_{\rm
2pt}(t_s,P_z)$, $i=$ SS or SP, we define the effective mass
\begin{equation}
   a E_{\rm eff}(t_s,P_z)=\ln\left(\frac{C^i_{\rm 2pt}(t_s/a,P_z)}{C^i_{\rm 2pt}(t_s/a+1,P_z)}\right).
\end{equation}
Our results for the effective masses are shown in Fig.~\ref{fig:meff} for the SP and
SS correlators.

The effective mass should approach a constant corresponding to the ground state
energy $E_0(P_z)$ at sufficiently large $t_s$. The momentum dependence of the ground
state energy is expected to be described by the dispersion relation
$E_0(P_z)=\sqrt{P_z^2+M^2}$, with $M$ being the nucleon mass. Therefore, in
Fig.~\ref{fig:meff}, we show the expected ground state energy at different $P_z$
obtained from the dispersion relation as horizontal lines at the right for
comparison. Along with the expected asymptotic values at large $t_s$, we also show
the $t_s$-dependence of the effective mass  based on an effective two-state fit
to the two-point function, as we will explain shortly. Indeed, we see that the
effective masses approach the corresponding values.  The effective masses
corresponding to the SP correlator reach a plateau at a slightly larger $t_s$ than
the SS correlators. On the other hand, at small $t_s$, the effective masses for the
SP correlators are smaller than those for SS correlators.  This implies that the
contribution of the excited states is smaller for the SP correlator, for which a
plausible reason could be that the different excited states contribute with different
signs to the correlator. Thus, even though the ground and the excited state energies
are the same in the SP and SS correlators, the two are affected differently by the
higher excited states, which we can take advantage of to obtain the excited state
spectrum reliably.

\begin{figure}[t]
  \centering
\includegraphics[scale=0.45]{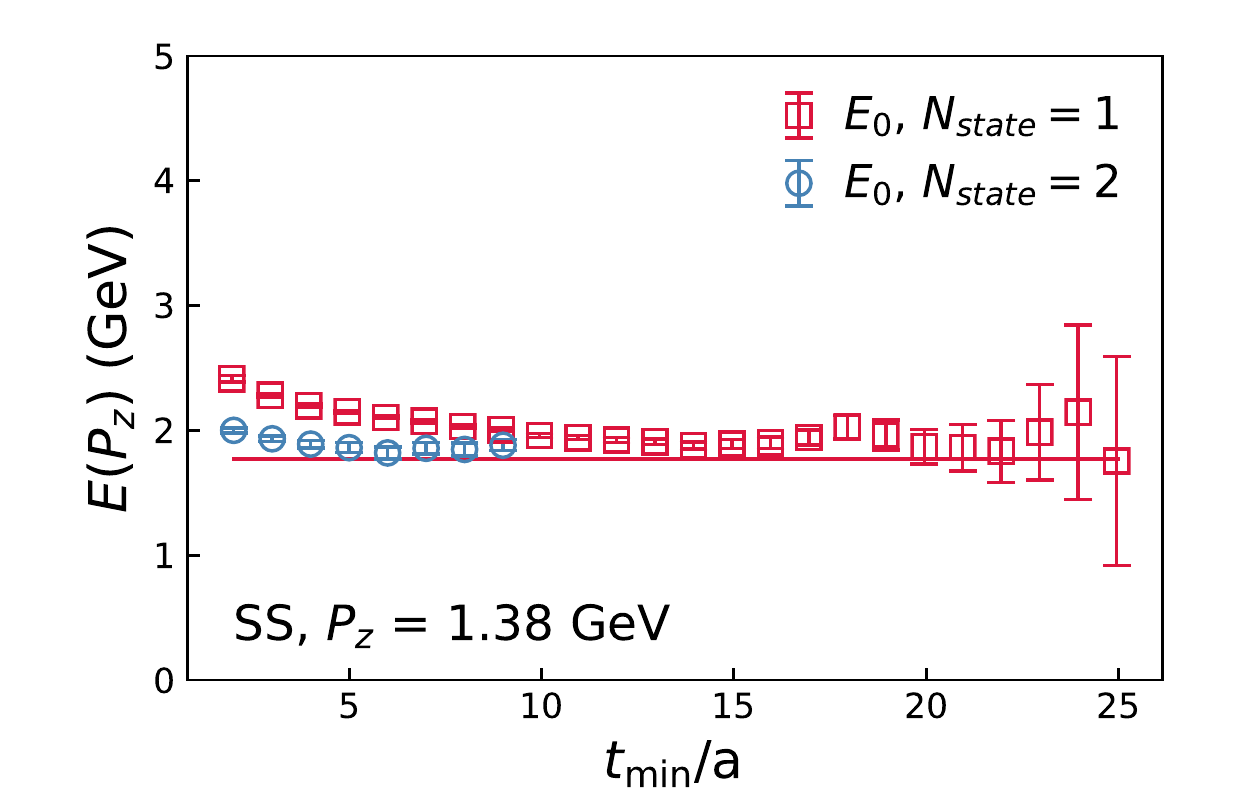}
\includegraphics[scale=0.45]{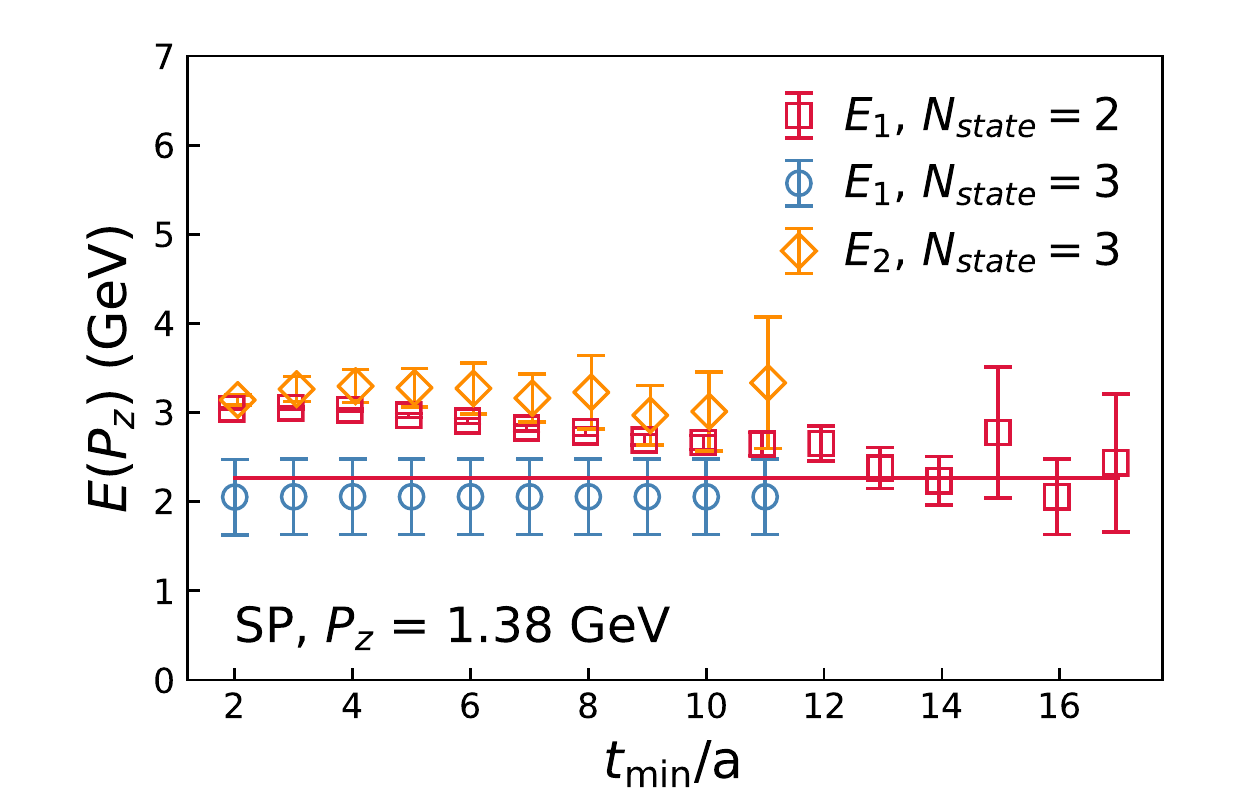}
\includegraphics[scale=0.45]{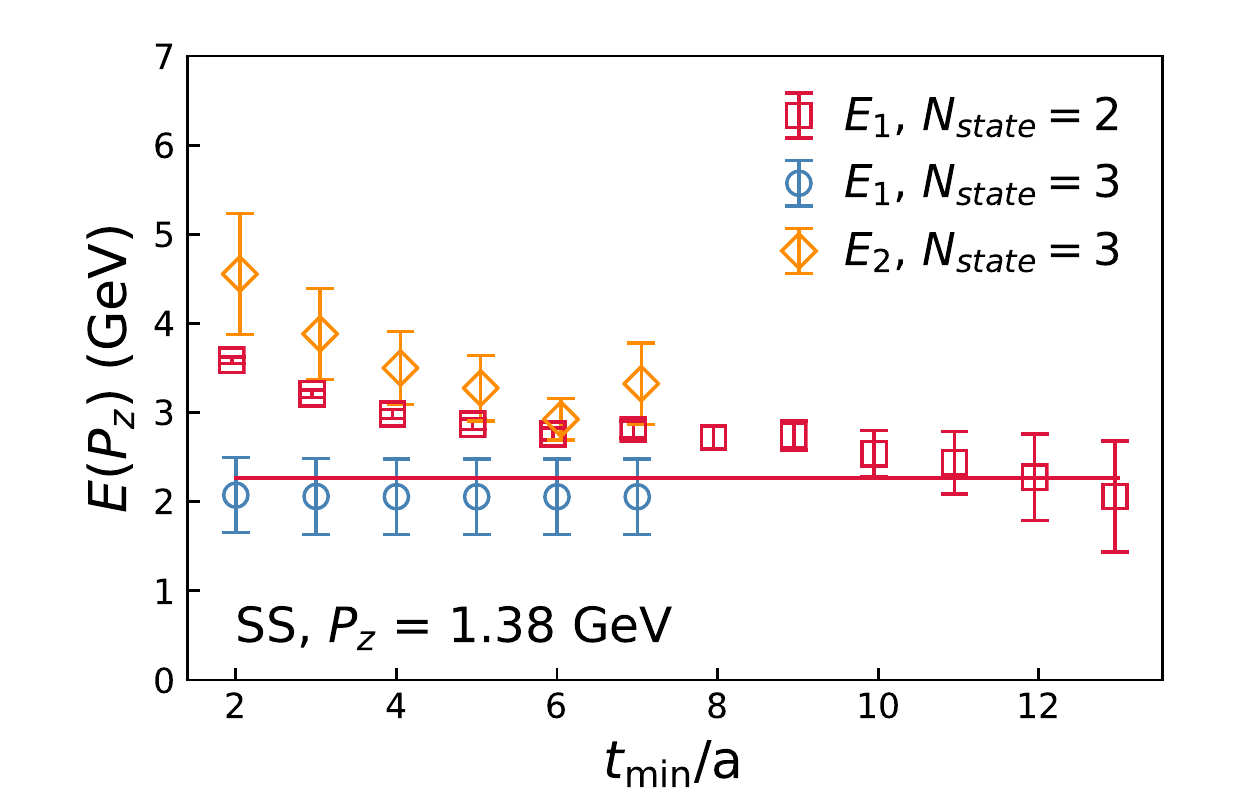}
\includegraphics[scale=0.45]{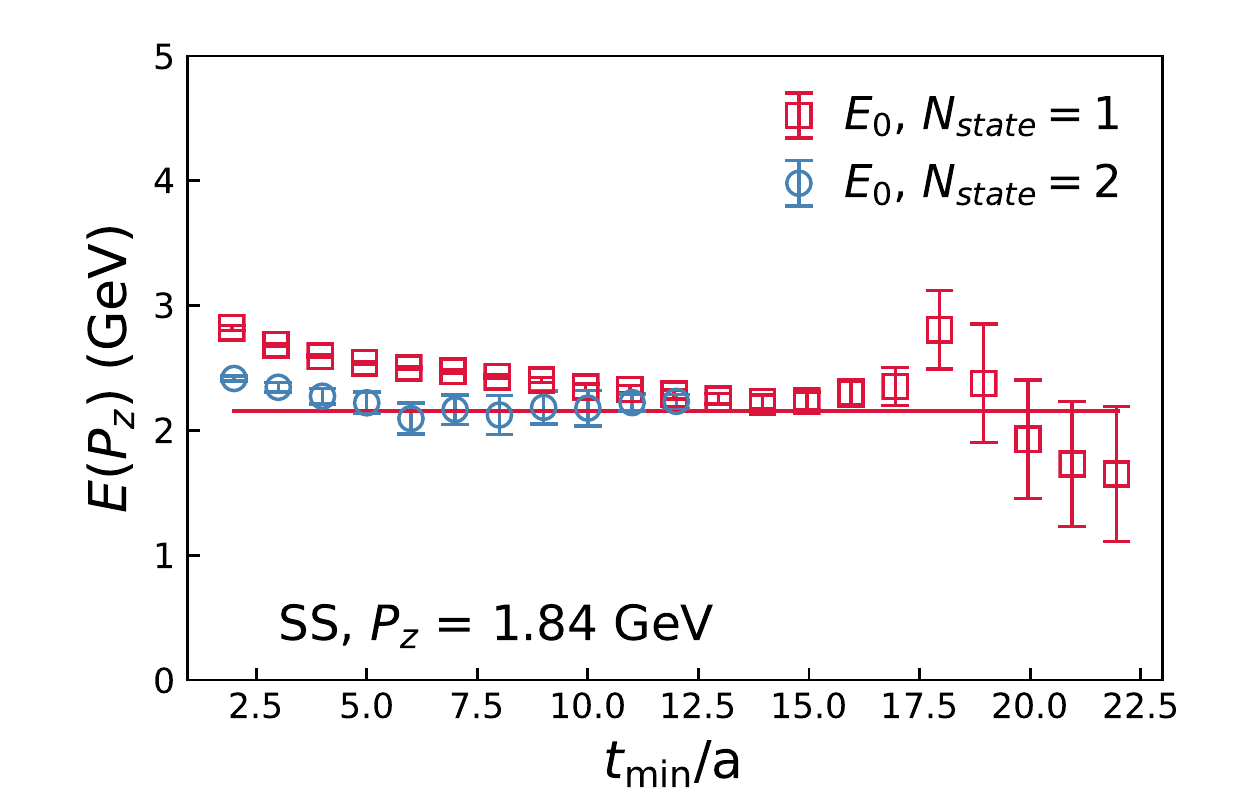}
\includegraphics[scale=0.45]{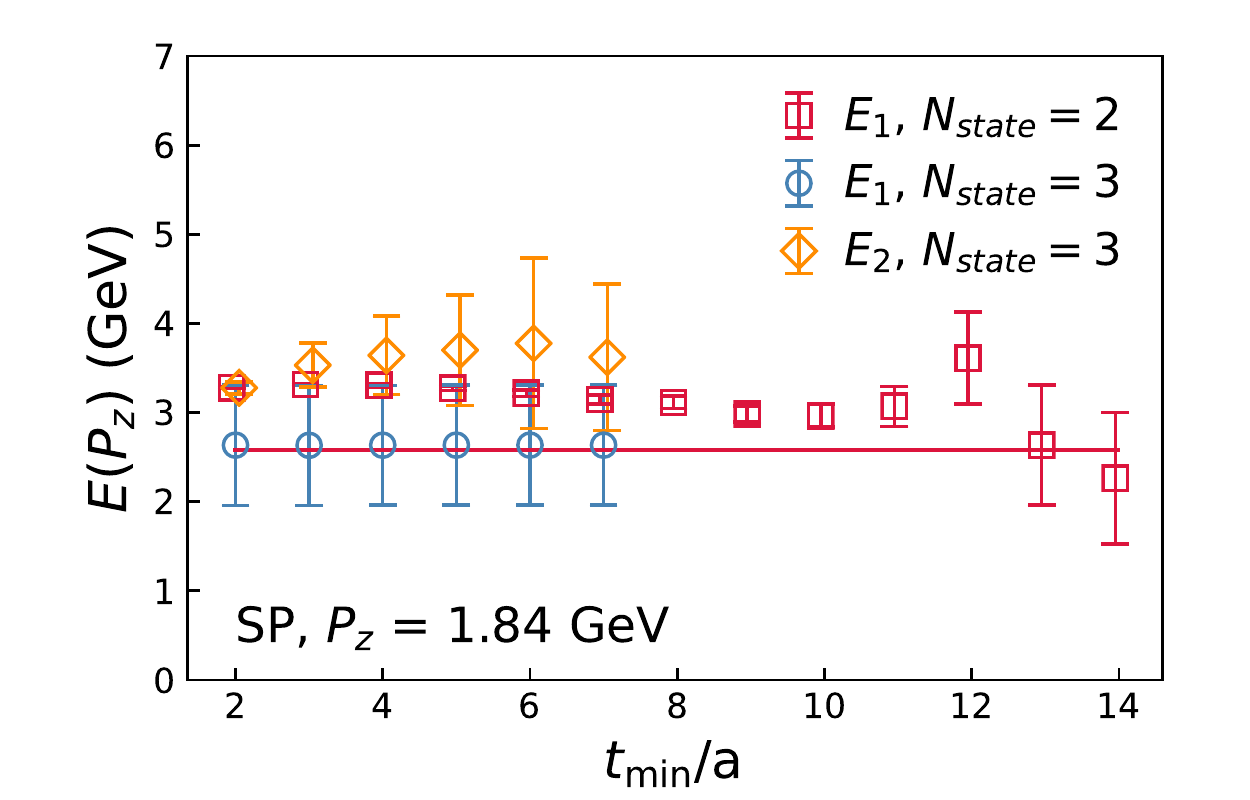}
\includegraphics[scale=0.45]{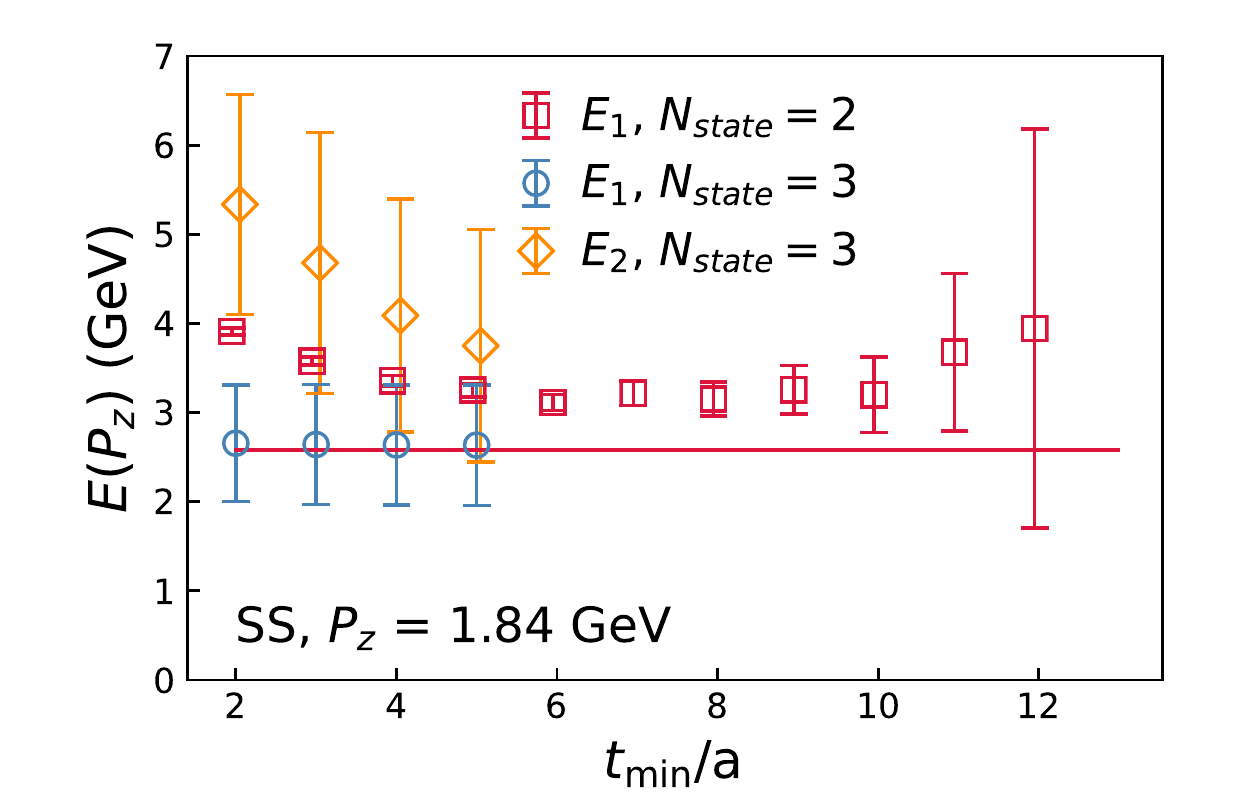}
\caption{Fit results for $n_z=3$ (up) and $n_z=4$ (down) the nucleon two-point
function. Left panels are for the ground state ($E_0$) from one-state and two-state
fits. Middle and right panels are for the first ($E_1$) and second ($E_2$) excited
states, determined by two-state and three-state prior-based fits (see text for
details). The horizontal lines are the values calculated from the dispersion relation.
}
\label{fig:fit2pt}
\end{figure}

In order to determine the energy levels, we fit the spectral decomposition of $C_{\rm 2pt}(t_s)$,
\begin{equation}
    C_{\rm 2pt}(t_s)=\sum_{n=0}^{N_{\rm state}-1} A_n e^{-E_n t_s},
\end{equation}
truncated at $N_{\rm state}$ to the two-point function data over a range of values of
$t_s$ between $[t_{\rm min},32 a]$. Since the lattice extent in the time direction is
192, we did not find any effect of lattice periodicity in this range of $t_s$ to be
important. We performed this fitting with one-state ($N_{\rm state}=1$), two-state
($N_{\rm state}=2$), and  three-state  ($N_{\rm state}=3$) Ans\"atze.  The ground state
energies, $E_0$ from the fits of SS correlators for $n_z=3$ and $4$ are shown in left
panels of Fig. \ref{fig:fit2pt} as function of $t_{\rm min}$, where $t_{\rm min}$
indicates that only $C_{\rm 2pt}(t_s>t_\mathrm{min})$ have been fitted.  Similar
results were obtained at the other values of the momenta.  The horizontal lines in
the figures correspond to the results from the dispersion relation for $E_0$.  The
single exponential fits give a good description of the SS correlator for $t_{\rm
min}>11a$, while two exponential fits give stable results for the ground state energy
already for $t_{\rm min}>5a$.

We found the determination of the excited state energies from the SS correlators to
be more problematic than from SP correlators. The excited state energy for SS is not
well-constrained by simple two exponential fits, and it is also not very stable with
respect to the variation of $t_{\rm min}$. Since the SP and SS correlators receive
different contributions from excited states, we performed a combined analysis of them
to obtain more reliable results for the excited state energies. Since we were able to
obtain the ground state energy $E_0$ reliably from one or two exponential fits to
both the SS and SP correlators and they agree with the expectation from the
dispersion relation well, we used  $E_0$ as a prior to performed more stable
two-exponential fits. The results from the two-state exponential fits, with $E_0$ as
prior, for $n_z=3$ and $n_z=4$ are shown in middle and right panels
Fig.~\ref{fig:fit2pt} for the SP and SS correlators, respectively. For the SP
correlators, the excited state energy $E_1$ seems to approach a plateau smoothly for
$t_{\rm min}>13a$.  It is interesting to note that, empirically, we observe the
values of the plateaus agree with the dispersion relation
$E_1(P_z)=\sqrt{P_z^2+E_1(P_z=0)^2}$, which are shown as the horizontal lines. While
being an interesting observation, such a stringent identification of this state is
not important to our analysis and requires further studies to rigorously establish
this. For the SS correlator, $E_1$ develops a pseudo-plateau for $5a < t_{\rm min} <
10a$ and it relaxes to the true plateau (i.e., as identified from the SP case) for
$t_{\rm min}>11a$.  For $n_z=4$, it is actually difficult to identify the true
plateau.  To model the excited state contributions to $R(\tau, t_s)$ in the range
$0<\tau<t_s/2$, with $t_s=16a, 18a, 20a$, one might consider using the
well-determined values of $E_0$ and $E_1$ from the SP correlator at large $t_s$.
However, as we will demonstrate now, such choices provide a less accurate description
of the SS two-point function in the range $5a < t_s < 10a$.  A better description of
the excited-state contributions to the $C_{\rm 2pt}(5a < t_s < 10a)$ can be obtained
by using the effective pseudo-plateau value of $E_1$ in the range of $5a < t_{\rm
min} < 10a$.

Since, we observe $E_1$ to be well-described by a particle-like dispersion relation
for sufficiently large $t_s$, we perform three-state fits for both SP and SS
correlators by imposing a prior on $E_1$ as well, using its best estimate from 2-state fit of SP
correlators with the corresponding  Jackknife errors\cite{Izubuchi:2019lyk}. The results are shown in middle and right panels in
Fig.~\ref{fig:fit2pt}.  We see that with the prior-based three-state exponential
fits, we can obtain stable results for the first excited state energy, $E_1(P_z)$,
already for relatively small $t_{\rm min}$ which agrees with the dispersion relation
value that we input via the prior. The value of the second excited state is also
shown in Fig.~\ref{fig:fit2pt} and it roughly agrees with the values of $E_1$ from
the two-exponential fit (with prior only on $E_0$) at smaller $t_{\rm min}$.  Since
the value of $E_2$ is quite large, the third exponential probably corresponds to a
combination of several excited states. In Fig.~\ref{fig:meff}, we show the 1-$\sigma$
bands for the effective mass corresponding to: (1) Two-state fit that uses values of
$E_0$ and the true value of $E_1$; (2) two-state fit obtained by setting $E_1$ to be
the effective value in the range from $5a < t_{\rm min} < 10a$; (3) three-state fit
that we described above. We find that the curves (2) and (3) agree quite well with
each other in the range of $5a < t_s < 10a$ and they extrapolate in the similar
fashion to the asymptotic value $E_0$.  However, the curve (1) fails in capturing the
data in the range $5a < t_s < 10a$. Since for our three-point calculations the
source-sink separations were chosen to be  $t_s=16a, 18a, 20a$, we must model the
effective excited state contributions to the three-point functions in the range
$0<\tau<t_s/2$. Thus, through this analysis on SP and SS correlators, we numerically
demonstrate the usage of an effective value of $E_1$ in the range of $5a < t_s< 10a$
that is higher than the true value of $E_1$ is justified, and is the best
extrapolation one could perform for the extraction of bare matrix elements in the
absence of enough data to perform a three-state fit.

\begin{figure}[t]
  \centering
\includegraphics[scale=0.6]{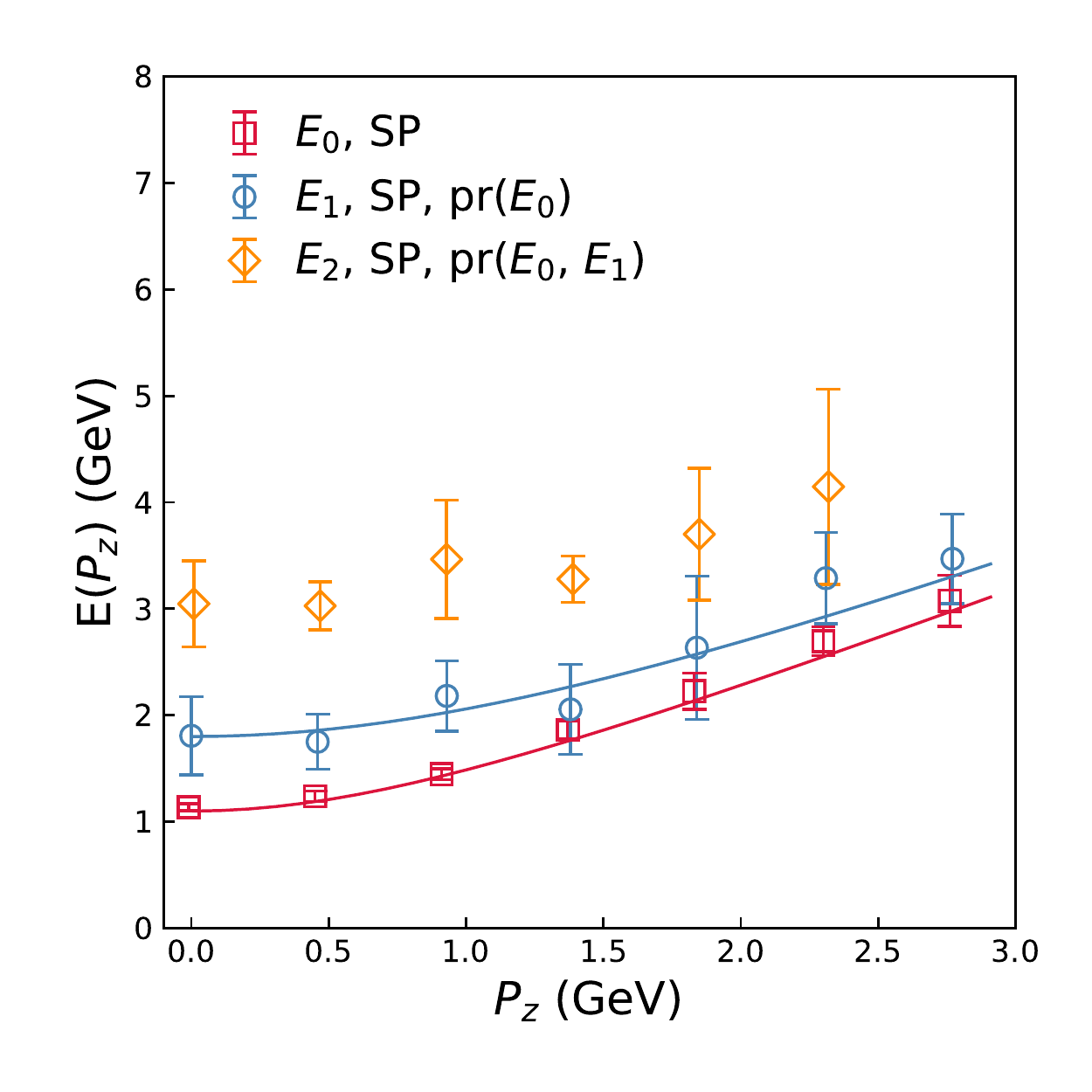}
\includegraphics[scale=0.6]{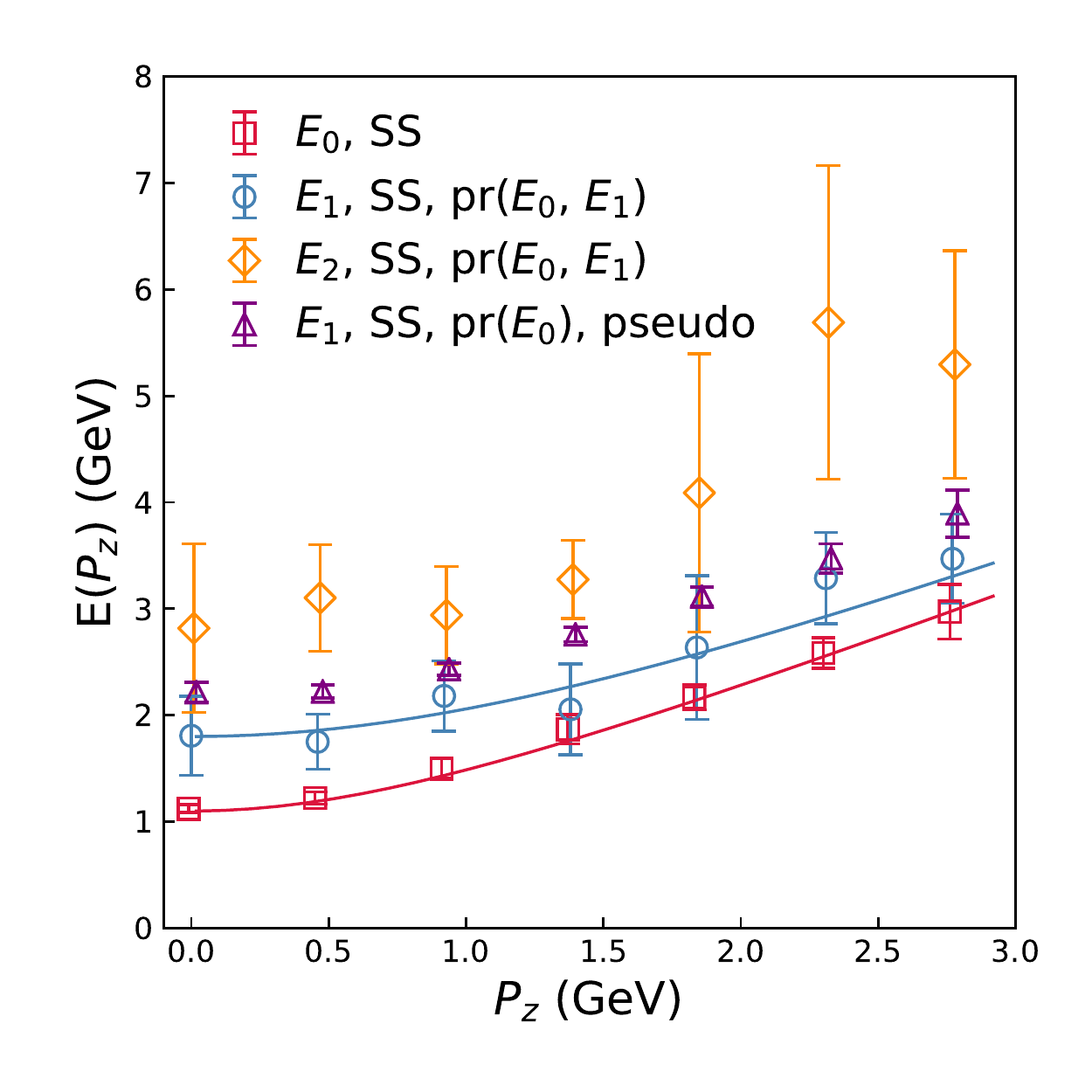}
\caption{The energies of different states as function of $P_z$. On the left panel,
the $P_z$ dependence of $E_0$, $E_1$ and $E_2$ for SP correlators are shown. The values of $E_1$ were
obtained from two-state fit with prior only on $E_0$, and $E_2$ from a three-state
fit with priors on both $E_0$ and $E_1$. On the right panel,
the $P_z$ dependence of true values (blue points) and effective values of $E_1$ (purple points) for SS correlators are both shown (see text for details). The lines indicate the corresponding continuum dispersion relations.
}
\label{fig:disprel}
\end{figure}

Let us now, summarize the analysis of the nucleon two-point function. Using boosted
Gaussian sources we were able to extract ground state energy levels up to momenta
$2.7$  GeV from SP and SS correlators. The ground state energy dependence on $P_z$
seem to follow the continuum dispersion relation. Using this fact, we performed
prior-based fits using the energy from the dispersion relation as a prior and
extracted the excited state energies as function of $P_z$. For SP correlator the
extracted value of $E_1$ agrees well with the one from the dispersion relation. We
show this in the left panel of Fig.~\ref{fig:disprel}. Furthermore, we were able to
extract an effective third energy level. These results are shown as blue points
($E_1$) and orange points ($E_2$) in the left panel of Fig.~\ref{fig:disprel}. Our
results for the energy levels obtained from SS as function of $P_z$ are summarized in
right panel of Fig. \ref{fig:disprel}. Here, the effective values of $E_1$ from the
pseudo-plateau and the true values are both showed. The main point of the elaborate
analysis is that even though a third excited state contributes in the relatively
shorter range of $t_s$ we use in the paper, it possible to describe the SS
correlator very well by a two state form with an effective value of $E_1$, which is
larger than the energy of the physical excited state. Further details on the analysis of the two-point functions are provided in Appendix~\ref{app:2pt}.

\section{Nucleon three-point correlators} \label{sec3pt}
In order to obtain the nucleon qPDF matrix element we consider the
ratio of the 3-point function to 2-point function, $R(z,P_z;t_s,\tau)$,
at different source sink separations, $t_s$, and operator
insertion, $\tau$. At fixed $(z, P_z)$ we are interested in fitting
the $(t_s,\tau)$-dependence as expected from the spectral decomposition
of $R$.  If only two states contribute to the correlation
functions, the dependence of this ratio on $\tau$ and $t_s$ is given
by the following form:
\begin{equation}
    R^{\rm fit}_3(t_s,\tau)=\frac{B_0+e^{-\Delta E t_s/2} \left( B_1 e^{-\Delta E (t_s/2-\tau)}+B_2 e^{\Delta E (t_s/2-\tau)} \right)
+B_3 e^{-\Delta E t_s}}{1+\frac{A_1}{A_0}e^{-\Delta E t_s}}.
\label{eq:R1}
\end{equation}
Here $B_0$ is the desired matrix element $h$, and $\Delta E=E_1-E_0$.
Generically, $B_1$ and $B_2$ are independent fit parameters, except
at $z=0$, where $B_1=B_2$. If we assume that the terms proportional
to $A_1$ are small, the denominator can be expanded to leading order
to obtain a simpler form
\begin{equation}
    R^{\rm fit}_2(t_s,\tau)=B_0+e^{-\Delta E t_s/2} \left( B_1 e^{-\Delta E (t_s/2-\tau)}+B_2 e^{\Delta E (t_s/2-\tau)} \right)
+B_3 e^{-\Delta E t_s}.
\label{eq:R2}
\end{equation}
Finally, if the term proportional to $B_3$ is also small compared to other terms,
we get an even simpler expression that depends
only on three parameters, $B_0$, $B_1$ and $B_2$:
\begin{equation}
    R^{\rm fit}_1(t_s,\tau)=B_0+e^{-\Delta E t_s/2} \left( B_1 e^{-\Delta E (t_s/2-\tau)}+B_2 e^{\Delta E (t_s/2-\tau)} \right).
    \label{eq:R3}
\end{equation}
For each $(z, P_z)$, we fitted the $(t_s,\tau)$-dependence of $R(z,P_z;t_s,\tau)$
to Eqs.~(\ref{eq:R1}, \ref{eq:R2}, \ref{eq:R3}) and determined $B_0$ in each case. In
all these fits we used a fixed value of $\Delta E(P_z)=E_1(P_z)-E_0(P_z)$, with the
pseudo-plateau values of $E_1(P_z)$ and the ground state energies $E_0(P_z)$
determined from the 2-point SS correlation function, as shown in
Fig.~\ref{fig:disprel}(right).

\begin{figure}[t]
  \centering
\includegraphics[scale=0.7]{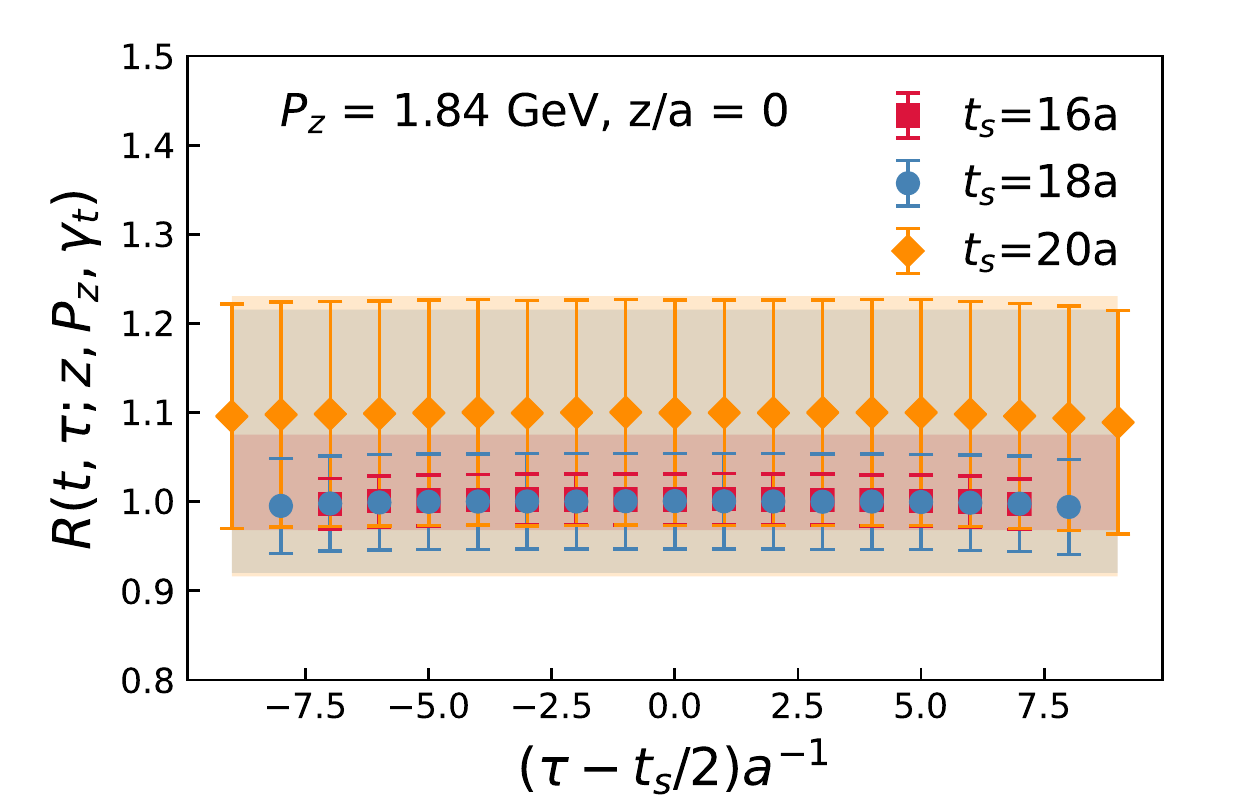}
\includegraphics[scale=0.7]{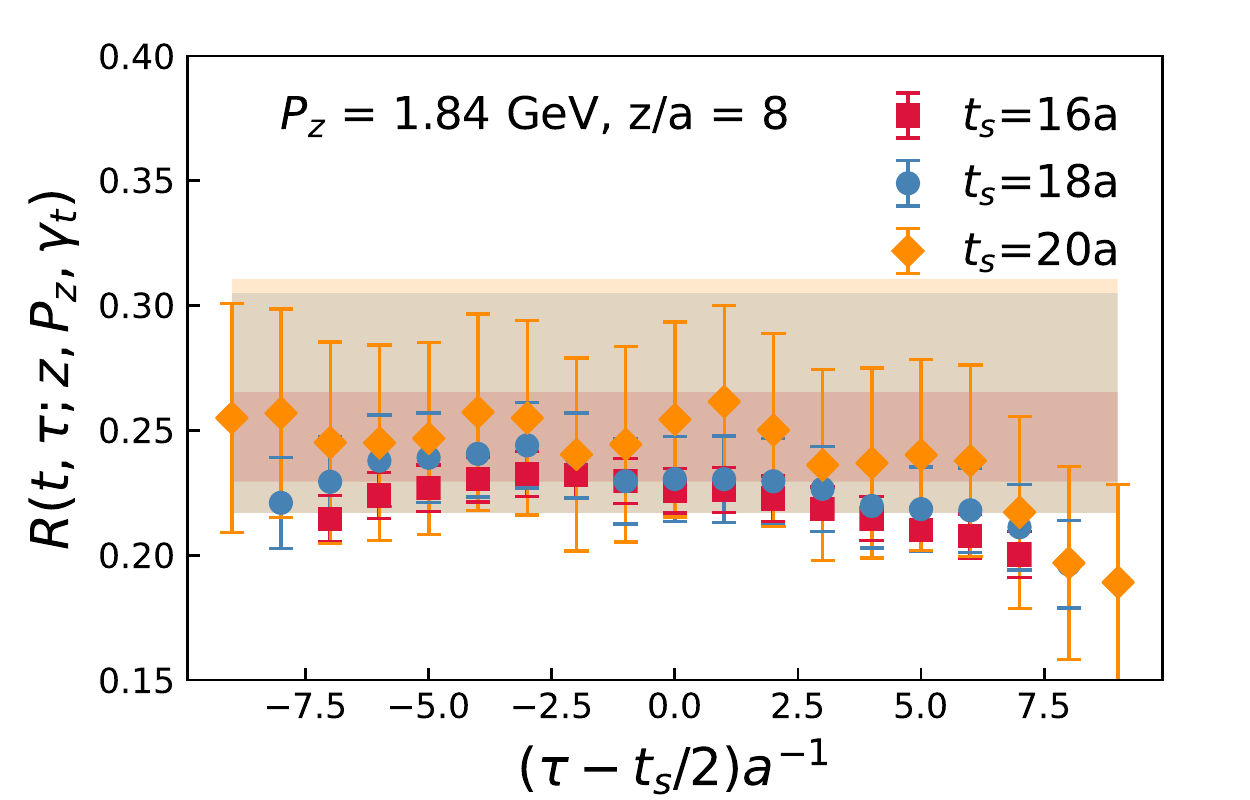}
\caption{The real part ratio of the three-point function to two-point function at $z=0$ (left) and $z=8a$ (right)
    as a function of $\tau$ and for different $t_s$. The red, orange and gray bands correspond to the bare matrix elements $B_0$ extracted from fits to $R^{\rm fit}_1$, $R^{\rm fit}_2$ and $R^{\rm fit}_3$, respectively.
}
\label{fig:rat_fit}
\end{figure}

\begin{figure}[t]
  \centering
\includegraphics[scale=0.7]{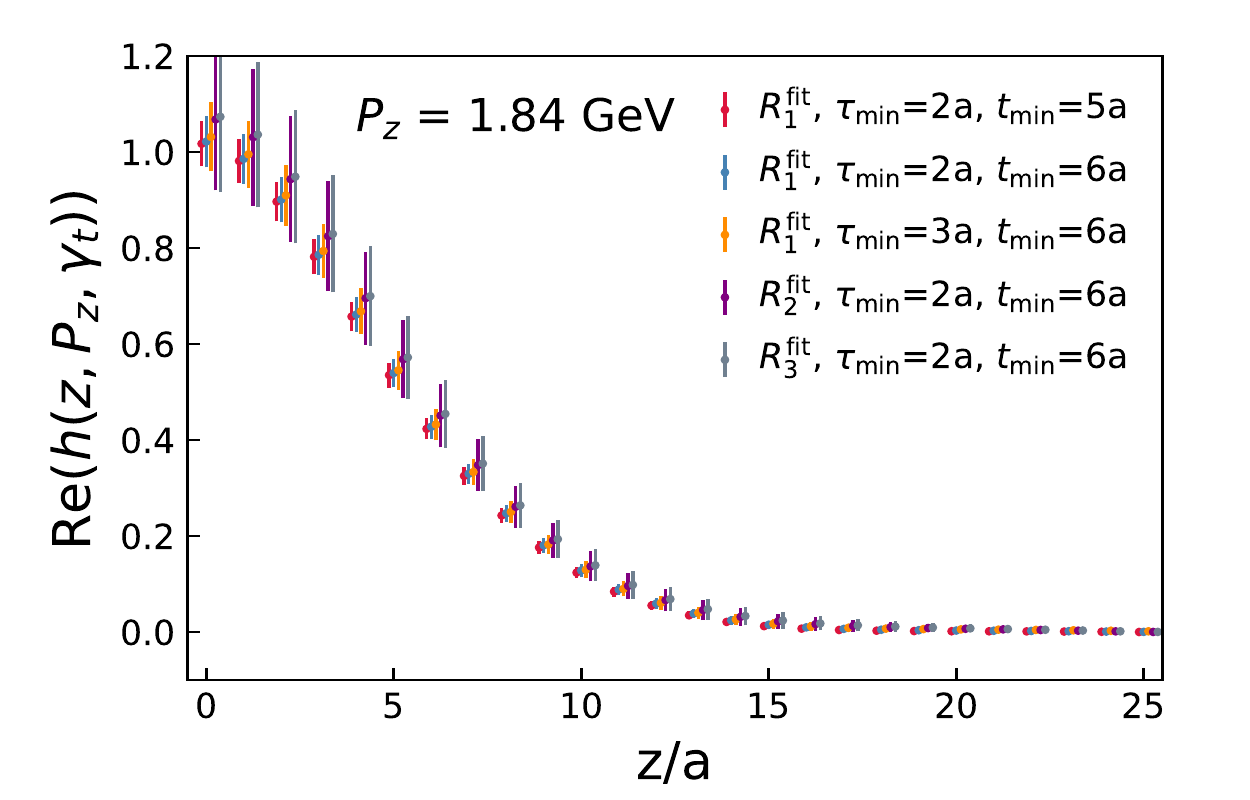}
\includegraphics[scale=0.7]{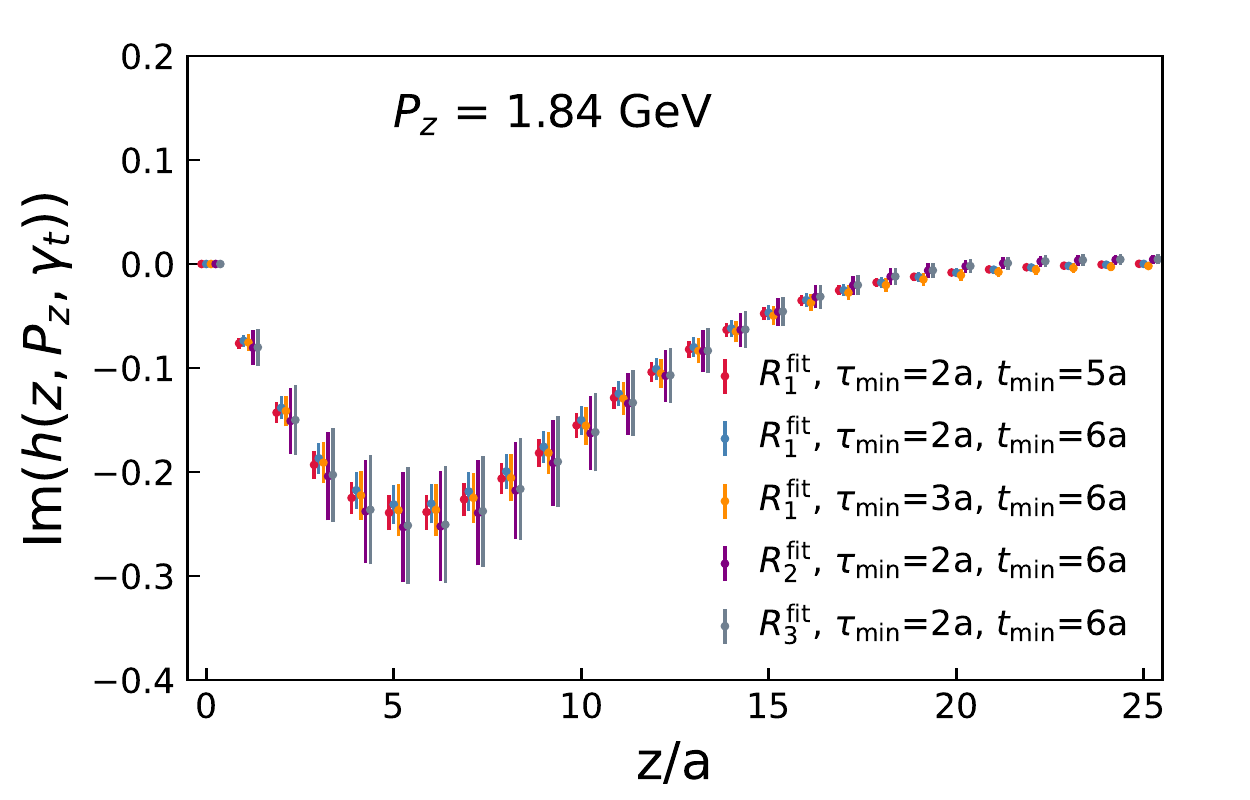}
\caption{The $z$ dependence of the bare matrix element for $n_z=4$. The left panel corresponds to the real
part and the right panel corresponds to the imaginary part. The different colors are the matrix elements 
    obtained by various extrapolation methods (denoted by $R_1^{\rm fit}, R_2^{\rm fit}$ and $R_3^{\rm fit}$), 
  the number of operator insertion points skipped near source and sink (denoted by $\tau_{\rm min}$), and 
  the $t_{\rm min}$ value in two-point function fit from which the excited state $E_1$ was obtained from. 
}
\label{fig:h_px4_zdep}
\end{figure}

In the following, we discuss the ratio of the 3-point function to 2-point function,
$R(z,P_z;t_s,\tau)$, and the corresponding fits for $\Gamma=\gamma_t$ and $n_z=4$. In
Fig.~\ref{fig:rat_fit}, we show the lattice data on this ratio, together with the fit
results for two representative values of $z$, namely $z=0$ and $z=8a$. The $t_s$
dependence of the lattice results is small compared to the statistical errors. In
particular, the difference between $t_s=12a$ and $t_s=16a$ data is quite small.  This
means that the contribution of the excited states is not large even though the
source-sink separation is below 1 fm.  Given that the $t_s$-dependence of the ratio
is small, it is natural to set $B_3=0$ since the term is suppressed by $e^{-t_s
\Delta E}$, and perform fits using $R^{\rm fit}_3(t_s,\tau)$. We performed fits of
the lattice results using the three different fit forms above and the value of
$\Delta E$ obtained from two-state fits of the 2-point functions with $t_{\rm
min}=6a$. We used operator insertion time $\tau \geqslant  \tau_{\rm min}=2a$ in the
fits.  The matrix elements, $B_0$, were obtained using the three fit functions agree
within errors. The real and imaginary part of the ratio $R$ should be symmetric and
antisymmetric with respect to $z=0$ at any fixed $t_s$ and $\tau$. In general, our
lattice data is compatible with this expectation. Hence, we  symmetrize and
antisymmetrize the real and imaginary part of the ratio with respect to $z=0$. The $z$-dependence of the bare matrix elements is shown in Fig.
\ref{fig:h_px4_zdep} for all three types of fits.  We see again that all three fits
give the consistent results within errors. Since $B_0$ obtained from $R^{\rm fit}_3$
and $R^{\rm fit}_2$ are consistent with that obtained from $R^{\rm fit}_1$, but with
larger errors, in the following we will focus on the results obtained from $R^{\rm
fit}_1$. We also carried out additional checks for any systematic effects, as
discussed below.

We performed several checks to understand the systematic effects in $R^{\rm fit}_1$.
First, we studied the dependence of the extracted matrix element on $\tau_{\rm min}$
and found no significant dependence on it.  Second, we performed the fits using only
a single source-sink separation $t_s$ and compared the corresponding results from the
three values of $t_s$. Interestingly, the matrix elements calculated for source sink
separation $t_s=16a,~18a$ and $20a$ agree within errors, though the $t_s=20a$ results
have very large errors. We also studied the variation of the extracting matrix
element on $\Delta E$ by using $E_1$ obtained using different values of $t_{\rm
min}$. We found no significant variation. Finally, we used the summation method to
obtain the matrix element. This determination has very large statistical errors but
it is still compatible with all other determinations. The above checks of systematic
effects are discussed further in the Appendix~\ref{app:3pt}. We performed a similar
analysis for the three-point functions corresponding to the helicity
qPDF, i.e. for $\Gamma=\gamma_z \gamma_5$. Details of those analysis also are
discussed in the Appendix~\ref{app:3pt}.

\section{Nonperturbative renormalization}
\begin{figure}[t]
\includegraphics[width=0.45\textwidth]{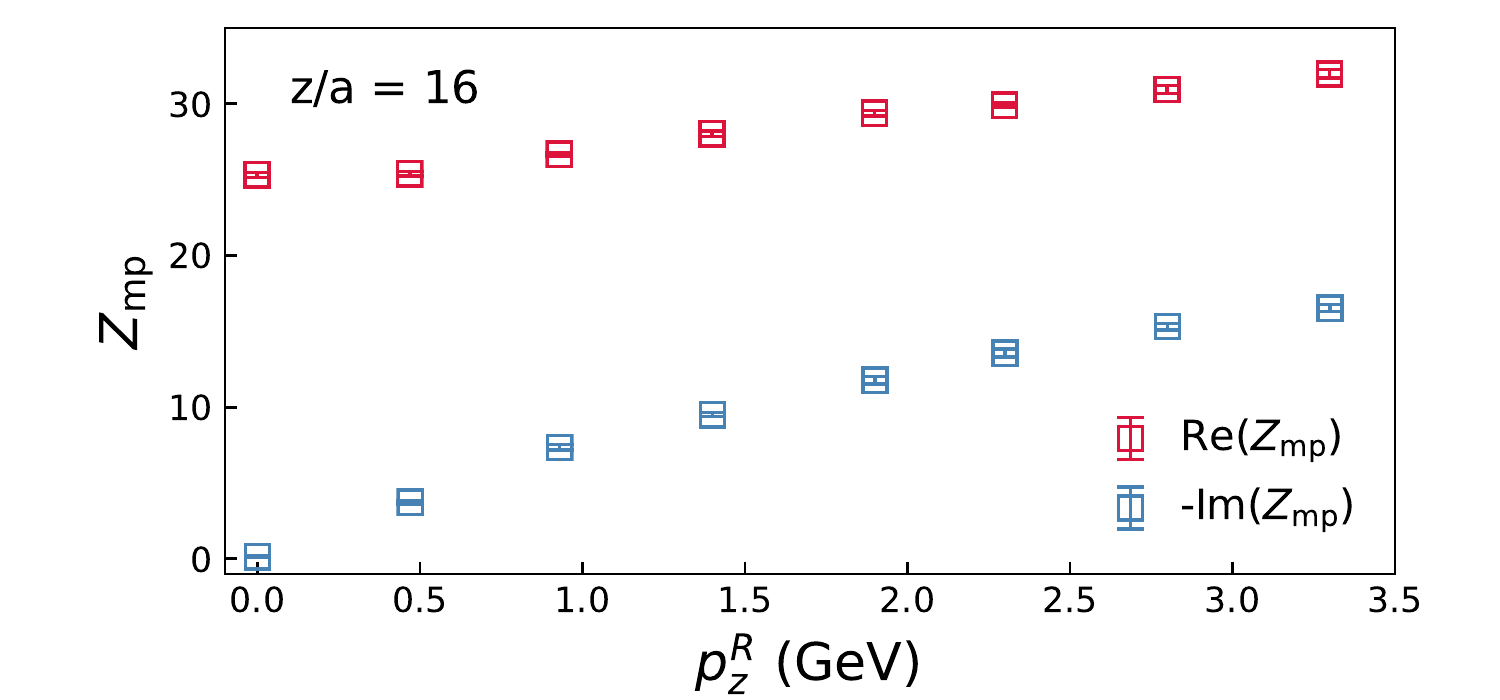}
\includegraphics[width=0.45\textwidth]{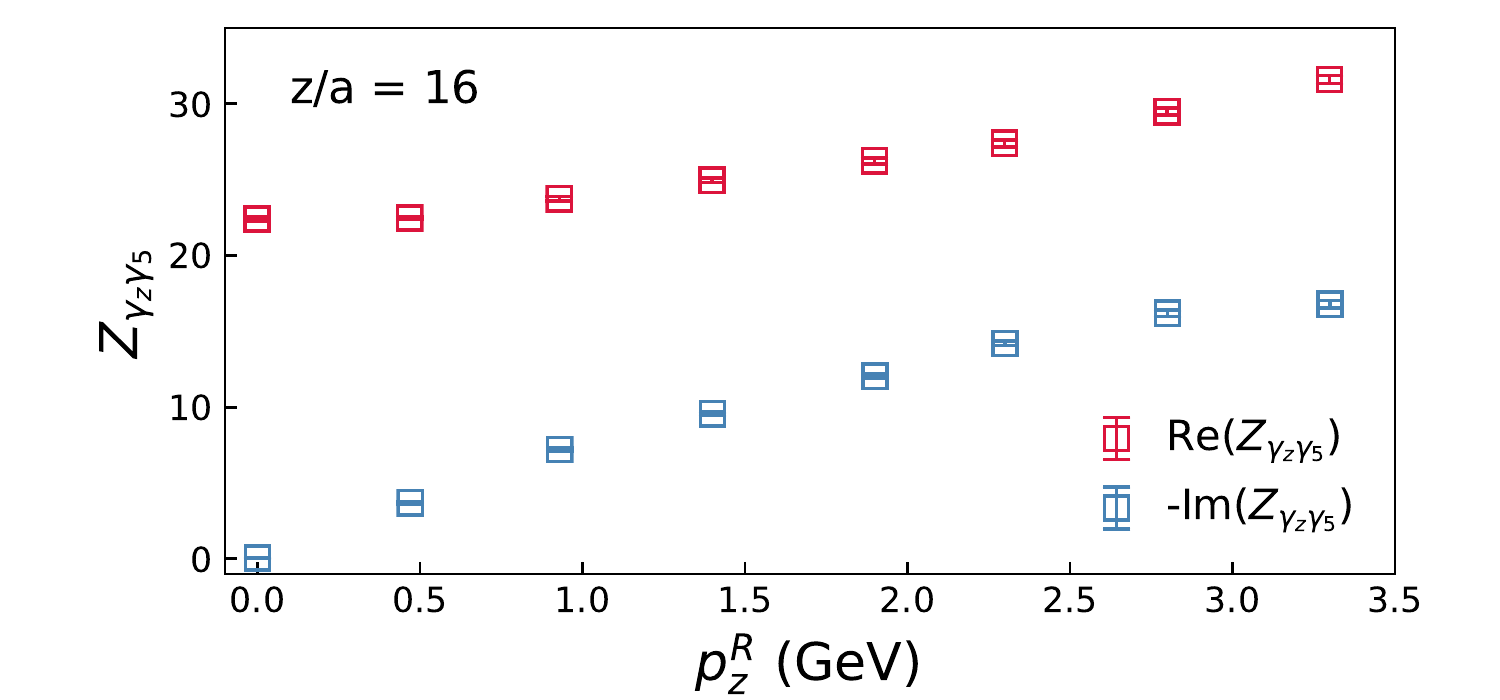}
\includegraphics[width=0.45\textwidth]{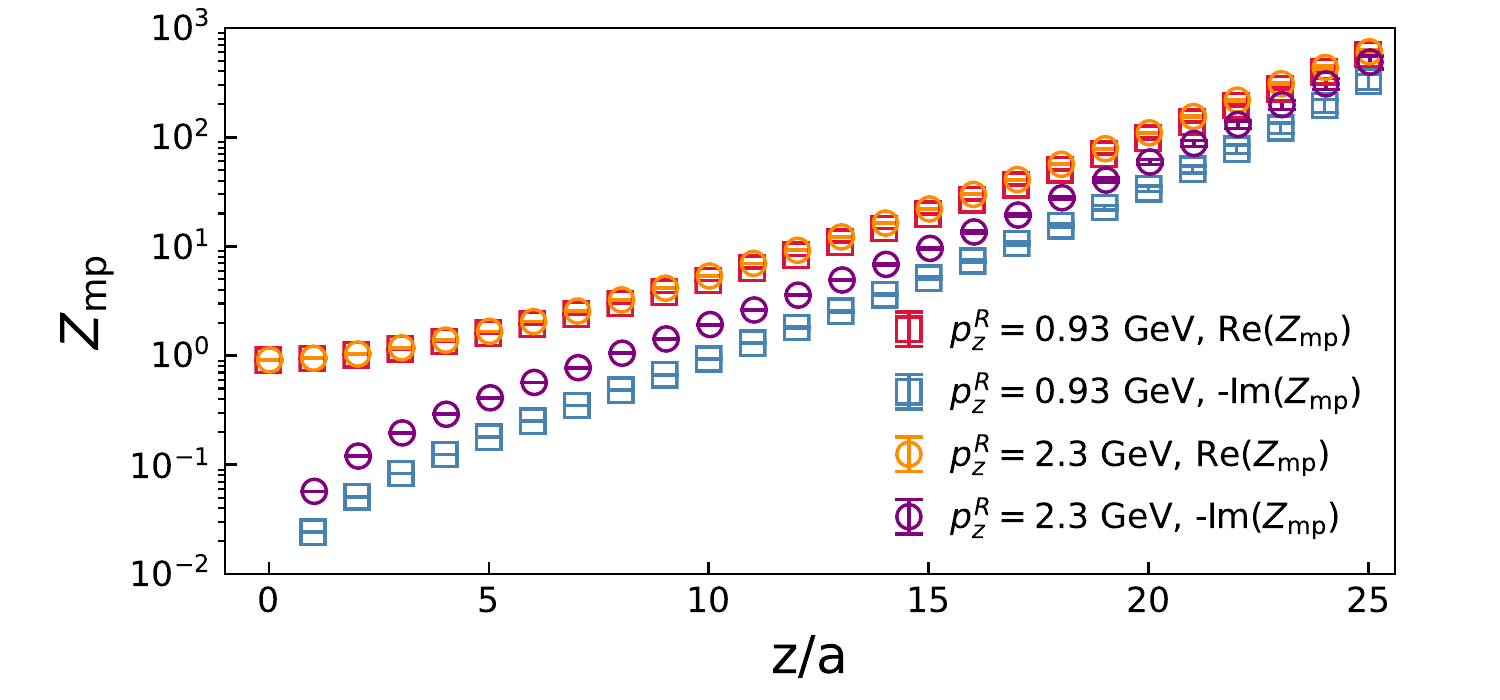}
\includegraphics[width=0.45\textwidth]{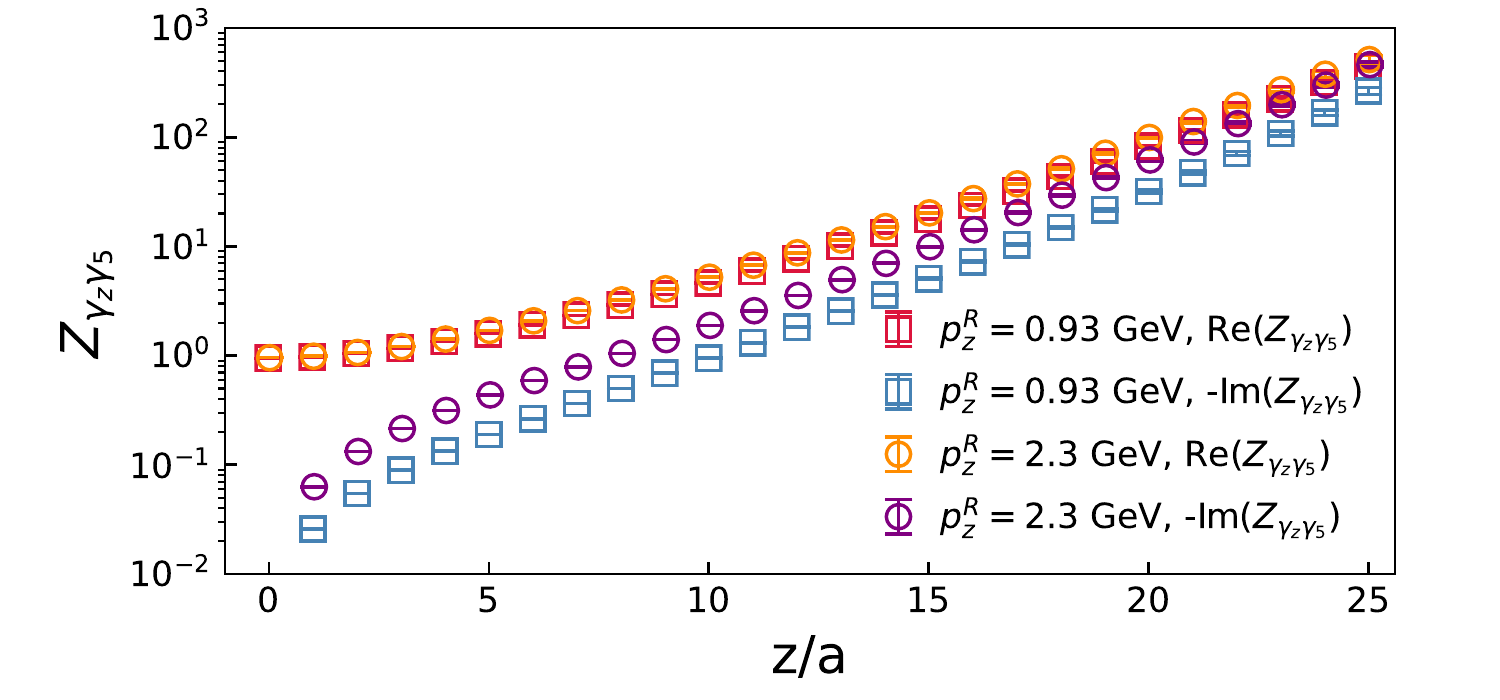}
\caption{The renormalization constant $Z_{\rm mp}$ and $Z_{\gamma_z \gamma_5}$ at $\mu_R=4~$GeV. The upper panels show $Z_{\rm mp}$ and $Z_{\gamma_z \gamma_5}$ as a function of $p^R_z$ at $z=16 a \approx 0.67$~fm. The lower panels show the $z/a$ dependence with $p_z^R$ = 0.93 GeV and 2.3 GeV.
}
\label{fig:RI_MOM}
\end{figure}

We calculate the nonperturbative renormalization of the qPDF
operator in the RI-MOM scheme using off-shell quark states in the Landau
gauge~\cite{Alexandrou:2017huk,Chen:2017mzz}.  The matrix element
of $O_{\Gamma}(z)$ in an off-shell quark state $|p\rangle$ is
\begin{equation} \label{eq:wallsource}
\Lambda(p,z,\Gamma)=\langle S(p)\rangle^{-1} \left \langle \sum_w {\gamma_5}S ^{\dagger}(p,w+zn){\gamma_5}\Gamma W_z(w+zn,w)S(p,w) \right \rangle \langle S(p)\rangle^{-1} \ ,
\end{equation}
where $n^\mu=(0,0,0,1)$ is the unit vector along the $z$ direction,
and the summation is over all lattice sites $w$. The quark propagators are defined as
\begin{equation}
S(p,x)=\sum_{y} e^{ipy}\langle \bar{\psi}(x) \psi(y)\rangle\ ,\ S(p)=\sum_{x} e^{-ipx}S(p,x),
\end{equation}
and $\gamma_5$ is inserted on both sides of $S^{\dagger}(p,w+zn)$ in Eq.~\ref{eq:wallsource}
to get the necessary propagator $\sum_{y} e^{-ipy}\langle \bar{\psi}(y) \psi(w+zn)\rangle$.

For the unpolarized qPDF, we use the RI-MOM renormalization constant defined via
\begin{equation}\label{RIMOMrenormcond}
Z_{\rm mp}(z,p^R_z,a^{-1},\mu_R)= \left. \frac{\mathrm{Tr} [{\cal P}\Lambda_{\rm tree}(p,z,\gamma_t)]}{\mathrm{Tr}[{\cal P}\Lambda(p,z,\gamma_t)]} \right\vert_{p^2=\mu_R^2,\ p_z=p^R_z},
\end{equation}
where $\Lambda_{\rm tree}(p,z,\gamma_t)=\gamma_te^{-izp_z}$ is the tree level matrix
element in the momentum space. Furthermore, ${\mathcal P}=\gamma_t-(p_t/p_x)\gamma_x$
is the projection operator corresponding to the so-called minimal projection, where
only the term with the Dirac structure proportional to $\gamma_t$ is kept
\cite{Stewart:2017tvs,Liu:2018uuj}. Hence, we use the subscript `mp' for the
renormalization constant. The renormalization constant $Z_{\rm
mp}(z,p^R_z,a^{-1},\mu_R)$ depends on the lattice spacing $a$, as well as the other
two scales $p^R_z$ and $\mu_R$.

We followed a similar procedure for the longitudinally polarized case, where the RI-MOM renormalization
constant is defined as
\begin{equation}
Z_{\gamma_z \gamma_5}(z,p^R_z,a^{-1},\mu_R)= \left. \frac{\mathrm{Tr} [{\cal P}\Lambda_{\rm tree}(p,z,\gamma_z \gamma_5)]}{\mathrm{Tr}[{\cal P}\Lambda(p,z,\gamma_z \gamma_5)]} \right\vert_{p^2=\mu_R^2,\ p_z=p^R_z},
\label{RIMOMrenormcond_pol}
\end{equation}
with $\Lambda_{tree}(z,p_z,\gamma_z \gamma_5)=\gamma_z \gamma_5 e^{-ip_z z}$.
The projection operator ${\cal P}$ in this case was chosen to be ${\cal P}=\gamma_5 \gamma_z/4$.

We calculated the non-perturbative RI-MOM renormalization constants in Landau gauge.
The calculations were performed using 14 gauge configurations.
The relative uncertainties of the renormalization constants for $z=0,16,32$ are 0.02\%, 1\% and 10\%, respectively.
Such precision is much better than that of our nucleon matrix elements with the same $z$, so it is enough at the present stage.
We used the following values of the momenta for the off-shell quark state:
$ap=\frac{2\pi}{L} (5,5,5,0)$, $\frac{2 \pi}{L} (6,2,1,17/3)$ and $\frac{2 \pi}{L} (7,4,3,1/3)$,
$L=64$ being the spatial size the of the lattice.
These momenta correspond to $\mu_R=|p|=3.99$~GeV, $3.94$~GeV and $3.97$~GeV,
i.e. to $\mu_R \sim 4$~GeV within $1.5\%$. Since all the spatial directions are equivalent, each of them could be considered
as the $z$-direction and, therefore, with the above choice of the three momenta we have
$p_z^R=0.46\times\{0,1,2,\ldots,7\}$~GeV.

The renormalization constant is plotted in Fig.~\ref{fig:RI_MOM}.
Due to the linear divergence, the renormalization constant can be far from 1 at a large $z\approx 0.67$~fm,
making the nonperturbative renormalization unavoidable.
Fig.~\ref{fig:RI_MOM} also shows that the renormalization constant will be sensitive to the value of $p^R_z$,
while such a dependence should be canceled by the matching in the continuum if we have the matching formula up to all orders, because the PDFs or the Mellin moments in $\overline{\rm MS}$ scheme have no dependence on $p^R_z$.
We will consider the residual $p^R_z$ dependence in the final PDF prediction as a systematic uncertainty.

\begin{figure}[t]
  \centering
    \includegraphics[scale=0.7]{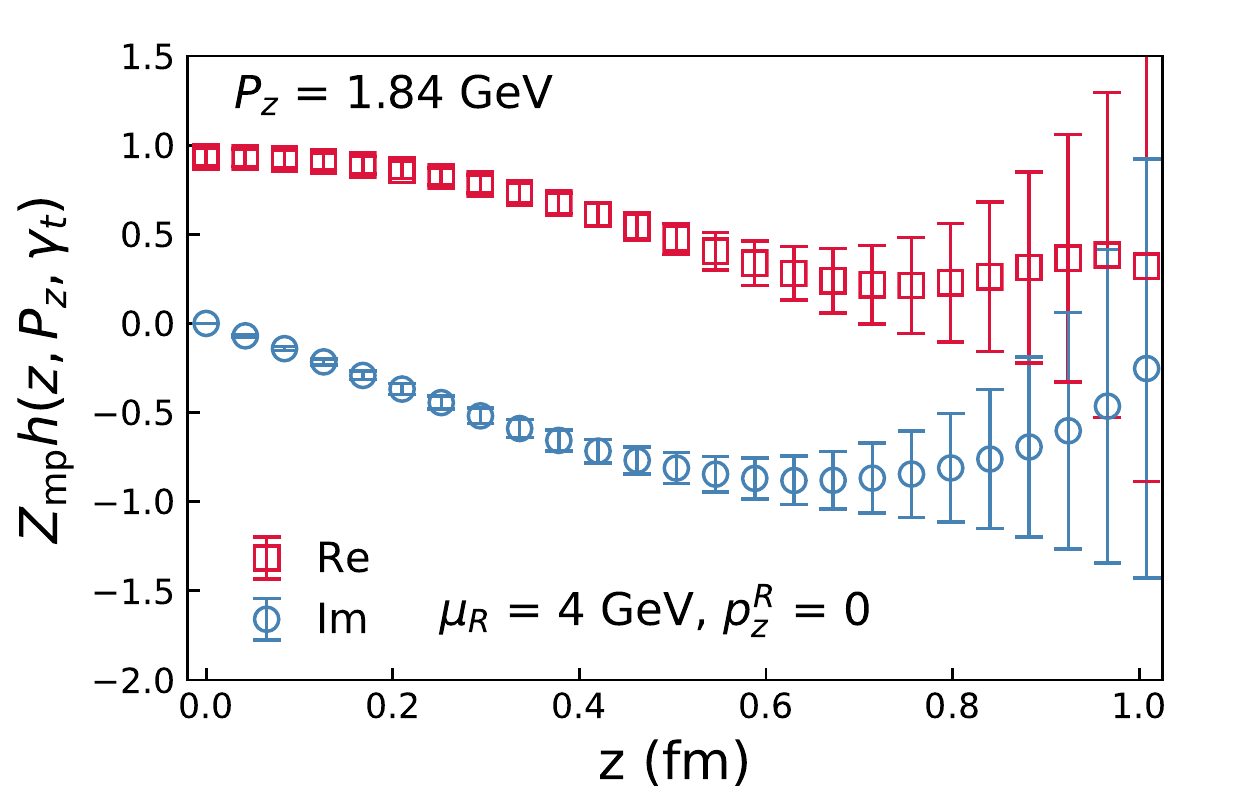}
        \includegraphics[scale=0.7]{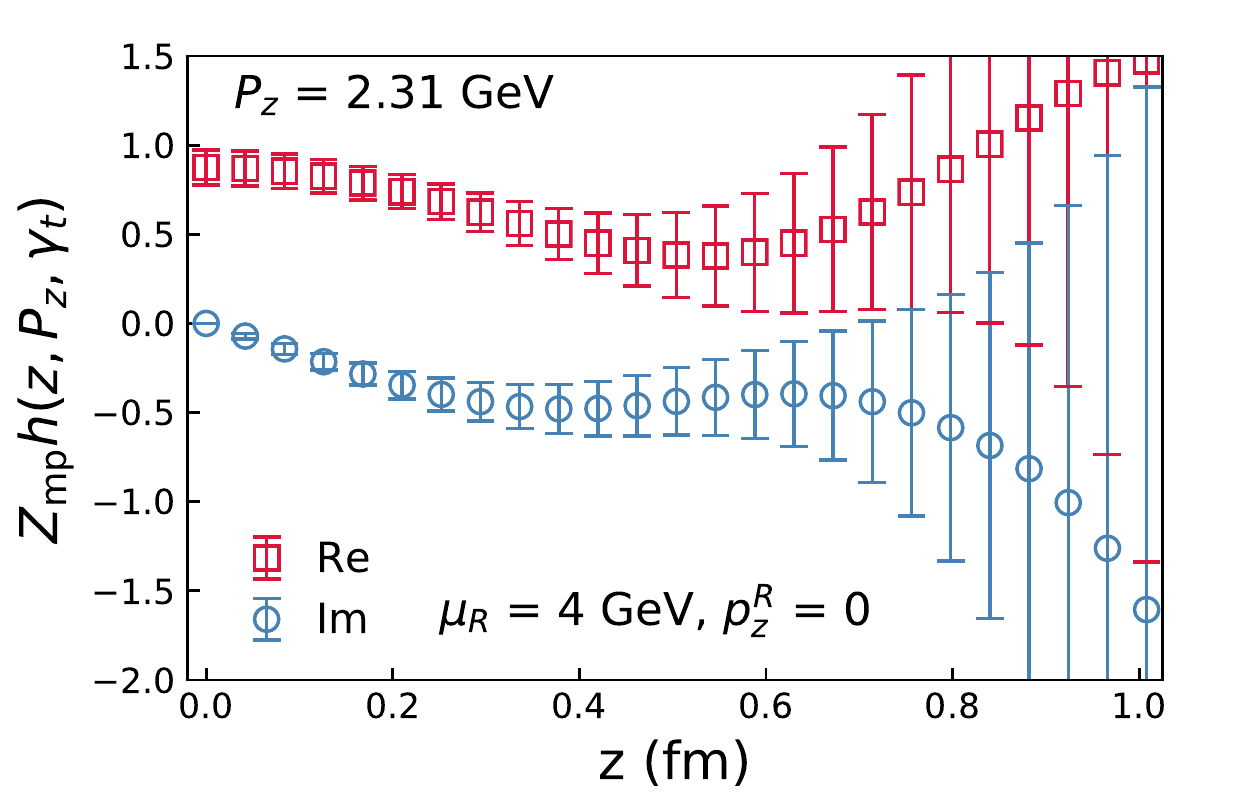}

    \includegraphics[scale=0.7]{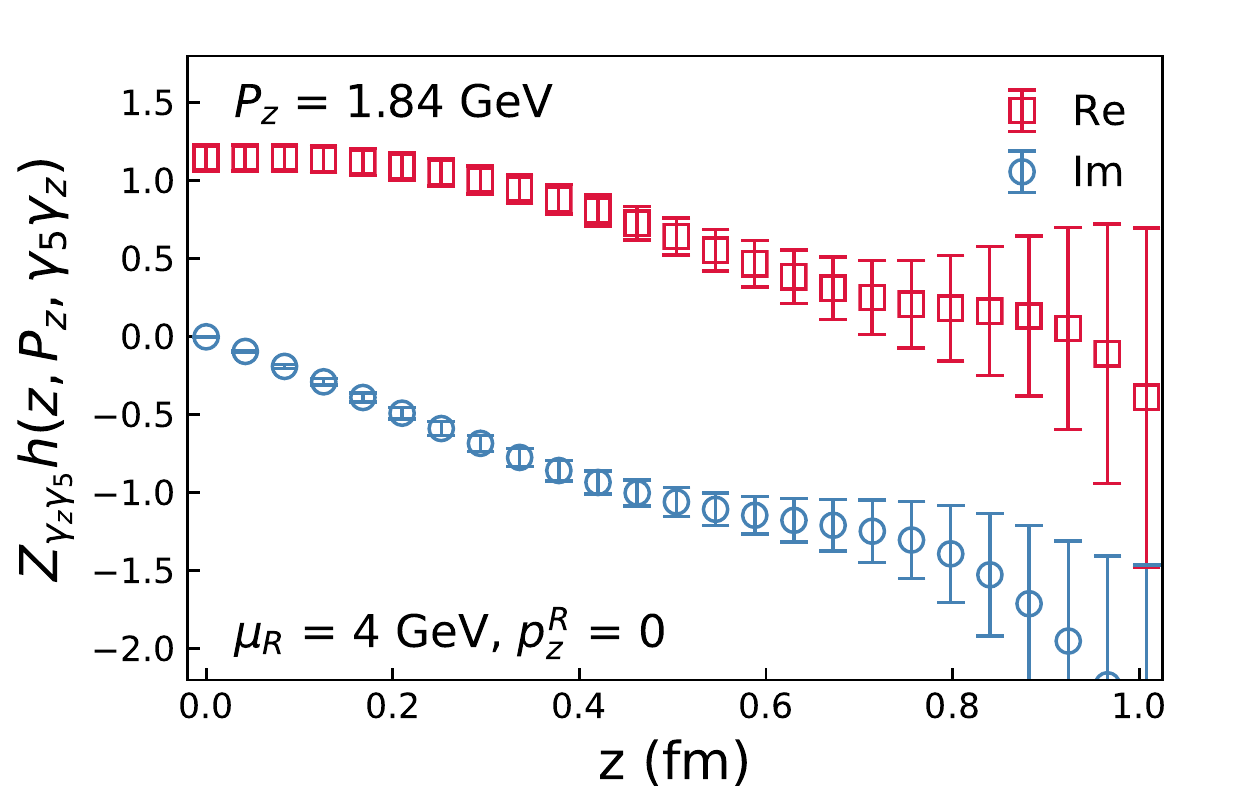}
        \includegraphics[scale=0.7]{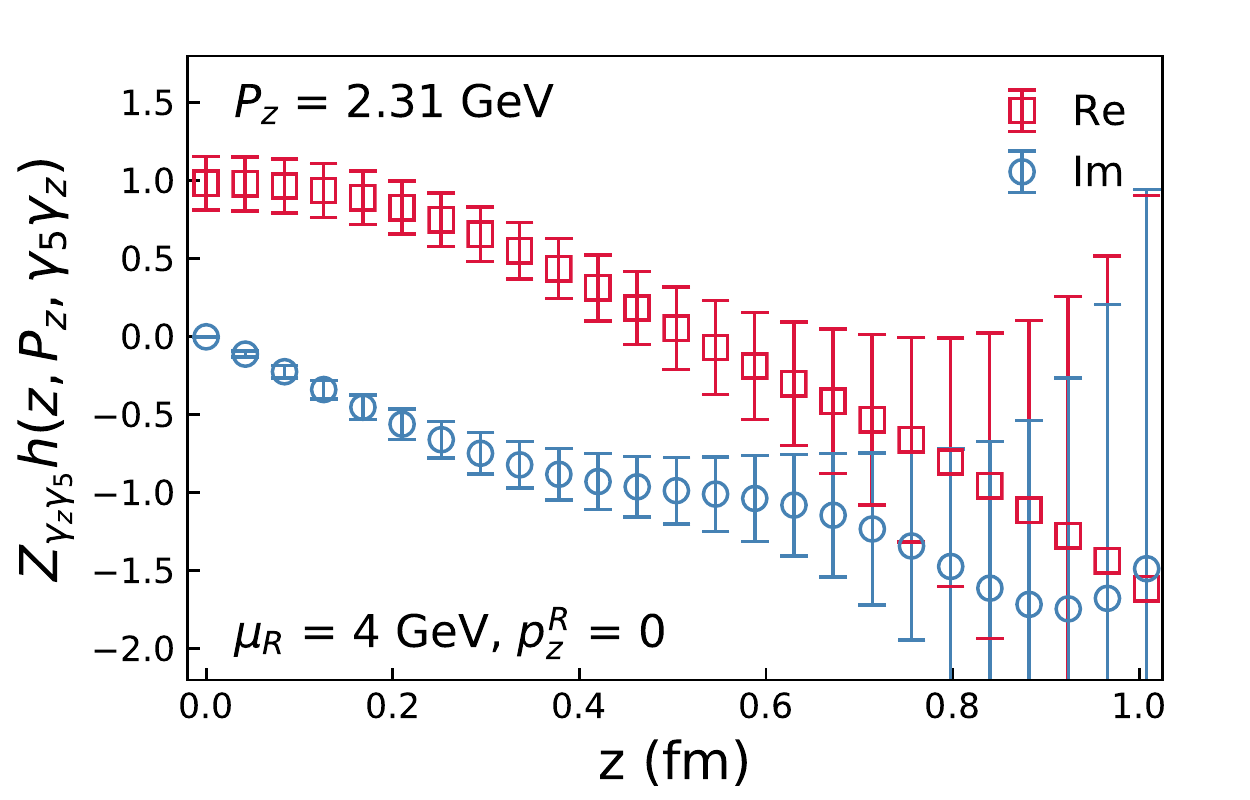}

\caption{Top panels: The $z$-dependence of the real and imaginary parts of
    the RI-MOM renormalized (modulo the wavefunction renormalization, $Z_q$) unpolarized qPDF matrix element for $P_z$ = 1.84 GeV (left) and 2.31 GeV (right).
    Bottom panels: Similar results for the real and imaginary parts of the helicity matrix element.
    }
\label{fig:unpolreME}
\end{figure}

Having determined the renormalization constants $Z_{\rm mp}$ and $Z_{\gamma_z \gamma_5}$ we obtained
the renormalized matrix elements, i.e., coordinate space qPDF.
For the unpolarized case,
\begin{equation}
h_R(z,P_z,\mu_R,p_z^R)=Z_q Z_{\rm mp}(z,p_z^R,a^{-1},\mu_R) h(z,P_z,\gamma_t),
\end{equation}
and for longitudinally polarized case,
\begin{equation}
\Delta h_R(z,P_z,\mu_R,p_z^R)=Z_q Z_{\gamma_z \gamma_5}(z,p_z^R,a^{-1},\mu_R) h(z,P_z,\gamma_z \gamma_5).
\end{equation}
In the above equations, $Z_q$ is the quark wavefunction renormalization factor.

In Fig.~\ref{fig:unpolreME} we show the renormalized matrix elements, modulo the
factor $Z_q$, in the RI-MOM scheme at $p_z^R=0$, $\mu_R=4$ GeV. We find that the
errors are large. We can achieve substantial error reductions at $z\ne 0$, by
redefining the renormalized matrix elements as
\begin{equation}
h_R(z,P_z,\mu_R,p_z^R) \equiv \frac{h_R(z,P_z,\mu_R,p_z^R)}{h_R(z=0,P_z,\mu_R,p_z^R)},
\qquad \mathrm{and} \qquad
\Delta h_R(z,P_z,\mu_R,p_z^R) \equiv \frac{\Delta h_R(z,P_z,\mu_R,p_z^R)}{\Delta h_R(z=0,P_z,\mu_R,p_z^R)}.
\end{equation}
The errors of the matrix elements for $z \neq 0$ are reduced due to the strong
correlations between $z \neq 0$ (particularly for for small $z$ close to $z=0$) and $z=0$ matrix elements for each gauge
configurations. The effectiveness of this procedure in can be seen from
Figs.~\ref{fig:comp2lat_z} and \ref{fig:comp2lat_pol_z}.  As one can see the error
reduction due to this division is very significant. In fact, with this method, the
errors are reduced enough that the $z$-dependence of the matrix element is well
constrained also for $n_z=5$.  Since for the extraction of the qPDF we are only
interested in the $z$-dependence of the matrix element, and we know that the
unpolarized isovector nucleon matrix element at $z=0$ is the isospin of the nucleon,
which is unity (in our convention, c.f. Eq. ~\ref{eq:isovector} ) after renormalization, we can consider the above improved ratio of
renormalized matrix elements.   However, the effect of taking this ratio is not
trivial in the case of the matrix element of the helicity qPDF--- the
value of the renormalized matrix element at $z=0$ should be $g_A\approx1.3$; this
procedure is equivalent to studying a helicity PDF with the
first moment normalized to unity, i.e., in a normalization where  $g_A=1$.

\section{Unpolarized PDF: perturbative matching and comparisons with $\mathbf{h_R(z,P_z)}$}
\begin{figure}[t]
  \centering
  \includegraphics[scale=0.7]{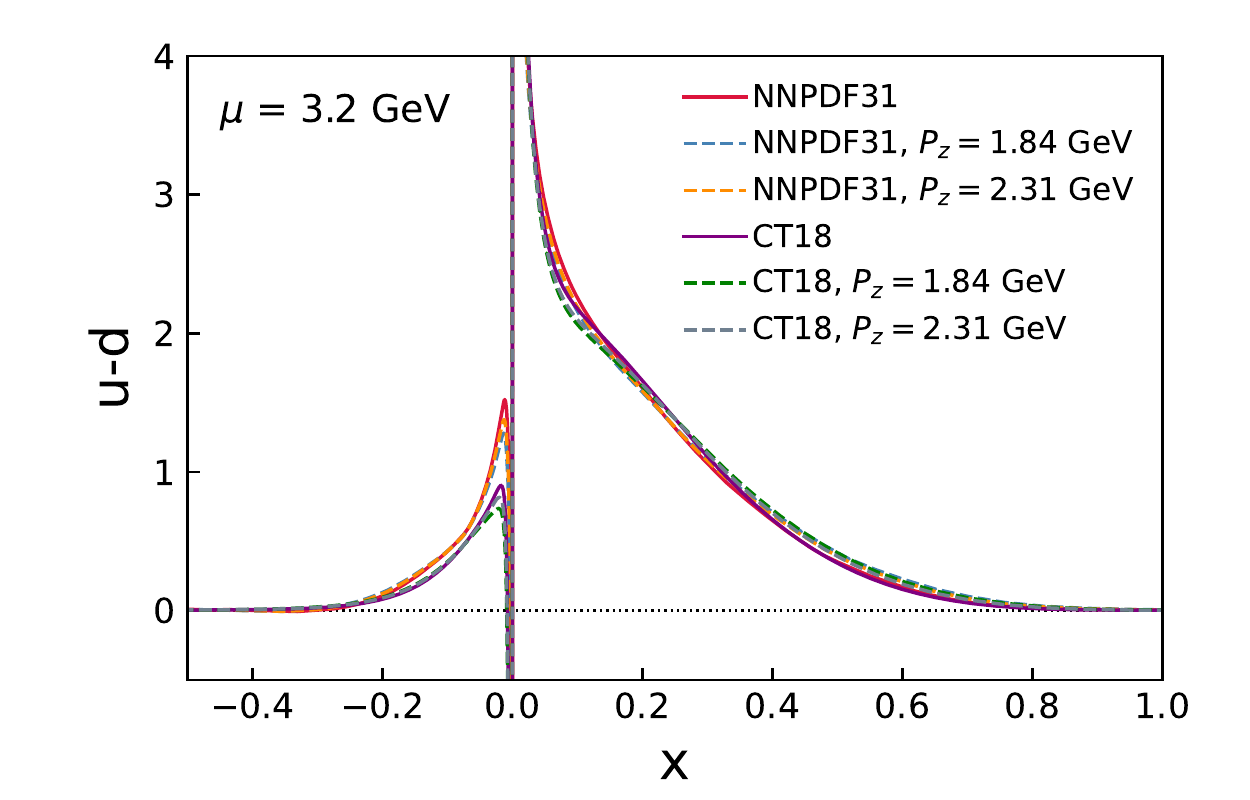}
  \caption{The NNLO isovector nucleon PDFs CT18~\cite{Hou:2019efy} and
  NNPDF3.1~\cite{Ball:2017nwa} (solid lines), and the corresponding target-mass
  corrected ones (dashed lines), at a scale $\mu=3.2$~GeV. See text for details.
  }
\label{fig:corrPDF}
\end{figure}

\begin{figure}[t]
  \centering
\includegraphics[scale=0.7]{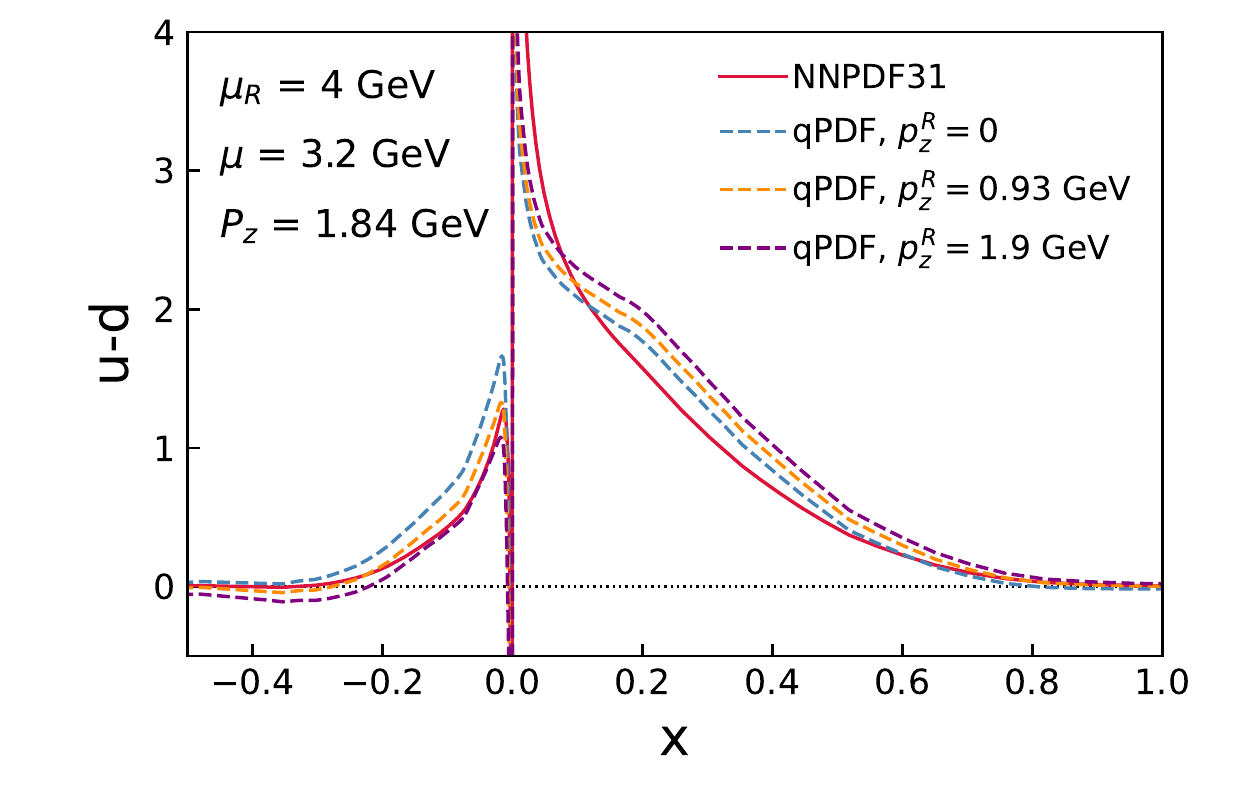}
\includegraphics[scale=0.7]{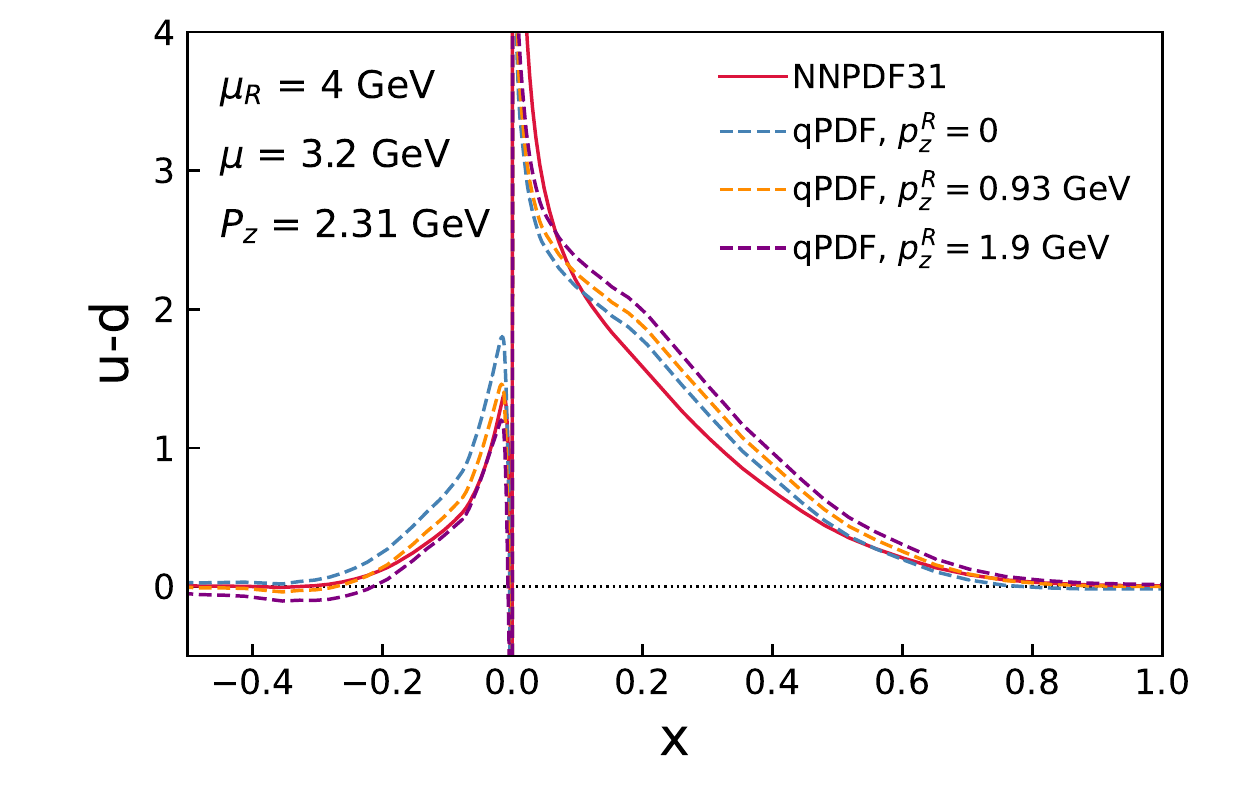}
\caption{qPDF corresponding to NNPDF3.1 for $P_z$=1.84~GeV (left) and $P_z$=2.3~GeV
(right) with $\alpha_s=0.25$ and 3 different RI-MOM renormalization condition.
}
\label{fig:match}
\end{figure}

\begin{figure}[t]
  \centering
\includegraphics[scale=0.7]{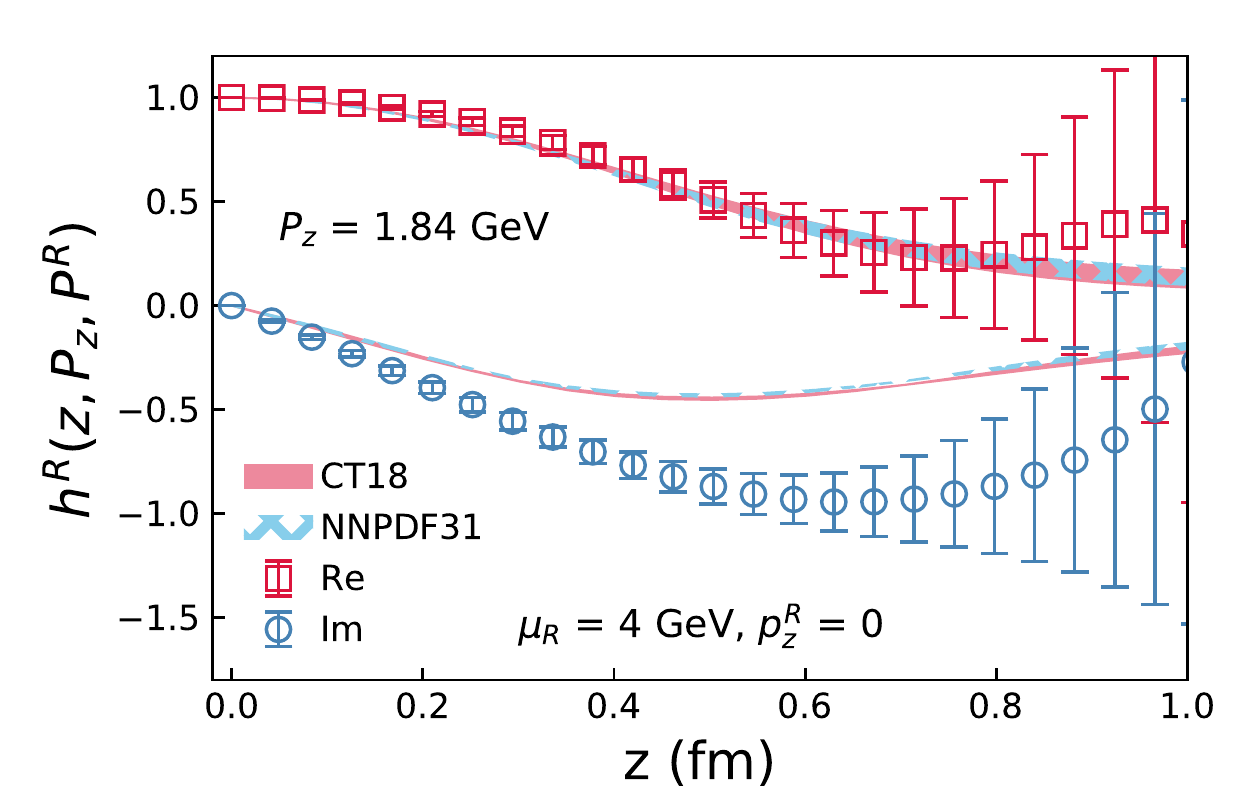}
\includegraphics[scale=0.7]{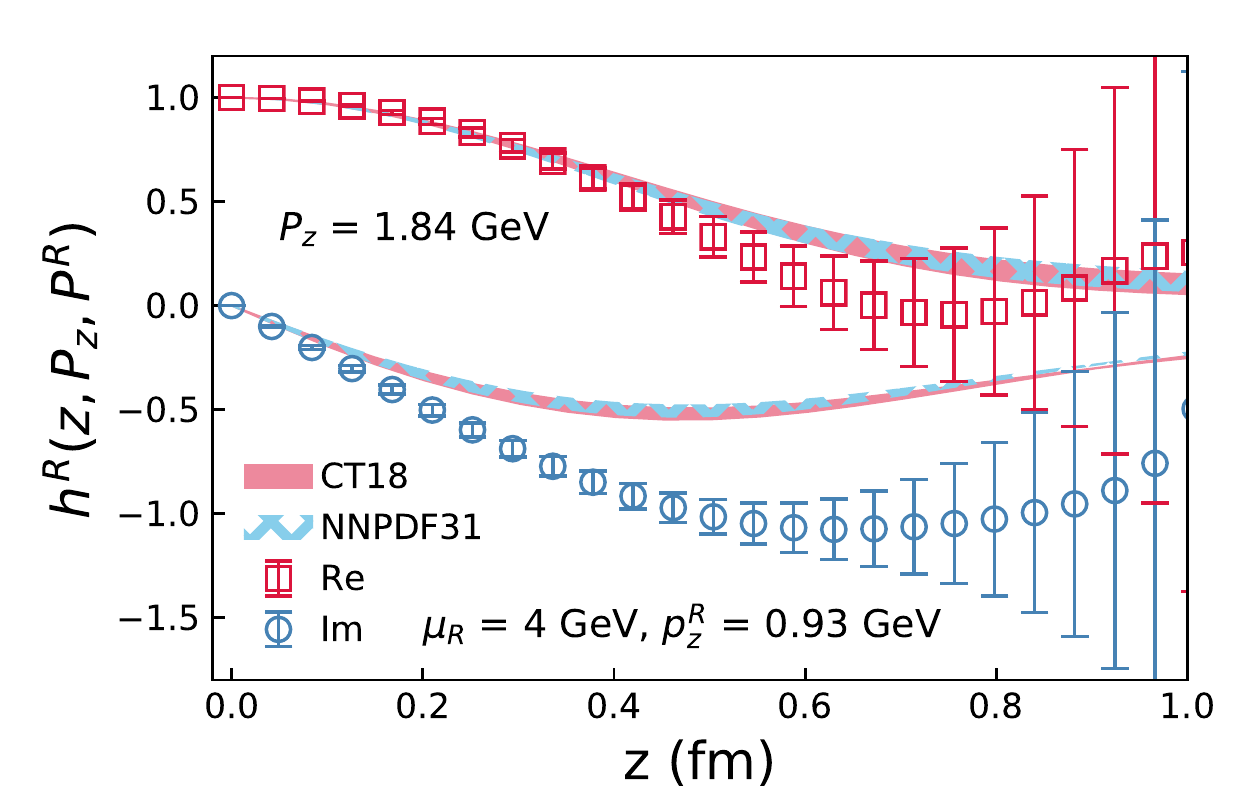}
\includegraphics[scale=0.7]{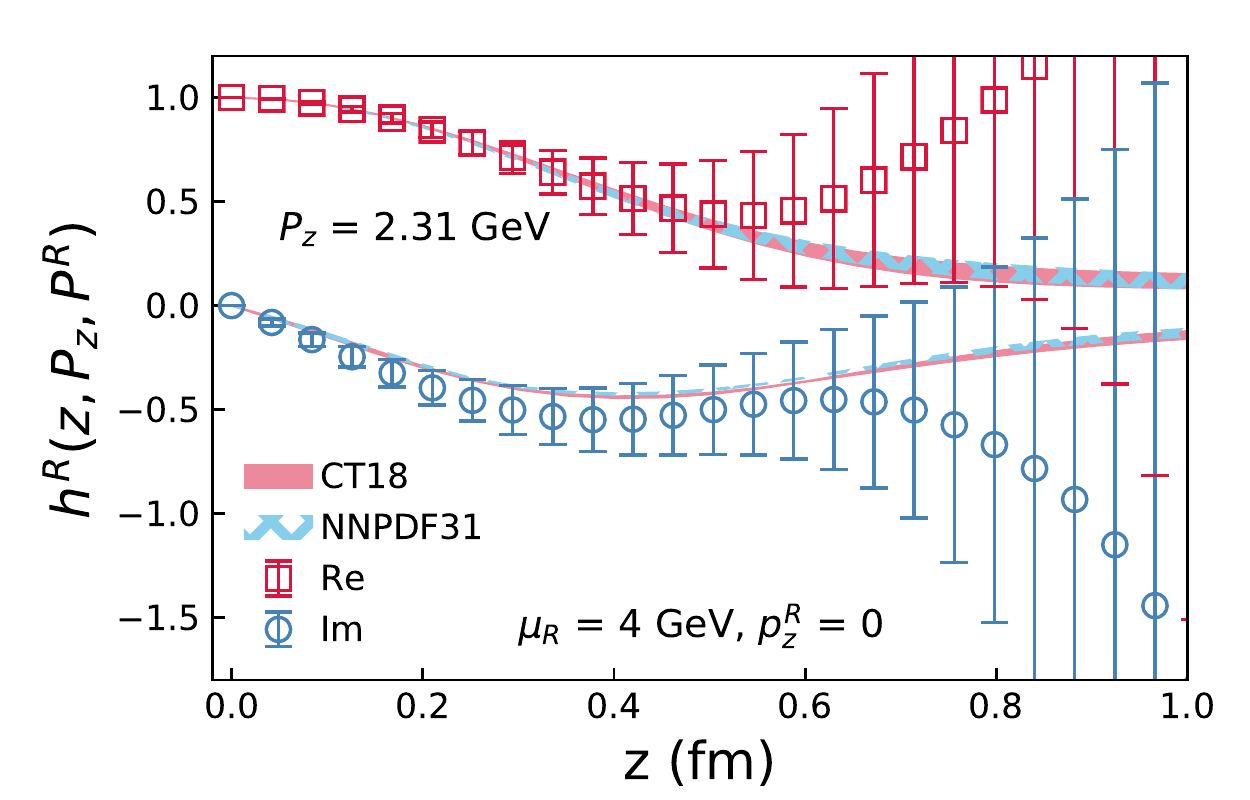}
\includegraphics[scale=0.7]{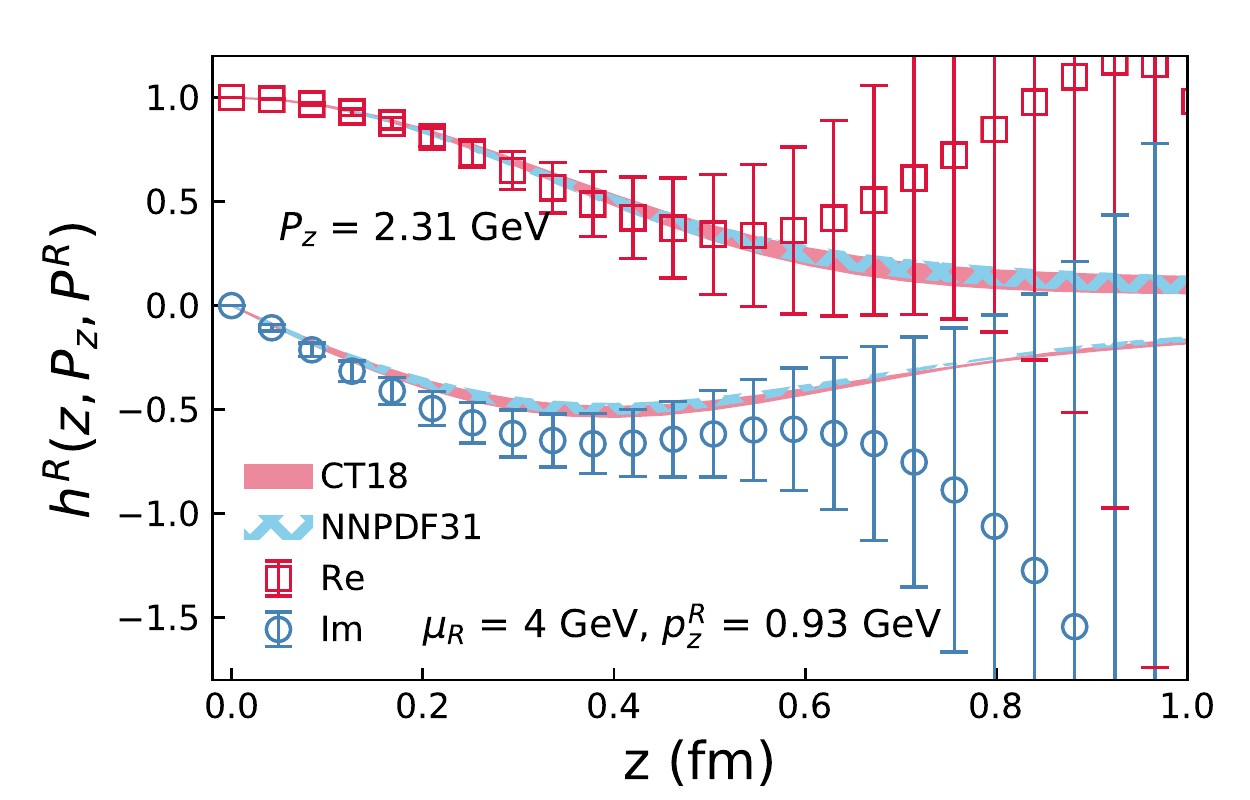}
\caption{Comparisons of the qPDF with the ones obtained from the global analysis for
two values of the RI-MOM renormalization scale, $p_z^R=0$~GeV (left row) and
$p_z^R=0.93$~GeV (right row), and for two values of the nucleon boost momenta,
$P_z=1.84$~GeV (upper column) and  $P_z=2.31$~GeV (lower column).
}
\label{fig:comp2lat_z}
\end{figure}

In this section, we will discuss how the renormalized coordinate-space qPDF,
$h_R(z,P_z,\mu_R,p_z^R)$, can be related and compared with phenomenological
unpolarized nucleon PDF, such as the CT18~\cite{Hou:2019efy} and
NNPDF3.1~\cite{Ball:2017nwa}, extracted from the global analysis of experimental
data. The unpolarized quark PDF in the valence region is well constrained through
global analysis. Therefore, it is natural to start from these phenomenological PDFs as
a function of Bjorken-$x$, use the perturbative matching to reconstruct the
corresponding coordinate-space qPDF as a function of $z$ for different $P_z$
values, and compare with our results for $h_R(z,P_z,\mu_R,p_z^R)$. The reason for
comparing in the $z$-space, rather than constructing the $x$-dependent PDF from our
$h_R(z,P_z,\mu_R,p_z^R)$ and then comparing with the phenomenological PDFs, is the
following: As can seen from Figs.~\ref{fig:comp2lat_z} and \ref{fig:comp2lat_pol_z},
$h_R(z,P_z,\mu_R,p_z^R)$ is quite noisy for $z\geqslant0.5$~fm. Thus, the Fourier
transformation which is needed to calculate the qPDF in $x$ -space is difficult to
perform.  Similar approach also had been used for pion PDF~\cite{Izubuchi:2019lyk}.

\begin{figure}[t]
  \centering
\includegraphics[scale=0.7]{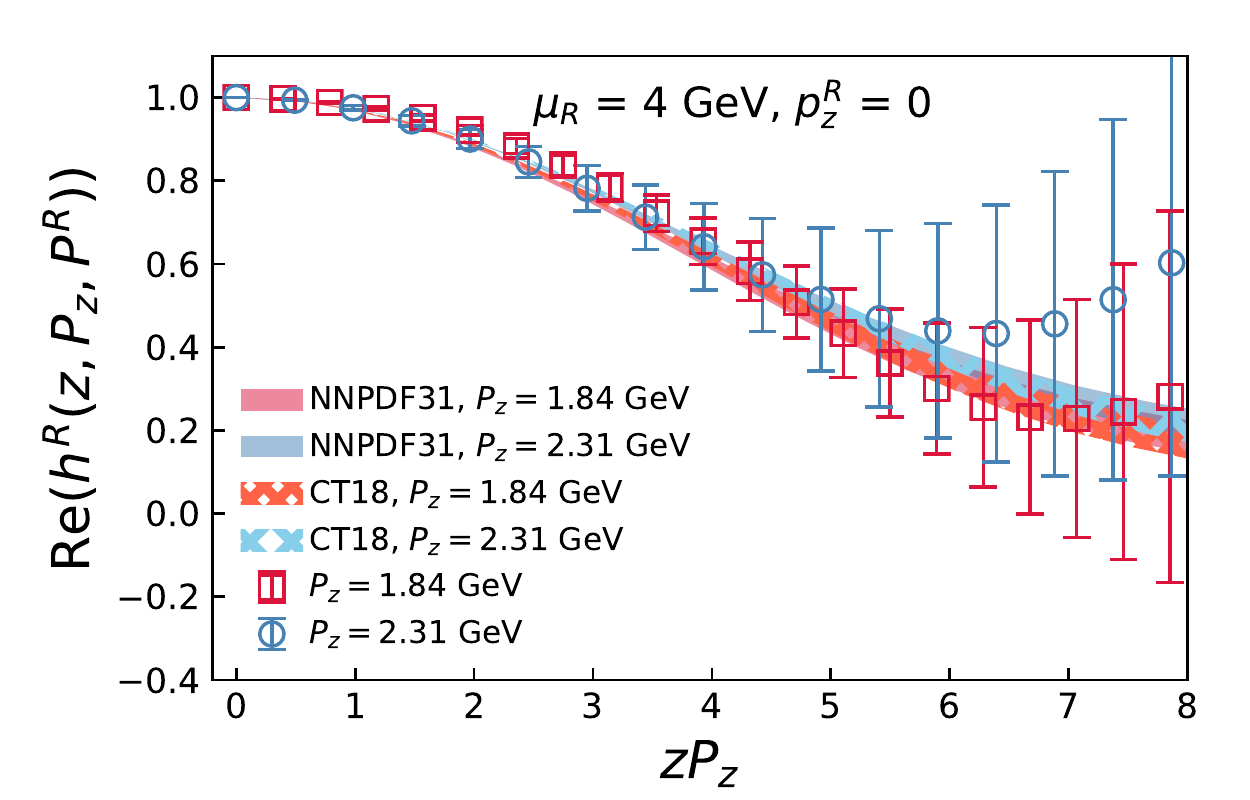}
\includegraphics[scale=0.7]{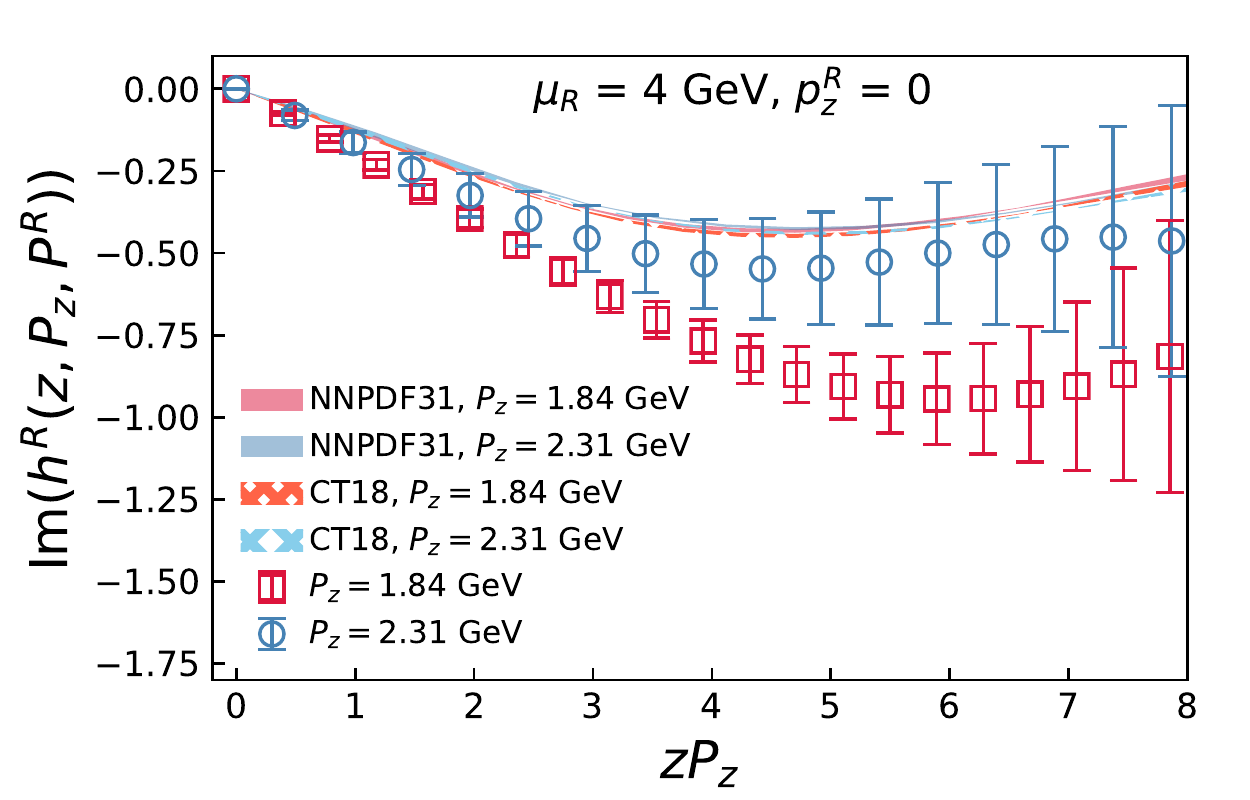}
\caption{Comparisons of the real (left panel) and imaginary (right panel) parts of
the ITDs, in the RI-MOM renormalization at the scales $p_z^R=0$ and $\mu_R=4$~GeV,
with that obtained from the CT18 and NNPDF3.1unpolarized isovector nucleon PDFs.
}
\label{fig:comp2lat}
\end{figure}

\begin{figure}[t]
  \centering
\includegraphics[scale=0.7]{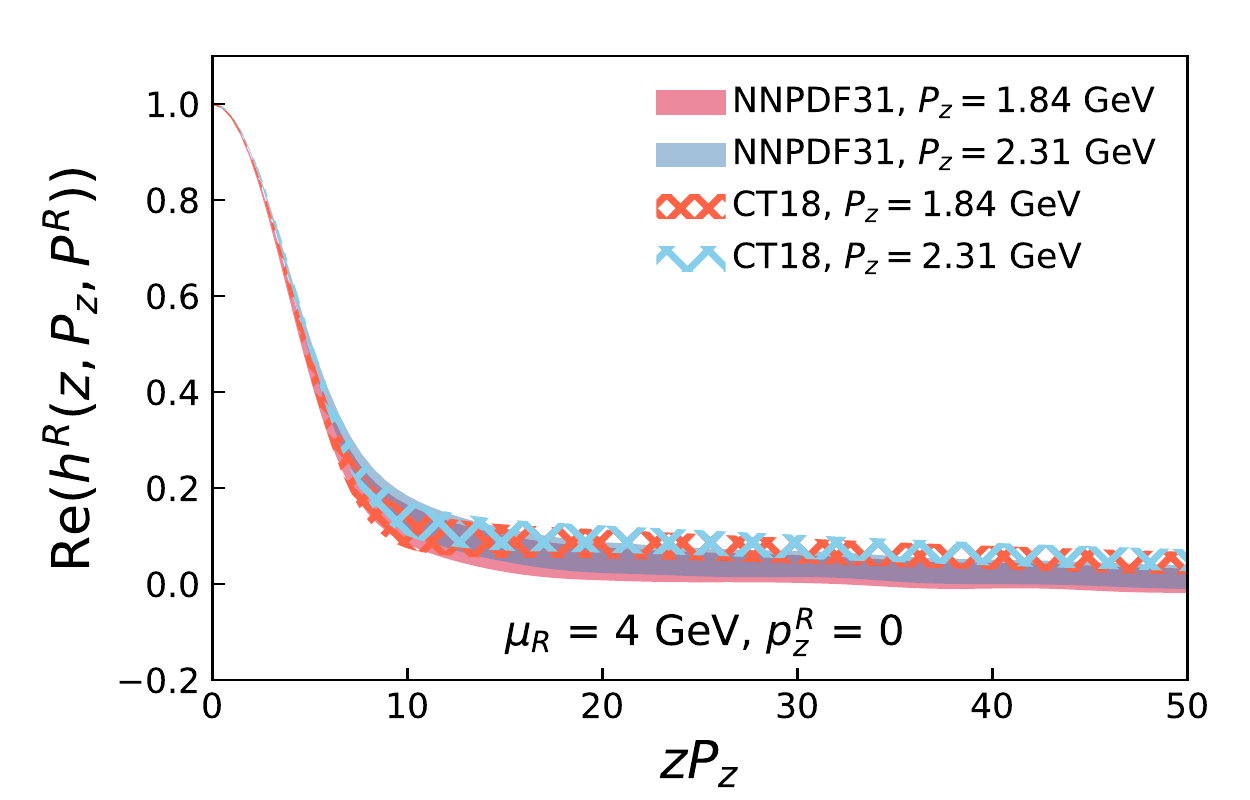}
\includegraphics[scale=0.7]{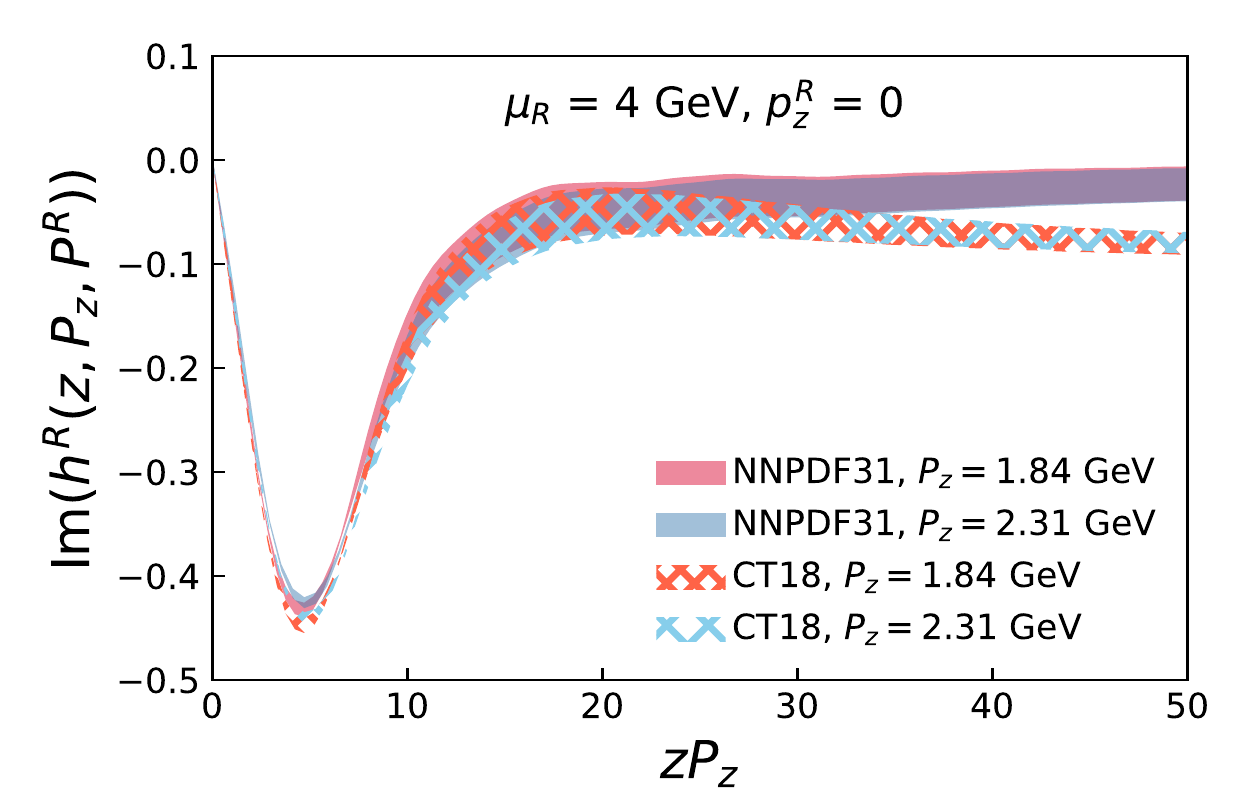}
\caption{
Real (left) and imaginary (right) parts of the ITDs corresponding to target-mass corrected  CT18 and NNPDF3.1 PDFs,
for $p_z^R=0$,  in an extended range of the Ioffe-time.
}
\label{fig:qPDFextnu}
\end{figure}

Even at the leading $\alpha_s^0$ order the  qPDF and the PDF differ due to the trace
term in the small $z$-expansion~\cite{Ji:2013dva,Izubuchi:2018srq}. This difference
was explicitly calculated in Ref.~\cite{Chen:2016utp}. In the context of DIS, such
corrections have been studied long ago~\cite{Nachtmann:1973mr}, and are known as
target-mass corrections. Following Ref.~\cite{Chen:2016utp}, we introduce the
target-mass corrected PDF
\begin{equation}
q^\prime(x, P_z) = \frac{1}{\sqrt{1+4c}} \left[ \frac{f_+}{2}
q\left(\frac{2x}{f_+}\right) - \frac{f_-}{2} q\left(\frac{-2x}{f_-}\right) \right],
\label{targetmass}
\end{equation}
where $c=M^2/(4P_z)^2$, $f_\pm=\sqrt{1+4c}\pm 1$, $q(x)$ is the usual PDF that
corresponds to $P_z \rightarrow \infty$.  In our analysis we use two sets of NNLO PDF
for the u and d quark and anti-quark distributions, the CT18~\cite{Hou:2019efy}, and
NNPDF3.1~\cite{Ball:2017nwa}, evaluated at scale $\mu=3.2$ GeV. If the matching was
known to all orders of perturbation theory, the prediction for real space qPDF should
have been independent of the value of $\mu$ at which the PDF was evaluated. Since
the matching only known to 1-loop order, we chose a scale $\mu=3.2$~GeV that is of
the same order of the other momentum scales used in our computations and, thereby,
avoided corrections due to large logarithms.  The lightcone quark PDF for $u$ quark
is calculated as $q_u(x)=u(x)$, $x>0$ and $q_u(x)=-\bar u(-x)$, $x<0$.  The isovector
nucleon PDF, $q_u^\prime(x)-q_d^\prime(x)$ is shown in Fig.~\ref{fig:corrPDF}.

In Fig.~\ref{fig:corrPDF}, we also show the target-mass corrected isovector nucleon
PDF for the two momenta used in our study, namely 1.84~GeV and 2.31~GeV. We see from
the figure that target-mass correction is small for the values of $P_z$ used in this
study. Using the target-mass corrected NNPDF3.1 isovector nucleon PDF obtained from
Eq.~(\ref{targetmass}) and the 1-loop matching to RI-MOM  we obtained the
corresponding qPDF for $P_z=1.84$~GeV and $P_z=2.31$~GeV, $\mu_R=4$ GeV, and
$p_z^R=0, 0.93, 1.9$ GeV. The functions $f_{1,\gamma_t}$ and
$f_{2,\gamma_t,\mathrm{mp}}$ in Eq.~(\ref{eq:C_f12}) for the 1-loop matching to
RI-MOM scheme with minimal projection were taken from Eq.~(28) Eq.~(31) of
Ref.~\cite{Liu:2018uuj}. Fig.~\ref{fig:match} shows comparisons of the NNPDF3.1 with
the corresponding qPDFs. In these comparisons $\alpha_s$ was evaluated at scale
$\mu=3.2$~GeV, which resulted in a value $\alpha_s=0.25$. We see significant
differences between the PDF and qPDF.  For large positive $x$, the qPDF is larger
than PDF, while for negative $x$ the qPDF can turn negative for some $P_z^R$.  The qPDF strongly depends
on the choice of the RI-MOM scales. It is possible to choose the RI-MOM scale such
that the qPDF is negative for $x<-0.2$, even though the PDF is positive.

By Fourier transforming the CT18 and NNPDF3.1 target-mass corrected qPDFs with
respect to $x$ we obtained the corresponding distributions as a function of the
so-called Ioffe-time, $zP_z$, i.e. the corresponding ITDs~\cite{Ioffe:1969kf}. Since
the matching is only up to 1-loop order, the scale entering $\alpha_s$ is not fixed.
We considered three choices of the scale for $\alpha_s$, namely $\mu/2, \mu, 2 \mu$.
The corresponding variations in the ITDs can be considered as estimates of the
perturbative uncertainties, and are shown as bands in Fig.~\ref{fig:comp2lat}. In the
same figure, also we compare with the lattice results for the ITDs in RI-MOM
renormalization, at the renormalization scales of $\mu_R=4$~GeV and $p_z^R=0$~GeV.
Albeit large errors, for both values of $P_z$ the real parts of the ITDs compare well
at least up to $zP_z\lesssim5$. However, lattice results for the imaginary parts of ITDs undershoot the
phenomenological ITDs even for $zP_z\gtrsim2$.

Albeit the significant difference between CT18 and NNPDF3.1 PDFs in the small-$x$
region, Fig.~\ref{fig:comp2lat} do not show any visible
difference in their corresponding ITDs. To understand this better,
Fig.~\ref{fig:qPDFextnu} we explore these ITDs in an extended range of Ioffe-time.
The difference between the PDFs in the negative-$x$ region is only reflected in
$<10\%$ difference in the imaginary part of the ITDs for $zP_z>25$, with essentially
showing no difference in the real part ITDs even up to $zP_z=50$.

To explore the dependence of the lattice results on the choice of RI-MOM scale
$p_z^R$ and the range of validity of the 1-loop matching, in
Fig.~\ref{fig:comp2lat_z} we show comparisons between the qPDFs as a function of $z$
obtained in the lattice calculations and from the global analysis of PDF for two
different choices of renormalization scale, namely $p_z^R=0, 0.93$~GeV. Very little
dependence on the $p_z^R$ was observed. While the real part of the qPDF obtained
from the global analysis agrees with the lattice results up to $z\sim1$~fm within
relative large errors, for the agreement is limited only for $z\lesssim0.2$~fm. For
$P_z=2.31$~GeV the agreements seem to extend to larger values of $z$, partly because
of larger errors. However, it is encouraging that the central value seems to shift
towards the global analysis results as $P_z$ is increased from $1.84$~GeV to
$2.31$~GeV. In any case, at large $z$, we see clear tension between the imaginary
part of the lattice qPDF lattice and the results of global analysis. This suggests
that the range of applicability of 1-loop matching is perhaps limited to
$z\lesssim0.2$~fm  in the case of the nucleon. It remains to be seen if this
agreement gets better with the addition of higher-loop corrections, or this observed
discrepancy arises because of contamination of higher-twist effects at larger $z$.
This observation has an important implication for our ability to described the
$x$-dependence of PDF within the LaMET framework. For example, if the 1-loop
perturbative matching works only for $z\simeq0.2$~fm,  reliable calculations of
nucleon PDF down to  $x\simeq0.1$ will need $P_z\gtrsim10$~GeV.

\section{Helicity PDF: perturbative matching and comparisons with $\mathbf{\Delta h_R(z,P_z)}$}
\begin{figure}
  \centering
  \includegraphics[scale=0.7]{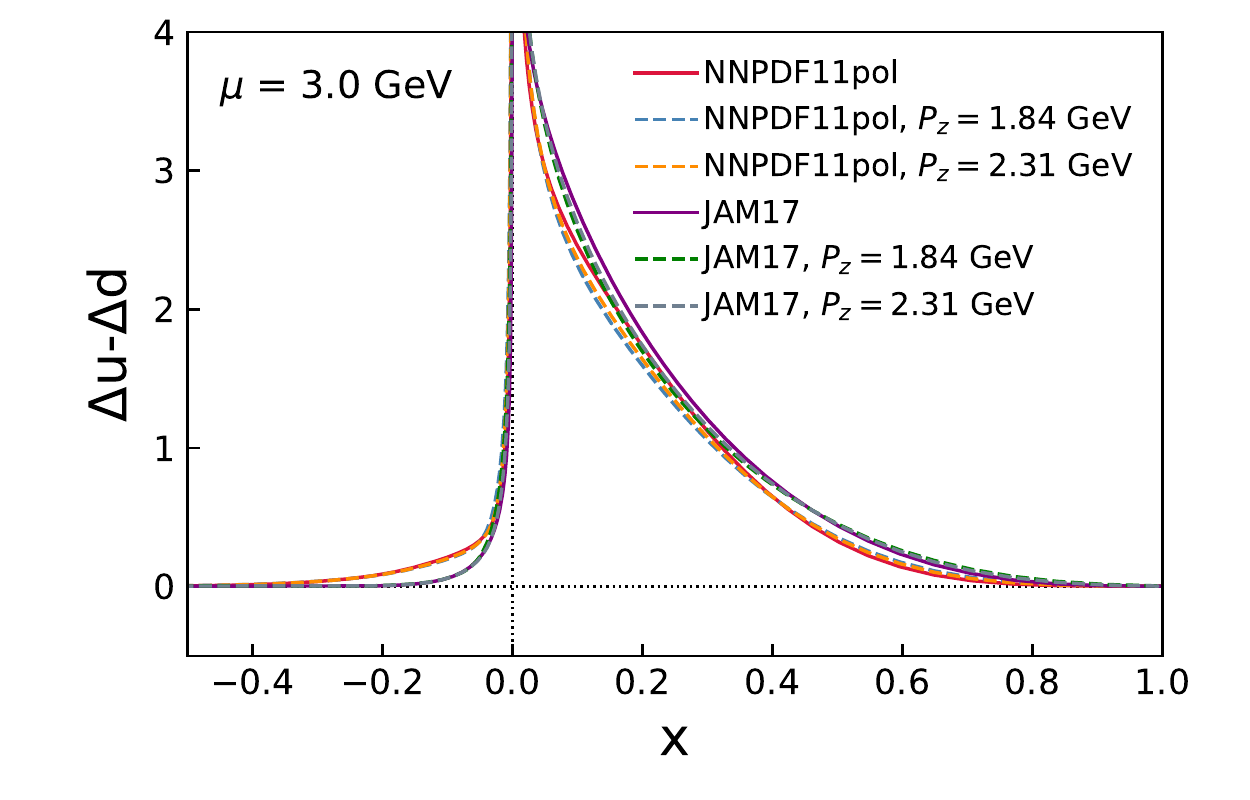}
\caption{NNPDFpol1.1 and JAM17 isovector helicity PDF at a scale $\mu=3$~GeV. Also,
shown are the corresponding target-mass corrected isovector helicity PDFs (dashed
lines) for $P_z=1.84$~GeV and $P_z=2.31$~GeV.
}
\label{fig:PDF_global_pol}
\end{figure}

\begin{figure}[t]
  \centering
\includegraphics[scale=0.7]{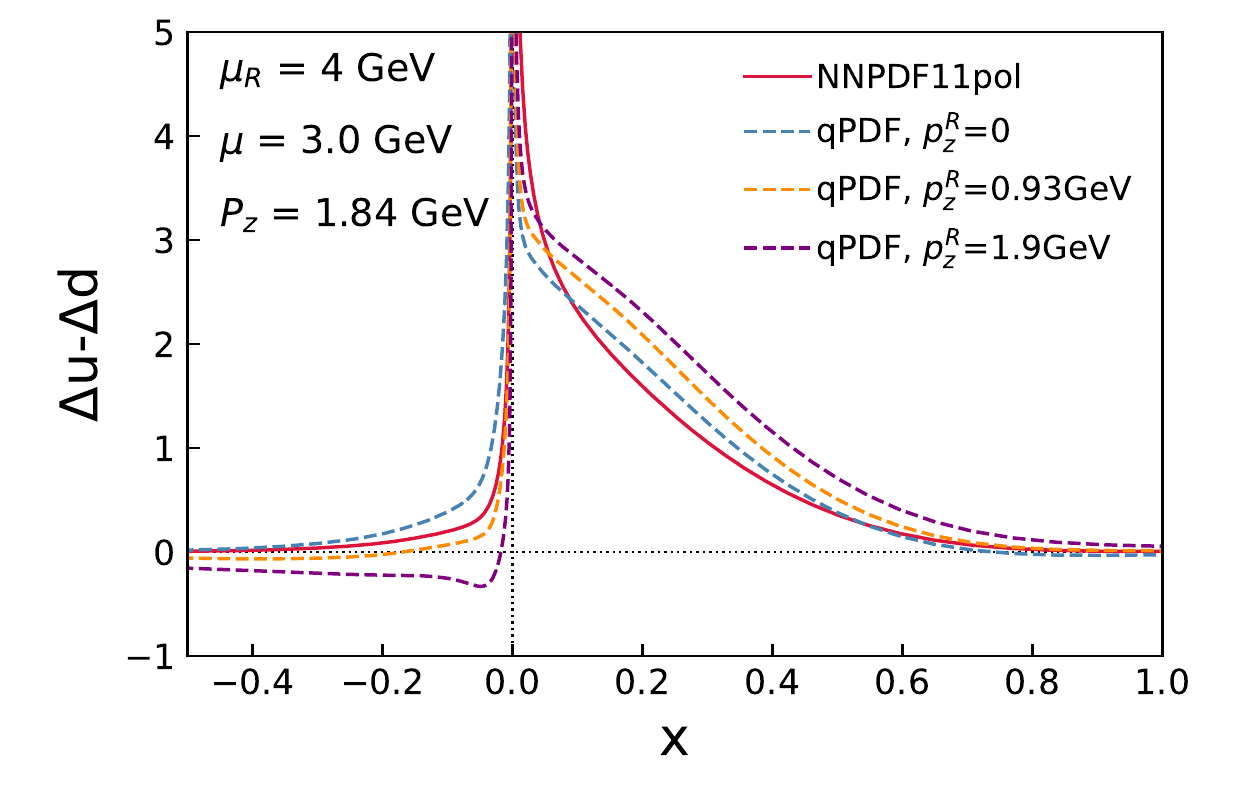}
\includegraphics[scale=0.7]{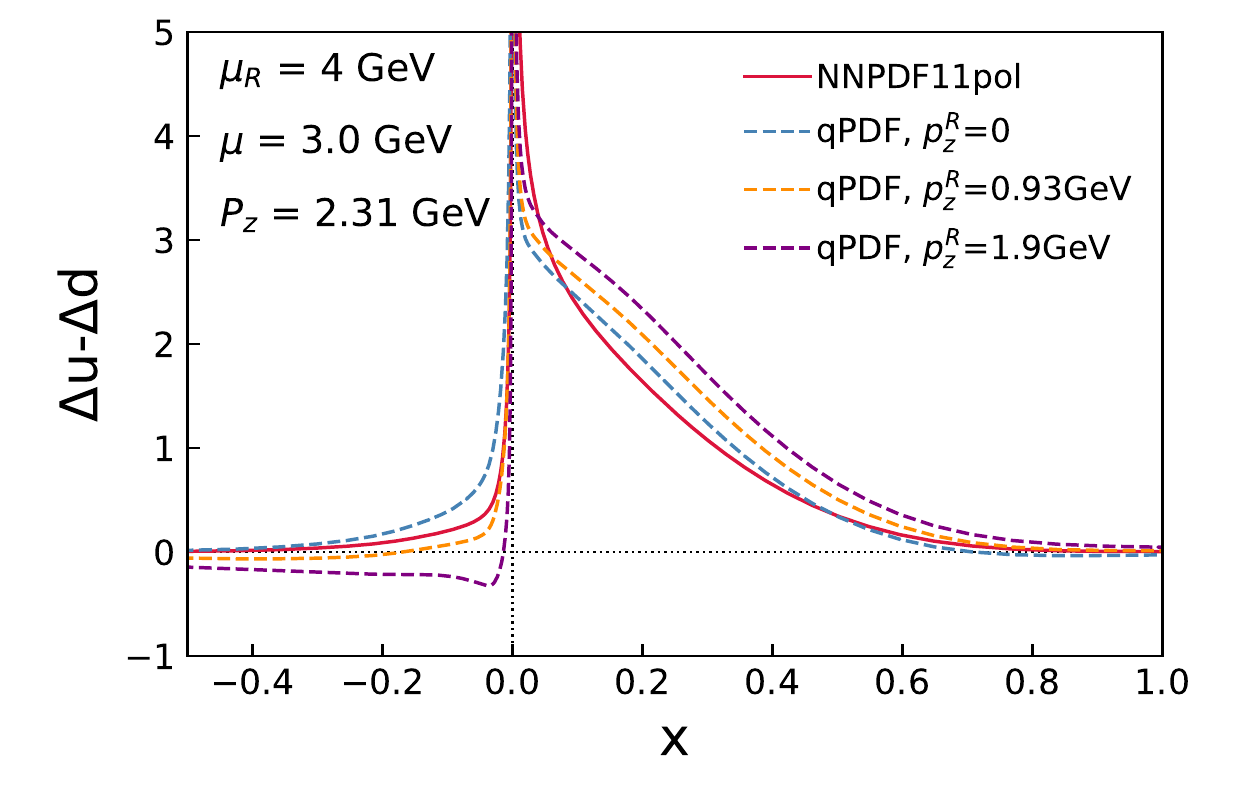}
\caption{qPDF corresponding to NNPDF1.1pol for $P_z$=1.84~GeV (left) and $P_z$=2.3~GeV
(right) with $\alpha_s=0.25$ and 3 different RI-MOM renormalization condition.
}
\label{fig:match_heli}
\end{figure}

Our analysis of helicity qPDF closely follows the analysis performed in the
unpolarized case, namely we start from the helicity PDF obtained in global analyses,
reconstruct the corresponding target-mass corrected qPDF, and then compare with the
lattice results. The helicity PDF have been extracted from the global analysis by
NNPDF collaboration using DIS, inclusive $W^{\pm}$ and jet production data from RHIC,
as well as the open charm data from COMPAS resulting in
NNPDFpol1.1~\cite{Nocera:2014gqa}. The JAM collaboration used the DIS and SIDIS data
in their global analysis, combined with $e^{+} e^{-}$ data to constrain the
fragmentation functions at NLO~\cite{Ethier:2017zbq}.  The resulting PDF
parameterization is called JAM17. In Fig.~\ref{fig:PDF_global_pol}, we show the
isovector helicity PDF $\Delta q_u-\Delta q_d$. The positive-$x$ region corresponds
to quark contribution, while the negative-$x$ region corresponds to anti-quark
region. The target-mass corrected helicity PDF, $\Delta q^\prime(x, P_z)$, was
obtained from helicity PDF, $\Delta q(x)$, following Ref.~\cite{Chen:2016utp}:
\begin{equation}
\Delta q^\prime(x, P_z) = \frac{1}{1+4c} \left[ \frac{f_+}{2}\Delta
q\left(\frac{2x}{f_+}\right) + \frac{f_-}{2}\Delta q\left(\frac{-2x}{f_-}\right)
\right] - \int^{x}_{\pm\infty} \frac{2c}{(1+4c)^{3/2}} \left[
\Delta q\left (\frac{2y}{f_+}\right ) + \Delta q\left(\frac{-2y}{f_-}\right ) \right],
\end{equation}
where $c=M^2/4P_z^2$, $f_\pm=\sqrt{1+4c}\pm 1$, and for the integration limits
$+\infty$ (-$\infty$) correspond to $x>0$ ($x<0$).

\begin{figure}
  \centering
 \includegraphics[scale=0.7]{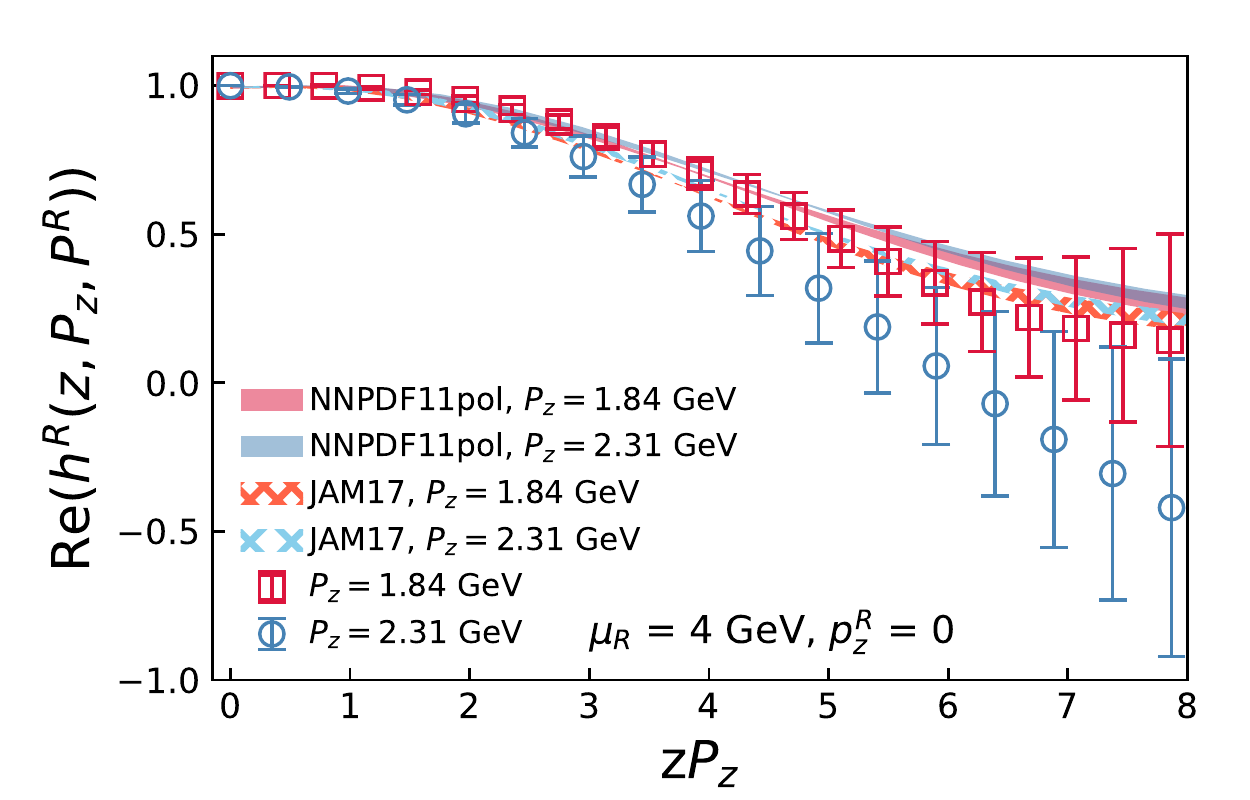}
 \includegraphics[scale=0.7]{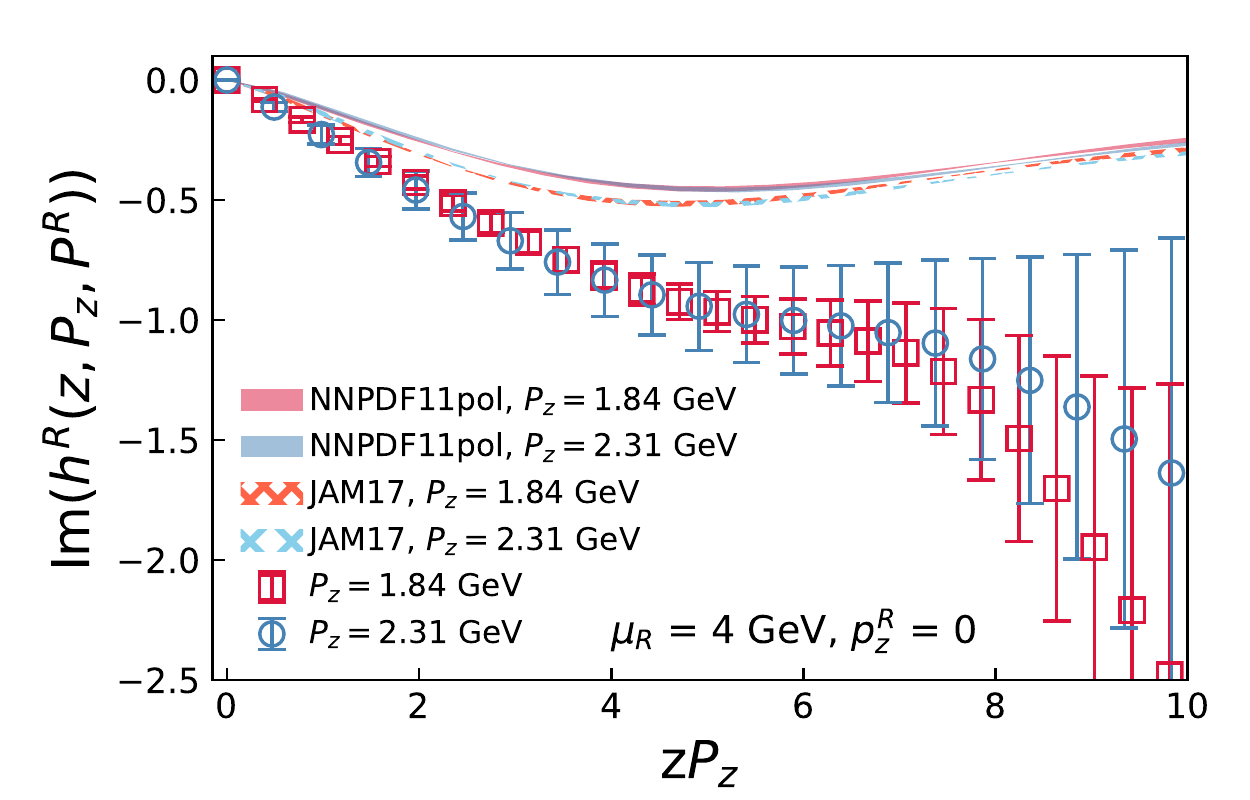}
\caption{Comparisons of the real (left panel) and imaginary (right panel) parts of
the isovector helicity ITDs, in the RI-MOM renormalization at the scales $p_z^R=0$
and $\mu_R=4$~GeV, with that obtained from the NNPDF1.1pol and JAM17.
}
\label{fig:comp2lat_pol}
\end{figure}

Although the matching for helicity qPDF has not been explicitly presented in the
literature before, it was straightforwardly deduced from the results presented in
Ref.~\cite{Liu:2018uuj}. The key observation here was the fact that, owing to the
chiral symmetry,  for a mass-less quarks in 1-loop perturbation theory
$\mathrm{Tr}[\gamma_5\gamma_z\Lambda(p,z,\gamma_z\gamma_5)] =
\mathrm{Tr}[\gamma_z\Lambda(p,z,\gamma_z)]$. Thus, the 1-loop matching of the
helicity qPDF in the RI-MOM scheme with minimal projection is same as that for the
unpolarized qPDF with $\Gamma=\gamma_z$ (instead of the $\Gamma=\gamma_t$ used
before), and with the RI-MOM renormalization condition corresponding to the
projection operator $\mathcal{P}=\gamma_z$ (instead of the minimal projection). The
1-loop matching for the  $\Gamma=\gamma_z$ operator is known for two different RI-MOM
projections, the minimal projection and the $\slashed{p}$ projection, corresponding
to $\mathcal{P}=\gamma_z-(p_z/p_x)\gamma_x$ and $\mathcal{P}=\slashed{p}/(4p_z)$,
respectively~\cite{Liu:2018uuj}. The function that depends on the RI-MOM projection
operator, i.e. $f_{2,\gamma_z,\gamma_z}$, entering the matching coefficient in
Eq.~\ref{eq:C_f12} was simply deduced from these known results. The Lorentz structure
of  $\Lambda(p,z,\gamma^{\alpha})$ for a general $\gamma_\alpha$, $\alpha=x,y,z,t$ is
given by
\begin{equation}
\Lambda^{(1)}(p,x,\gamma_{\alpha}) =  \gamma_{\alpha} \left[ \tilde f_t(x,\rho) \right]_{+}
+ \gamma_z \frac{p_{\alpha}}{p_z} \left[ \tilde f_z(x,\rho) \right]_{+}
+  \frac{\slashed{p} p_{\alpha}}{p^2} \left[ \tilde f_p(x,\rho) \right]_{+} \,
\end{equation}
and $f_{2,\gamma_z,\mathrm{mp}}=\tilde f_t+\tilde f_z$ and $f_{2,\gamma_z,
\slashed{p}}=\tilde f_t+\tilde f_z+\tilde f_p$~\cite{Liu:2018uuj}. Here, the
subscript `+' refers to the standard plus-prescription and $\rho=-p^2/p_z^2$. The
functions $\tilde f_t$, $\tilde f_z$ and $\tilde f_p$ have been calculated in
Ref.~\cite{Liu:2018uuj}, and we use the same notations here. Therefore, for the case
of $\mathcal{P}=\gamma_z$ the RI-MOM projection-dependent function is given by
\begin{equation}
f_{2,\gamma_z,\gamma_z} =
\tilde f_t + \tilde f_z +  (p_z^2/p^2) \tilde f_p =
f_{2,\gamma_z,\mathrm{mp}} + \left( f_{2,\gamma_z,\slashed{p}} - f_{2,mp} \right) / r .
\label{f2gammaz}
\end{equation}
Thus, for the helicity qPDF the 1-loop matching RI-MOM function in the minimal
projection scheme is the same as in Eq.~\ref{eq:C_f12}, but with
$f_{2,\gamma_z,\mathrm{mp}}$ given by Eq.~\ref{f2gammaz}, and $f_{1,\gamma_z}$,
$f_{2,\gamma_z,\mathrm{mp}}$ and $f_{2,\gamma_z,\slashed{p}}$ are given by
Eqs.~(A6-A8) of  Ref. ~\cite{Liu:2018uuj}.

\begin{figure}
  \centering
  \includegraphics[scale=0.7]{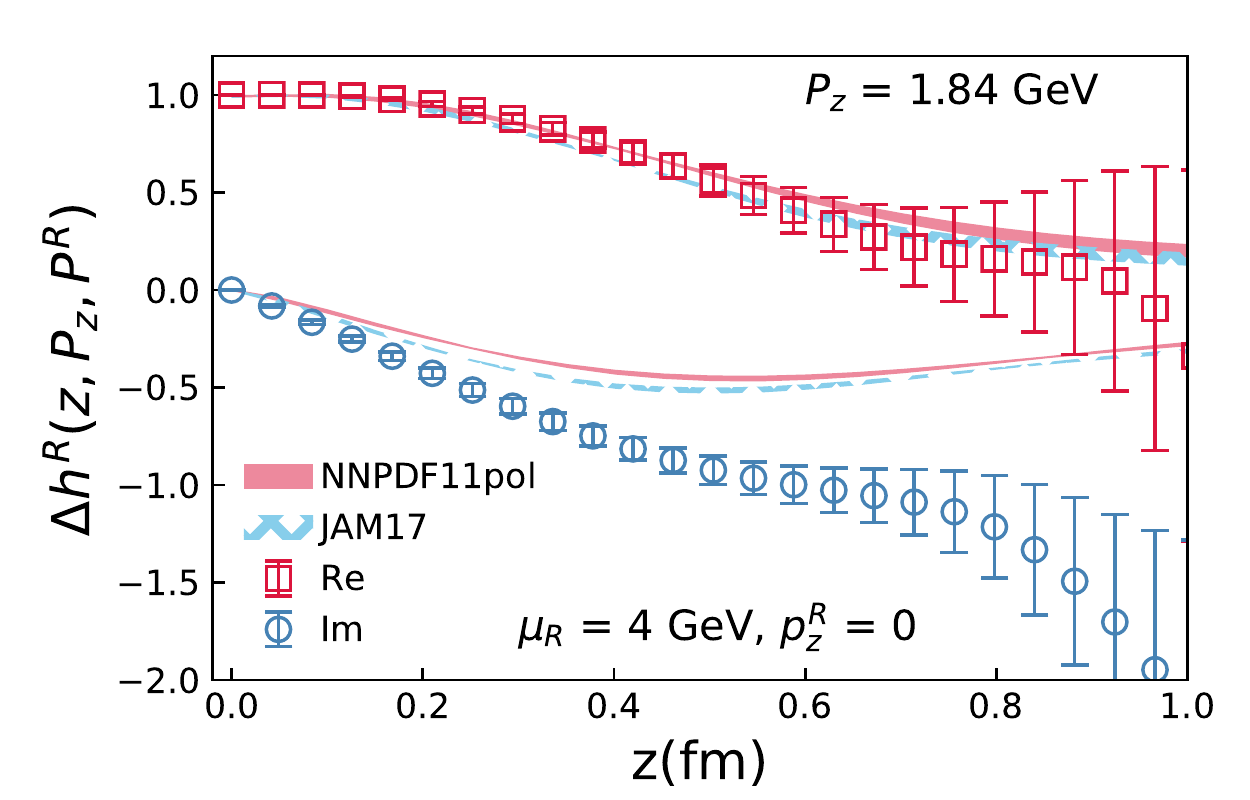}
  \includegraphics[scale=0.7]{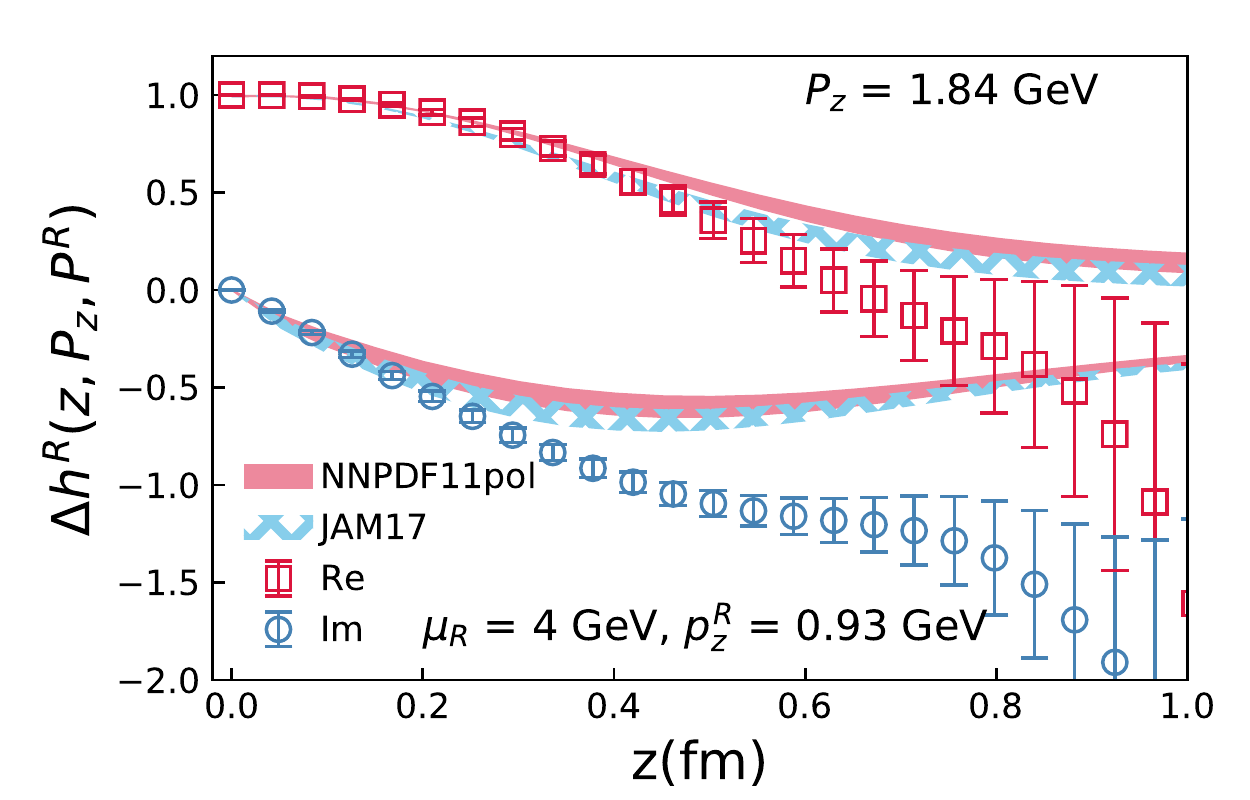}
  \includegraphics[scale=0.7]{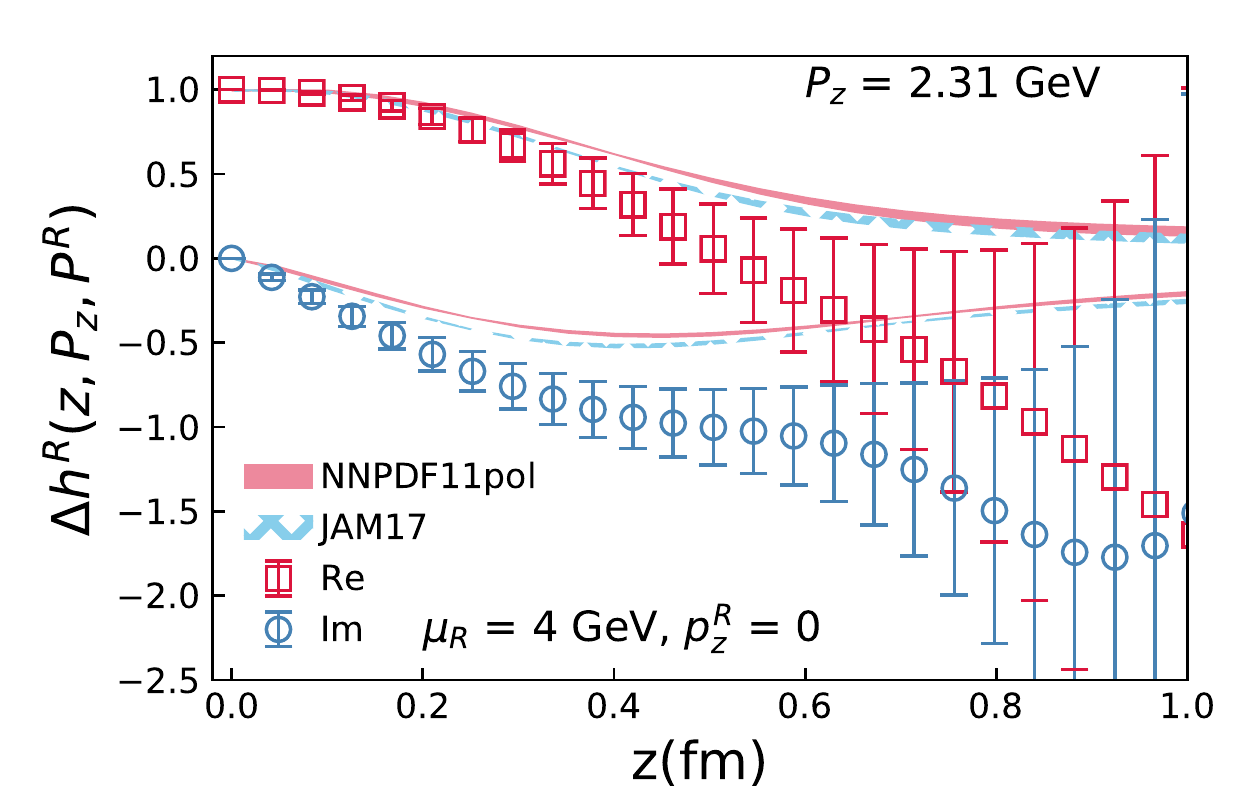}
  \includegraphics[scale=0.7]{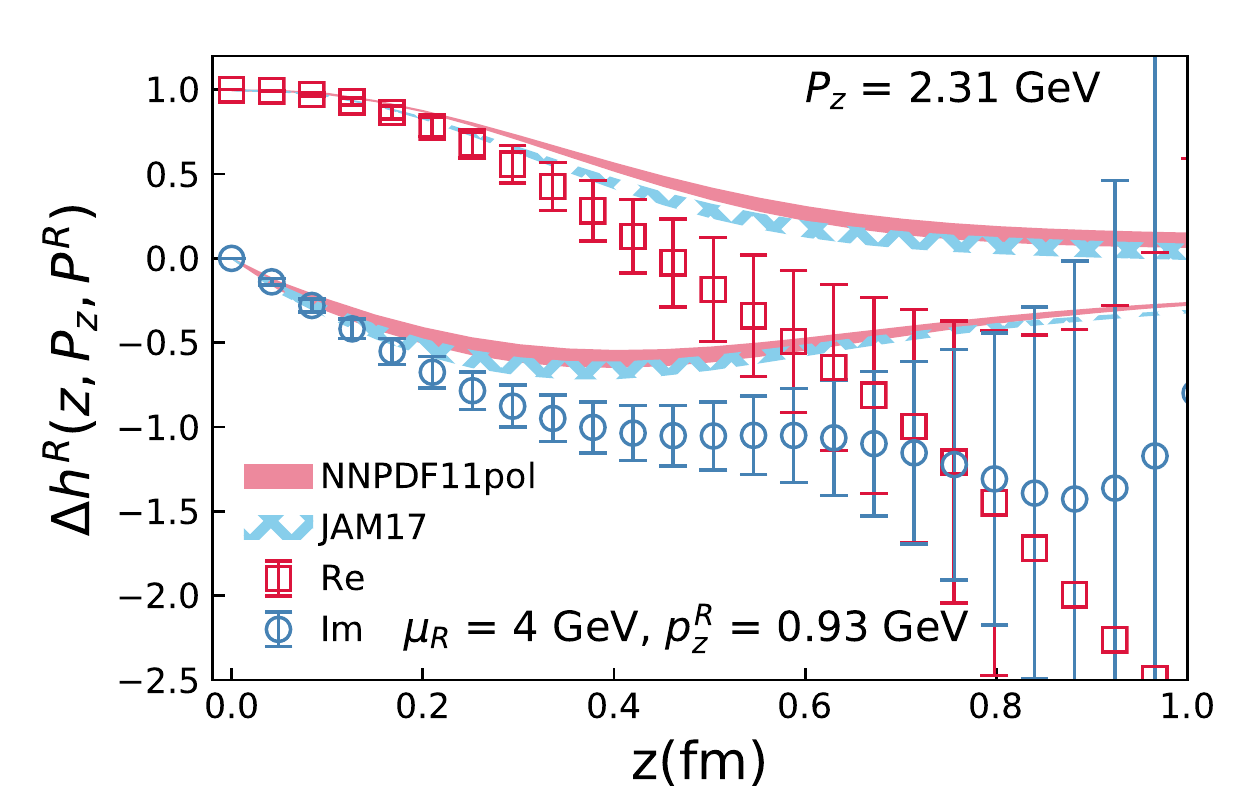}
  \caption{
  Comparisons of the isovector helicity qPDF with the ones obtained from the global
  analysis for two values of the RI-MOM renormalization scale, $p_z^R=0$~GeV (left
  row) and $p_z^R=0.93$~GeV (right row), and for two values of the nucleon boost
  momenta, $P_z=1.84$~GeV (upper column) and  $P_z=2.31$~GeV (lower column).
}
  \label{fig:comp2lat_pol_z}
\end{figure}

Using the matching discussed above, we can obtain the isovector helicity qPDF from
the target-mass corrected NNPDFpol1.1 and JAM17. As before, the 1-loop matching we
used $\alpha_s$ evaluated at scale $\mu=3.0$~GeV, and the scale was varied between $\mu/2$ to $2\mu$ to
estimate the scale uncertainty. We found noticeable difference between the
isovector helicity PDFs and the corresponding qPDFs in Fig. \ref{fig:match_heli}. By Fourier transforming the qPDFs we obtained
the isovector helicity ITDs and compared it with our lattice results in Fig.
\ref{fig:comp2lat_pol}. Since we normalized our lattice results by the value of
matrix element at $z=0$, we normalized the phenomenological ITDs by dividing with
$g_A=1.25$. Within the large statistical errors, we did not find a significant $P_z$
dependence of the lattice results. While the real parts of the lattice results agree
with that obtained from the phenomenological PDFs up to $zP_z\lesssim3$,  the
imaginary parts do not agree quantitatively but also have larger errors.
We also explored the dependence of our result on the choice of RI-MOM scales.  In
Fig. \ref{fig:comp2lat_pol_z}, we compare the qPDFs for $\mu_R=4$~GeV, and
$p_z^R=0$~GeV and $p_z^R=0.93$~GeV. From the figure, we see that the comparison
between the results of lattice calculation, as well as the global analyses are not
sensitive to the choice of the renormalization scales. For both values of $P_z$, the
agreement between the lattice and the global analyses extends to values of $|z|$ of
about $0.3$~fm for the real parts, but not for the imaginary parts. In the  next section we
will discuss how these disagreements show up in the moments of the PDFs.

\section{Moments of PDF from ratio of Ioffe-time distributions}\label{secratio}
In the previous sections, we analyzed the boosted nucleon matrix
matrix elements renormalized in the RI-MOM scheme and matched it
to the PDFs in the $\overline{\rm{MS}}$ scheme. Due to the multiplicative
renormalizability of $h(z,P_z,\gamma_t)$ and $h(z,P_z,\gamma_z\gamma_5)$,
we can form well-defined renormalized quantities by taking the
ratios of matrix elements at two different momenta $P_z$ and $P^\prime_z$ as
\begin{equation}
{\cal M}(z,P_z,P^\prime_z,\Gamma) = \frac{h(z,P_z,\Gamma)}{h(z,P^\prime_z,\Gamma)}
\frac{h(0,P^\prime_z,\Gamma)}{h(0,P_z,\Gamma)}.
\label{defratio}
\end{equation}
The second factor on the right hand side of the above definition normalizes the $z=0$
matrix element to unity, as we did in the case of the RI-MOM scheme. The choice
$P'_z=0$ in the ratio is usually referred to as the reduced Ioffe-time
distributions~\cite{Radyushkin:2017cyf}, and one should think of $P^\prime_z\ne 0$ as
a generalization of this choice. Here, we take $P_z=2.31$~GeV and $P'_z=1.84$~GeV,
respectively. Since both $P_z,P^\prime_z > \Lambda_\mathrm{QCD}$ and the nucleon
mass, we expect this ratio to be simply described by the leading twist
expression~\cite{Izubuchi:2018srq},
\begin{equation}
    {\cal M}(z,P_z,P^\prime_z,\Gamma)=\frac{\sum_{n=0}\frac{c_n(\mu z)}{c_0(\mu
    z)}\frac{(-i z P_z)^{n}}{n!}\xm{n}_{P_z}(\mu)} {\sum_{n=0}\frac{c_n(\mu z)}{c_0(\mu
    z)}\frac{(-i z P^\prime_z)^{n}}{n!}\xm{n}_{P'_z}(\mu)}.
\label{ope}
\end{equation}
Following Ref.~\cite{Chen:2016utp}, the target-mass corrected unpolarized PDF moments $\xm{n}$ can be obtained by relation:
\begin{equation}
\frac{\xm{n}_{P_z}}{\xm{n}_{~~~}}=\sum_{i=0}^{\lfloor (n+1)/2 \rfloor}C_{n-i+1}^ic^i
\end{equation}
and for the helicity case,
\begin{equation}
\frac{\xm{n}_{P_z}}{\xm{n}_{~~~}}=\sum_{i=0}^{\lfloor n/2 \rfloor}\left( \frac{n-i+1}{n+1} \right) C_{n-i}^ic^i
\end{equation}
where $C_n^i$ is the binomial function, $c=M^2/4P_z^2$. In Eq.~\ref{ope}, $c_n(\mu z)$ is the 1-loop order Wilson coefficients in the
$\overline{\mathrm{MS}}$ scheme. The Wilson coefficients describes the $z$ dependence
of the twist-2 local operator associated with the $n^\mathrm{th}$ moment of the PDF,
$\xm{n}(\mu)$, in the $\overline{\mathrm{MS}}$ scheme and at a factorization scale
$\mu$. As in our RI-MOM analysis, we will use $\mu=3.2$~GeV for the unpolarized case
and $\mu=3$~GeV for the
helicity case in the following analysis.

\begin{figure}[t!]
  \centering
\includegraphics[scale=0.7]{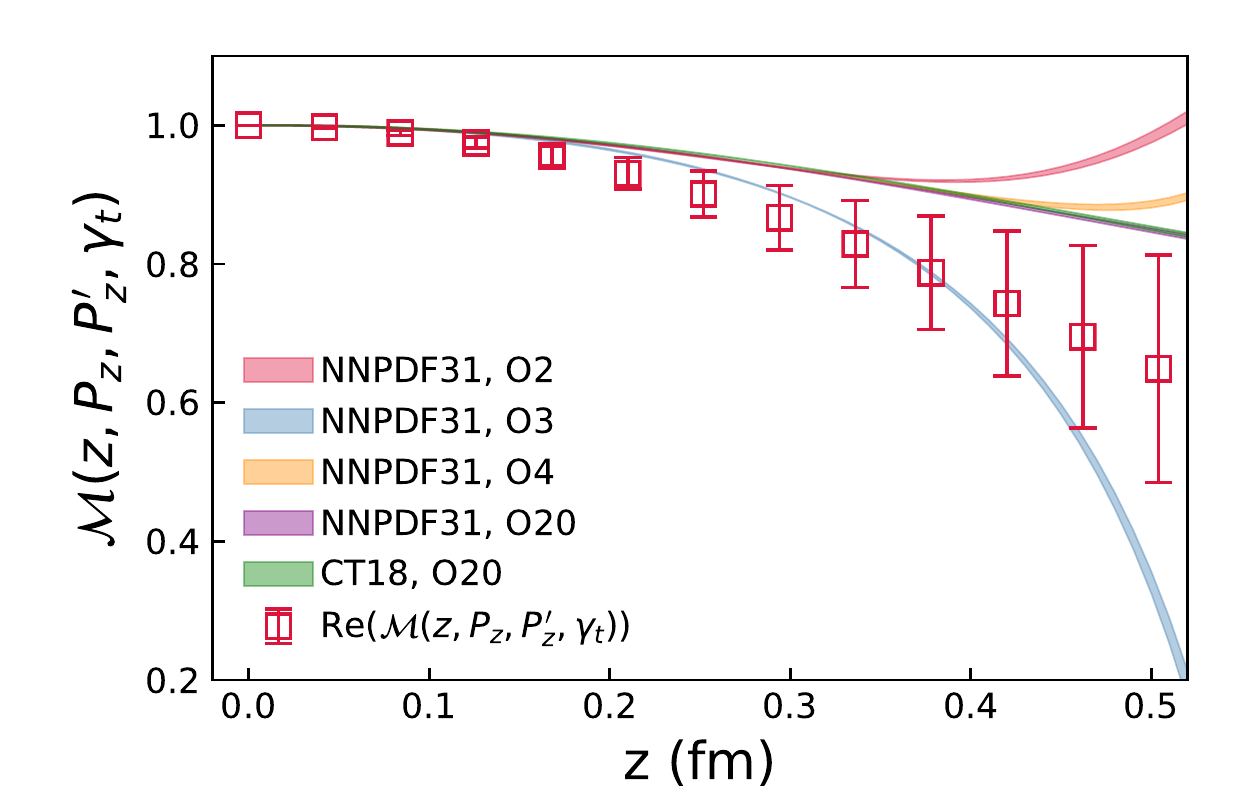}
\includegraphics[scale=0.7]{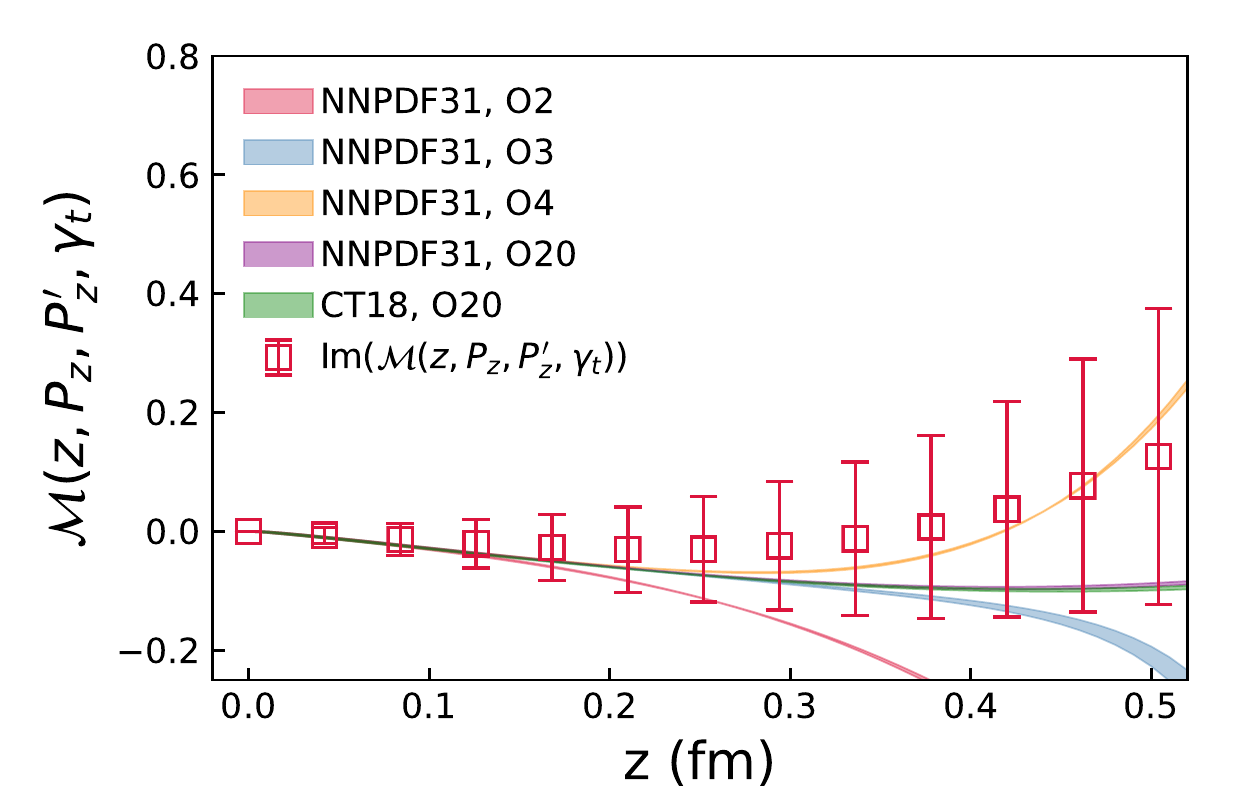}
\caption{
The real (left) and imaginary(right) parts of ${\cal M}(z,P_z,P'_z,\gamma_t)$ is
shown for $P_z=2.31$~GeV and $P^\prime_z=1.84$~GeV. The data points are from our
lattice calculations, whereas the various colored bands are the corresponding results
from the isovector  unpolarized PDFs from NNPDF3.1 and CT18. The band in these
phenomenological expectations arise due to variations of $\alpha_s(\mu)$ within the
scale scale $\mu/2$ to $2\mu$. For NNPDF3.1, we also show results by truncating the
expansion in Eq.~\ref{ope} at various orders, $n=n_\mathrm{max}$, in the PDF moments;
these results are denoted by `$\mathrm{O}n_\mathrm{max}$'.
}
\label{unpol_ratio_pheno}
\end{figure}

\begin{figure}[t!]
  \centering
\includegraphics[scale=0.7]{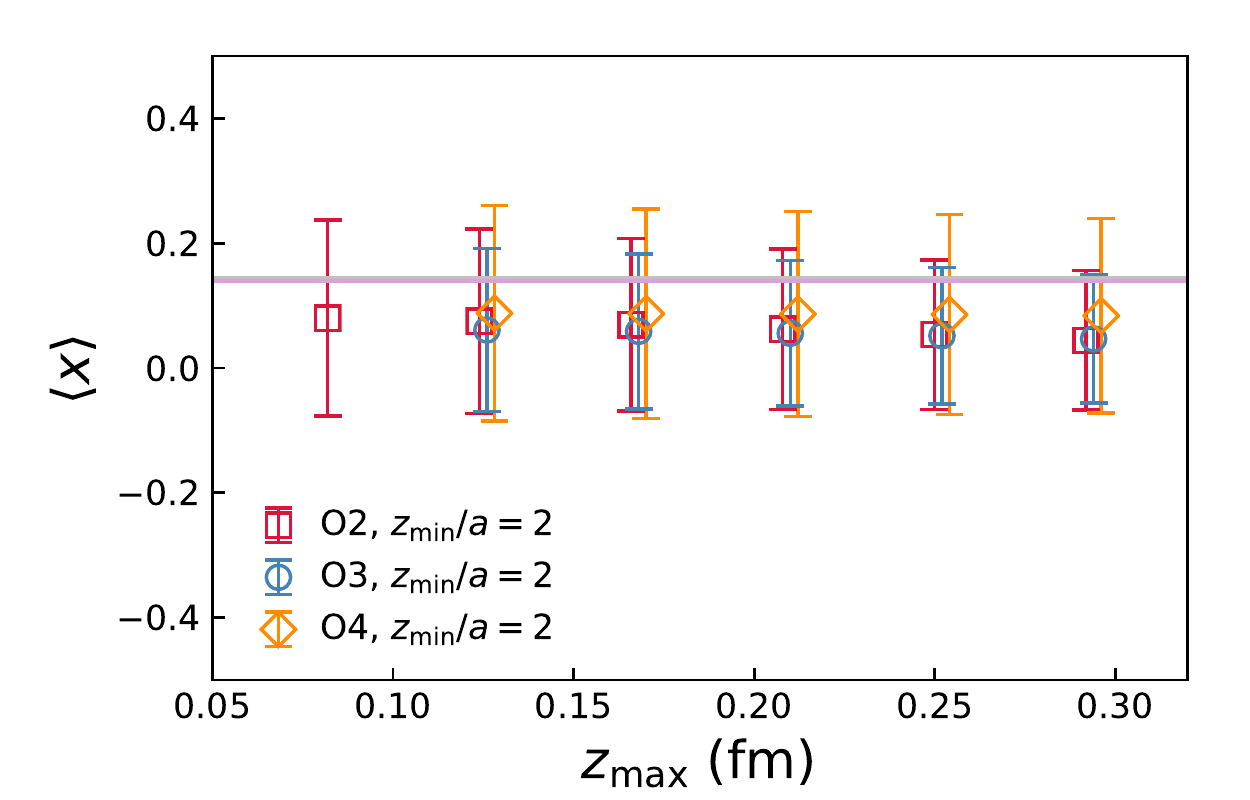}
\includegraphics[scale=0.7]{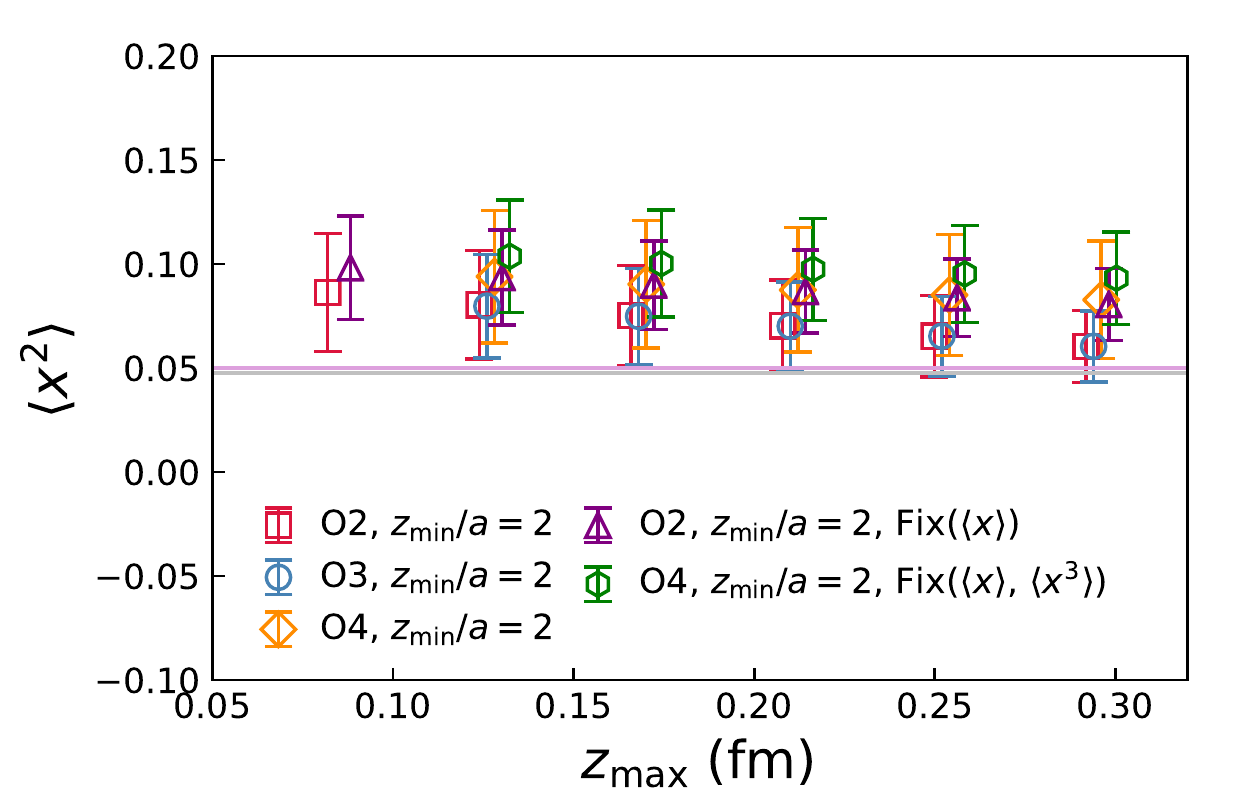}
\includegraphics[scale=0.7]{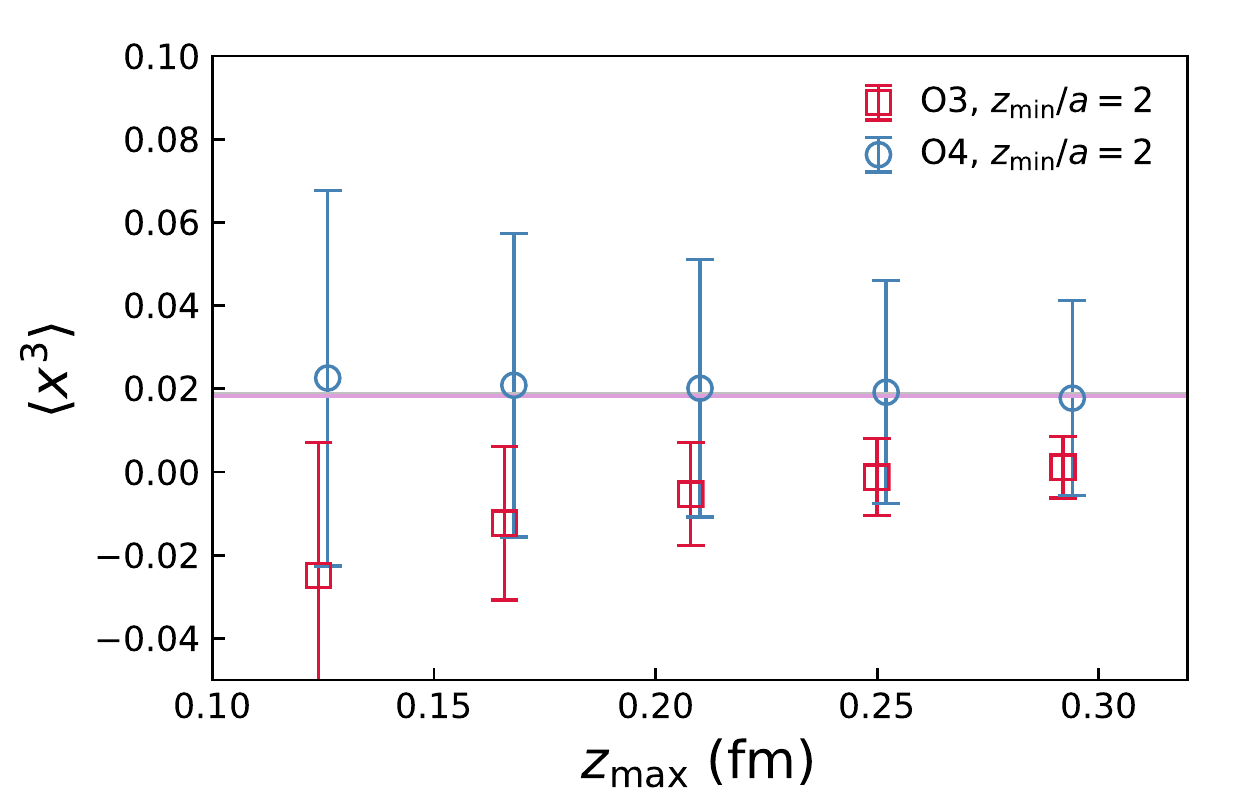}
\includegraphics[scale=0.7]{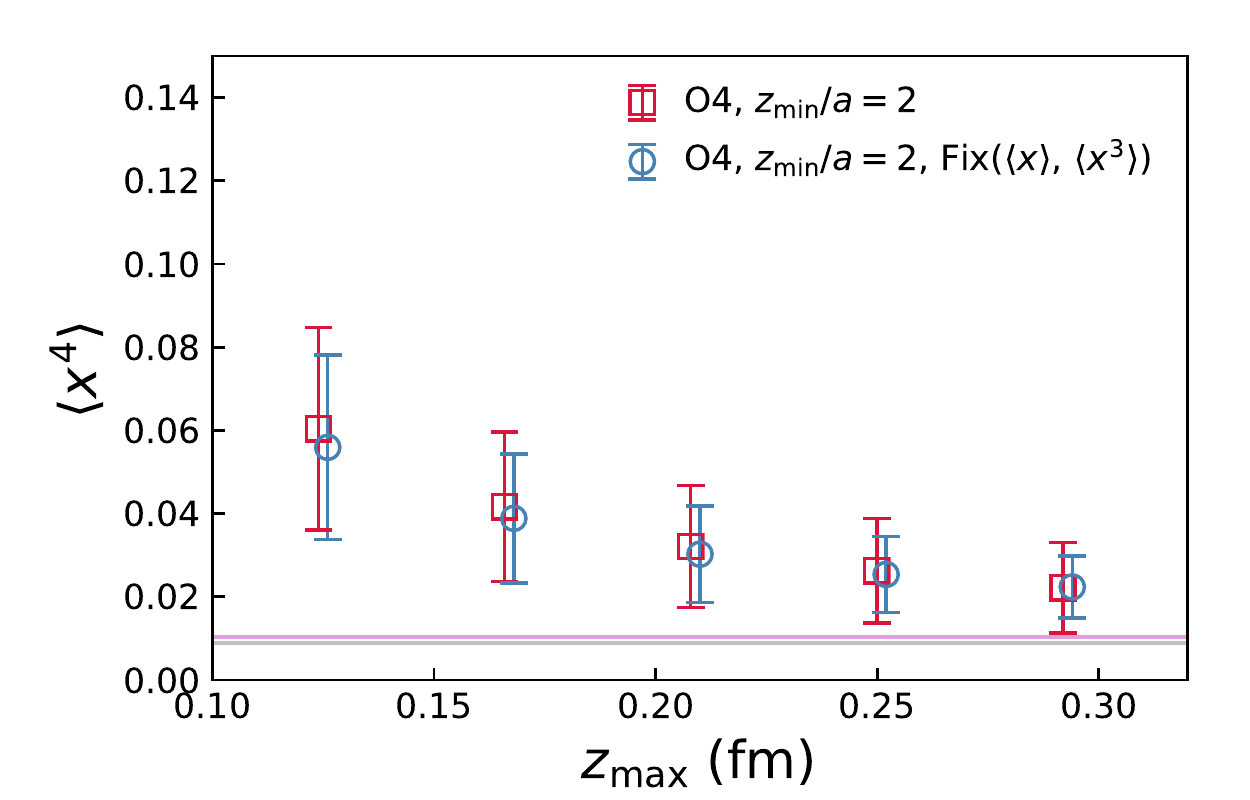}
\caption{
The moments of isovector unpolarized PDF, $\xm{1}$ (top-left), $\xm{2}$ (top-right),
$\xm{3}$ (bottom-left), and $\xm{4}$ (bottom-right),  that best describes the ratio
${\cal M}(z,P_z,P^\prime_z,\gamma_t)$ with $P_z=2.31$~GeV and $P'_z=1.84$~GeV. In
each of the panels, the moment $\xm{n}$ is shown as a function of $z_{\rm max}$ of
the fit using the functional form in Eq.~\ref{ope} over a range $[2a,z_{\rm max}]$ of
the data.  The results from fits using only moments up to $n=2$ as free parameters in
Eq.~\ref{ope} are labeled `O2', and those up to $n=4$ are labeled `O4'. The results
from fits that fix the moment $\xm{1}$, or $\xm{1}$ and $\xm{3}$, to their global fit
values are also shown.  For comparisons, results from CT18 and NNPDF3.1 are shown as
the horizontal lines.
}
\label{unpol_moments}
\end{figure}

Now, we can perform an independent analysis that avoids the usage of RI-MOM procedure
completely and compare the outcome to the prediction for ${\cal
M}(z,P_z,P^\prime_z,\Gamma)$ from the knowledge of NNPDF and CTEQ PDF moments. We
perform such a comparison in Fig.~\ref{unpol_ratio_pheno}. For this, we used the
values of $\xm{n}(\mu)$ up to an order $n=n_\mathrm{max}$ for NNPDF31 in
Eq.~\ref{ope},  , and the complete result for  CT18, to obtain the
phenomenological expectation for the ratio ${\cal M}(z,P_z,P^\prime_z,\gamma_t)$. The
results obtained by using the truncation order $n_\mathrm{max}=2,3,4,20$ using the
NNPDF31 values for $\xm{n}$ are shown as different colored bands in
Fig.~\ref{unpol_ratio_pheno}. It is clear that inclusion of up to $n_\mathrm{max}=20$
moments is sufficient for convergence to the correct PDF within $z\leqslant0.5$~fm.
For $z<0.3$~fm, which is where the lattice data has a good signal to noise ratio, we
find that $N=4$ is sufficient to describe the lattice results. This gives us
an idea of which moments are being probed by our lattice data at different $z$. We
observe some discernible differences between the phenomenological expectations and
our lattice ${\cal M}(z,P_z,P^\prime_z,\gamma_t)$ for $z>0.2$~fm, as we also observed
in the case of RI-MOM scheme in Fig.~\ref{fig:comp2lat}. To understand this, we
estimate the values of the moments $\xm{n}$ that best describe our lattice data. To
avoid overfitting the data, we truncate the expansion in Eq.~\ref{ope} at most by
$n=4$.  In order to avoid lattice artifacts that might be present for $z$ of the
order of lattice spacing, we fit the data only from $z=2a$ to a value $z_{\rm max}$.
The variation of the best fit values of $\xm{n}$ with $z_{\rm max}$ is a source of
systematic error. In Fig.~\ref{unpol_moments}, we show the $z_{\rm max}$ dependence
of our estimates for $\xm{1}, \xm{2}, \xm{3}$ and $\xm{4}$. From
Fig.~\ref{unpol_ratio_pheno}, we note the noisy determination of the imaginary part
of ${\cal M}$. As a consequence, we find our estimates of $\xm{1}$ and $\xm{3}$ to be
noisy as well.  On the contrary, we were able to determine $\xm{2}$ and $\xm{4}$
reasonably well. In addition to $z_{\rm max}$ dependence, we also studied whether our
determination of the moments is affected by the order of truncation used in
Eq.~\ref{ope}. We observe no significant variations with truncation.  For comparison,
the NNPDF and CT18 values of these moments are shown by the horizontal lines.
Further, when we fix the values of $\xm{1}$ and $\xm{3}$ from NNPDF to reduce the
number of fit parameters, we find the estimates for $\xm{2}$ to be slightly elevated
in value and in the direction away from NNPDF,CT18 value. To a small extent, this is
seen in $\xm{4}$ as well. Thus, the observed difference between our lattice result
and the NNPDF, CT18 results could be attributed to this tendency for our lattice
values of $\xm{2},\xm{4}$ to be slightly higher than the corresponding
phenomenological values.

\begin{figure}
  \centering
\includegraphics[scale=0.7]{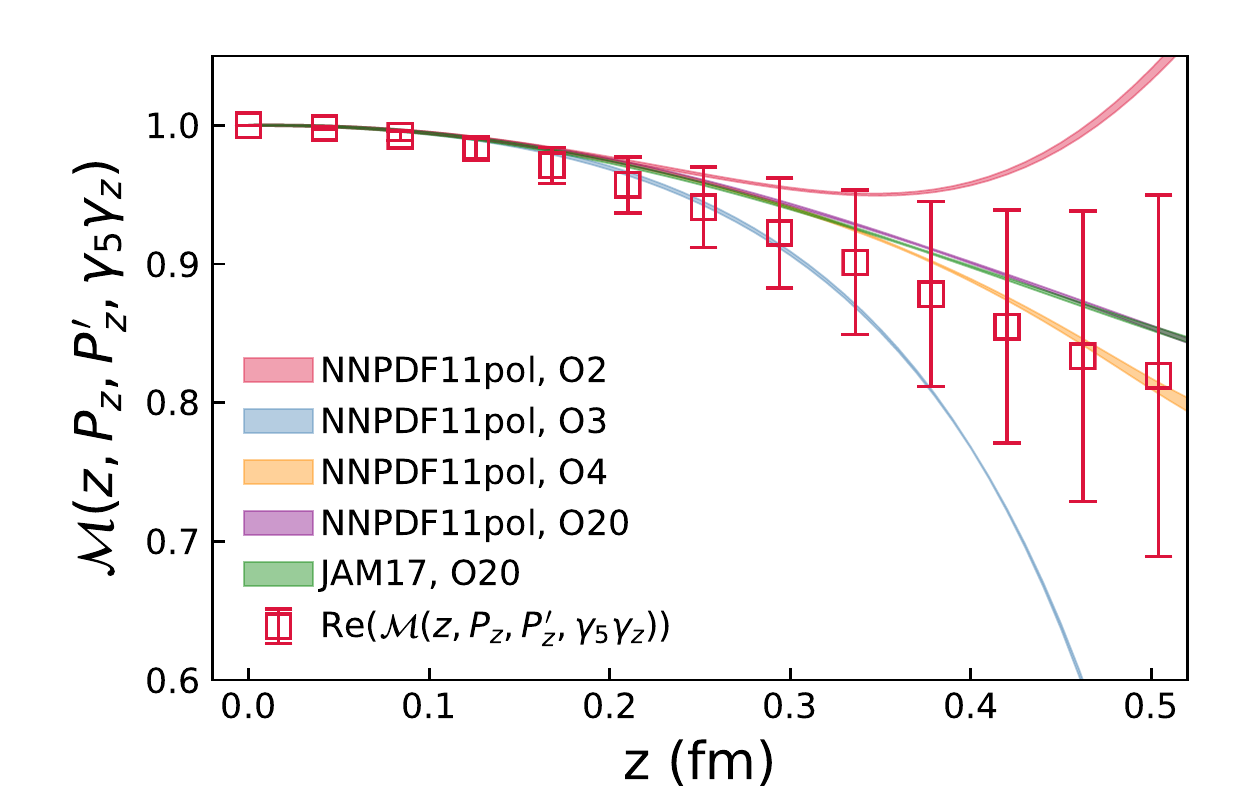}
\includegraphics[scale=0.7]{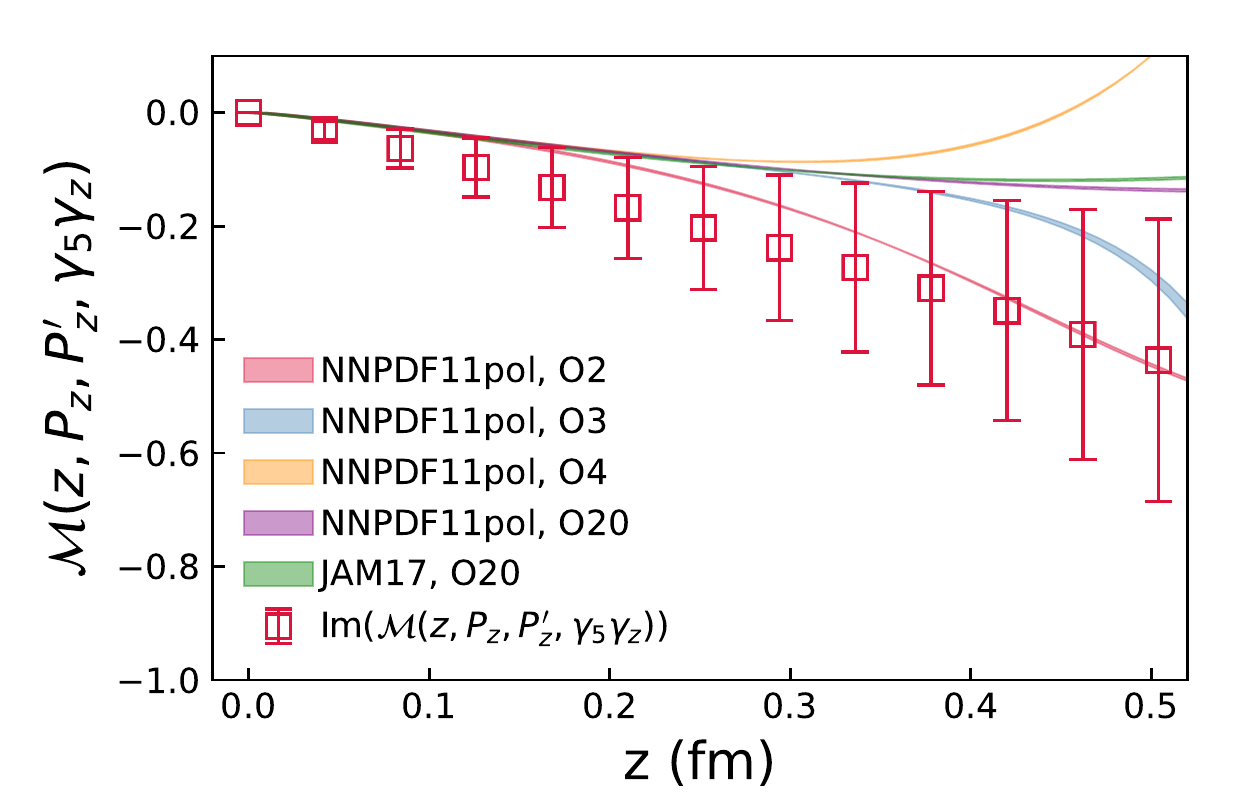}
\caption{
The real (left) and imaginary(right) parts of ${\cal M}(z,P_z,P'_z,\gamma_5\gamma_z)$
is shown for $P_z=2.31$~GeV and $P^\prime_z=1.84$~GeV. The data points are from our
lattice calculations, whereas the various colored bands are the corresponding results
from the isovector helicity PDFs from NNPDF1.1pol and JAM17.  For NNPDF1.1pol, we
also show results by truncating the expansion in Eq.~\ref{ope} at various orders,
$n=n_\mathrm{max}$, in the PDF moments; these results are denoted by
`$\mathrm{O}n_\mathrm{max}$'.
}
\label{pol_ratio_pheno}
\end{figure}

\begin{figure}
  \centering
\includegraphics[scale=0.7]{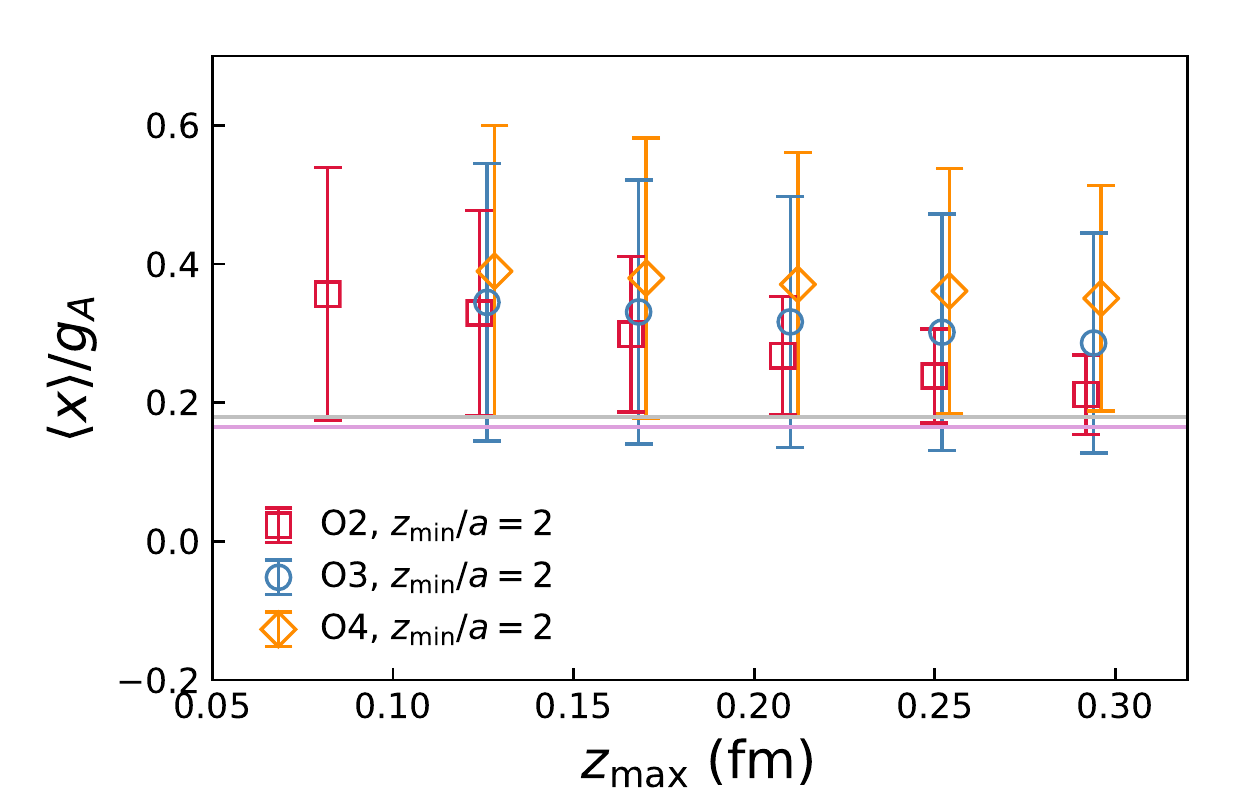}
\includegraphics[scale=0.7]{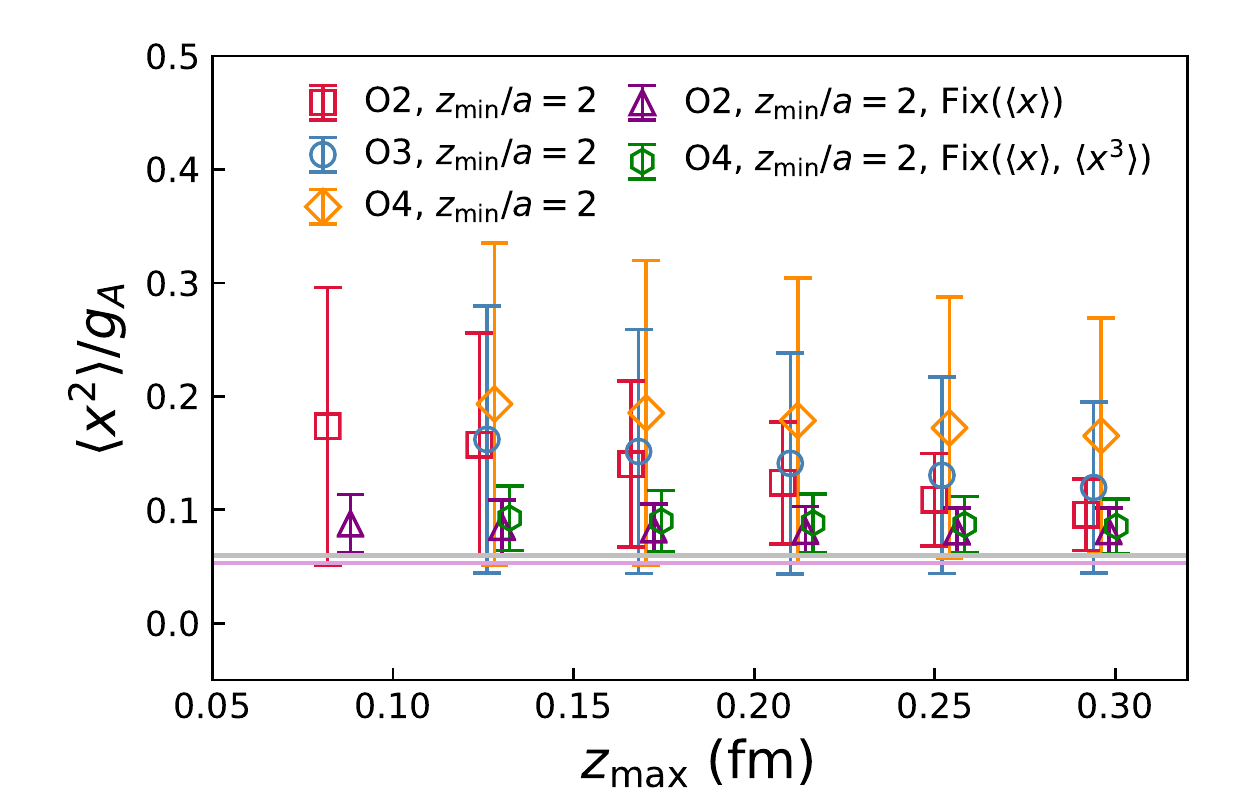}
\includegraphics[scale=0.7]{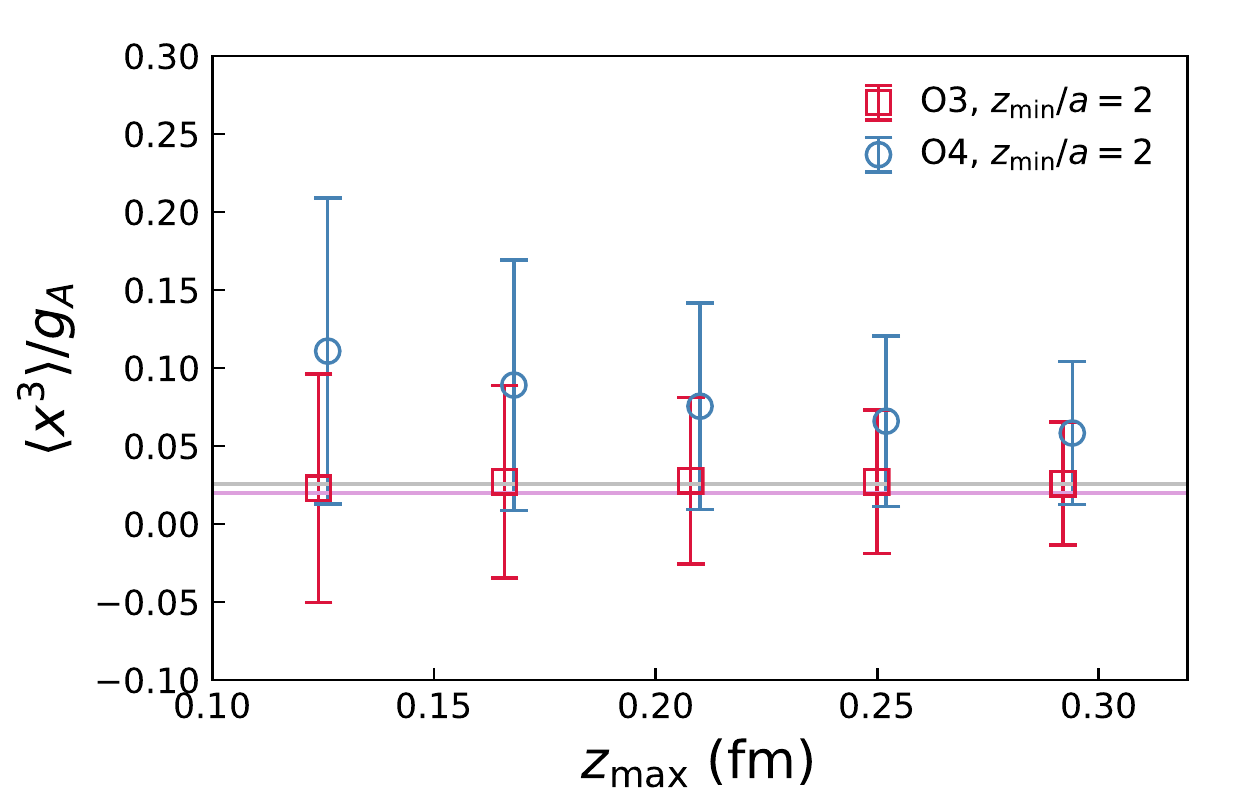}
\includegraphics[scale=0.7]{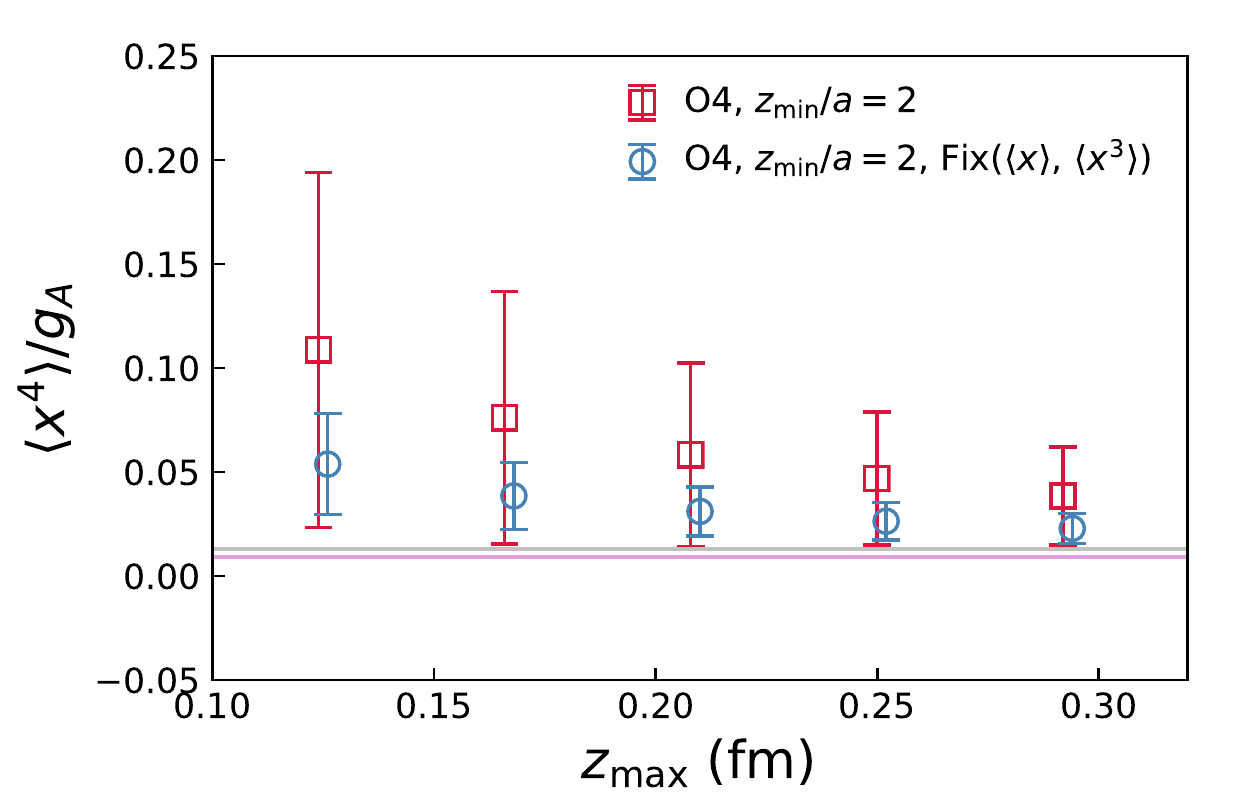}
\caption{
The moments of helicity PDF, $\xm{1}$ (top-left), $\xm{2}$ (top-right), $\xm{3}$
(bottom-left), and $\xm{4}$ (bottom-right),  that best describes the ratio ${\cal
M}(z,P_z,P^\prime_z,\gamma_5\gamma_z)$ with $P_z=2.31$~GeV and $P'_z=1.84$~GeV. In
each of the panels, the moment $\xm{n}$ is shown as a function of $z_{\rm max}$ of
the fit using the functional form in Eq.~\ref{ope} over a range $[2a,z_{\rm max}]$ of
the data.  The results from fits using only moments up to $n=2$ as free parameters in
Eq.~\ref{ope} are labeled `O2', and those up to $n=4$ are labeled `O4'. The results
from fits that fix the moment $\xm{1}$, or $\xm{1}$ and $\xm{3}$, to their global fit
values are also shown.  For comparisons, results from JAM17 and NNPDF1.1pol are shown
as the horizontal lines. 
}
\label{pol_moments}
\end{figure}

We repeated similar analysis for the helicity matrix element,
$\Gamma=\gamma_5\gamma_z$. In this case, the Wilson coefficients $c_n(\mu z)$ are the
same as in the case of unpolarized case with $\Gamma=\gamma_z$. Since we are setting
the value of the matrix elements at $z=0$ to be 1 through the ratio, we only obtain
the values of $\xm{n}/\xm{0}$ in the expansion Eq.~\ref{ope}, with $\xm{0}=g_A$. In
Fig.~\ref{pol_ratio_pheno}, we compare the results corresponding to the NNPDF11pol
and JAM17 with the lattice result for the ratio. As in the case of the unpolarized
matrix element, we also test the dependence of this comparison on the truncation
order $n_\mathrm{max}$. The sensitivity to higher moments is a bit more than that for
the unpolarized case, and we find convergence at only $n_\mathrm{max}=6$ at $z<0.3$
fm. Surprisingly, the global fit expectation agrees quite well with our lattice
result even though there is a little tension in the imaginary parts. As explained
above, we also obtain the best fit values of $\xm{1}/g_A, \xm{2}/g_A, \xm{3}/g_A$
and $\xm{4}/g_A$ that describe our lattice data via Eq.~\ref{ope} truncated at most
by $4^\mathrm{th}$ order.  In Fig.~\ref{pol_moments}, we show the results as a
function of the largest $z$ used in the fits, $z_{\rm max}$. Like the unpolarized PDF
case,  $\xm{1}/g_A$ is noisy, but seems agree with the global fit results. The more
precisely determined value of $\xm{2}/g_A$ is quite robust to various ways of fitting
the data and agrees nicely with the global fit values. To compare with other lattice caculations, we truncate the expansion in Eq.~\ref{ope} at
$n=2$, and estimate $\xm{1}/g_A$ at $\mu$ = 2 GeV with the $z$ in range [2a, 0.3 fm]. Our result $\langle x \rangle /g_A$ = 0.219(56) is compatible with the ETMC result~\cite{Abdel-Rehim:2015owa} 0.229(30)/1.242(57) within the error.

\section{Summary and conclusions}
In this paper we studied isovector unpolarized and helicity PDFs of proton using the
LaMET approach. The lattice calculations have been performed for an unphysically
large pion mass of $310$~MeV. On the other hand, our lattice study was carried out
using lattice spacing $a=0.042$~fm, which is the smallest lattice spacing used in
such studies. We argued that such small lattice spacing is essential for the validity of 1-loop
perturbative matching between PDF and qPDF, which is a key ingredient of LaMET.

Extracting the nucleon matrix elements for such large momenta and small lattice
spacing is challenging because of poor signal to noise ratio. To deal with this
problem we performed a detailed study of the nucleon two-point function with momentum
smeared source and sink, as well as with momentum smeared source and point sink to
better control the excited state contributions. We demonstrated that the ground state
can be reliably isolated up to the highest momenta used in this study. Furthermore,
for the Euclidean time separations used that are relevant for our lattice analysis
the two-point function is very well described by the ground state and and an
'effective' excited state contribution, with the energy that is larger than the true
excited state energy. Therefore, we argued that the two-state Ans\"atze is sufficient
to describe the dependence of the 3-point function on the source-sink separation and
on the operator insertion time. We showed that the qPDF matrix elements can be
extracted in this way, and the results do no depend on the choices of the fit
interval used in our study, demonstrating the robustness of our analysis procedure.

After non-perturbative RI-MOM renormalizations we compared the lattice calculations
of the spatial, $z$, dependence of  qPDFs with that from the phenomenological PDFs,
obtained from the global pQCD-based  analyses of pertinent experimental data
performed by different collaborations. Working in $z$-space allowed us to test the
LaMET approach. The comparisons showed that there is a rough agreement between the
lattice results and the results of global analysis, but only at quite small
distances. Even for the very small lattice spacing used in this study, there was not
enough data points to constrain the $x$-dependence of the PDFs. Instead, to translate
our $z$-space comparisons to $x$-dependence, we introduced a new ratio-based
renormalization scheme for the Ioffe-time distributions. Using our lattice
calculations for Ioffe-time distributions, renormalized via this new ratio-based
scheme, we determined the first moments of the isovector unpolarized and helicity
PDFs of proton, and compared these moments with that from the corresponding
phenomenological PDFs.

\section*{Acknowledgments}
This work was supported by: (i) The U.S. Department of Energy, Office of Science,
Office of Nuclear Physics through the Contract No. DE-SC0012704; (ii) The U.S.
Department of Energy, Office of Science, Office of Nuclear Physics and Office of
Advanced Scientific Computing Research within the framework of Scientific Discovery
through Advance Computing (SciDAC) award Computing the Properties of Matter with
Leadership Computing Resources; (iii) The Brookhaven National Laboratory’s Laboratory
Directed Research and Development (LDRD) project No. 16-37. (iv) The work of ZF, RL,
HL, YY and RZ are supported by the US National Science Foundation under grant PHY
1653405 “CAREER: Constraining Parton Distribution Functions for New-Physics
Searches”. (v) The work of XG is partially supported by the NSFC Grant Number 11890712. (vi) SS also acknowledges support by the RHIC Physics Fellow Program of the RIKEN BNL Research Center and  by the National Science Foundation under CAREER Award PHY-1847893.

This research used awards of computer time provided by the INCITE program at Oak
Ridge Leadership Computing Facility, a DOE Office of Science User Facility operated
under Contract No. DE-AC05-00OR22725.

We are grateful to Yong Zhao for advice on perturbative matching of the helicity
qPDF.

\bibliographystyle{apsrev4-1.bst}
\bibliography{ref}

\newpage
\appendix
\section{Analysis of the nucleon two point function} \label{app:2pt}

\begin{figure}
        \includegraphics[width=0.3\textwidth]{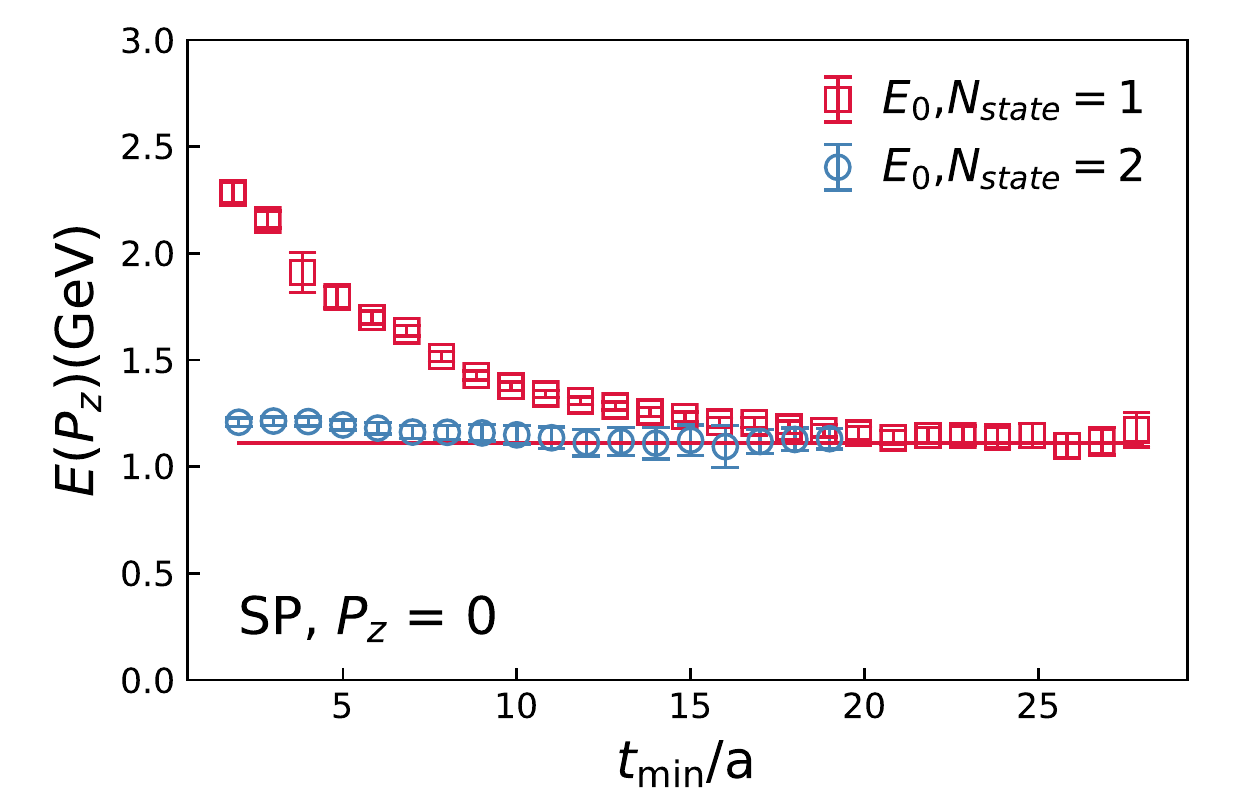}
        \includegraphics[width=0.3\textwidth]{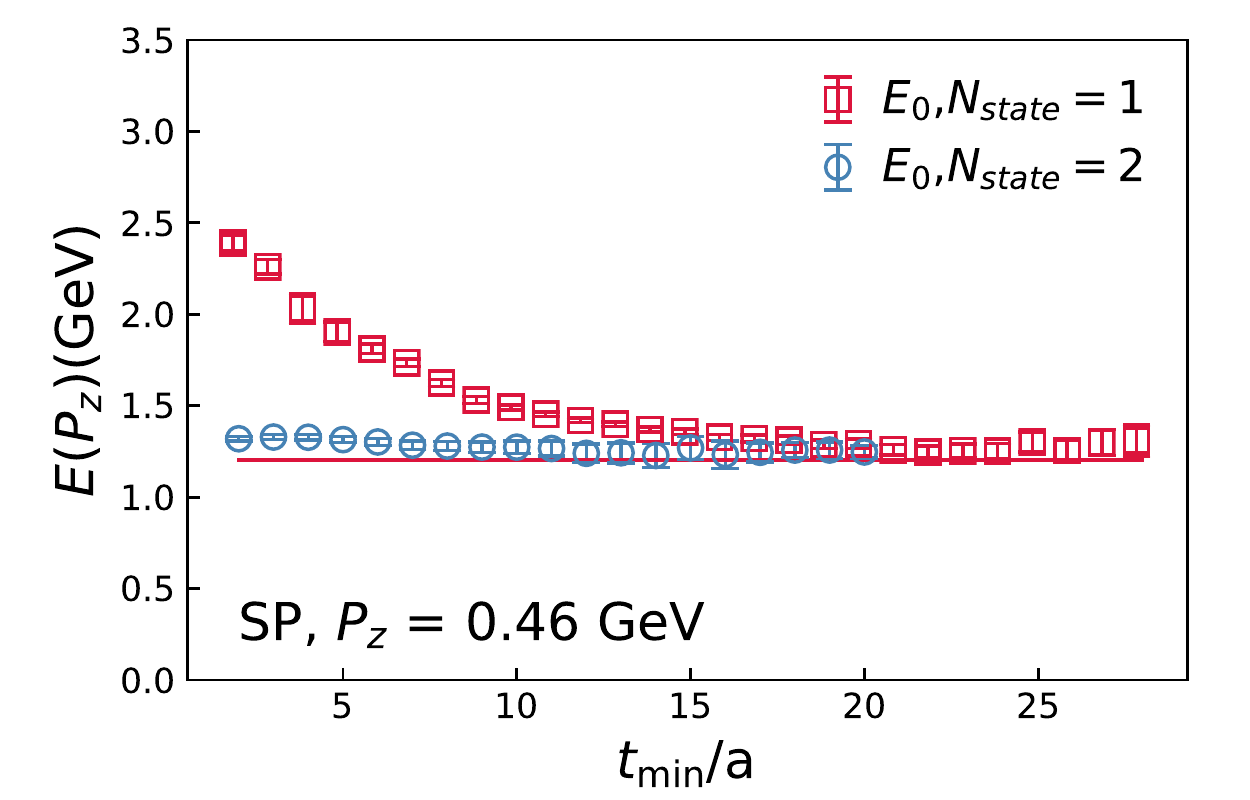}
        \includegraphics[width=0.3\textwidth]{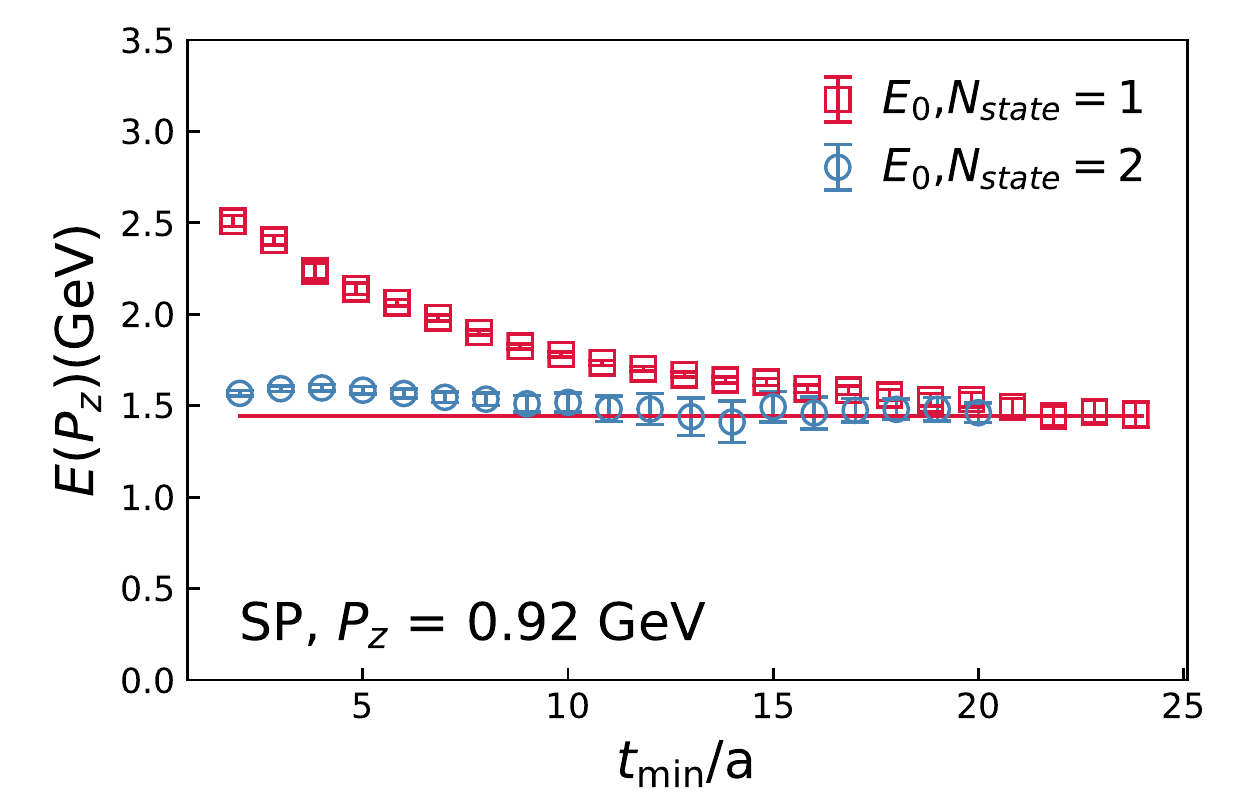}
        \includegraphics[width=0.3\textwidth]{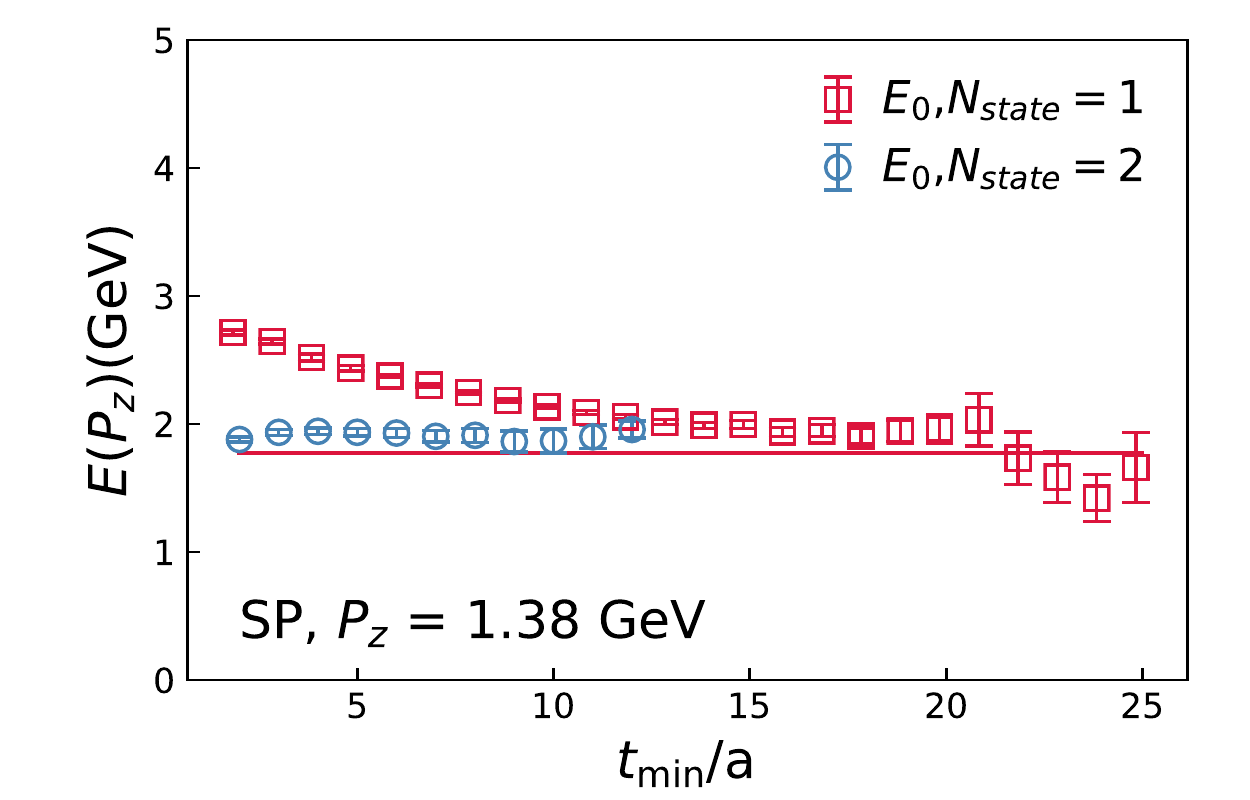}
        \includegraphics[width=0.3\textwidth]{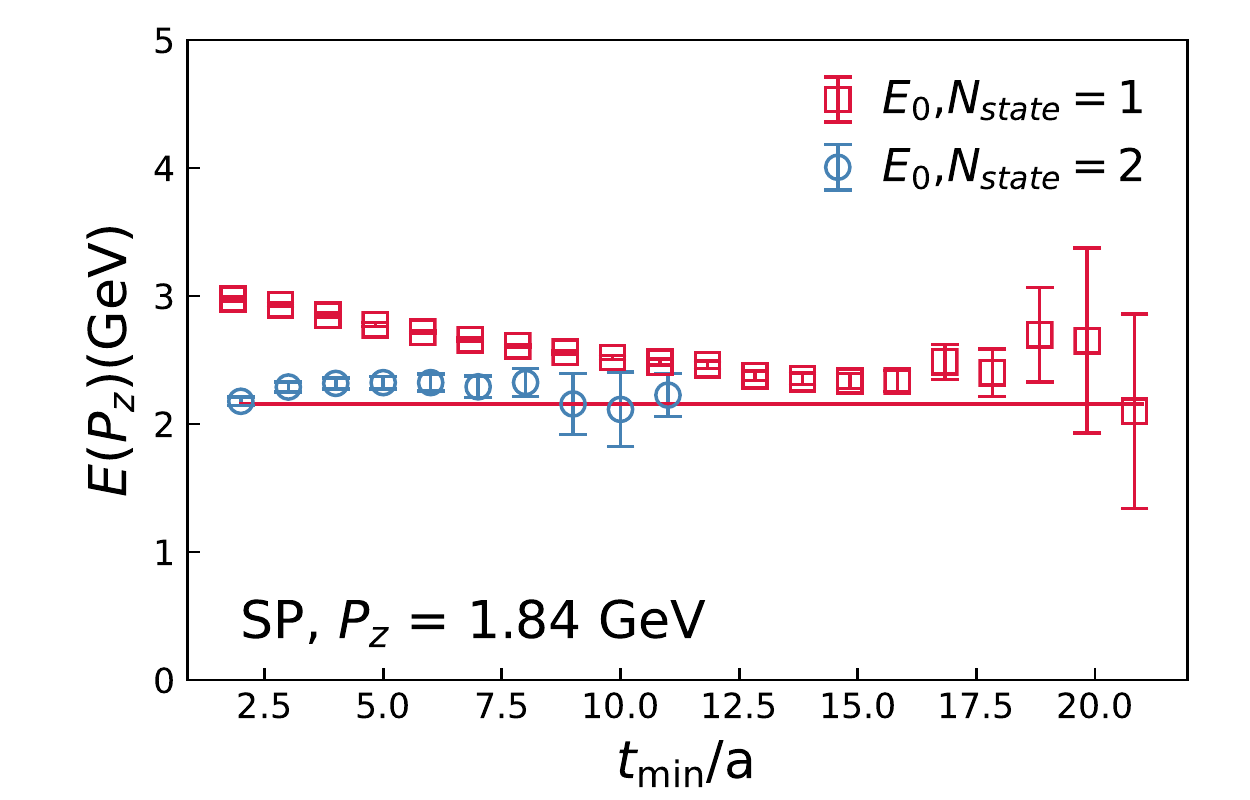}
        \includegraphics[width=0.3\textwidth]{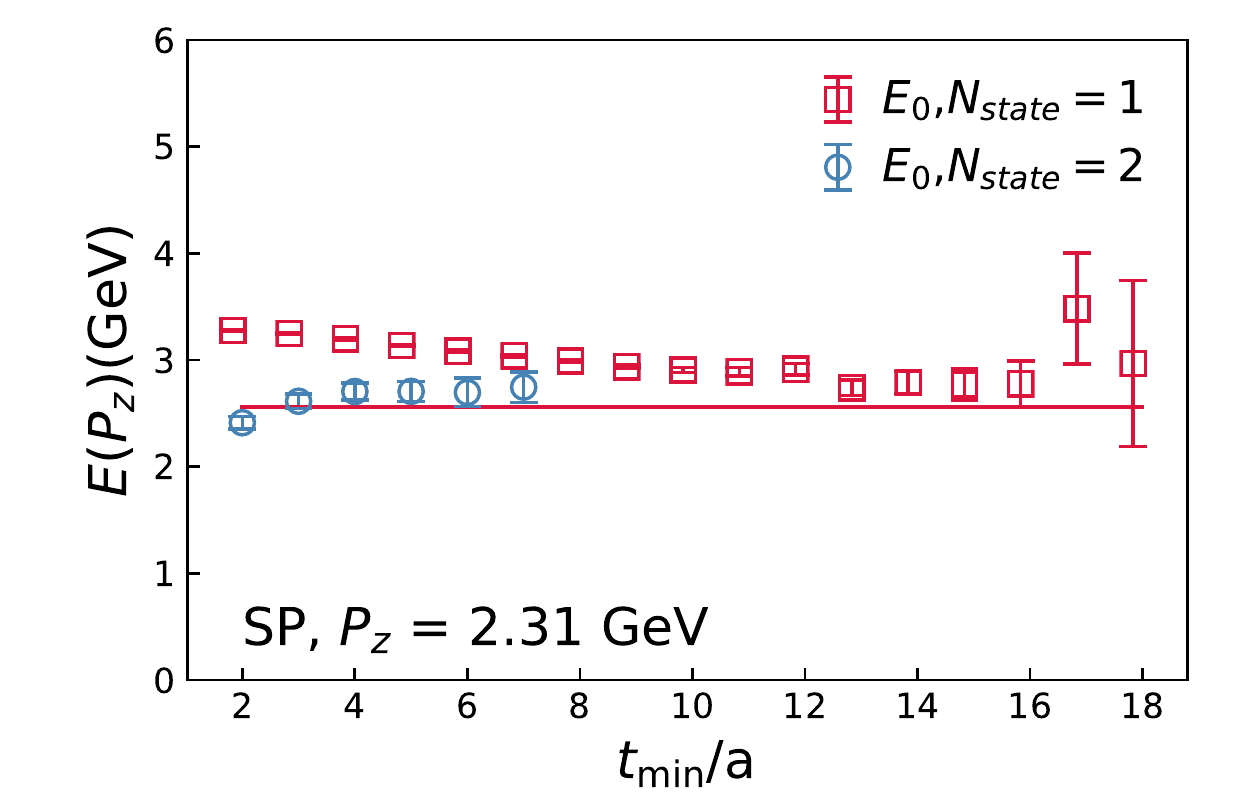}
        \caption{Ground state energy from unconstrained one state fit and two state fit of the SP correlators.}
\label{fig:fi2pt_E0_SP}
\end{figure}

In this appendix we discuss some details of the analysis of the SP and SS two point correlators.
In Fig. \ref{fig:fi2pt_E0_SP} we show the ground state energy from one and two exponential fits
of SP correlators as function of $t_{\rm min}$. Contrary to the fits of the SS correlators stable result
for the ground state energy, $E_0$ is only obtained for $t_{\rm min} \ge 20$.

\begin{figure}
        \includegraphics[width=0.3\textwidth]{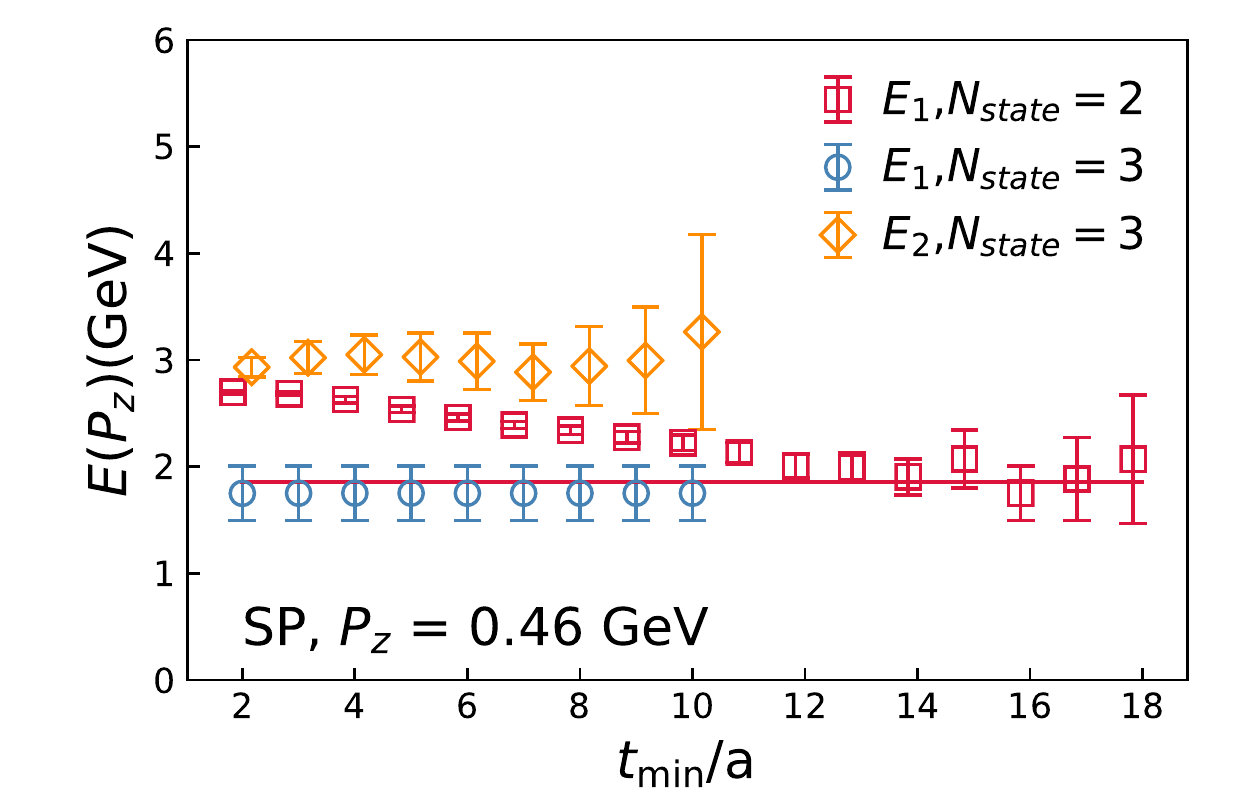}
        \includegraphics[width=0.3\textwidth]{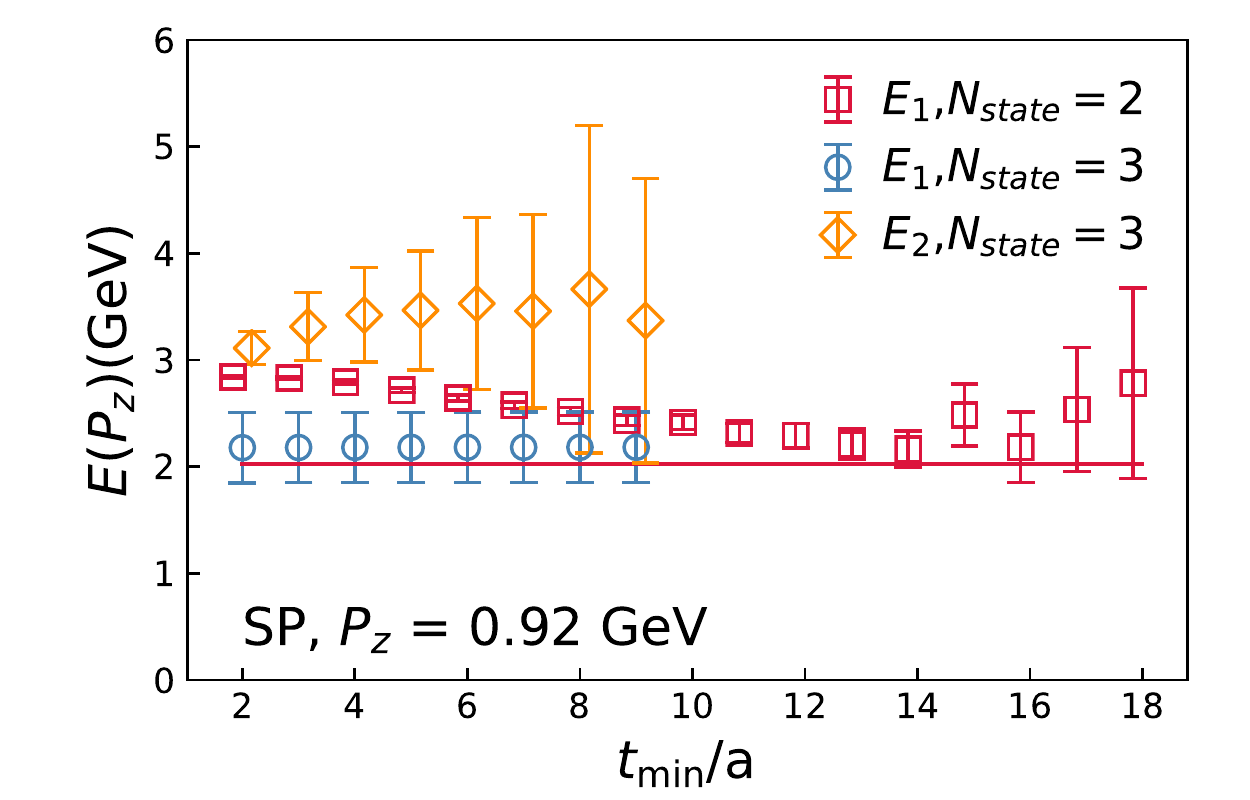}
        \includegraphics[width=0.3\textwidth]{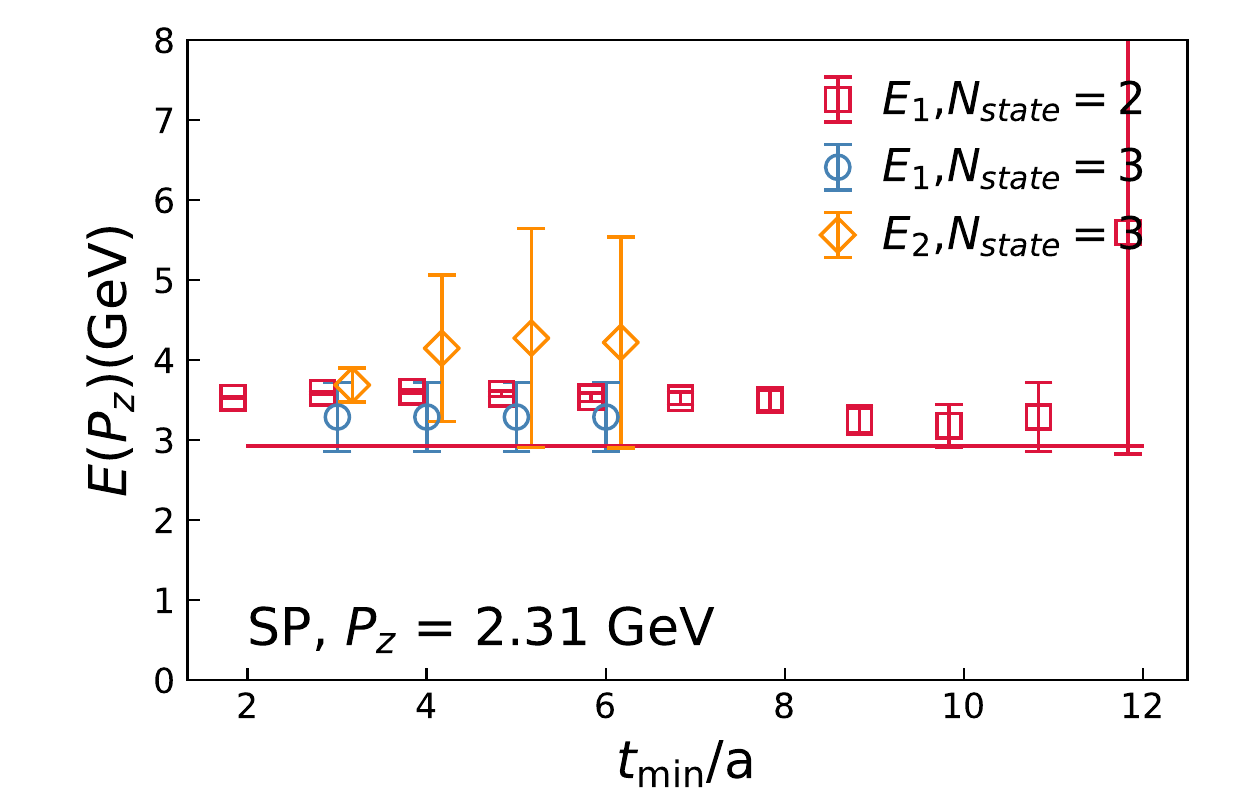}
\caption{The energies of the first ($E_1$) and second ($E_2$) excited states from constrained two-state and three-state fits of SP correlator for $n_z=1$ (left), $n_z=2$ (middle) and $n_z=5$ (right). The horizontal line is the values calculated from the dispersion relation.}
\label{fig:fit2pt_E1_SPadd}
\end{figure}

As discussed in the main text we performed prior-based fits of SP and SS correlators
for all values of $p_z$.
In Fig. \ref{fig:fit2pt_E1_SPadd} we show the results on $E_1(p_z)$ for $n_z=1,2$ and $5$
for prior-based fits of the SP correlator.

\begin{figure}
        \includegraphics[width=0.3\textwidth]{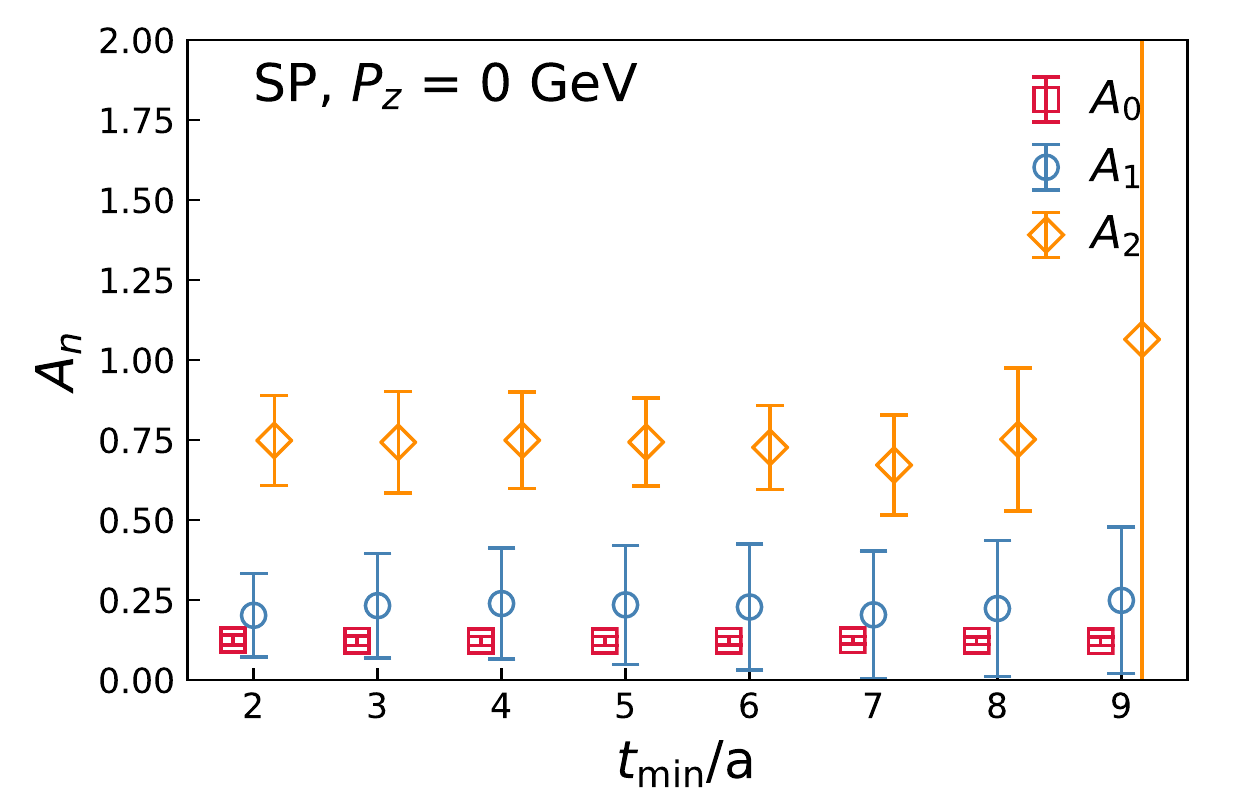}
        \includegraphics[width=0.3\textwidth]{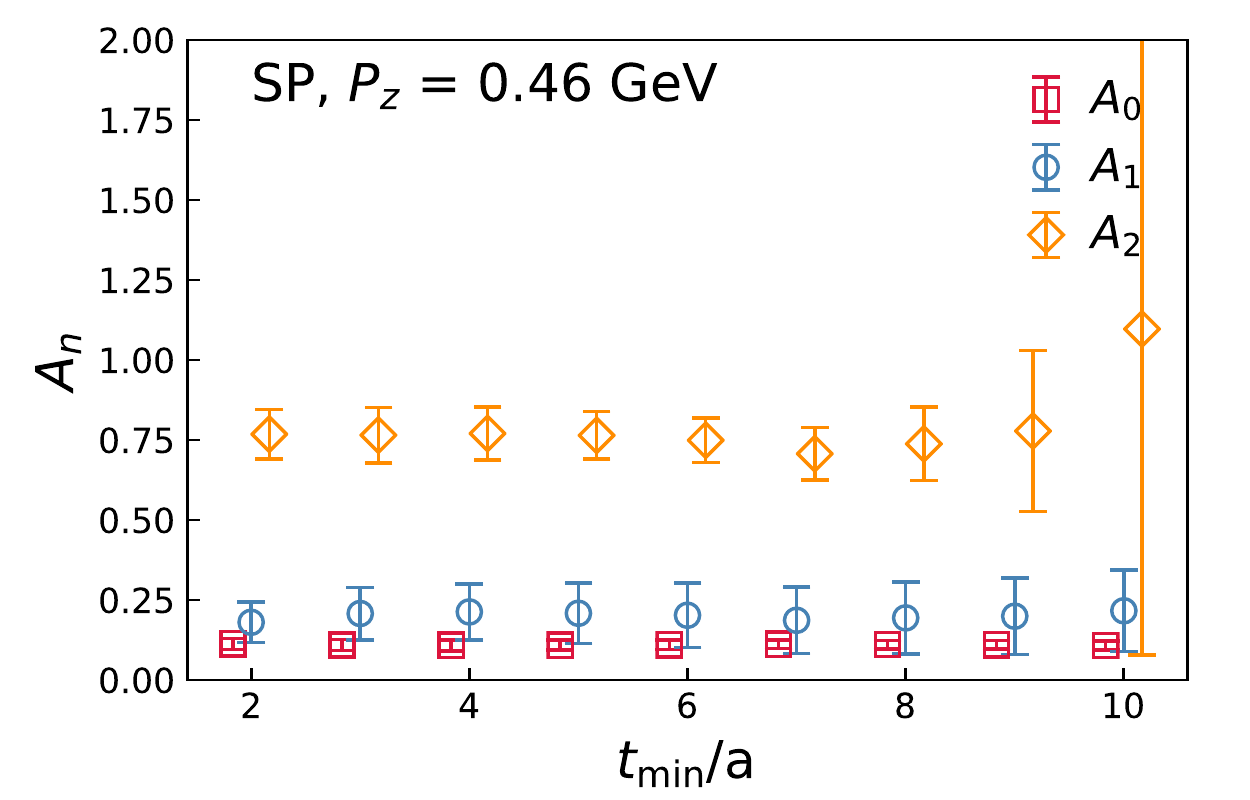}
        \includegraphics[width=0.3\textwidth]{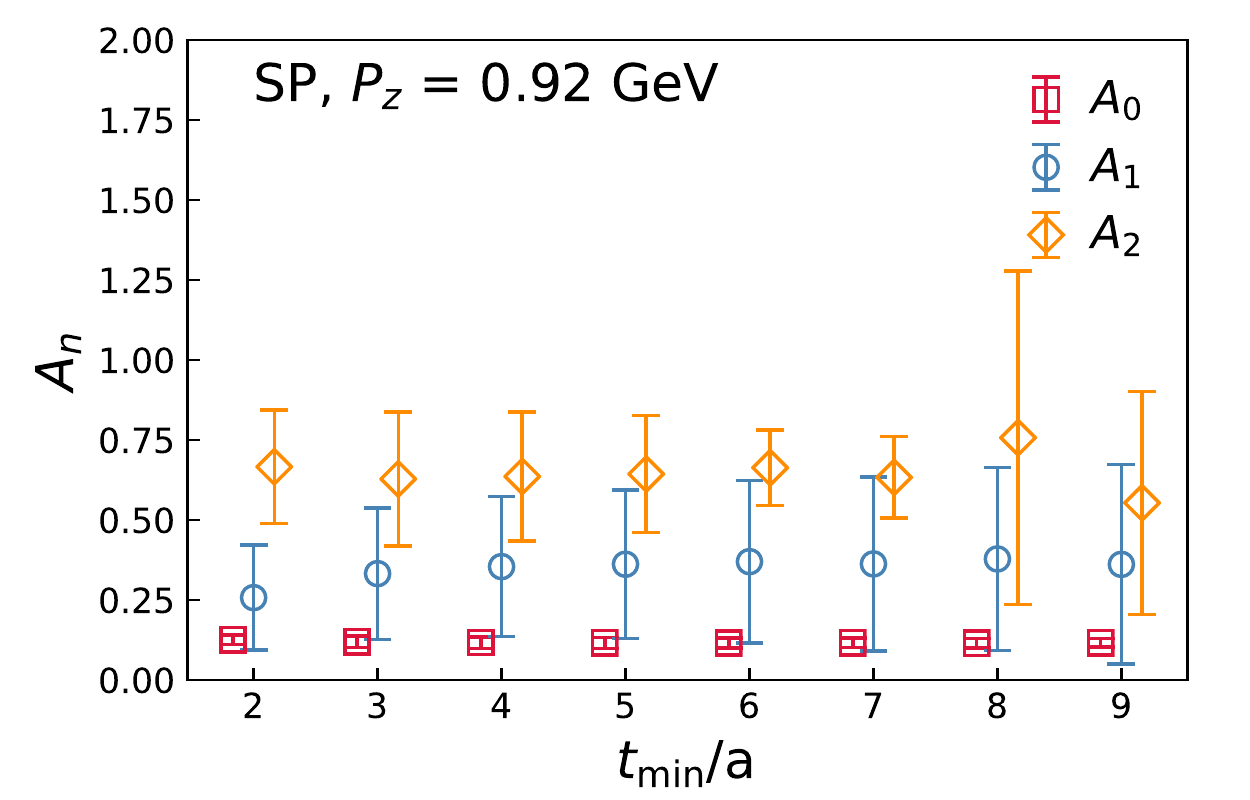}
        \includegraphics[width=0.3\textwidth]{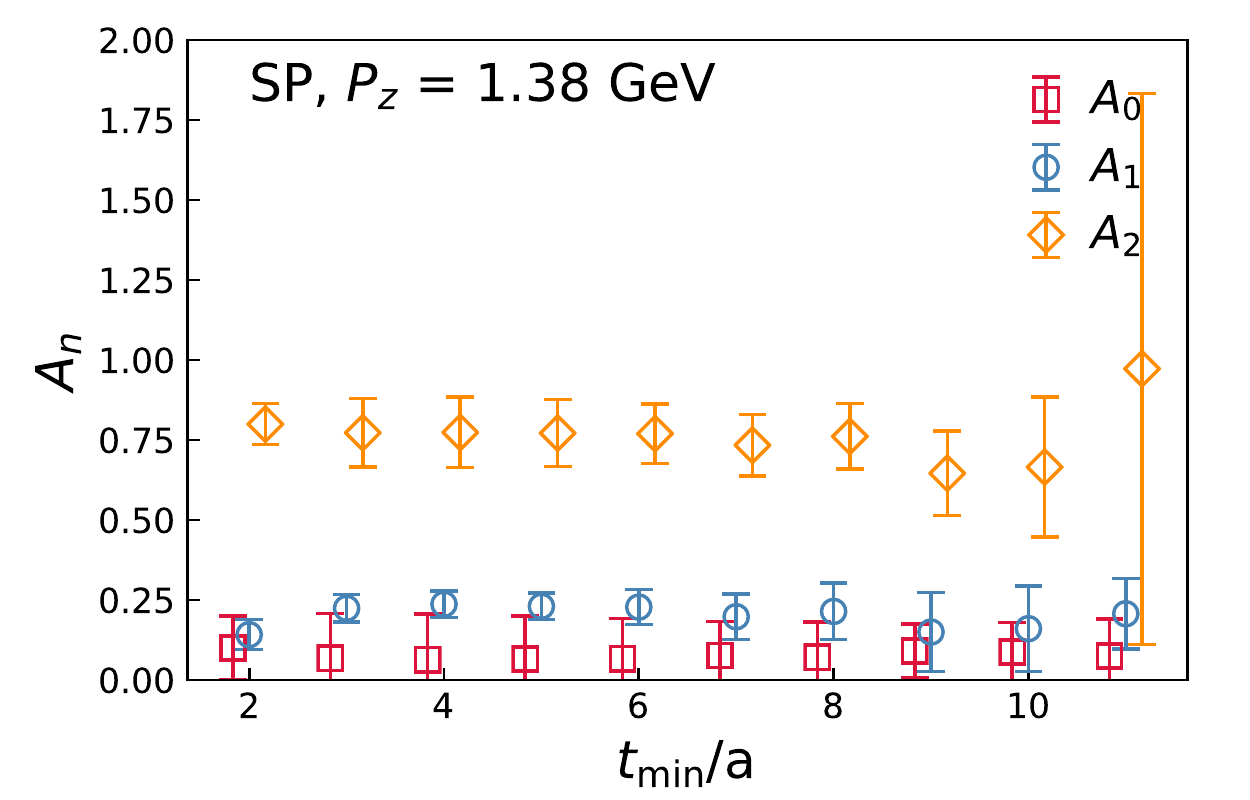}
        \includegraphics[width=0.3\textwidth]{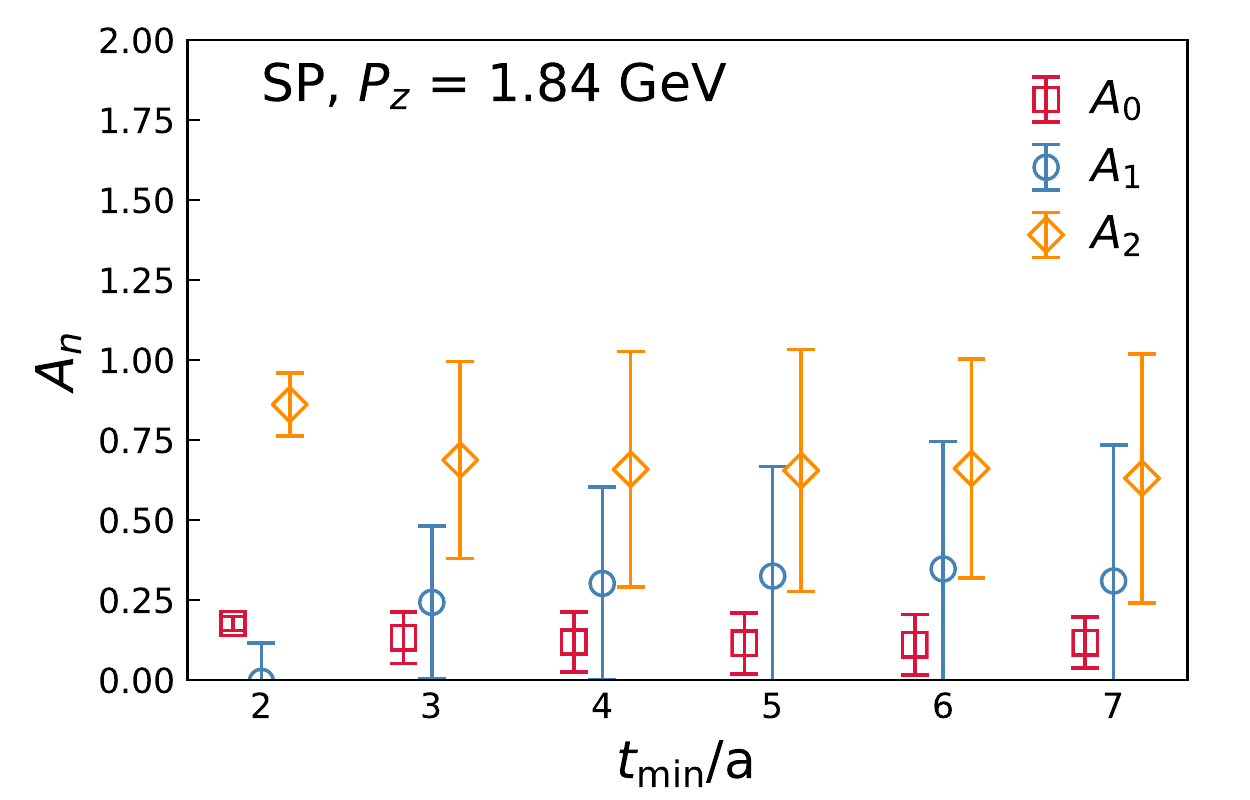}
        \includegraphics[width=0.3\textwidth]{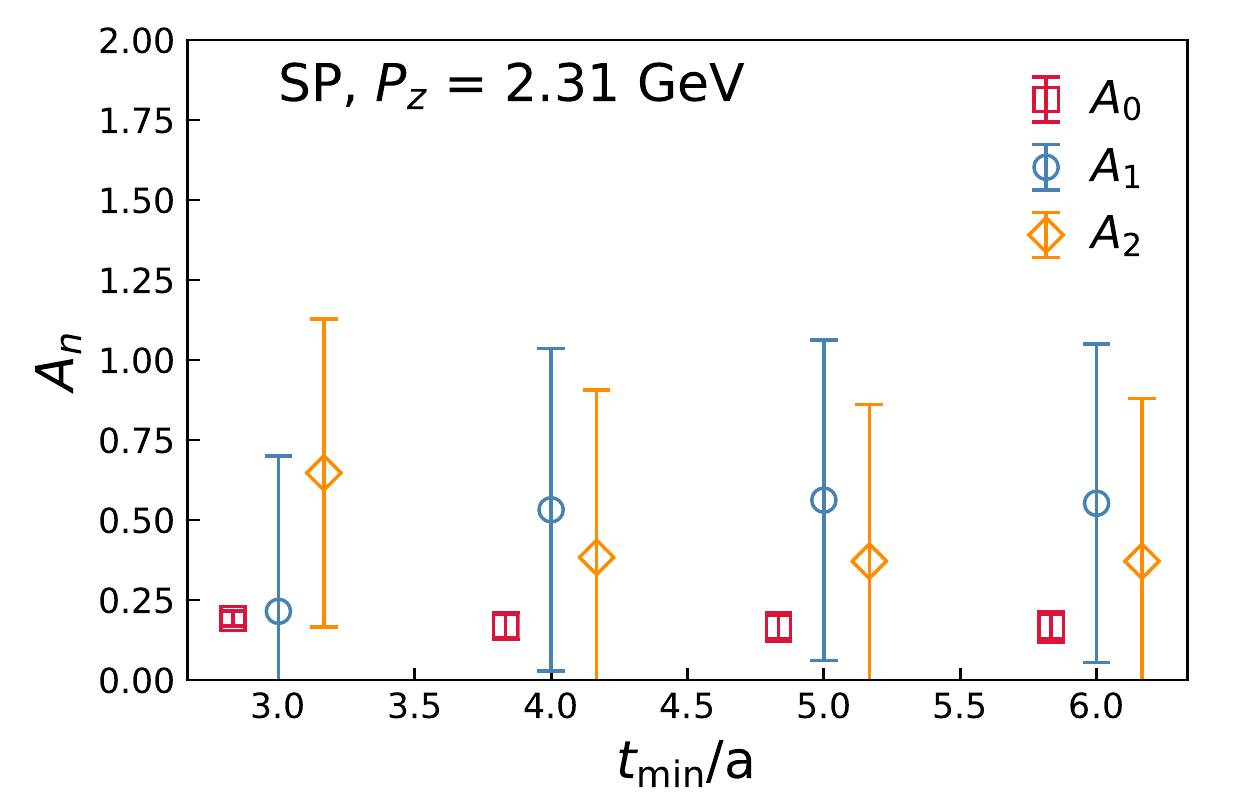}
\caption{The amplitudes of different states obtained from constrained 3-state fit of SP correlator and normalized
by $C_{2pt}^{SP}(t=0)$ as function of temperature.}
\label{fig:fit2pt_A_SP}
\end{figure}

We see clearly that $E_1$ approaches the value expected from the dispersion relation for
$t_{\rm min}>11$ if two exponential fit is used. For  constrained three exponential fits the same value
is approached for $t_{\rm min}=2$. In Fig. \ref{fig:fit2pt_A_SP} we show the amplitudes, $A_i,~i=1,2,3...$,
of different states normalized by the value of the two-point correlator at $t=0$, which
by definitions is equal to $\sum_i A_i$. We see that $A_1$ is slightly higher than $A_0$, while
$A_2$ is significantly larger than either $A_0$ or $A_1$.

\begin{figure}
\includegraphics[width=0.3\textwidth]{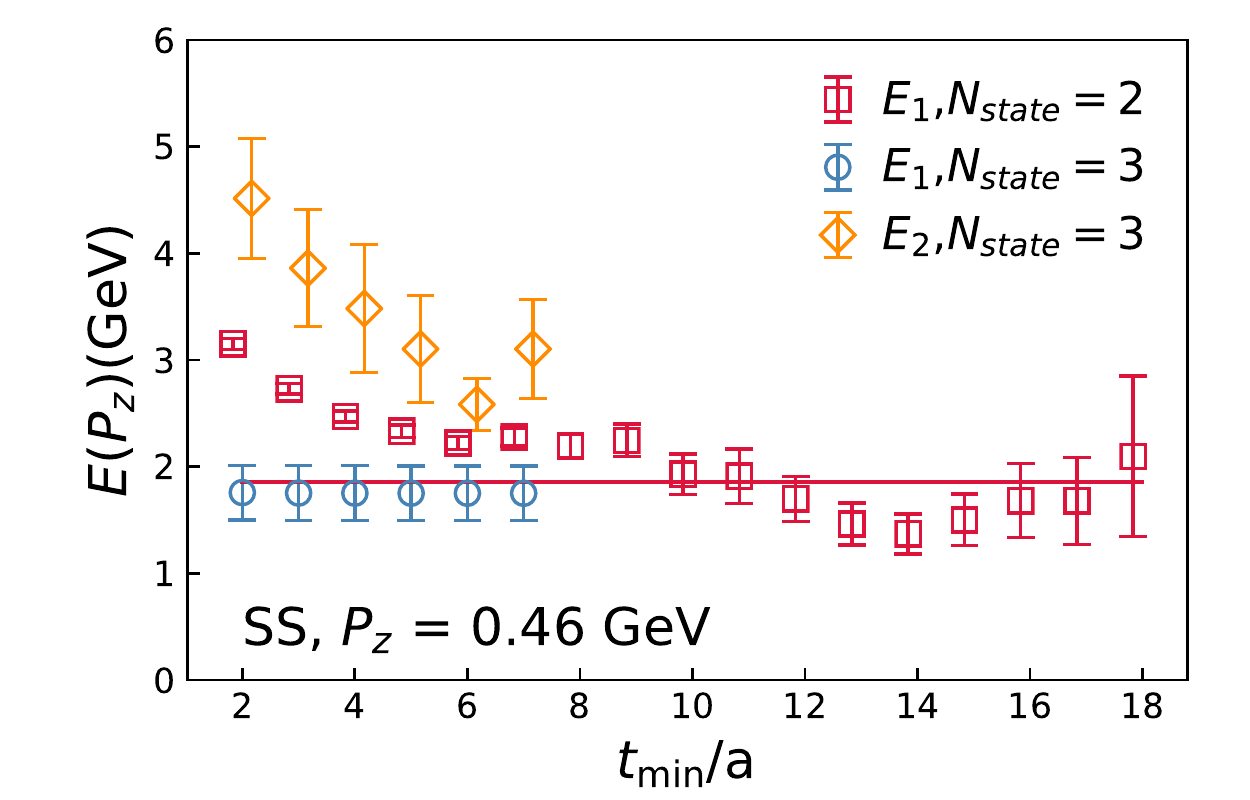}
\includegraphics[width=0.3\textwidth]{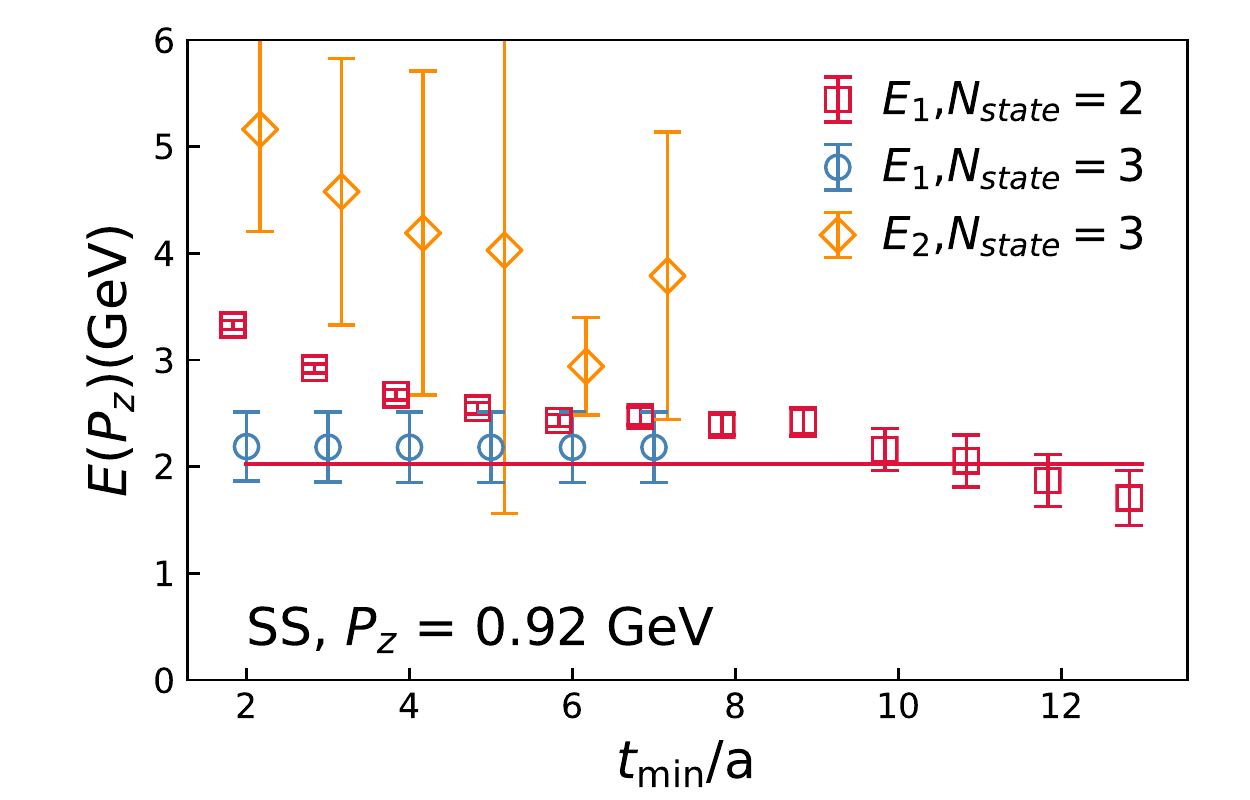}
\includegraphics[width=0.3\textwidth]{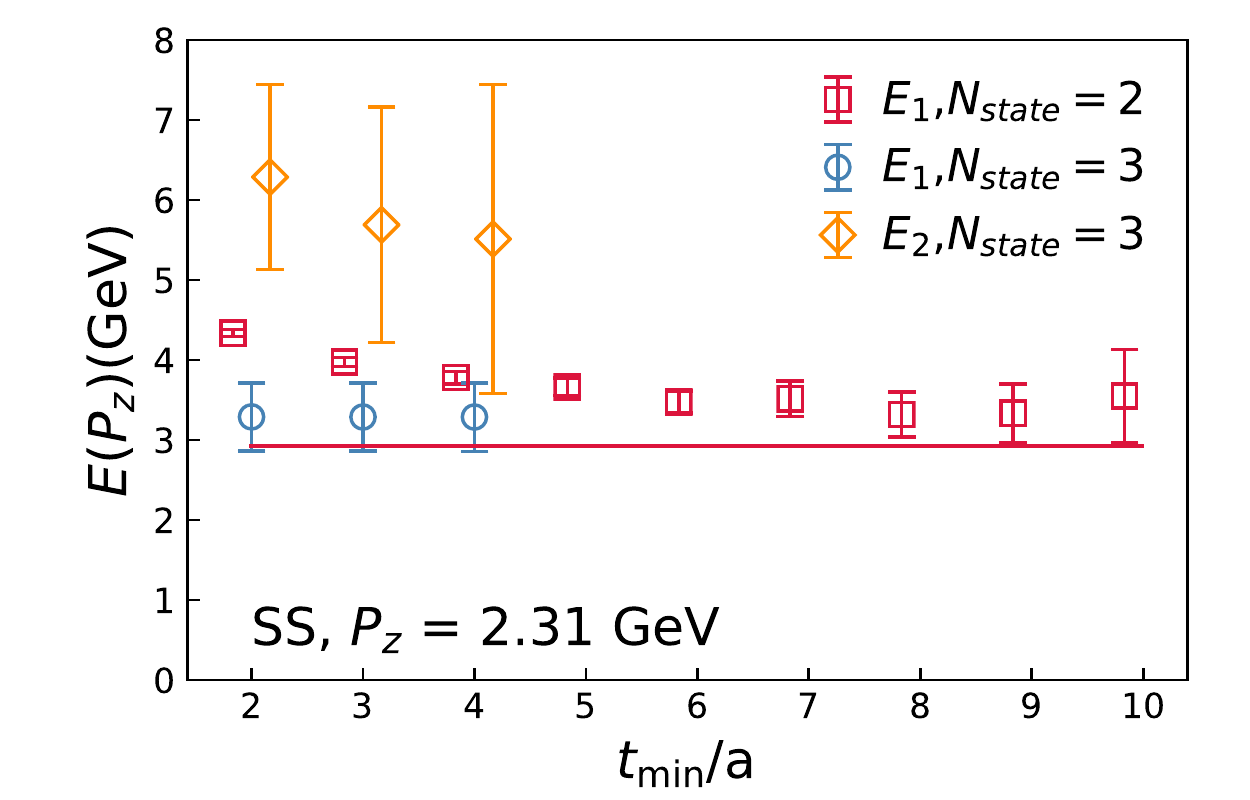}
\caption{The energies of the first ($E_1$) and second ($E_2$) excited states from constrained two-state and three-state fits of SS correlator for $n_z=1$ (left), $n_z=2$ (middle) and $n_z=5$ (right). The horizontal line is the values calculated from the dispersion relation.}
\label{fig:fit2pt_E1_SSadd}
\end{figure}

\begin{figure}
        \includegraphics[width=0.3\textwidth]{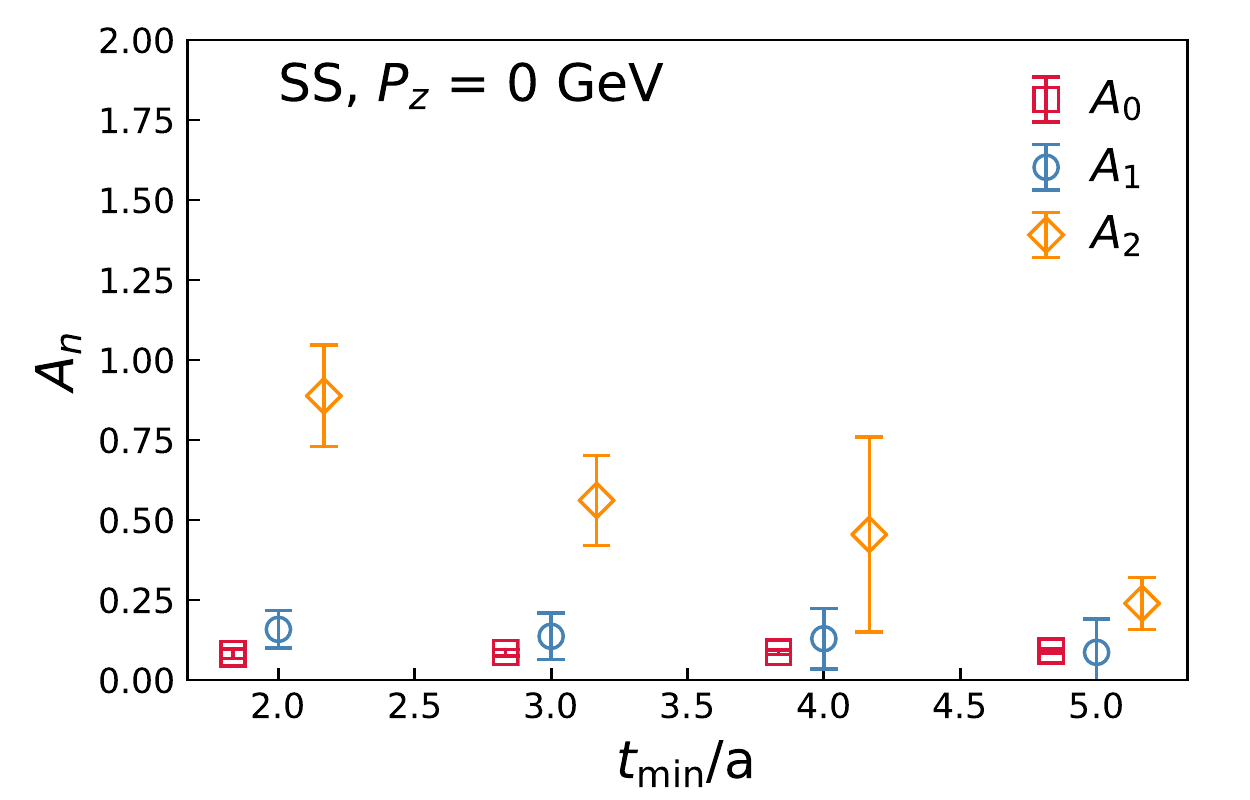}
        \includegraphics[width=0.3\textwidth]{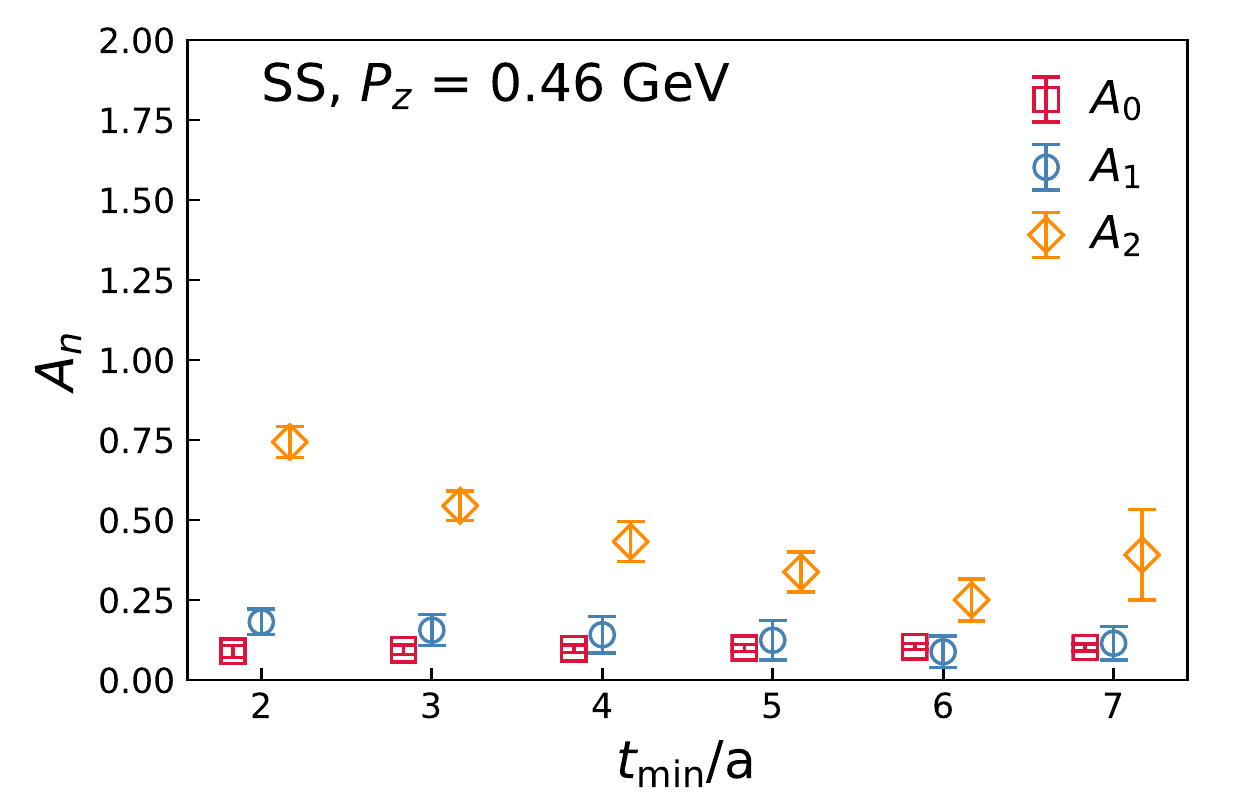}
        \includegraphics[width=0.3\textwidth]{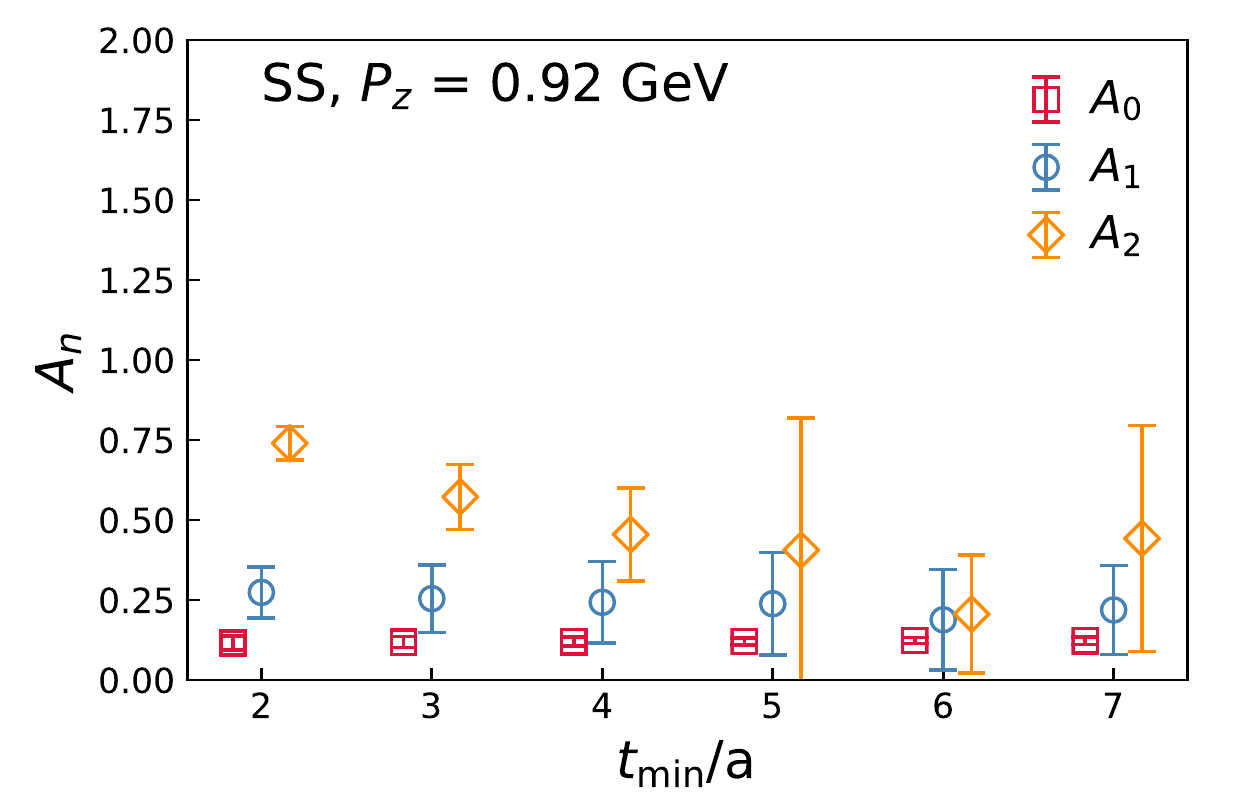}
        \includegraphics[width=0.3\textwidth]{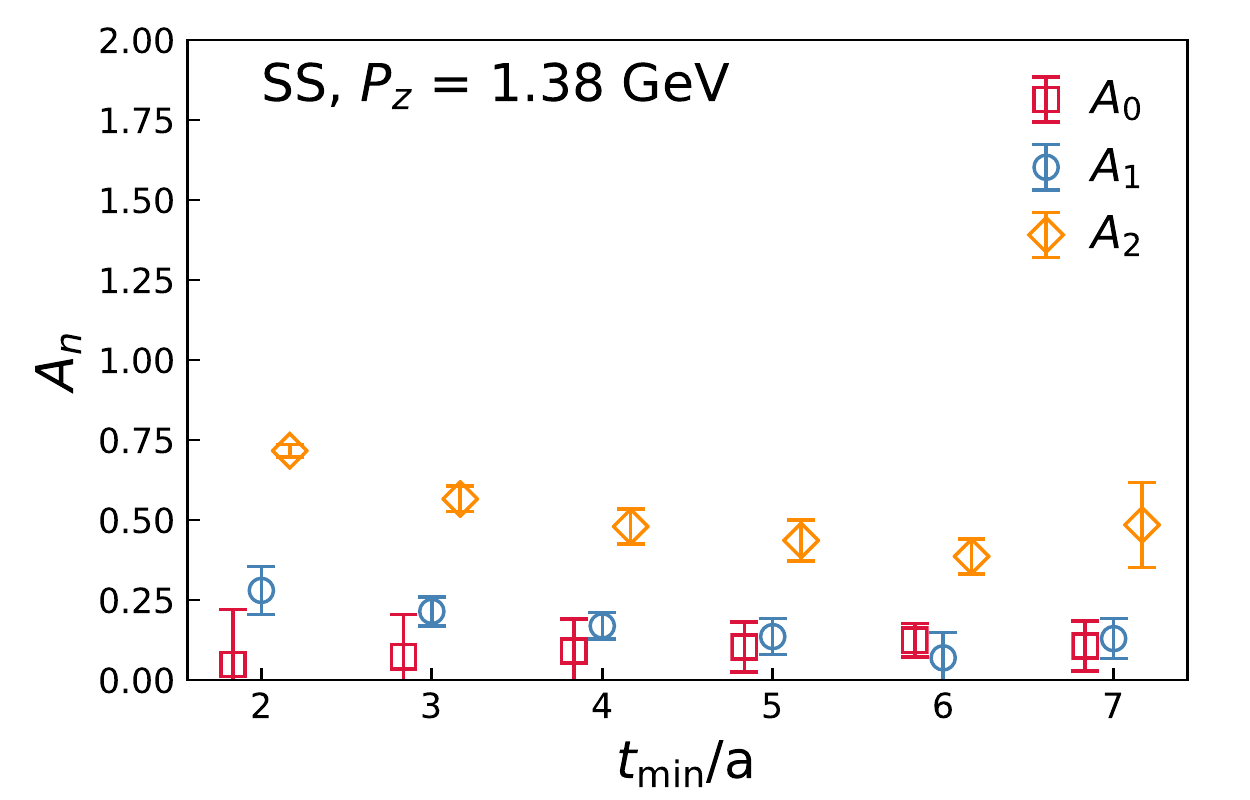}
        \includegraphics[width=0.3\textwidth]{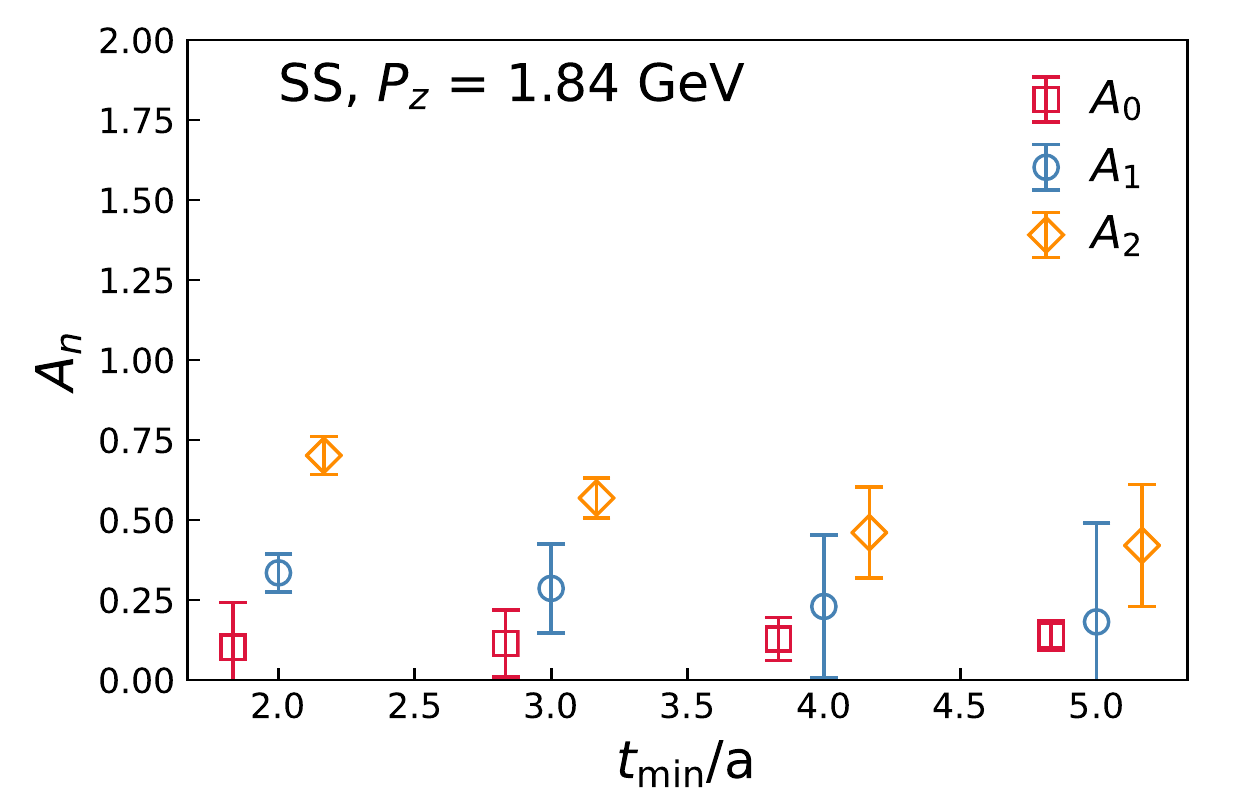}
        \includegraphics[width=0.3\textwidth]{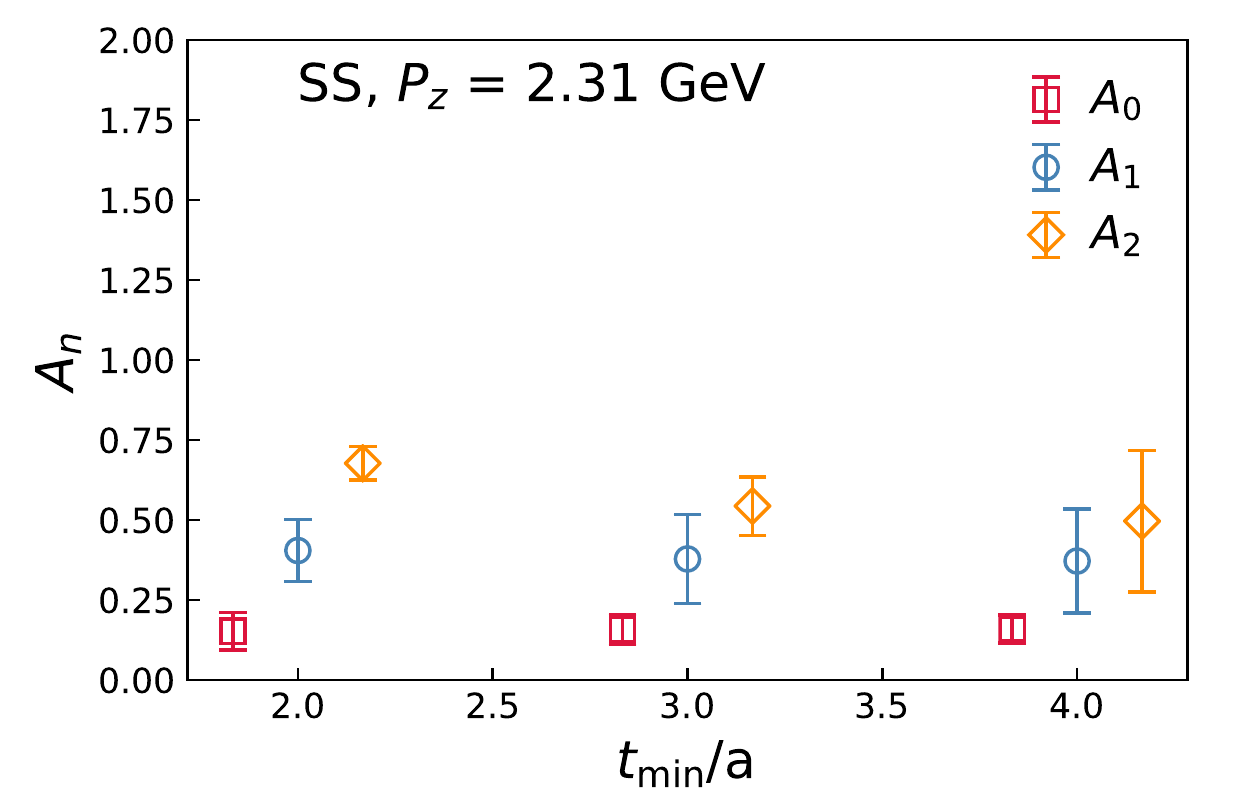}
\caption{The amplitudes of different states obtained from constrained 3-state fit of SS correlator and normalized by $C_{2pt}^{SS}(t=0)$ as function of temperature.}
\label{fig:fit2pt_A_SS}
\end{figure}

Similar analysis was performed
for SS correlators and the results for the excited state energies and amplitudes
are shown in Fig. \ref{fig:fit2pt_E1_SSadd} and Fig. \ref{fig:fit2pt_A_SS}, respectively.
From these figures we see that a pseudo-plateau develops for the first excited states
for $5<t_{\rm min}<10$ of 2-state fit. We see that $A_0$ and $A_1$ are similar in this case, and $A_2$ decreases as $t_{\rm min}$ increasing.

\section{Analysis of the three point function} \label{app:3pt}

\begin{figure}
\includegraphics[width=0.3\textwidth]{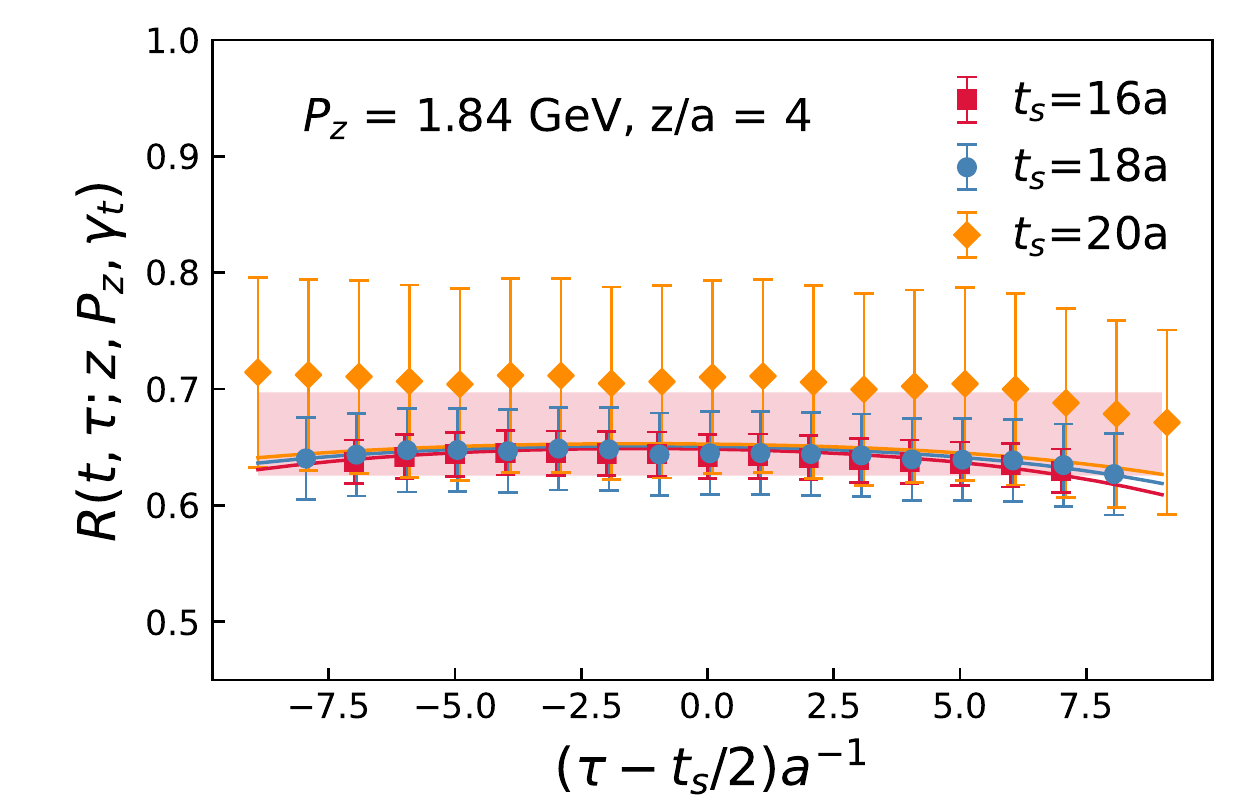}
\includegraphics[width=0.3\textwidth]{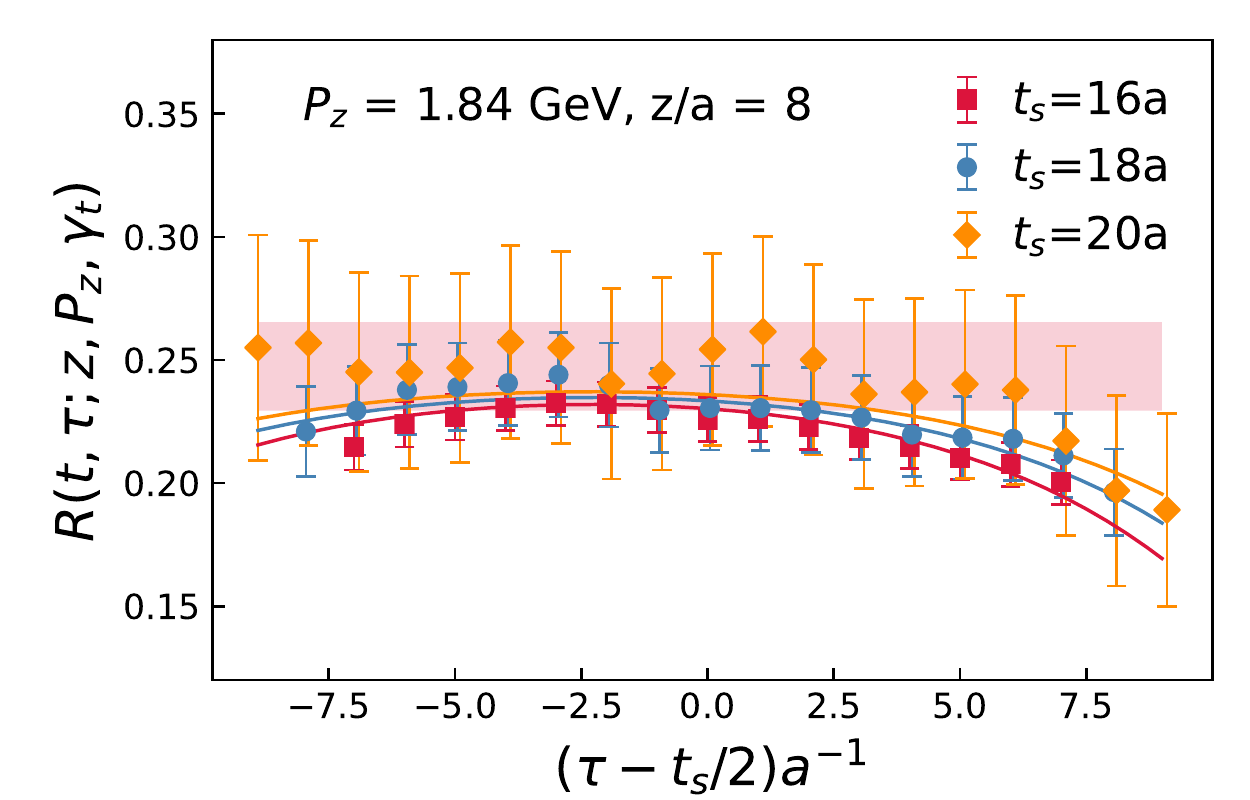}
\includegraphics[width=0.3\textwidth]{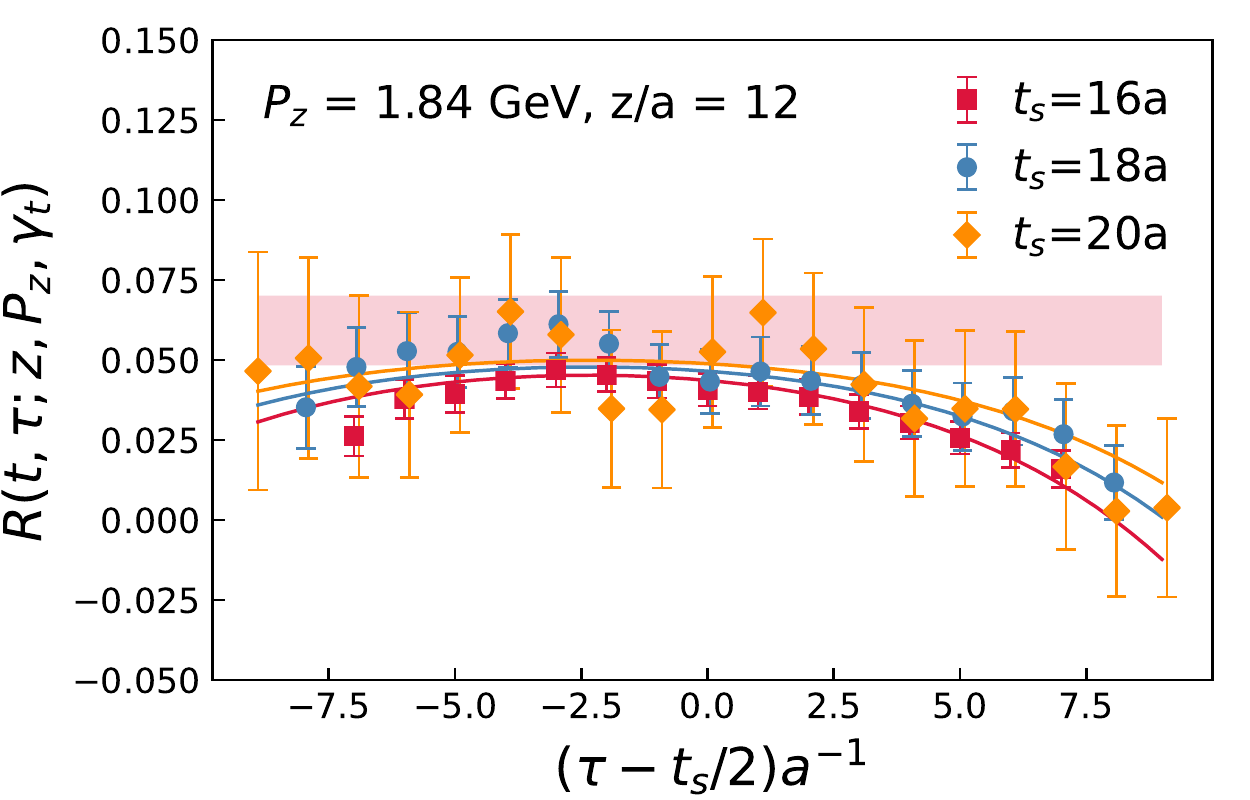}
\includegraphics[width=0.3\textwidth]{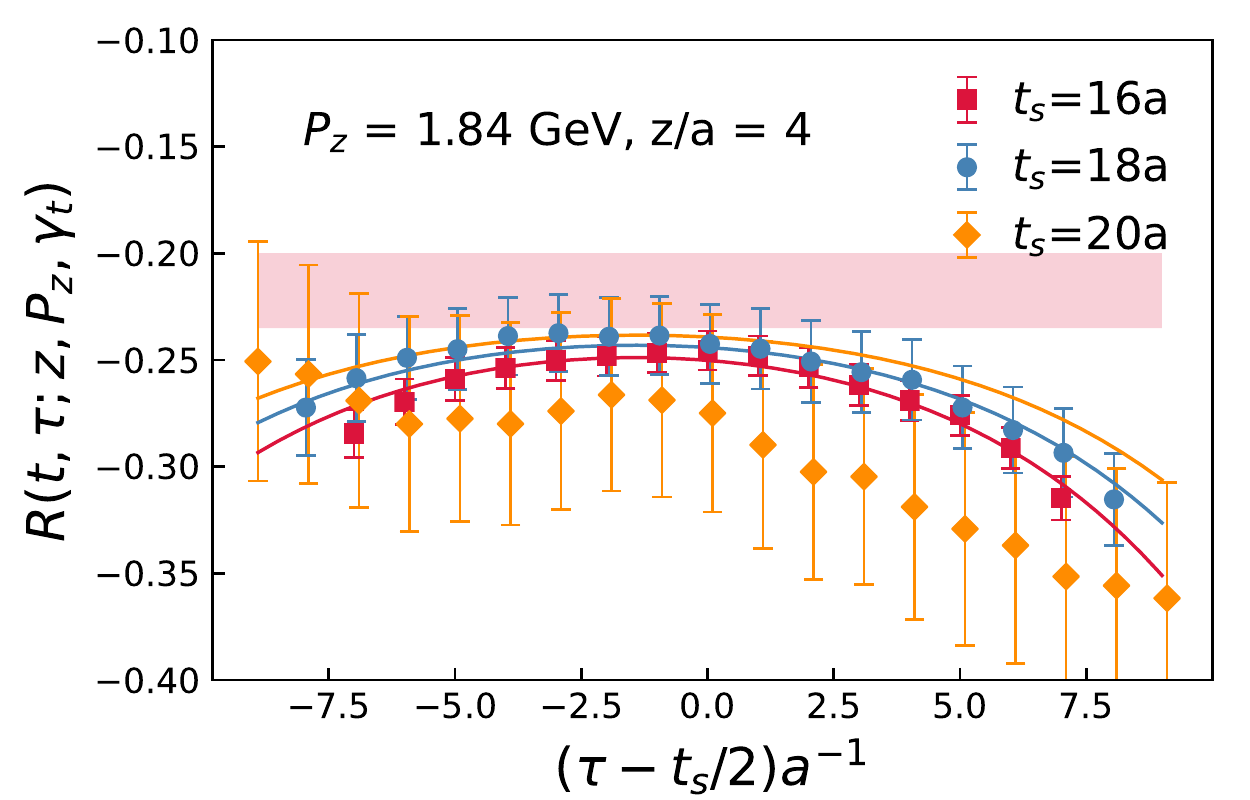}
\includegraphics[width=0.3\textwidth]{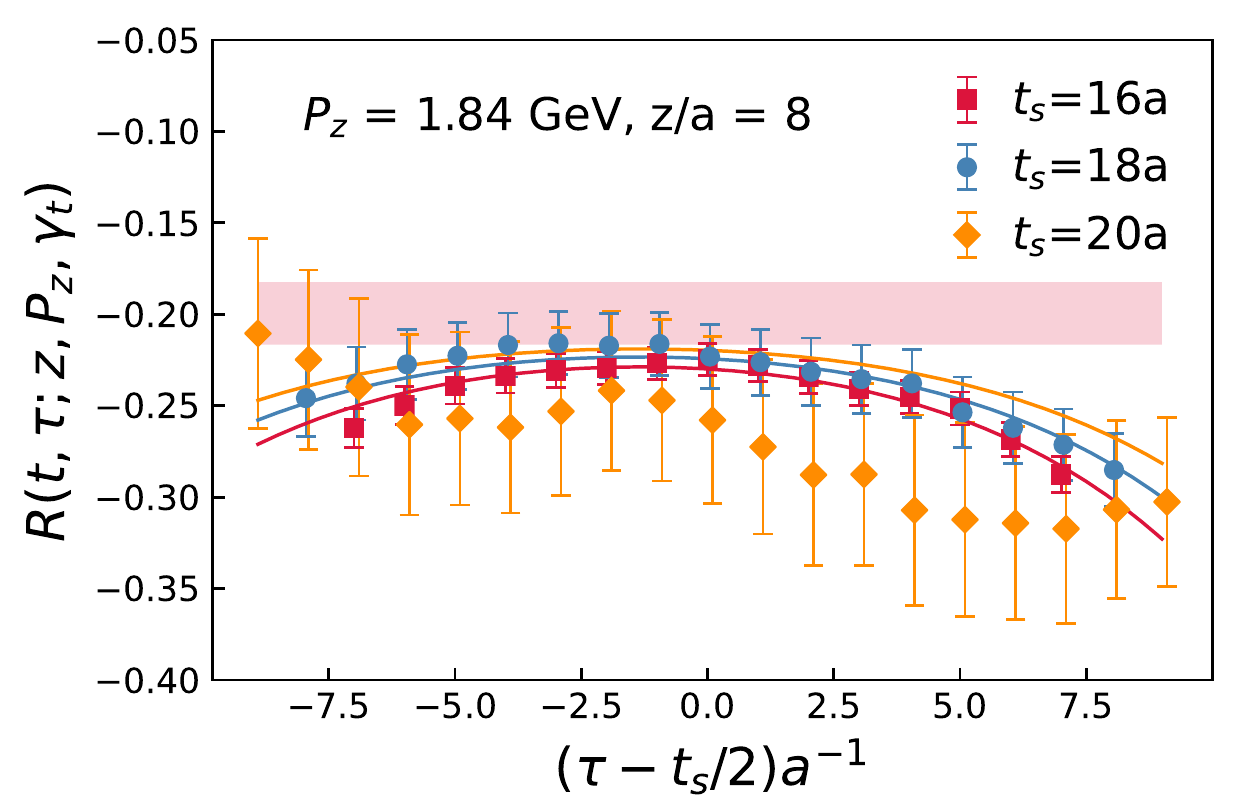}
\includegraphics[width=0.3\textwidth]{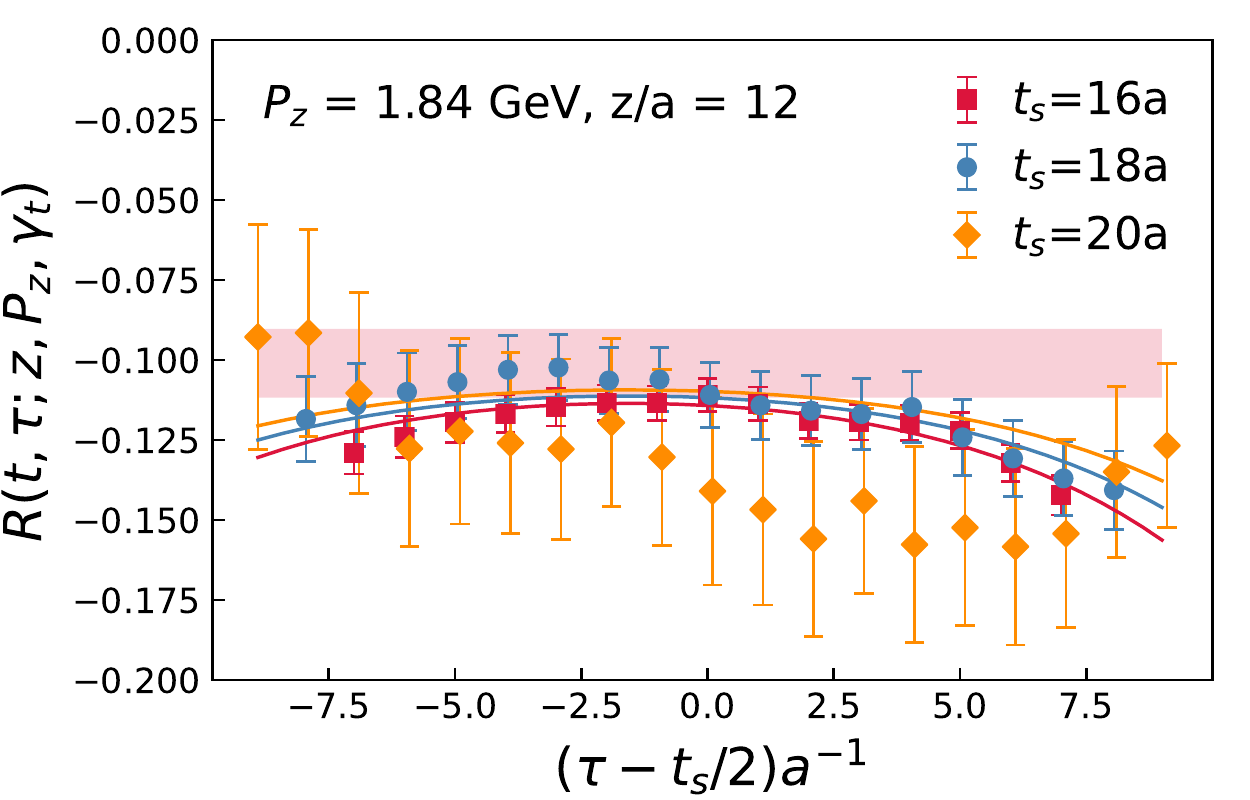}
\caption{
The ratio of the 3-point function to the 2-point function for z=4,8,12 and $n_z=4$. The upper panels show
the real part, while the imaginary part is shown in the lower panels. The results of $R^{fit}_1$
are shown as lines.}
\label{fig:samplefits_nz4}
\end{figure}

\begin{figure}
\includegraphics[width=0.3\textwidth]{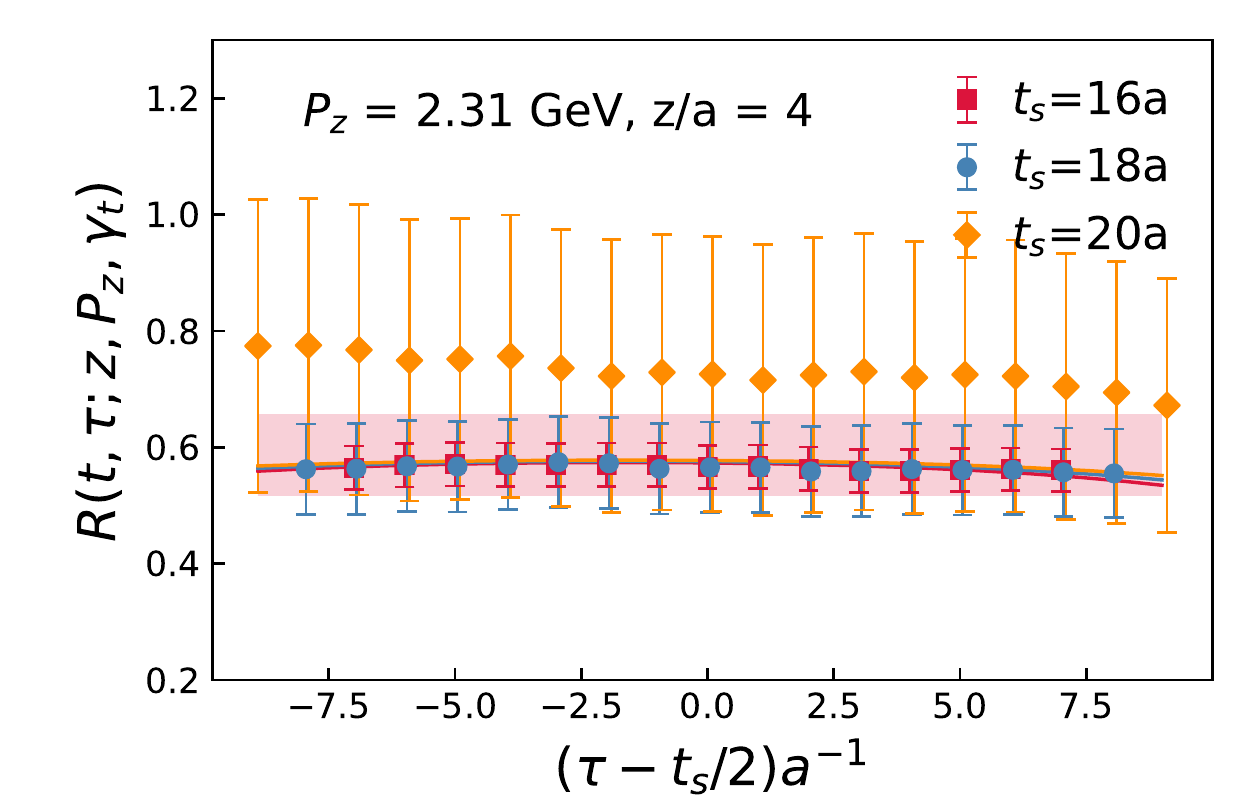}
\includegraphics[width=0.3\textwidth]{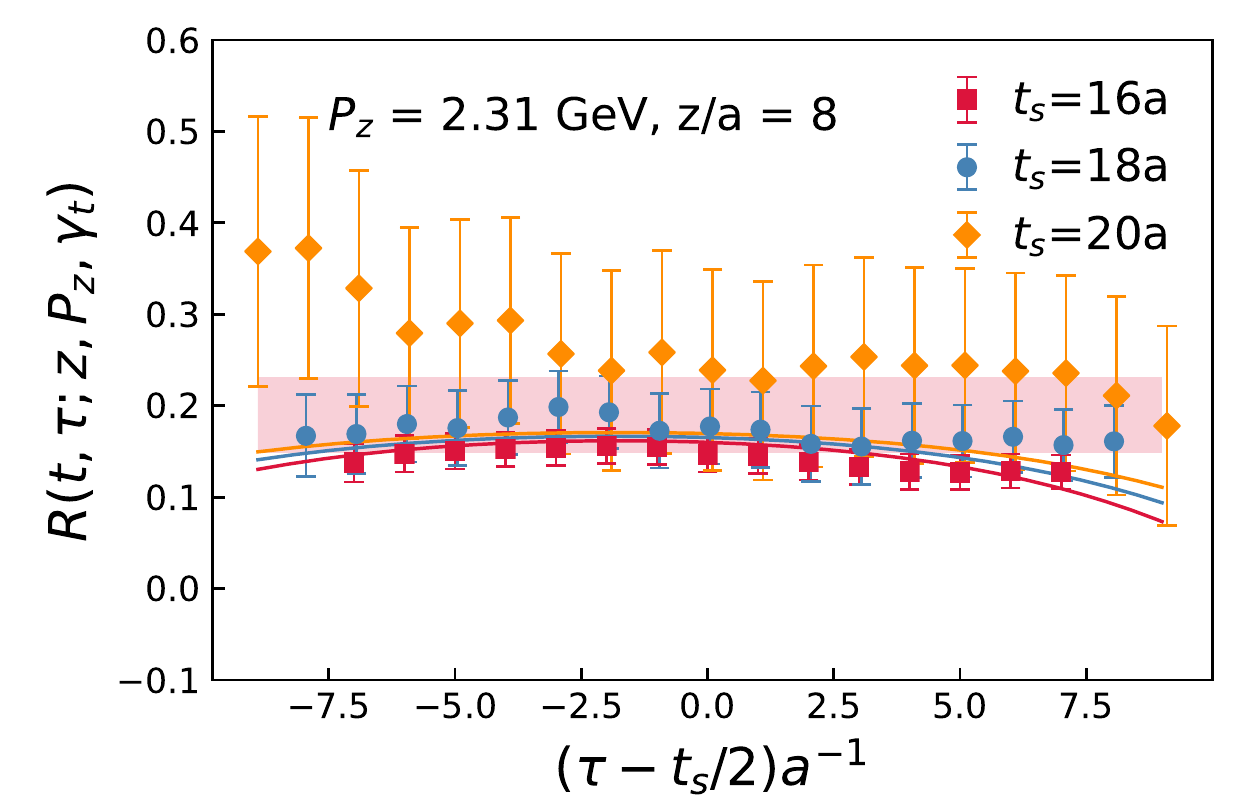}
\includegraphics[width=0.3\textwidth]{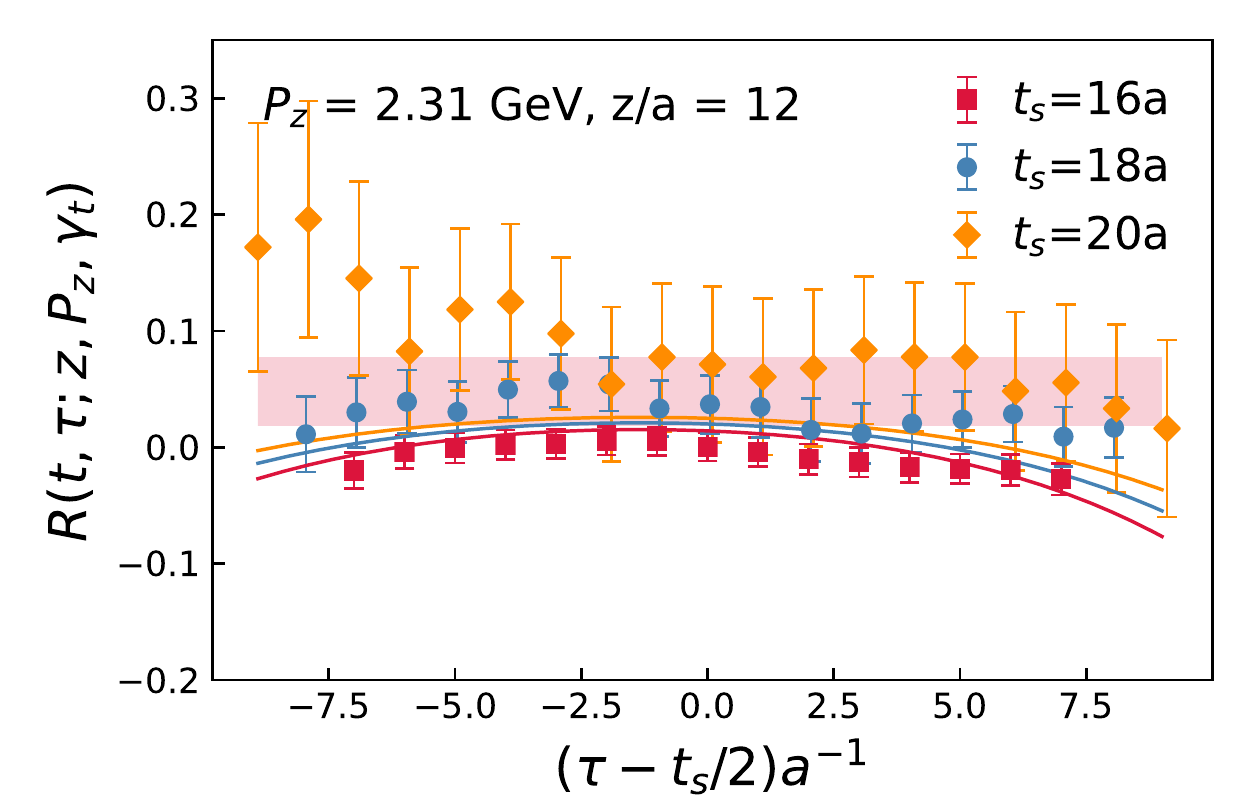}
\includegraphics[width=0.3\textwidth]{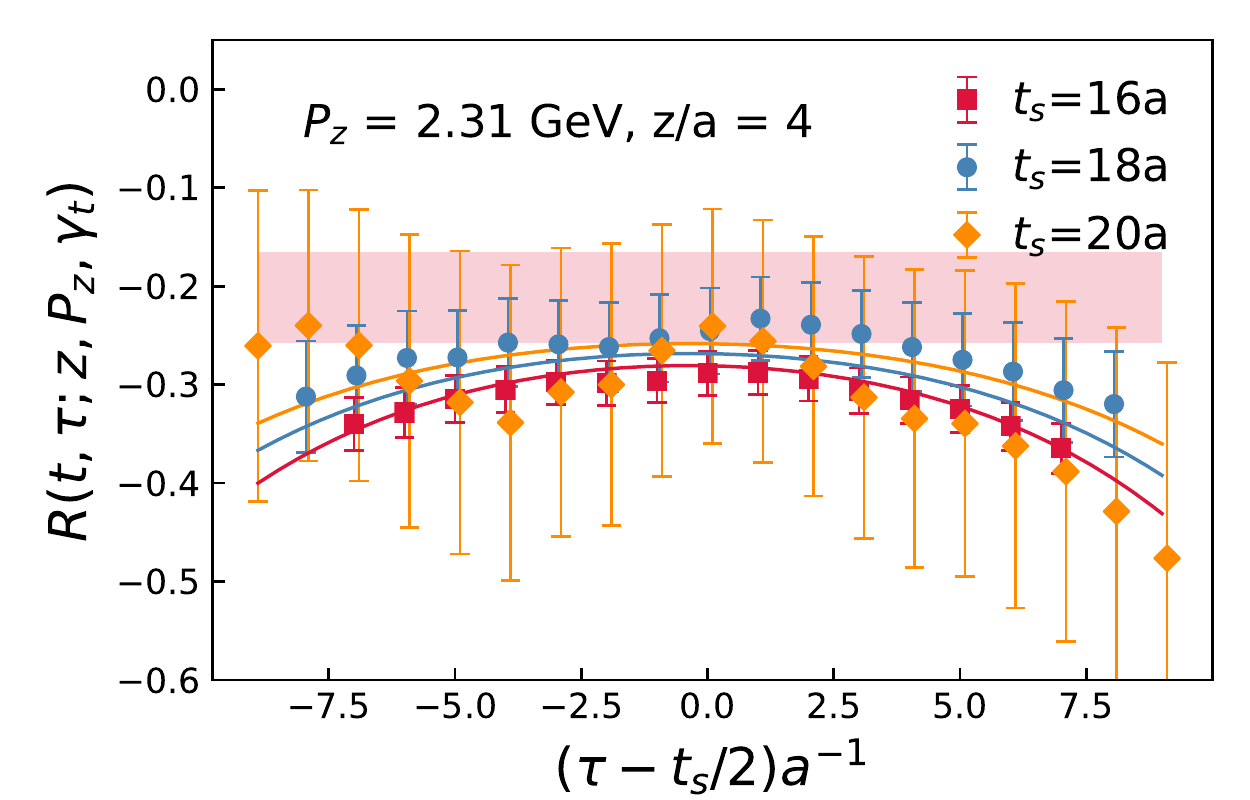}
\includegraphics[width=0.3\textwidth]{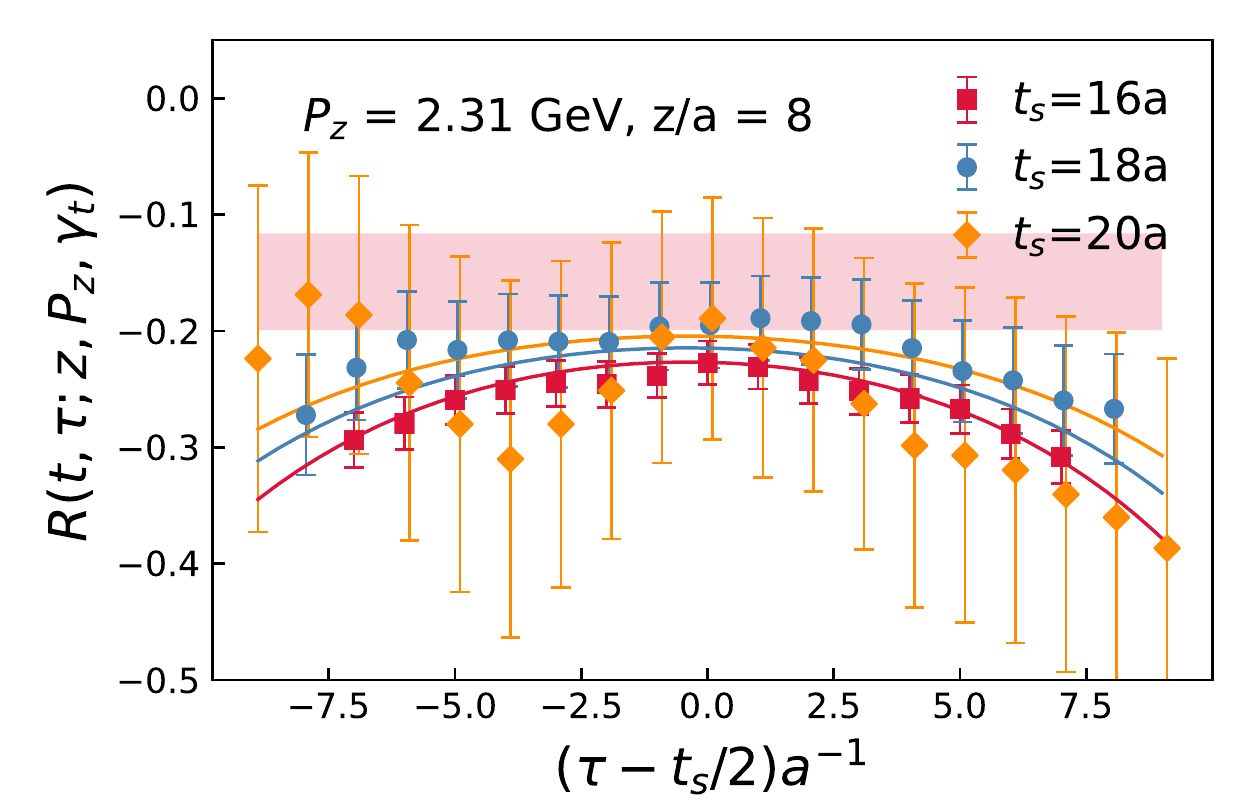}
\includegraphics[width=0.3\textwidth]{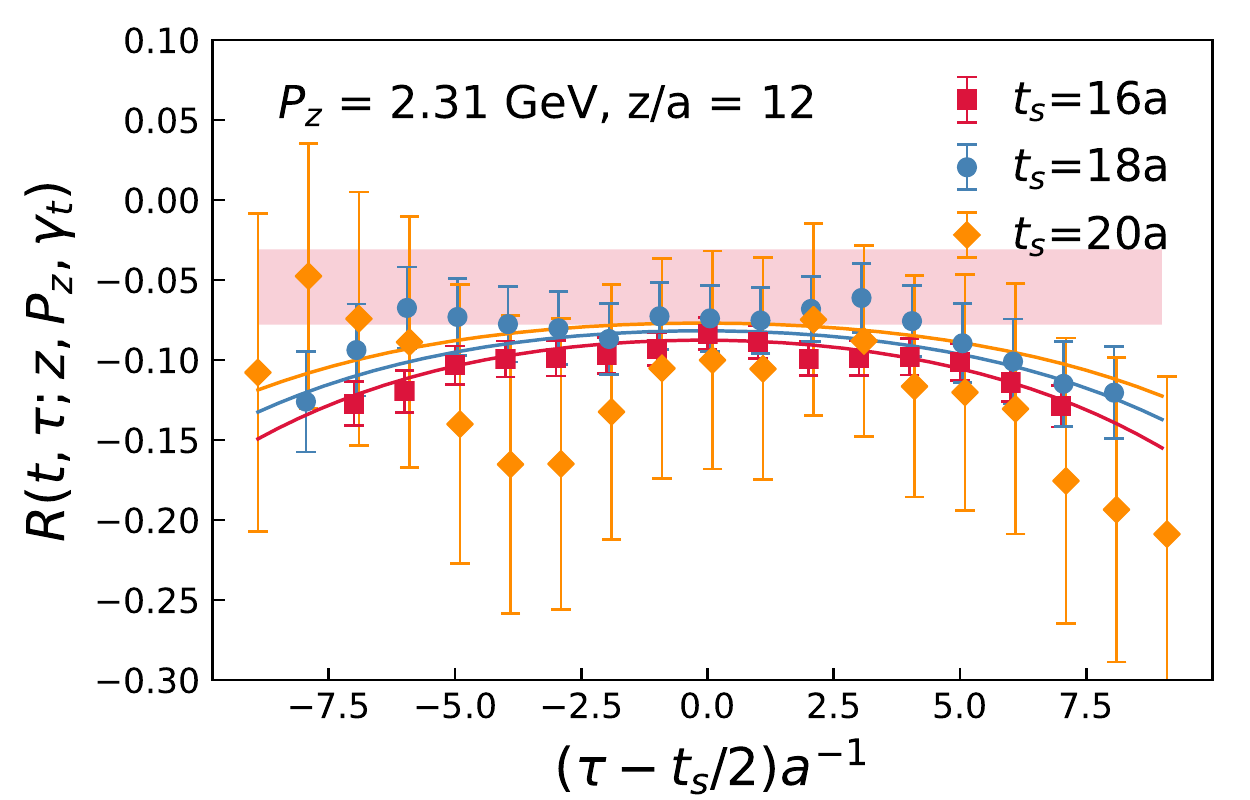}
\caption{
The ratio of the 3-point function to the 2-point function for z=4,8,12 and $n_z=5$. The upper panels show
the real part, while the imaginary part is shown in the lower panels. The results of $R^{fit}_1$
are shown as lines.}
\label{fig:samplefits_nz5}
\end{figure}

In this appendix we discuss further details of the extraction of the bare matrix element of
the qPDF operator. First, we show our results for the ratio of the 3-point function to two point function
for different source sink separation and different values of $z$ as function of the operator
insertion time $\tau$  in Fig. \ref{fig:samplefits_nz4} for $n_z=4$. In this figure we also show
the results for $R^{fit}_1$. As one can see from the figure $R^{fit}_1$ can describe the data well for all values of
$t$. In Fig. \ref{fig:samplefits_nz5} we show the same analysis but for $n_z=5$.

\begin{figure}
        \includegraphics[width=0.45\textwidth]{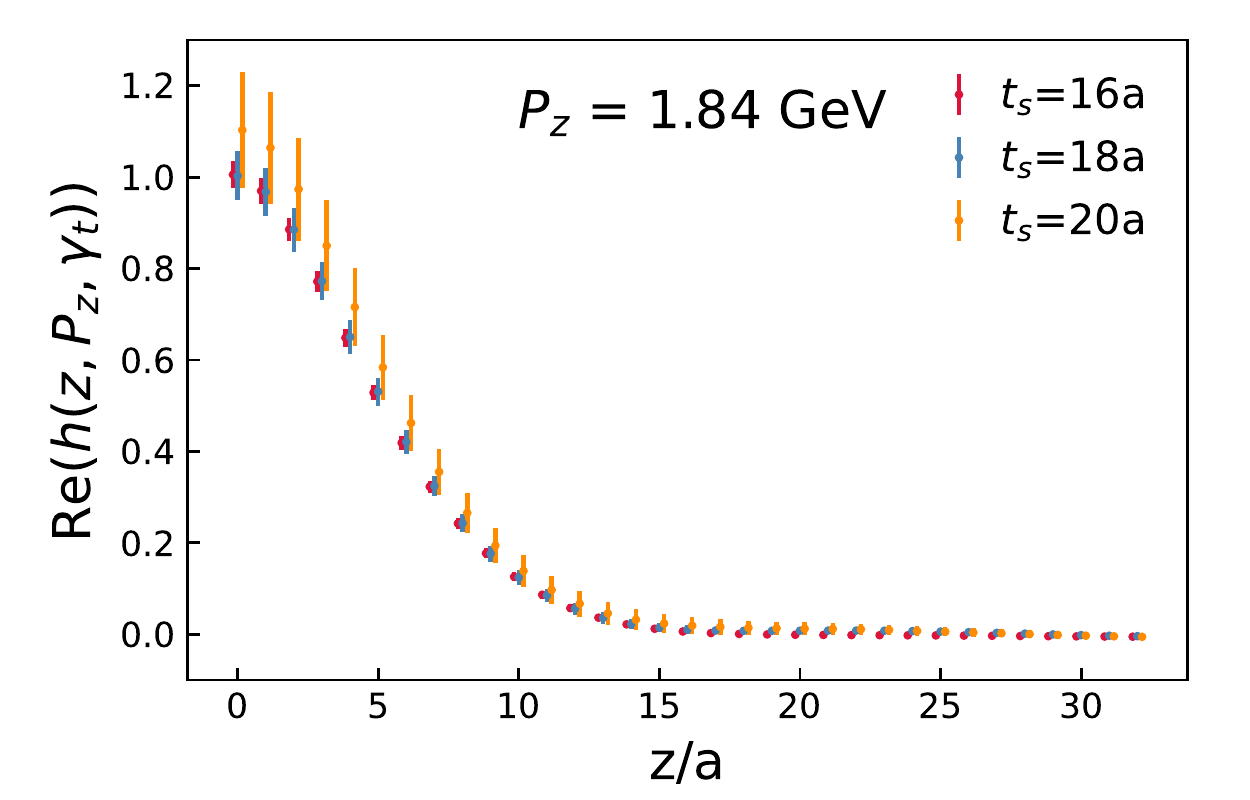}
        \includegraphics[width=0.45\textwidth]{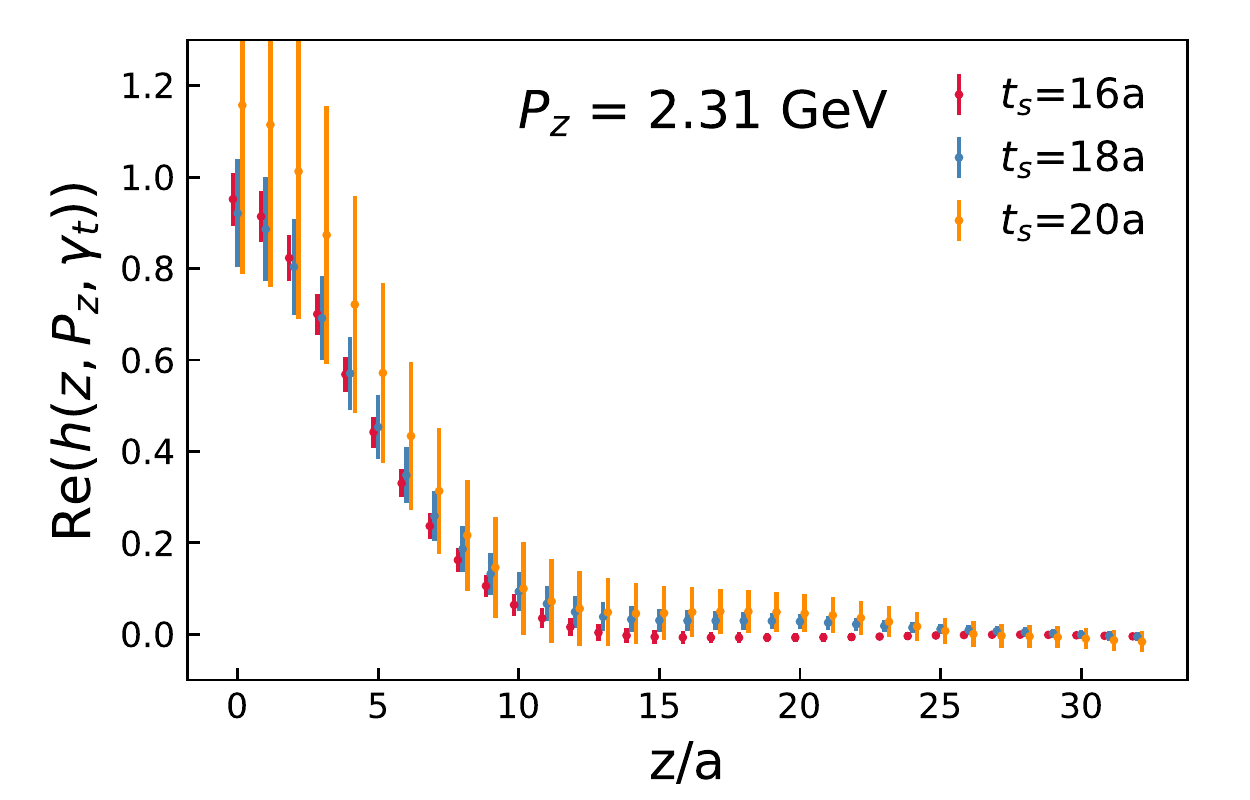}
        \caption{The $z$-dependence of the qPDF matrix element obtained using $R^{fit}_1$ with single value of
the source sink separation for $n_z=4,5$.}
\label{fig:single_t_fit}
\end{figure}

\begin{figure}
        \includegraphics[width=0.45\textwidth]{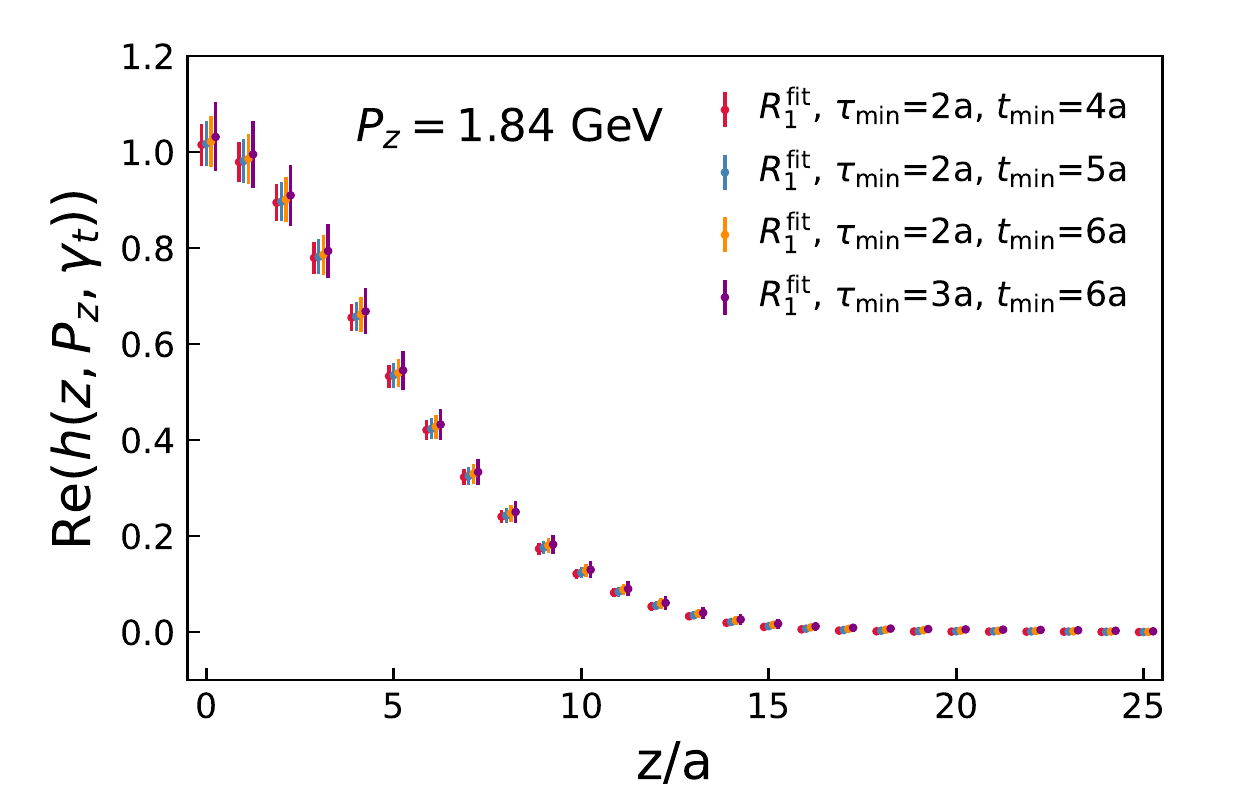}
        \includegraphics[width=0.45\textwidth]{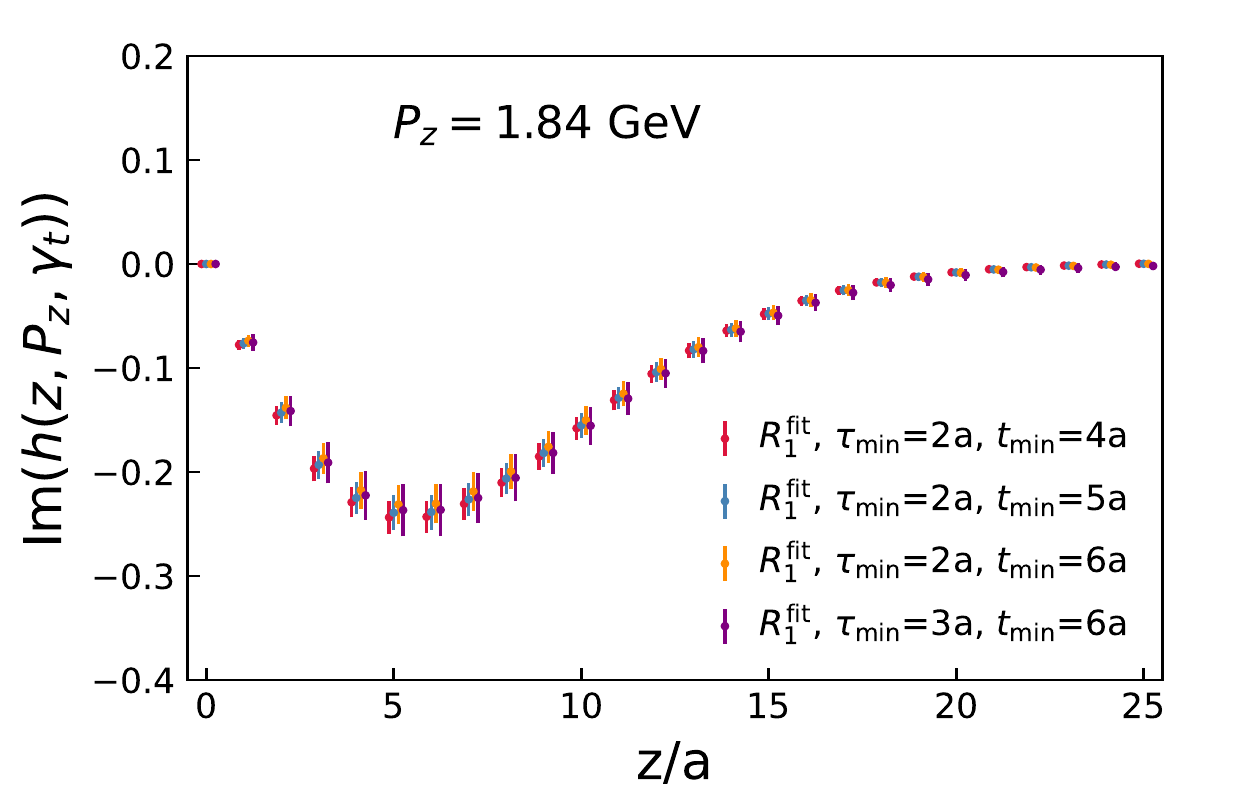}
        \includegraphics[width=0.45\textwidth]{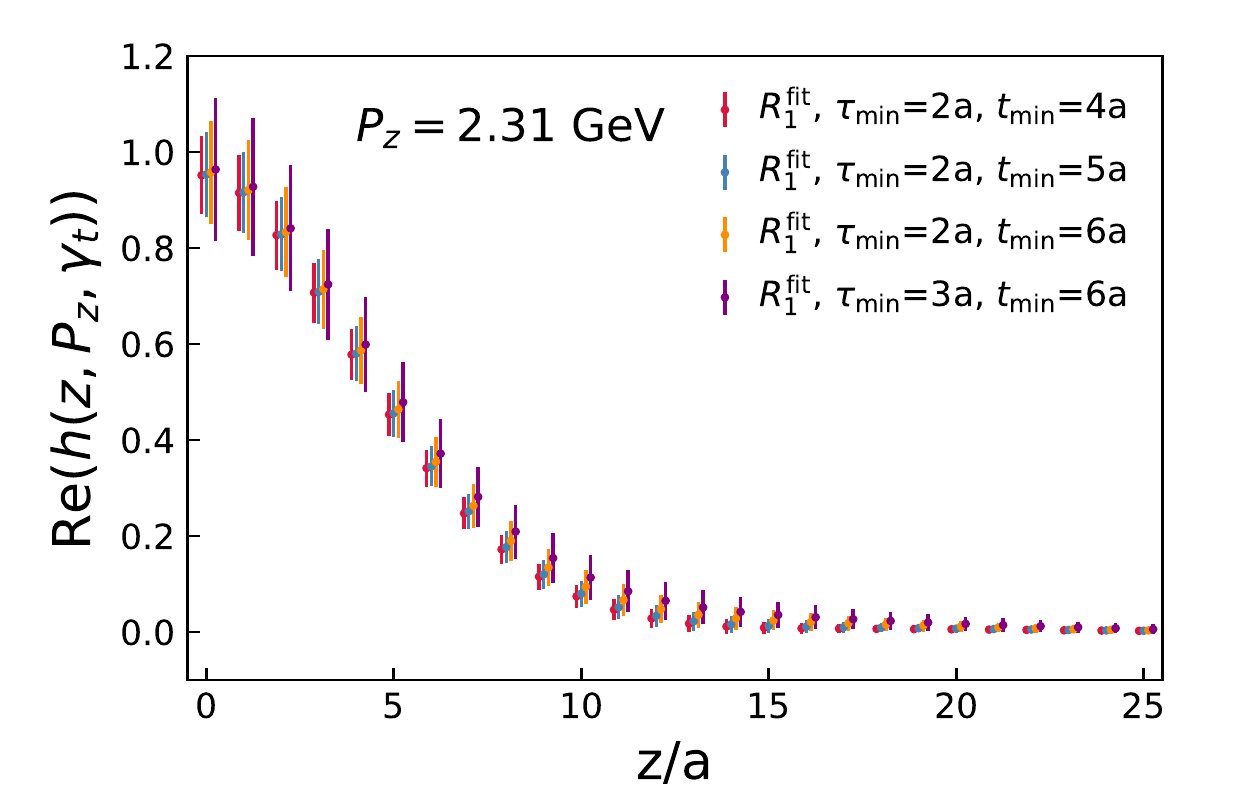}
        \includegraphics[width=0.45\textwidth]{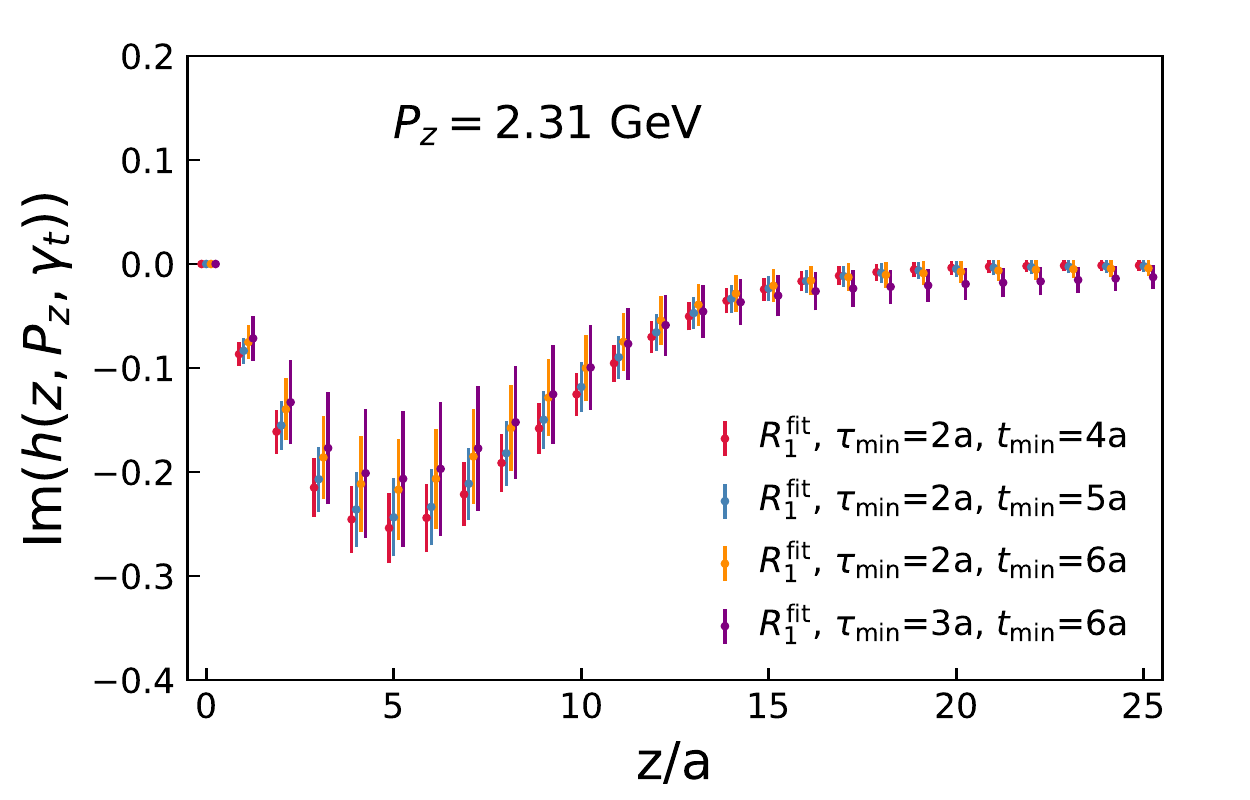}
\caption{Real (left) and imaginary (right) parts of the bare matrix as function of $z$.
The top panel show the result for $n_z=4$, the bottom panel show the results for $n_z=5$.
The results for different choices of $\tau_{\rm min}$ and $t_{\rm min}$ in the 2-point function fits
are shown. }
\label{fig:tmin_tau_min_dep}
\end{figure}

As discussed in the main text we performed $R^{fit}_1$ using single value of source sink separation.
The results are shown in Fig. \ref{fig:single_t_fit} for the real part of
the matrix element. As one can see from the figure the results
obtained from this fit for $t=16$, $18$ and $20$ agree within errors.
We performed fits using the form $fit_1$ with
$\tau>\tau_{\rm min}$ and taking the value of $E_1$ from the 2-point function fit with $t>t_{\rm min}$.
The results are shown in Fig. \ref{fig:tmin_tau_min_dep}. We see no significant
dependence on $\tau_{\rm min}$ and $t_{\rm min}$.

\begin{figure}
\includegraphics[width=0.41\textwidth]{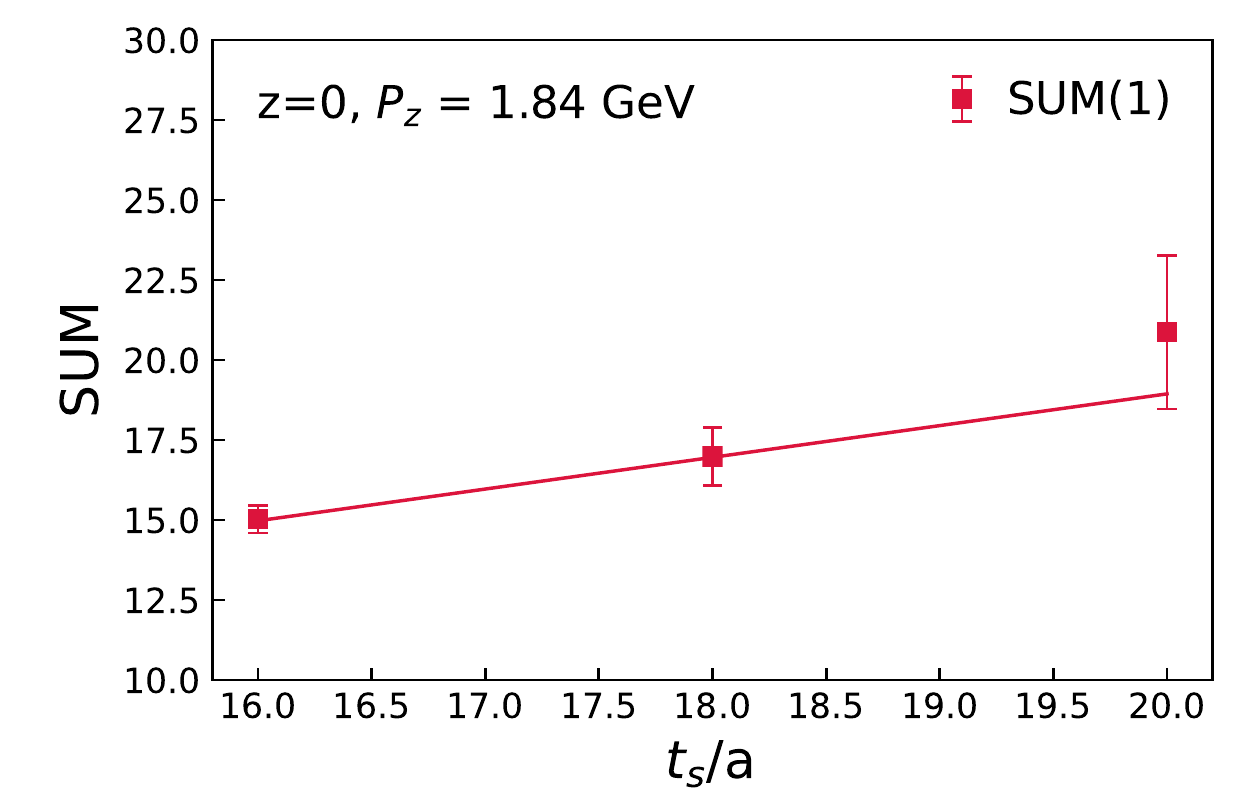}
\includegraphics[width=0.41\textwidth]{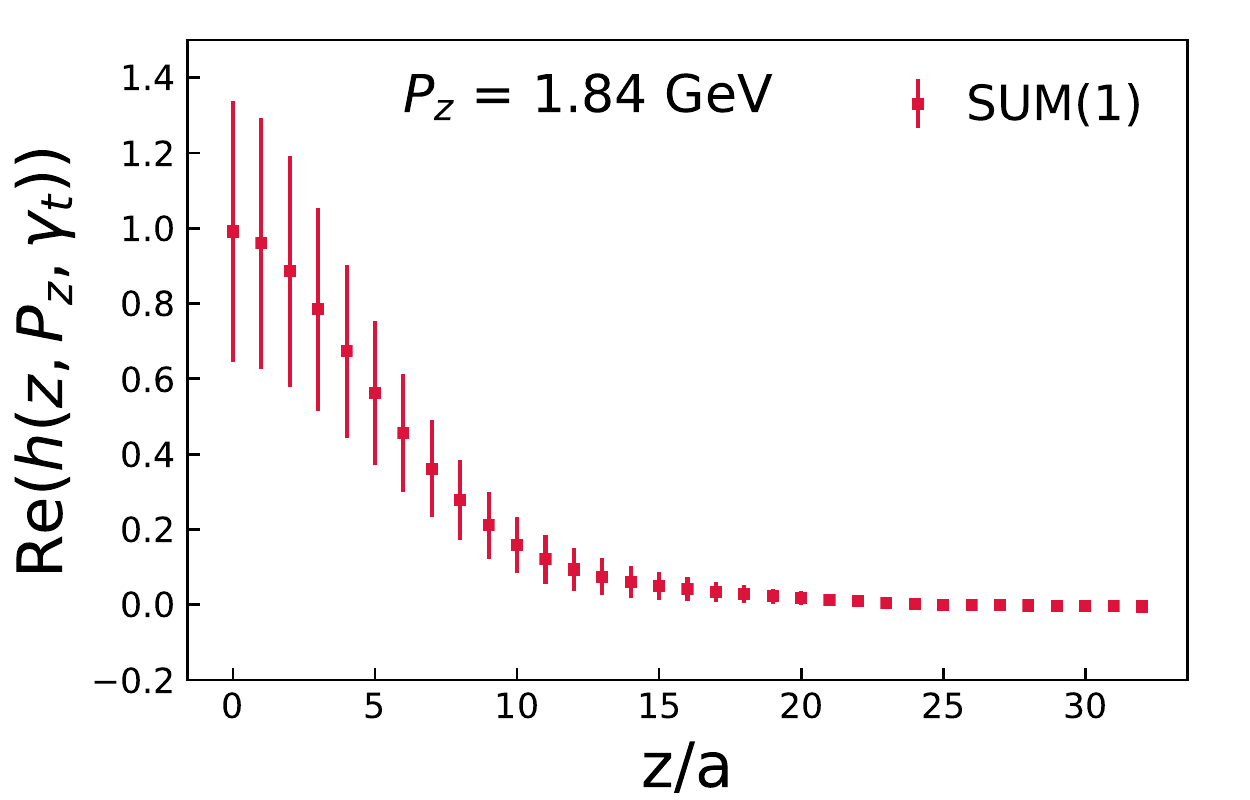}
\caption{The $t$ dependence of the sum of the ratio of the three point function to two point function (left) and the
$z$ dependence of the matrix element extracted from the summation method (right). SUM(n) means summation fit with n skipped time insertion.}
\label{fig:sum}
\end{figure}

\begin{figure}
\includegraphics[width=0.41\textwidth]{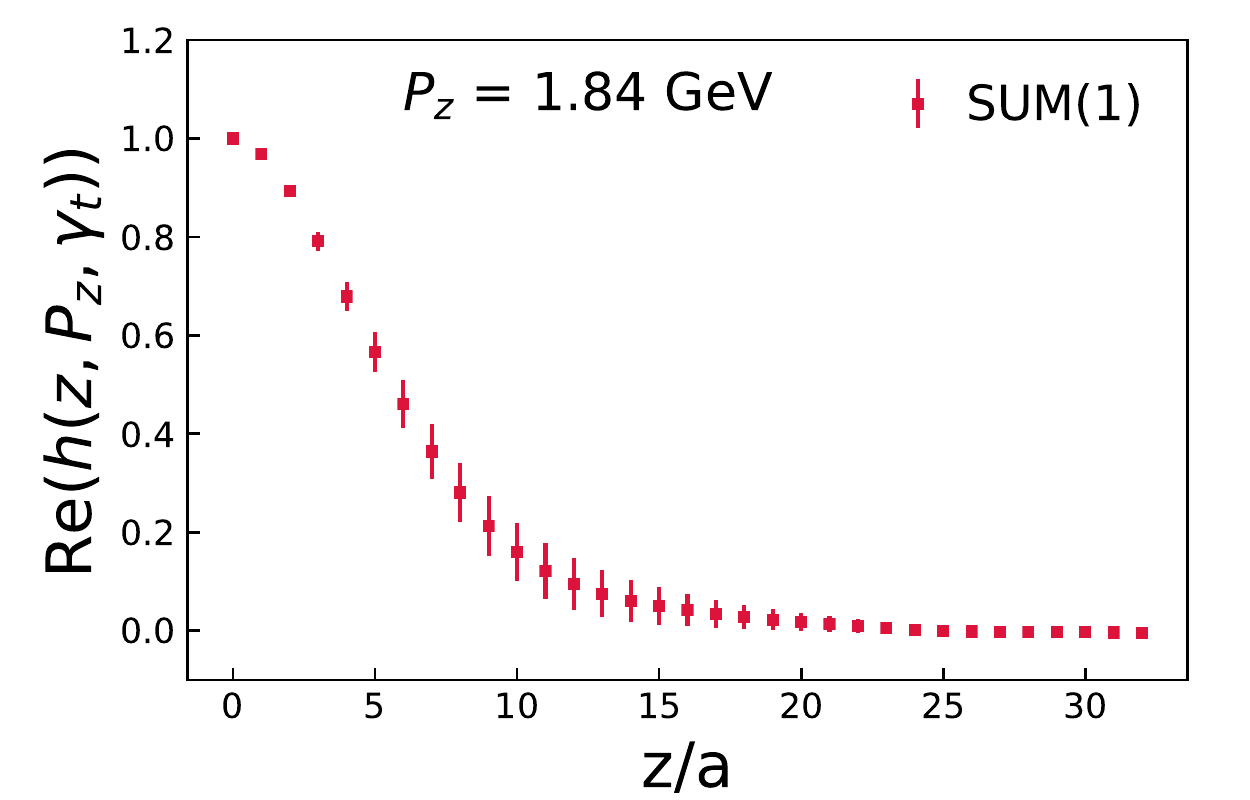}
\caption{The $z$-dependence of the real part of the bare qPDF matrix element obtained
by summation method after division by the matrix element for $z=0$ at $n_z=4$. SUM(n) means summation fit with n skipped time insertion.}
\label{fig:rath_nz4_sum}
\end{figure}

Another way to obtain the matrix element is to use the summation method. The summation method is illustrated in Fig. \ref{fig:sum}
for $n_z=4$. The results obtained from the summation method agree with those from $R^{fit}_1$ but have much larger errors.
The statistical errors of the $n_z=5$ data are too large to use the summation method.
Furthermore, we could also reduce the error in the summation method by dividing by
the matrix element at $z=0$ as can be seen in Fig. \ref{fig:rath_nz4_sum}

\begin{figure}
        \includegraphics[width=0.3\textwidth]{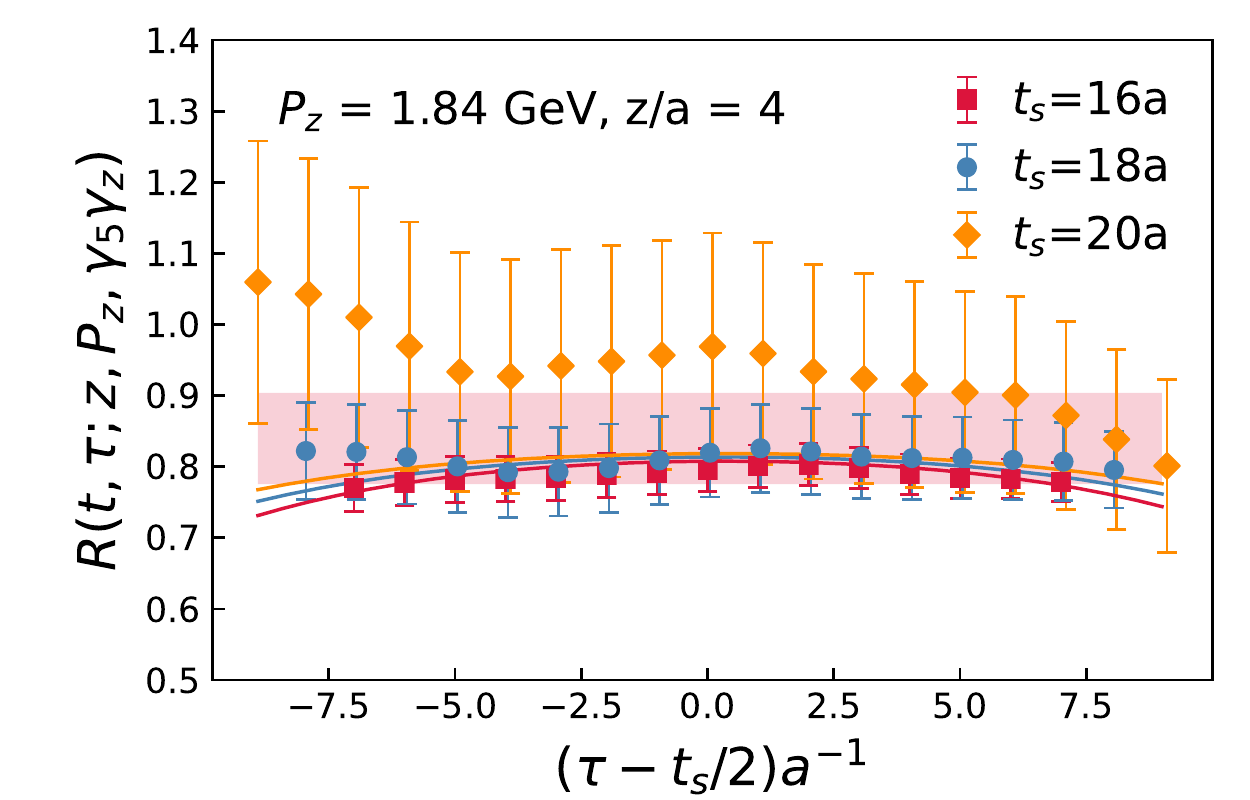}
        \includegraphics[width=0.3\textwidth]{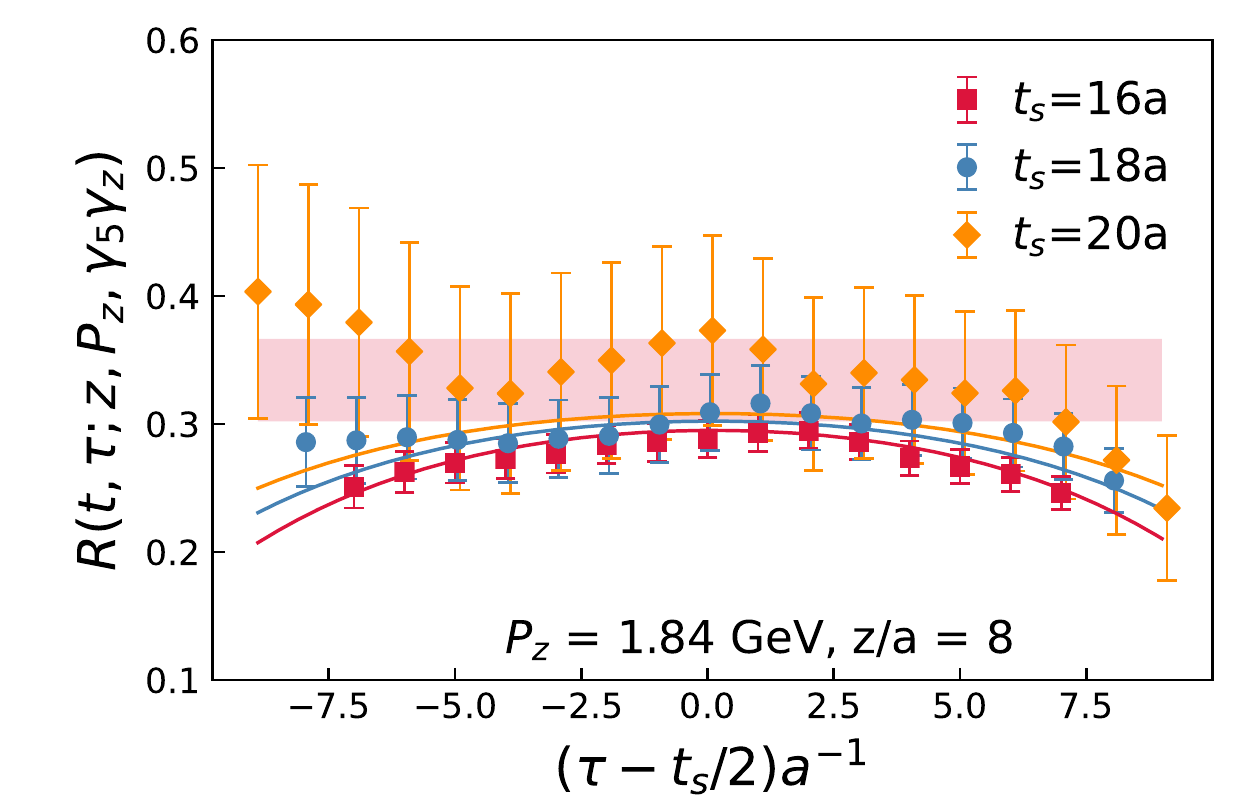}
        \includegraphics[width=0.3\textwidth]{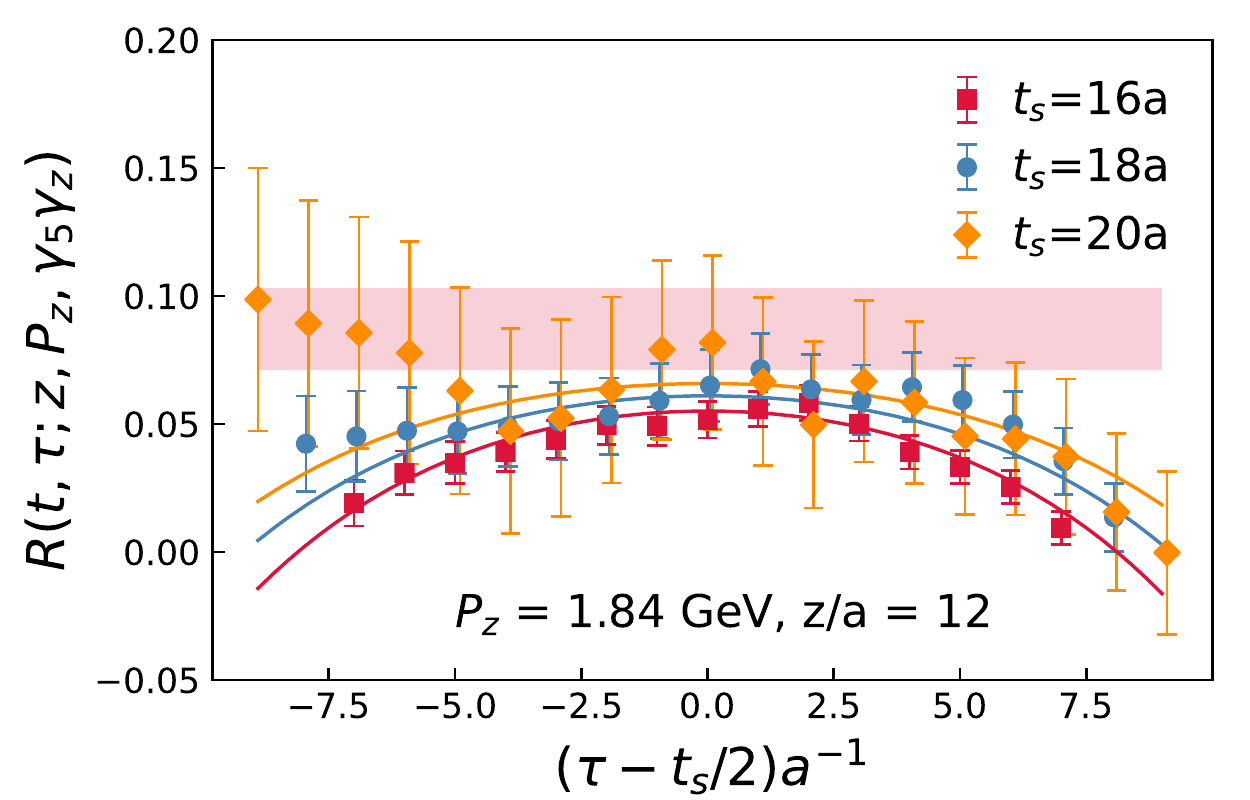}
        \includegraphics[width=0.3\textwidth]{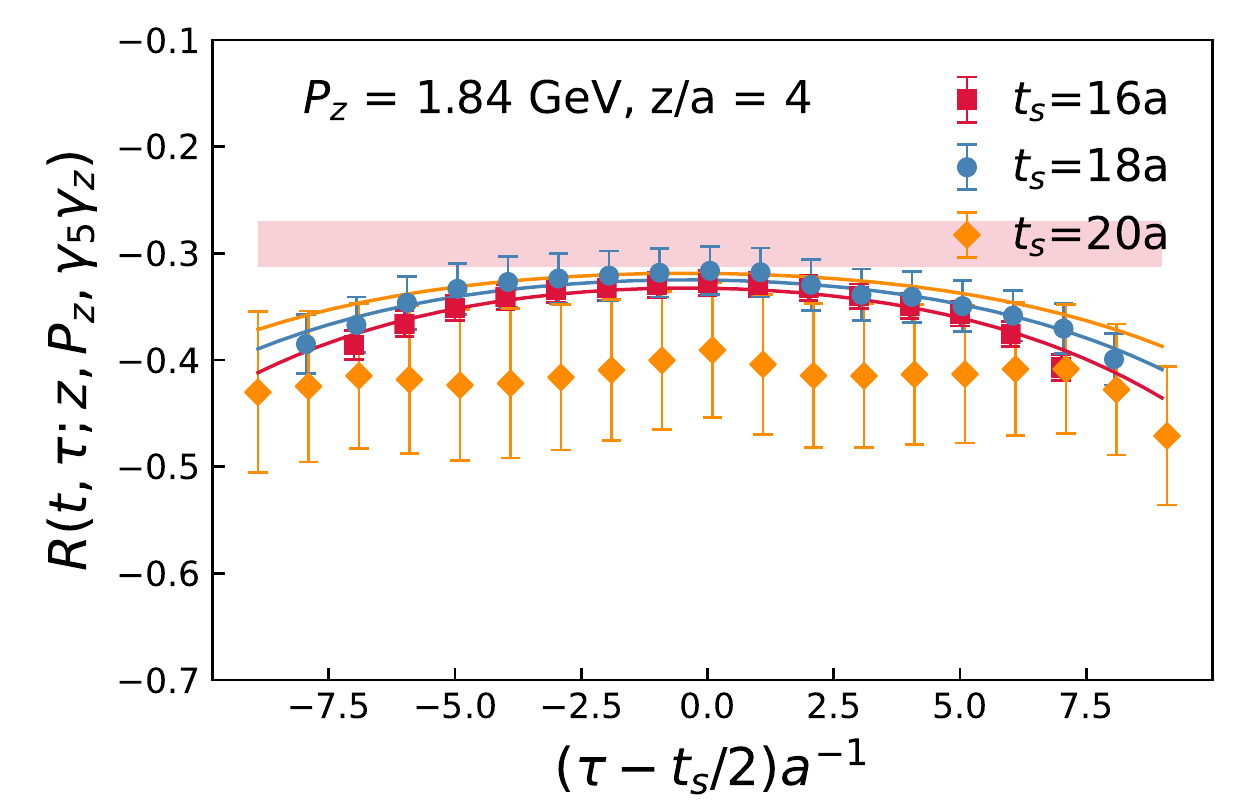}
        \includegraphics[width=0.3\textwidth]{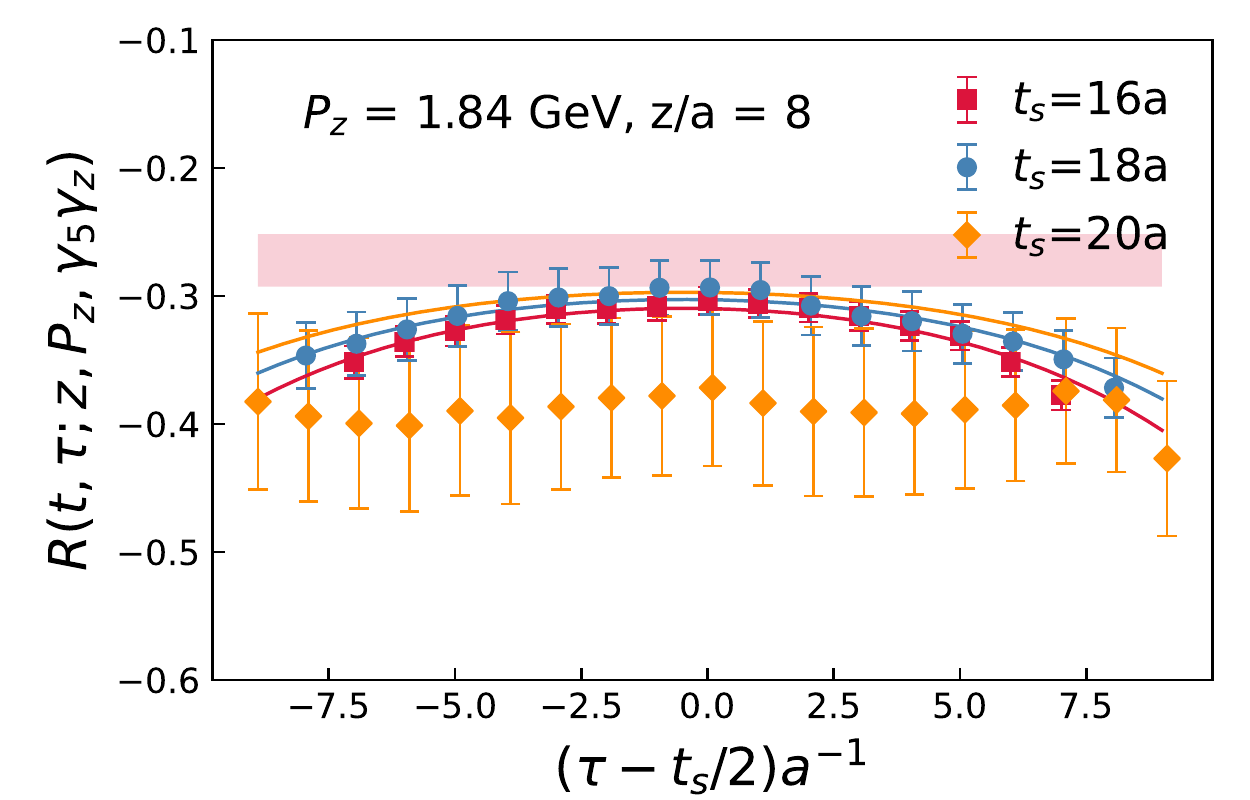}
        \includegraphics[width=0.3\textwidth]{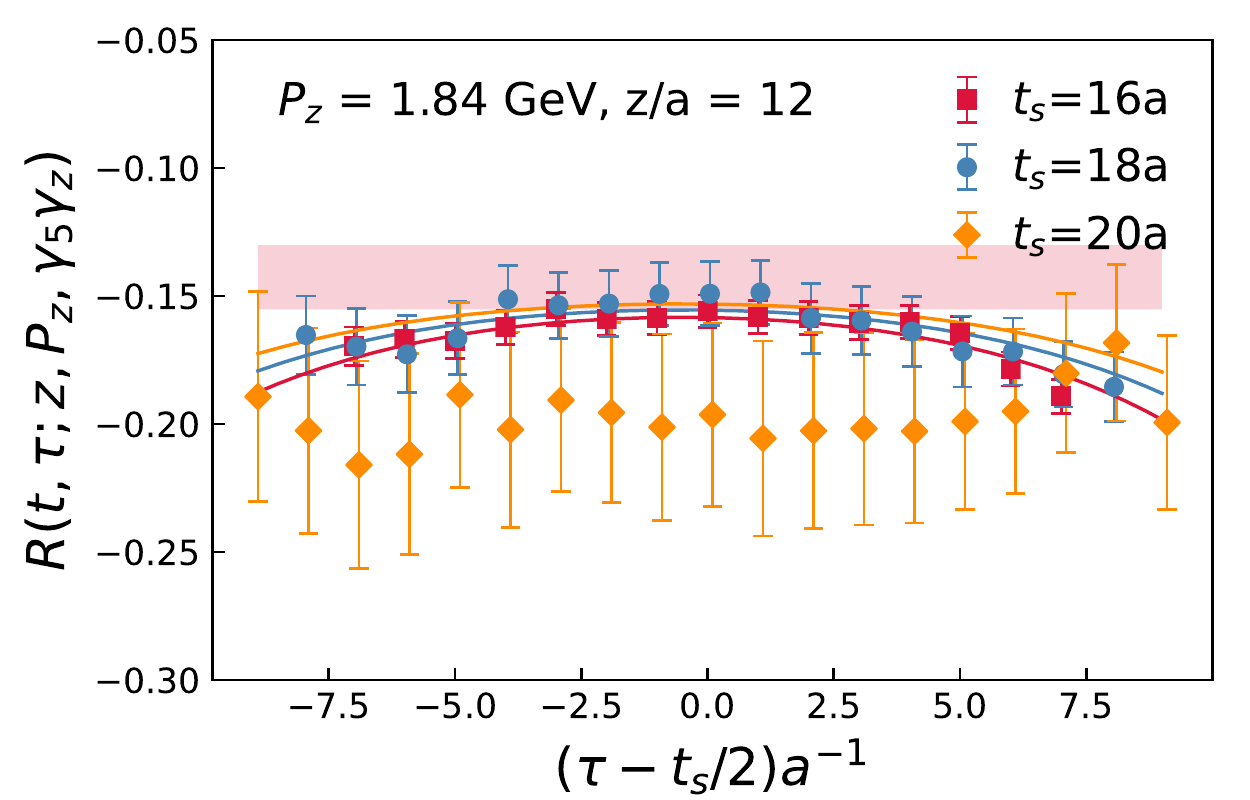}
\caption{
The ratio of the 3-point function to the 2-point function
corresponding to helicity qPDF
for z=4,8,12 and $n_z=4$. The upper panels show
the real part, while the imaginary part is shown in the lower panels. The results of $R^{fit}_1$
are shown as lines.}
\label{fig:samplefits_pol_nz4}
\end{figure}

\begin{figure}
        \includegraphics[width=0.3\textwidth]{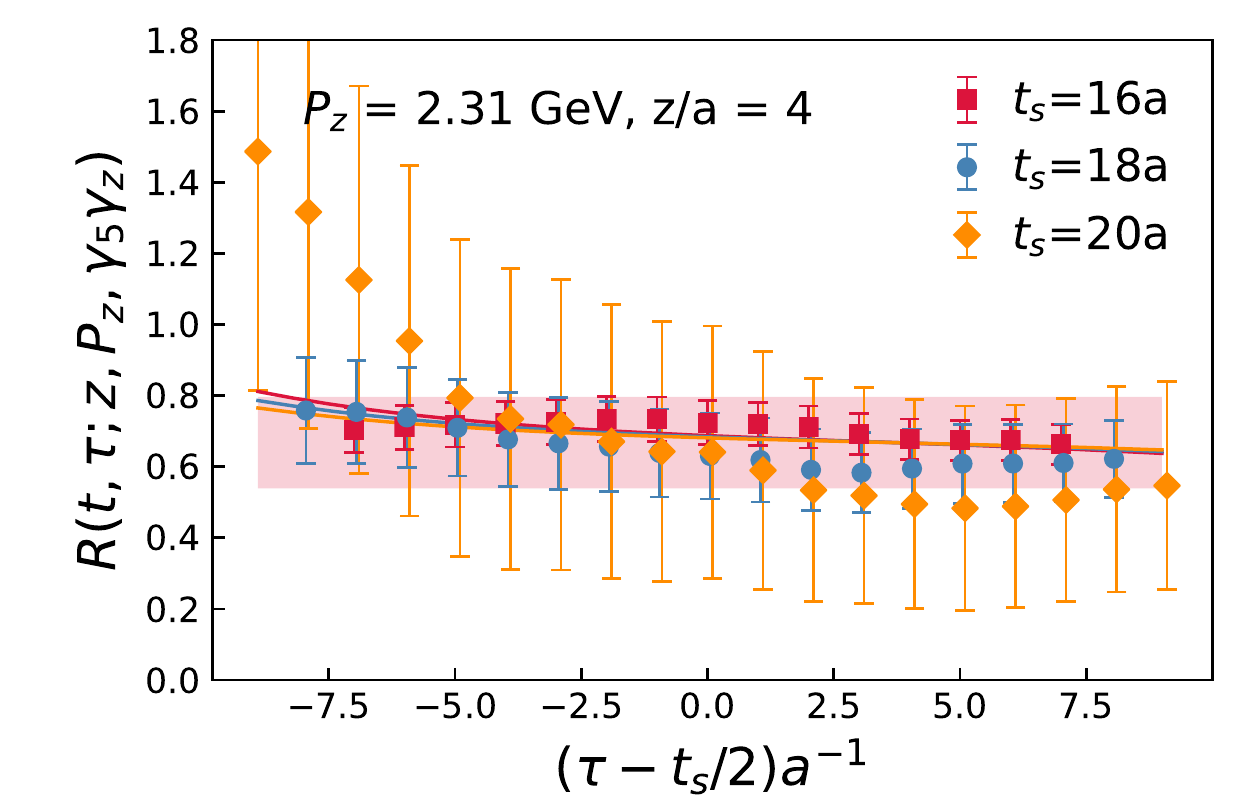}
        \includegraphics[width=0.3\textwidth]{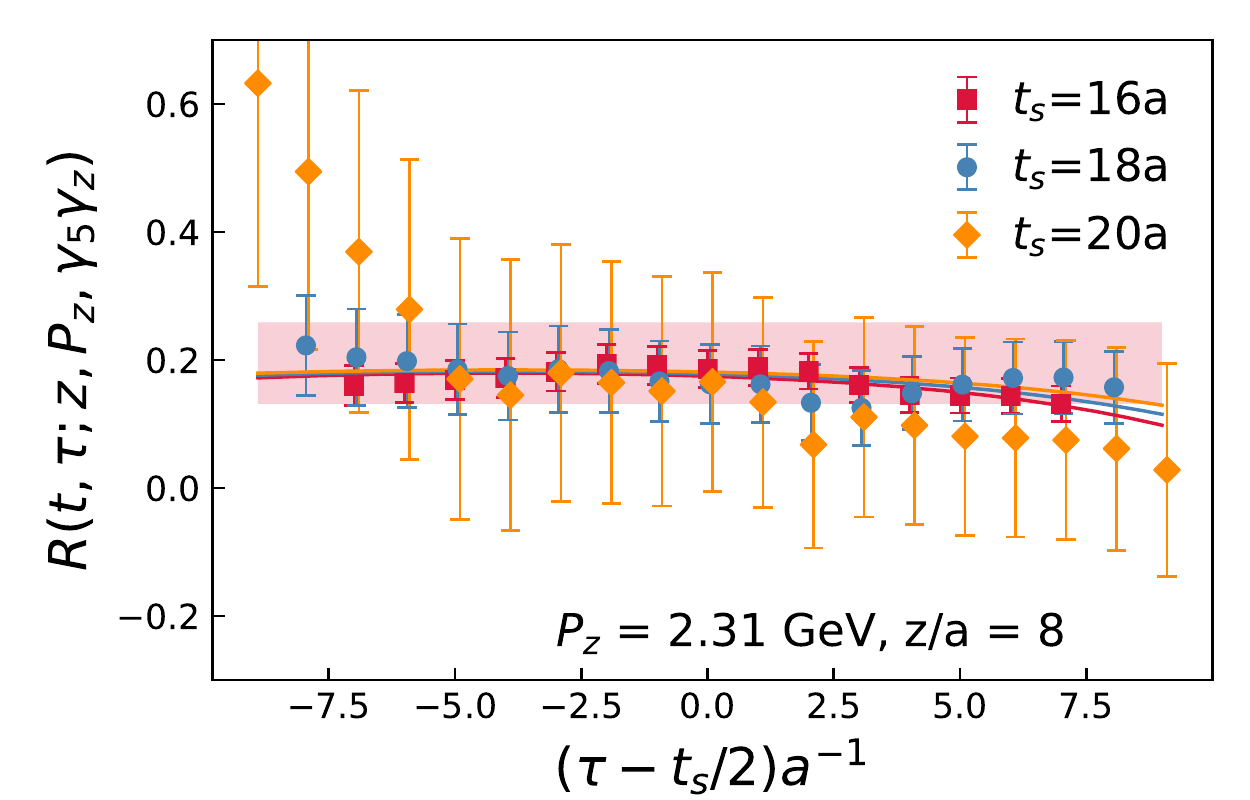}
        \includegraphics[width=0.3\textwidth]{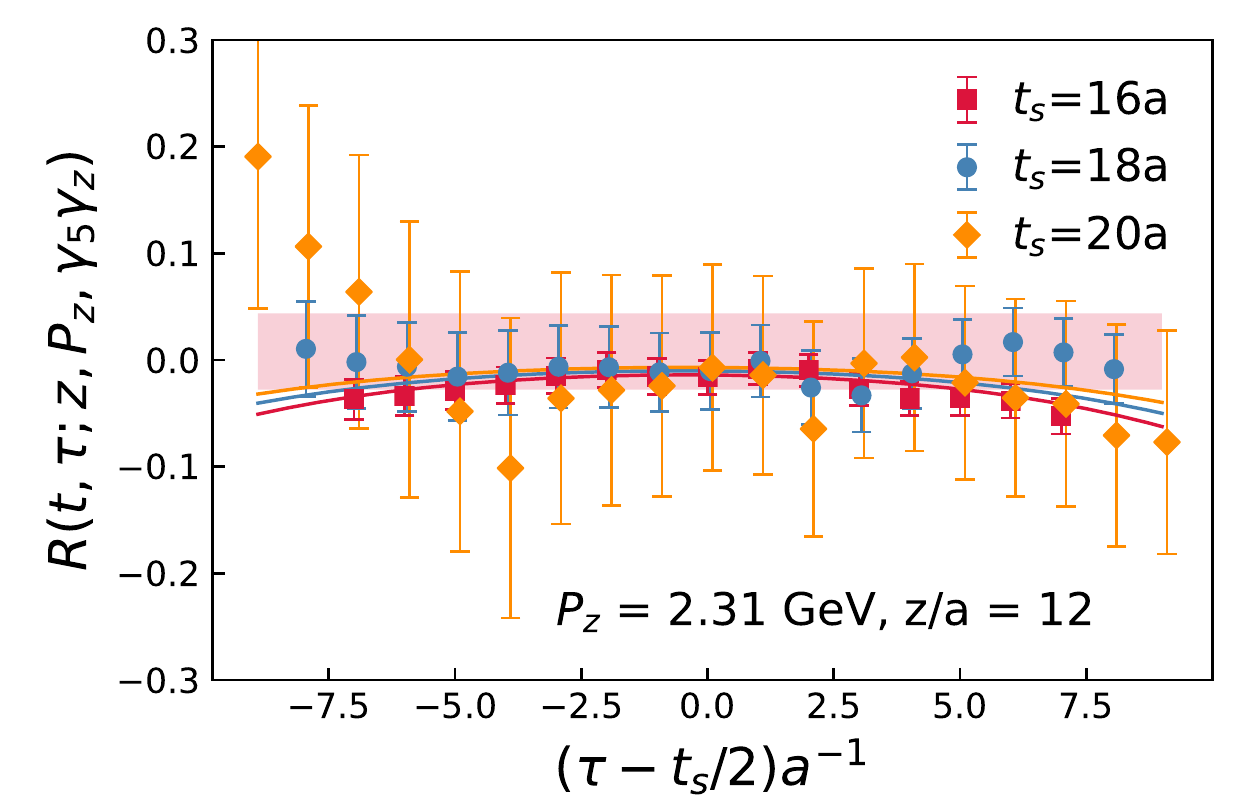}
        \includegraphics[width=0.3\textwidth]{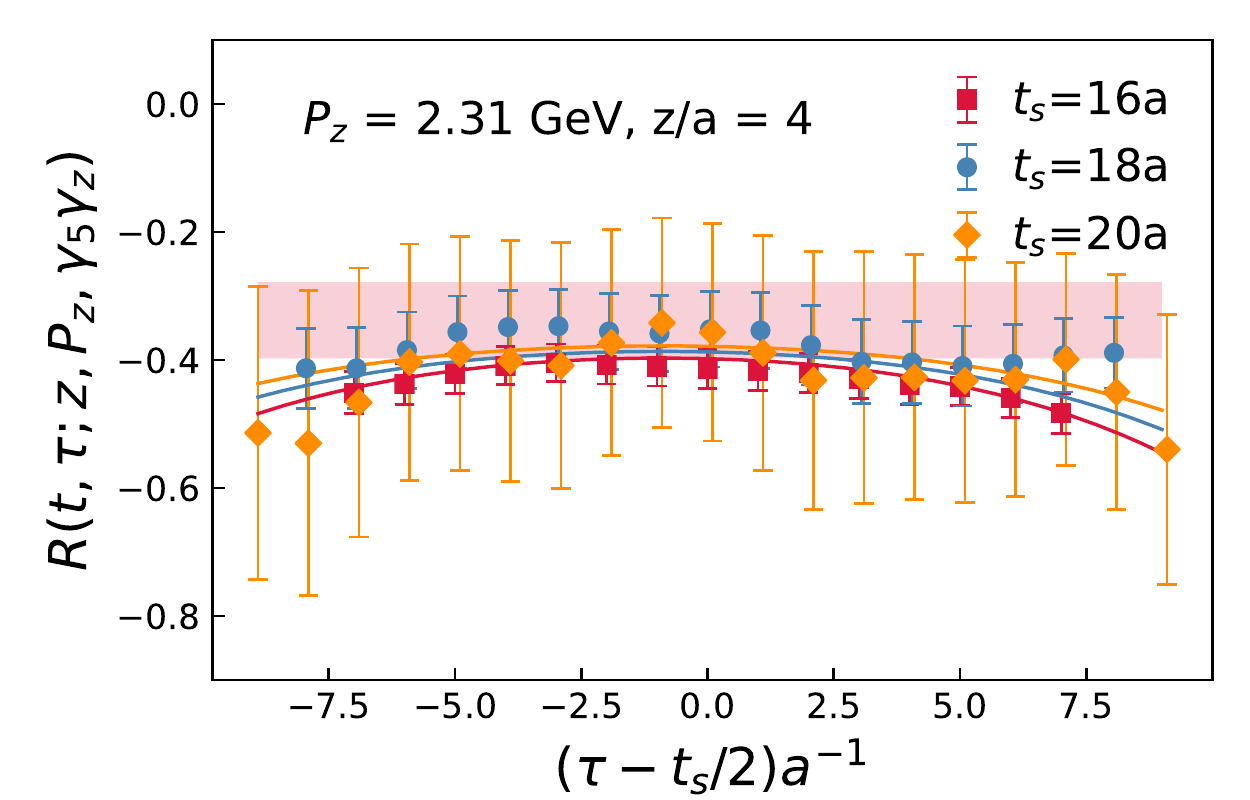}
        \includegraphics[width=0.3\textwidth]{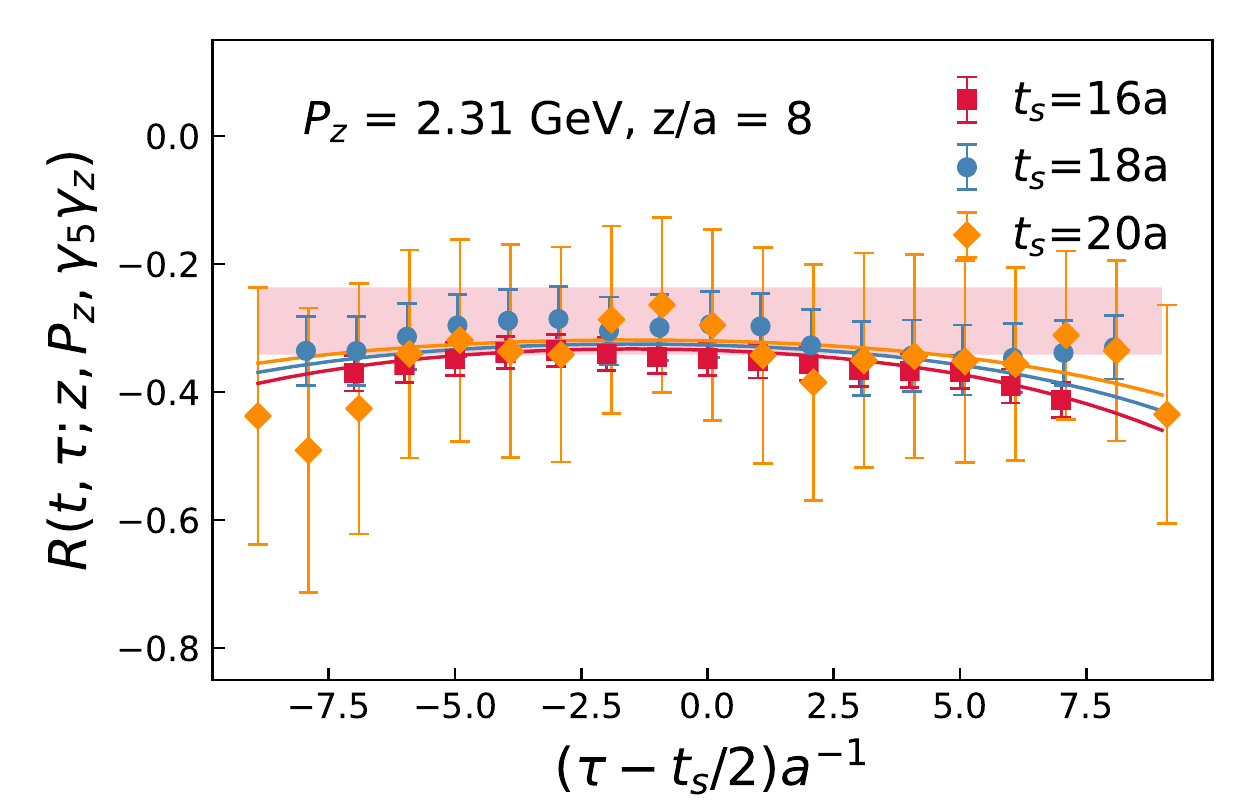}
        \includegraphics[width=0.3\textwidth]{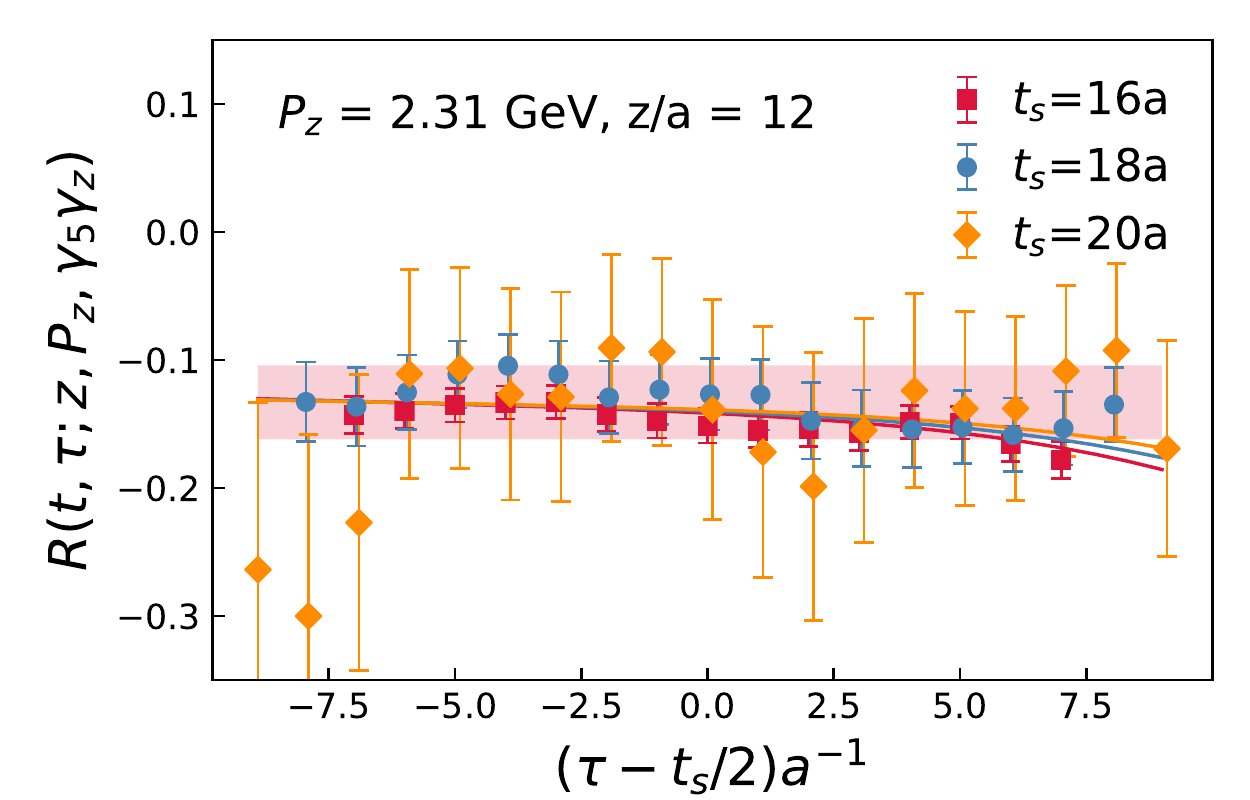}
\caption{
The ratio of the 3-point function to the 2-point function
corresponding to helicity qPDF
for $z=4,8,12$ and $n_z=5$. The upper panels show
the real part, while the imaginary part is shown in the lower panels. The results of $R^{fit}_1$
are shown as lines.}
\label{fig:samplefits_pol_nz5}
\end{figure}

\begin{figure}
        \includegraphics[width=0.45\textwidth]{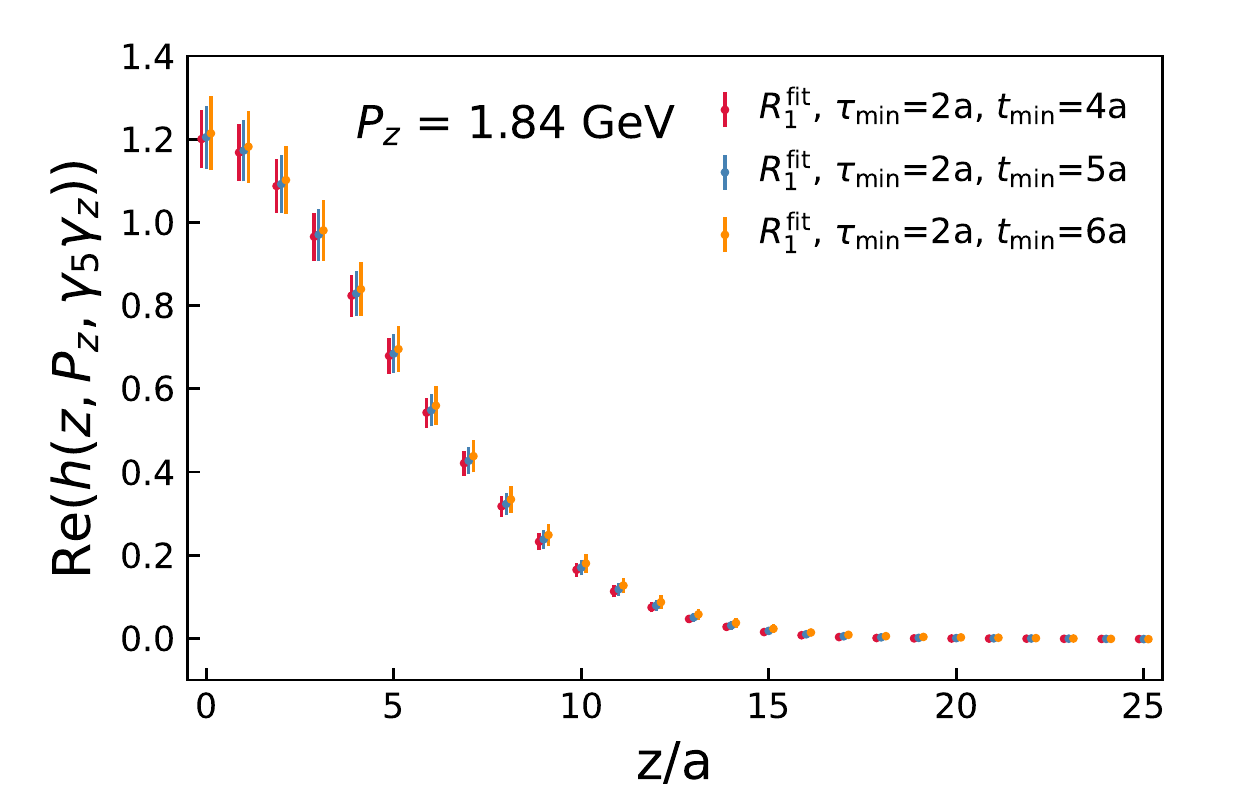}
        \includegraphics[width=0.45\textwidth]{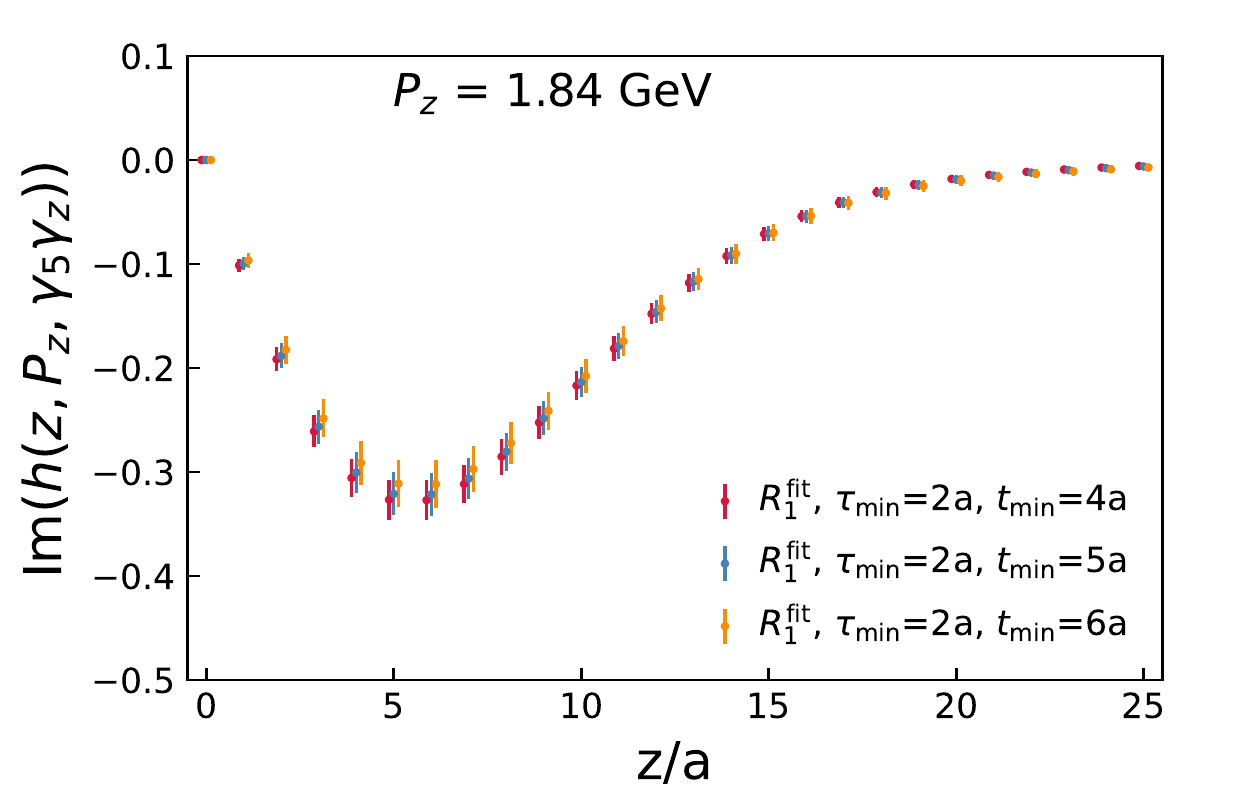}
        \includegraphics[width=0.45\textwidth]{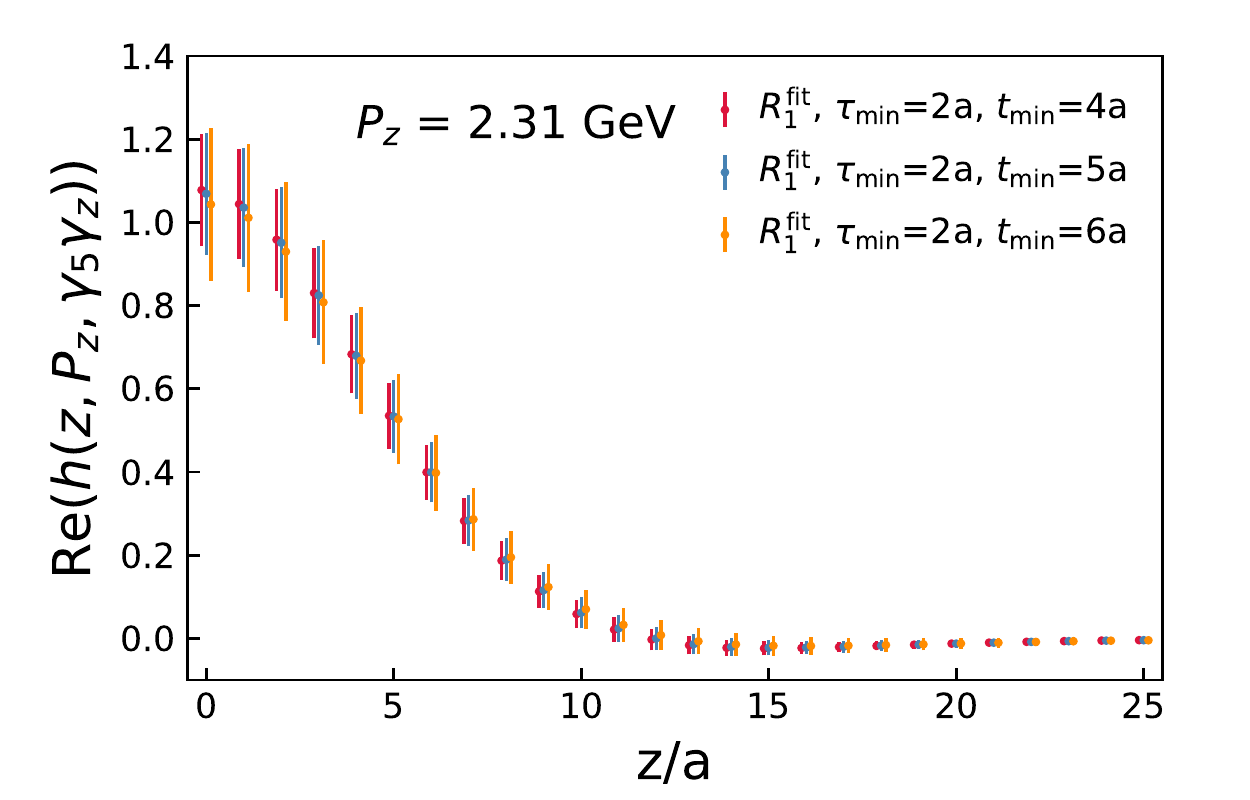}
        \includegraphics[width=0.45\textwidth]{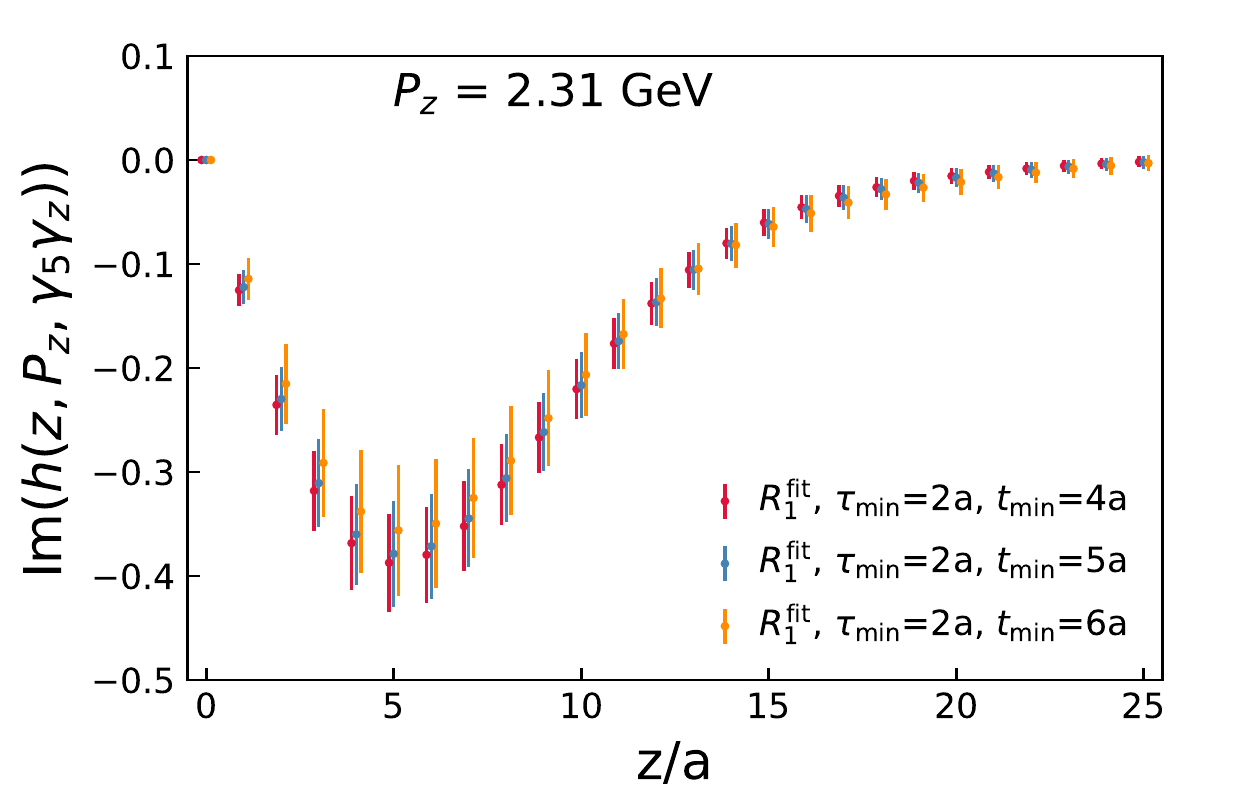}
\caption{The real (left) and the imaginary (right) parts of the bare matrix corresponding to helicity qPDF.
The upper panels correspond to $n_z=4$, while the lower panels correspond to $n_z=5$.
See text for further details.}
\label{fig:tmin_tau_min_dep_pol}
\end{figure}

Similar analysis of the ratio of the three point function to two point function was carries
out for longitudinally polarized qPDF operator.
The results are summarized in Figs. \ref{fig:samplefits_pol_nz4}, \ref{fig:samplefits_pol_nz5}
\ref{fig:tmin_tau_min_dep_pol}.

To take the advantage of correlation between different $z$ and cancel the field renormalization factor,
we divided the bare matrix elements by the matrix element at $z=0$.
The errors are much smaller after this division as discussed in the main text.

\end{document}